# MAGNESIUM BASED MATERIALS FOR HYDROGEN BASED ENERGY STORAGE: PAST, PRESENT AND FUTURE

V.A. Yartys[1*], M.V. Lototskyy[2], E. Akiba[3], R. Albert[4], V.E. Antonov[5],
J.-R. Ares[6], M. Baricco[7], N. Bourgeois[8], C.E. Buckley[9], J.M. Bellosta von Colbe[10],
J.-C. Crivello[8], F. Cuevas[8], R.V. Denys[1], M. Dornheim[10], M. Felderhoff[4],
D.M. Grant[11], B.C. Hauback[1], T.D. Humphries[9], I. Jacob[12], T.R. Jensen[13],
P.E. de Jongh[14], J.-M. Joubert[8], M.A. Kuzovnikov[15], M. Latroche[8],
M. Paskevicius[9], L. Pasquini[16], L. Popilevsky[17], V.M. Skripnyuk[17], E. Rabkin[17],
M. V. Sofianos[9], A. Stuart[11], G. Walker[11],
Hui Wang[18], C.J. Webb[19] and Min Zhu[18]

[1] Institute for Energy Technology, Kjeller NO-2027, Norway;
[2] HySA Systems, University of the Western Cape, Bellville 7535, South Africa;
[3] Kyushu University, Japan;
[4] Max-Planck-Institut für Kohlenforschung, Germany;
[5] Institute of Solid State Physics RAS, Russia;
[6] Universidad Autonoma de Madrid, Spain;
[7] Department of Chemistry and NIS, University of Torino, Italy;
[8] Université Paris Est, ICMPE (UMR7182), CNRS, UPEC, F-94320 Thiais, France;
[9] Physics and Astronomy, Curtin University, Australia;
[10] Helmholtz-Zentrum Geesthacht, Germany;
[11] University of Nottingham, U.K.;
[12] Ben-Gurion University of the Negev, Israel;
[13] Aarhus University, Denmark;
[14] Utrecht University, Netherlands;
[15] Max Planck Institute for Chemistry, Germany;
[16] Department of Physics and Astronomy, University of Bologna, Italy;
[17] Technion - Israel Institute of Technology, Israel;
[18] South China University of Technology, China;
[19] Qld Micro- and Nanotechnology Centre, Griffith University, Australia

---

[*] Corresponding author, volodymyr.yartys@ife.no

## Abstract

Magnesium hydride owns the largest share of publications on solid materials for hydrogen storage. The "Magnesium group" of international experts contributing to IEA Task 32 "Hydrogen Based Energy Storage" recently published two review papers presenting the activities of the group focused on magnesium hydride based materials and on Mg based compounds for hydrogen and energy storage. This review article not only overviews the latest activities on both fundamental aspects of Mg-based hydrides and their applications, but also presents a historic overview on the topic and outlines projected future developments. Particular attention is paid to the theoretical and experimental studies of Mg-H system at extreme pressures, kinetics and thermodynamics of the systems based on $MgH_2$, nanostructuring, new Mg-based compounds and novel composites, and catalysis in the Mg based H storage systems. Finally, thermal energy storage and upscaled H storage systems accommodating $MgH_2$ are presented.

## Content

# INTRODUCTION AND BACKGROUND

The hydrogen economy is an alternative to the current energy landscape based on fossil fuel consumption that creates enormous economic and environmental problems. In this context, the development of a safe, effective and economical way to store hydrogen is a necessary step to become more competitive with respect to other fuels. Besides gas and liquid storage, the storing of hydrogen into a solid has been considered a viable alternative since it is possible to contain more hydrogen per unit volume than liquid or high pressure hydrogen gas [1]. This was previously proposed by Hofman et al. in the early 70's [2] and since then, a zoo of materials has been deeply developed as summarised in numerous reviews [3-5]. However, as Table 1 shows, these compounds are only partially able to fulfil the different requirements (capacity, reversibility, and price) required for most applications.

Magnesium started to be investigated as a means to store hydrogen around 50 years ago, since it has the advantage of fulfilling the "natural" targets of (i) high abundance [6] (2% of earth surface composition and virtually unlimited in sea water), (ii) non toxicity and (iii) relative safety of operation as compared to other light elements and their hydrides that quickly and exothermically oxidize in air. Moreover, magnesium is produced by a well-established technology and its raw materials cost is relatively low. This, as well as its high volumetric (0.11 kg H/l) and gravimetric (7.6 mass% H) capacities, places magnesium as a feasible material to store hydrogen and it has attracted huge attention during recent years, as Figure 1 shows.

Magnesium and magnesium alloys have been intensively studied as hydrogen storage materials since the late 1960s. A rather comprehensive, although not complete, review of the related works published before 1985 was presented in [7]. A brief review covering a period up to 1997 was given in [8]. During the first decade of 2000s, several reviews on Mg-based hydrogen storage materials were published [9-14].

*Table 1. Qualitative analysis of main criteria of solid hydrogen storage families according to DOE 2020 targets for on-board applications. (Color code: Red = deficient ; Yellow = Fair ; Green = Good)*

| Compound families | Gravimetric capacity | Volumetric capacity | Minimum and maximum delivery temperature | Absorption / desorption rates | Toxicity , abundancy |
|---|---|---|---|---|---|
| Metallic hydrides (AB$_2$, AB$_5$..) | 🔴 | 🟢 🟡 | 🟢 | 🟢 | 🔴 🟡 |
| Magnesium hydride and alloys | 🟢 🟡 | 🟡 | 🔴 🟡 | 🟡 | 🟢 |
| Complex hydrides (alanates, borohydrides) | 🟢 | 🟡 🟢 | 🟡 🔴 | 🟡 | 🔴 🟡 |
| Chemical hydrides (amides, aminoboranes..) | 🟢 | 🟡 | 🟡 🔴 | 🔴 🟡 | 🔴 🟡 |
| Adsorbent materials (nanocarbon, MOFS) | 🟢 🟡 | 🟡 | 🔴 | 🟡 | 🟢 🟡 |

The latter review articles were focused on specific aspects of magnesium-based hydrogen storage materials including rare earth–Mg–Ni-based hydrogen storage alloys for electrochemical applications (2011 [15]); catalyst/additive-enhanced MgH$_2$ (2015 [16]; 2017 [17]); nanostructuring and size effects (2015 [18]; 2017 [19]); interrelations between composition, structure, morphology and properties of the Mg-based hydrides (IEA Task 32 report, 2016 [20]); optimisation of MgH$_2$ through the use of catalytic additives, incorporation of defects and an understanding of the rate-limiting processes during absorption and desorption (IEA Task 32 report, 2016 [21]).

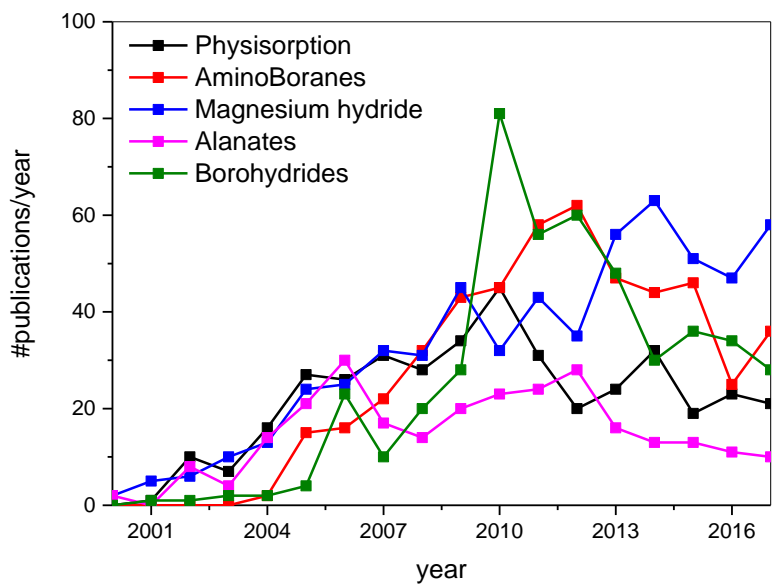

*Figure 1. Number of articles published during the last years 2000-2017 having "hydrogen storage" and the "name" of the respective compound in the **title**, **abstract** and **keyword** fields of the publication. Source: Scopus.*

Clearly, the progress achieved on magnesium as a hydrogen storage material over the last 20 years has been enormous. This is due to the use of a variety of synthesis methods, from high energy milling to magnesium cluster intercalation, which are able to tune the different kinetic and thermodynamic properties of MgH$_2$ via alloying, doping, nanosizing and nanoconfinement, strain-effects, etc.

The "Magnesium group" of international experts contributing to IEA Task 32 "Hydrogen Based Energy Storage" recently published two review papers presenting the activities of the group focused on Mg based compounds for hydrogen and energy storage [20] and on magnesium hydride based materials [21]. In the present review, the group gives an overview of the most recent developments in synthesis and hydrogenation properties of Mg-based hydrogen storage systems, highlighting the importance of magnesium based research on hydrogen storage materials for the future.

The first chapter (by V.A. Yartys, M.V. Lototskyy, J.R. Ares and C.J. Webb) gives a general review of R&D activities in the field of magnesium-based hydrogen storage materials including main properties and features of the hydrides, as well as a historical overview. It is followed by four chapters that deal with preparation routes and properties of nanostructured hydrogen storage materials based on Mg, associated with the most promising directions in the field which dynamically developed during the last decade. This part starts with analysis of the effects of nanostructuring of Mg-based materials as related to their hydrogen storage performance (L. Pasquini and P.E. de Jongh), followed by consideration of Mg-based nanomaterials prepared by mechanical alloying and reactive ball milling (M.V. Lototskyy, R.V. Denys and V.A. Yartys), and in-depth consideration of the mechanochemistry of magnesium during ball milling in H$_2$ gas (F. Cuevas, M. Latroche). The next chapter (M. Zhu, H. Wang, M.V. Lototskyy, V.A. Yartys, L. Popilevsky, V.M. Skripnyuk and E. Rabkin) presents overview of various catalysts, which improve hydrogen sorption / desorption performance of nanostructured materials on the basis of Mg.

Furthermore, this review considers important fundamental aspects including hydrogen sorption kinetics (M. Baricco), experimental (V.E. Antonov and M.A. Kuzovnikov) and theoretical (N. Bourgeois, J.-C. Crivello and J.-M. Joubert) studies of H–Mg system under high pressures and structural features of mixed transition metal – Mg complex hydrides (B.C. Hauback) and new ternary intermetallic hydride (I.Jacob, R.V. Denys and V.A. Yartys).

Finally, the review considers aspects related to the application of magnesium-based hydrides, including non-direct thermal desorption methods (J. Ares), the effects observed during the cycling of H–Mg system at high temperatures (A. Stuart, D. Grant, and G. Walker), as well as the analysis of application potential of hydride forming magnesium compounds for thermal energy storage (T.D. Humphries, M.V. Sofianos, M. Paskevicius, C.E. Buckley, R. Albert and M. Felderhoff) and an overview of developments of Mg-based hydrogen and heat storage systems (J.M. Bellosta von Colbe, M. Dornheim, M.V. Lototskyy and V.A. Yartys).

Future prospects of research and development in the field of magnesium based materials for hydrogen based energy storage are outlined in the final chapter of this review with contributions from all co-authors.

## Mg-Based Hydrides: Main Properties and Features

α-magnesium dihydride (MgH$_2$) can be synthesized directly from magnesium metal and hydrogen gas as a product of reversible interaction:

$$Mg + H_2 \rightleftarrows MgH_2 \qquad\qquad (1)$$

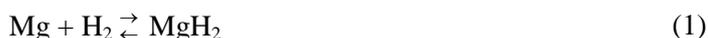

and by chemical methods, starting from magnesium-organic compounds.

In contrast to the chemically synthesized MgH$_2$ being a very active material which self-ignites when exposed to air and intensively reacts with water, magnesium hydride obtained by interaction of metallic Mg with H$_2$ gas is relatively inert and safe during handling [22]. MgH$_2$ is a stoichiometric hydride having a mixed ionic-covalent type of chemical bond [23]. It has a tetragonal, rutile-type crystal structure (space group #136; $a$=4.517Å, $c$=3.020 Å) [24,25]. The transformation of hexagonal magnesium (space group #194; $a$= 3.209 Å, $c$= 5.211 Å; bulk density 1.74 g/cm$^3$) to tetragonal MgH$_2$ (bulk density of 1.45 g/cm$^3$) is accompanied by a 20% increase in volume, while the volumetric hydrogen density in magnesium hydride is high, about 0.11 g/cm$^3$, more than 50 % greater than that of liquid hydrogen.

Under pressures up to 8 GPa and temperatures up to 900 °C, α-MgH$_2$ transforms to γ-MgH$_2$ with an orthorhombic α-PbO$_2$ type structure and octahedral H-coordination for the Mg atoms. The reverse transformation γ-MgH$_2$ → α-MgH$_2$ begins at T~350 °C at atmospheric pressure [26] and at T~250 °C in vacuum [27]. At pressures higher than ~8 GPa, γ-MgH$_2$ transforms to denser polymorphs: pyrite-type cubic β-MgH$_2$ with the $Pa\overline{3}$ space group [28]; orthorhombic HP1 with the $Pbc2_1$ space group [29] (previously identified as δ' with the $Pbca$ space group [28]) and cotunnite-type orthorhombic HP2, space group $Pnma$ [29].

The crystal and electronic structures of the α, β and γ allotropic modifications of MgH$_2$ will be considered in this review later; in the chapter "Modelling the MgH$_2$ hydride within high pressure model".

We note that occasionally the α modification of MgH$_2$ is called β-MgH$_2$, while the letter "α" is reserved for the primary solid solutions of hydrogen in hcp Mg. For convenience, from here on, we shall keep the notation from the original publications. However, the above-mentioned classification should be taken into account to avoid misunderstanding.

The hydrogenation of Mg-containing intermetallic compounds was first studied in 1967 [30], and the reaction of Mg$_2$Cu with hydrogen gas proceeded according to the following scheme:

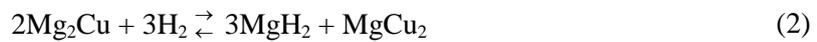

$$2Mg_2Cu + 3H_2 \rightleftarrows 3MgH_2 + MgCu_2 \tag{2}$$

Reaction (2) proceeds rapidly at $T$ ~300 °C and $P$(H$_2$) ~20 bar. It results in a hydride disproportionation of the starting intermetallic compound Mg$_2$Cu, but unlike the majority of the reactions of this class being irreversible processes, reaction (2) allows reversible absorption of up to 2.6 wt% H at temperatures lower than for the Mg–H systems (Reaction (1)).

Later, Reilly and Wiswall [31] showed that interaction of hydrogen with another Mg-based intermetallic compound, Mg$_2$Ni, results in the reversible formation of the ternary hydride, Mg$_2$NiH$_4$:

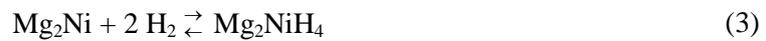

$$Mg_2Ni + 2 H_2 \rightleftarrows Mg_2NiH_4 \tag{3}$$

Reaction (3) proceeds rapidly at $T$ ~325 °C and $P$(H$_2$) ~20 bar allowing reversible absorption of up to 3.7 wt% H. In contrast with conventional intermetallic hydrides, such as AB$_5$H$_x$, AB$_2$H$_x$, etc., the formation of Mg$_2$NiH$_4$ is accompanied by essential changes in the crystal structure (see, for example, [32]). In Mg$_2$NiH$_4$, hydrogen atoms are placed in the vertices of [NiH$_4$]$^{4-}$ tetrahedra forming strong bounds with the central nickel atom. Therefore, Mg$_2$NiH$_4$ is considered as a complex hydride, rather than an interstitial intermetallic hydride.

Some magnesium-based complex hydrides (Mg$_2$FeH$_6$, Mg$_2$CoH$_5$, Mg$_3$MnH$_7$) similar to Mg$_2$NiH$_4$ were synthesized from parent metals and H$_2$ gas using special synthesis procedures (ball milling, sintering in hydrogen atmosphere under increased H$_2$ pressure, or under GPa-level hydrostatic pressure) [32–34]. The hydrogen-richest compound Mg$_2$FeH$_6$ can reversibly desorb / absorb up to 5.4 wt% of hydrogen at $T$=370–500 °C [33].

Among Mg-containing intermetallic compounds, the numerous group of magnesium-rich intermetallics with rare-earth metals, RE (e.g. REMg$_2$, RE$_2$Mg$_{17}$, RE$_5$Mg$_{41}$, etc), are of practical importance. Independent of their composition and structure, these intermetallics disproportionate during their interaction with hydrogen gas to yield a homogeneous mixture of magnesium hydride and rare-earth hydrides [35]:

$$RE_xMg_y + (1.5x + y)H_2 \rightarrow yMgH_2 + xREH_3 \tag{4}$$

The products of Reaction (4) are able to reversibly desorb up to 5.5–6 wt% H (depending on the amount of the rare-earth metal), according to the scheme [36]:

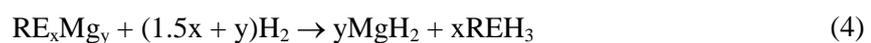

$$y\ MgH_2 + x\ REH_3 \rightleftarrows y\ Mg + x\ REH_z + [y+ x\ (3-z)\ /2]\ H_2,\ z=2.6\text{--}2.8 \rightleftarrows LaMg_{12} + H_2 \tag{5}$$

When starting from LaMg$_{11}$Ni, the following processes take place depending on conditions [37,38]:

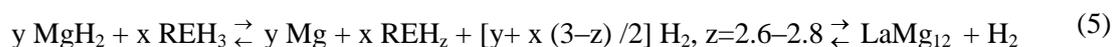
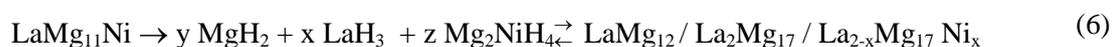

$$LaMg_{11}Ni \rightarrow y\ MgH_2 + x\ LaH_3\ + z\ Mg_2NiH_4 \rightleftarrows LaMg_{12}\ /\ La_2Mg_{17}\ /\ La_{2-x}Mg_{17}\ Ni_x \tag{6}$$

With the exception of Mg-containing intermetallic compounds with rare-earth and transition metals (e.g. REMg$_2$Ni$_9$ [20]), the products of hydrogenation of Mg-based H storage materials mainly contain either magnesium dihydride, MgH$_2$, and/or ternary complex hydrides, such as Mg$_2$NiH$_4$. The material composition is limited to elemental Mg, single-phase solid solution alloys and intermetallics, as well as Mg-based multiphase alloys and compositions. Consequently, there exists only a limited possibility for variation of the thermodynamic properties of these materials in systems with hydrogen gas.

Normally, MgH$_2$ adopts the rutile-type tetragonal phase following around a 20% expansion of the lattice of the initial magnesium metal. It exhibits a high absolute value for the enthalpy of decomposition (75 kJ/molH$_2$) [39], which is attributed to the ionic-covalent nature of the Mg–H bond [40]. Therefore, the plateau pressure of Mg/MgH$_2$ at room temperature is very low and subsequently, temperatures above 275 ºC are required to release hydrogen from MgH$_2$ under standard conditions of pressure, which is incompatible with most practical applications [14].

Figure 2 shows typical hydrogen desorption isotherms for Mg (Mg alloy) – hydrogen systems in comparison with a conventional "low-temperature" AB$_5$-type hydrogen storage intermetallic compound (compiled from the data published in [41,42]). It can be seen that although the hydrogen weight capacity of MgH$_2$ is more than 5 times greater than for the AB$_5$-based hydride, its dissociation requires a much higher temperature (~300 °C to provide a hydrogen pressure of 1 bar), than for the "low-temperature" interstitial type metal hydrides (<20–100 °C at $P$(H$_2$) = 1 bar). At the same temperature, dissociation of hydrides of Mg-containing intermetallics (e.g. Mg$_2$Ni) takes place at higher hydrogen pressures than for MgH$_2$. Unfortunately, this rather modest hydride destabilisation is associated with a notable decrease in gravimetric hydrogen storage capacity by a factor of ~2.

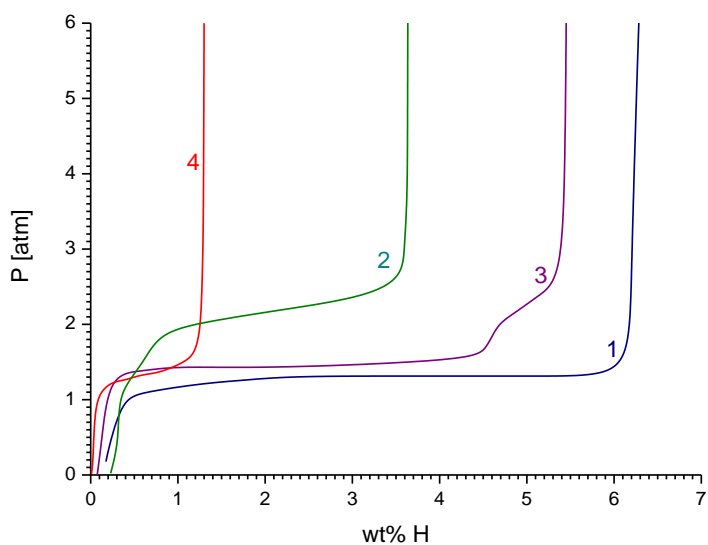

*Figure 2. Pressure – composition isotherms (H desorption) for systems of H$_2$ gas with: 1 – Mg-based nanocomposite, T=300 °C [41]; 2 – Mg$_2$Ni, T=300 °C [41]; 3 – Mg – Ni – Mm eutectic alloy , T=300 °C [42]; 4 – MmNi$_{4.9}$Sn$_{0.1}$, T=22 °C [41]. Mm is lanthanum rich mischmetal.*

Another disadvantage of magnesium as a hydrogen storage material is the low rate of its interaction with hydrogen gas, particularly during the first hydrogenation reaction. According to the data collected by different authors, reviewed in [43], the first hydrogenation of Mg at $T$=340–350 °C and H$_2$ pressure up to 30 bar can be completed within 6 to 336 hours. Such great discrepancies are caused by the high sensitivity of the reaction rate induced by the purity, particle size and surface state of the parent metal, activation conditions, purity of hydrogen gas, etc. The reaction can be accelerated by increasing the temperature to 400–450 °C and the hydrogen pressure to 100–200 bar, in combination with some gas-phase catalysts, such as iodine vapours, CCl$_4$, etc.

Mechanical treatment (ball milling) of the charge during hydrogenation also improves hydrogen absorption performance [44]. Historically, the ball milling of Mg powders was the first method shown to accelerate their hydrogenation kinetics [44,45]. Later, some other methods of plastic deformation of Mg and its alloys were also shown to have a beneficial effect on the hydrogenation kinetics. Particularly, the equal channel angular pressing (ECAP) of bulk Mg alloys with their subsequent dispersion into powder turned out to be as efficient as the prolonged ball milling in accelerating the kinetics of hydrogenation [46-50]. The positive effect of high pressure torsion (HPT) on both the kinetics of hydrogenation and the decrease in the temperature of the onset of hydrogenation has also been demonstrated [50]. The problem with the ECAP and HPT methods is their relatively high cost and up-scalability. Applying the low-cost plastic deformation methods such as cold rolling and forging the powders leads to a significant refinement of the microstructure and an improvement in the hydrogenation kinetics, comparable to those caused by the high energy ball milling, with better oxidation stability [51,52].

Typical results of early studies of hydrogen absorption kinetics by several magnesium-based materials are presented in Figure 3 and accelerated by introduction of conventional gas-phase catalysts, like $CCl_4$ (curve 2), into the hydrogen gas. However, hydrogen absorption remains too slow for practical hydrogen storage purposes. The rate of hydrogen absorption can also be increased when some elements (Al, Ga, In) are alloyed with Mg within the limits of diluted solid solution (curve 3). It was noted that when the concentration of the alloying element in such a solution is increased, the effect of improving the kinetics of hydrogen sorption by magnesium disappears [53]. This behaviour was explained by assuming that the improvement of hydrogen sorption properties of Mg is caused by facilitation of internal H diffusion due to the increased concentration of lattice defects within the solid. For oversaturated Mg-based solid solutions containing hydride-forming components, improvement of the hydrogen sorption kinetics appears again, in an even more pronounced form (curve 4). The presence of easily hydrogenated intermetallics results in further improvement of hydrogenation kinetics of Mg. This is illustrated by curve 5 which corresponds to a compacted mixture of Mg powder with 20 wt% $LaNi_5$; in such a composition practically complete hydrogenation of Mg in 30–60 minutes was observed [54,55]. The most probable mechanism of improvement of hydrogen sorption kinetics in this case was attributed to presence of the atomic hydrogen on the surface of $LaNi_5$ – a similar mechanism was supposed for hydrogenation of $Mg – Mg_2Cu$ alloys [56]. This mechanism is illustrated in [22] which describes the synthesis of $MgH_2$ from a charging of pure metal, when active hydrogen was generated by the interaction of zinc with hydrochloric acid. In this case the hydrogenation takes place even at room temperature and atmospheric pressure. The best hydrogen sorption performance is observed for multiphase Mg alloys containing rare-earth metals (RE). The alloys are represented by magnesium-rich intermetallics ($REMg_2$, $RE_2Mg_{17}$, $REMg_{12}$, etc.) which disproportionate yielding $MgH_2$ and $REH_3$ in the course of their interaction with hydrogen gas (Reaction (4)) [57]. Curve 6 in Figure 2 represents H sorption by $REMg_{12}$ intermetallics. In other cases, this group can be represented by alloys containing, apart from magnesium-based solid solution (or above-mentioned Mg-rich intermetallics), the hydride-forming intermetallic phases (e.g. $Mg_2Ni$). Such compositions have the best dynamics of hydrogen sorption (curve 7) which takes place even at room temperature. Kinetics and mechanism of the hydrogen absorption-desorption processes were characterised by in situ SR XRD (synchrotron radiation X-ray diffraction) [38].

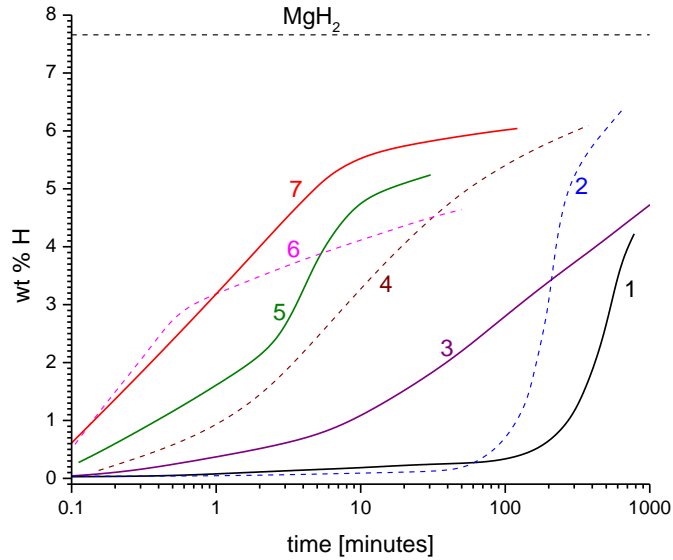

*Figure 3. Hydrogen absorption by magnesium and magnesium-based alloys / composites. The legend describing compositions and hydrogenation conditions of the samples is presented in Table 2.*

*Table 2. Sample compositions and hydrogenation conditions for hydrogen absorption curves presented in Figure 3*

| Curve # | Sample composition | | Hydrogenation conditions | | Notes |
|---|---|---|---|---|---|
| | Components | Phases | T, °C | P, bar | |
| 1 | Mg | Mg | 410 | 40 | |
| 2 | Mg | Mg | 410 | 40 | $H_2$ with the admixture of $CCl_4$ (~2%) |
| 3 | $Mg_{0.99}In_{0.01}$ | Mg (solid solution) | 270 | 80 | Alloy |
| 4 | Mg 94.11 Zn 4.01 La 1.24 Cd 0.52 Zr 0.12 (wt%) | Mg (solid solution) Traces of intermetallic phases | 340 | 30 | Industrial Mg alloy |
| 5 | Mg 80 LaNi$_5$ 20 (wt%) | Mg + LaNi$_5$ | 345 | 30 | Compacted mixture of powdered Mg and LaNi$_5$ |
| 6 | REMg$_{12}$ (RE = La, Ce) | REMg$_{12}$ (starting alloy and product of vacuum heating at T>450 °C) MgH$_2$+REH$_3$ (hydride) Mg + REH$_2$ (dehydrogenated sample) | 325 | 30 | Alloys |
| 7 | Mg$_{75}$Y$_6$Ni$_{19}$ | Mg+Mg$_2$Ni+YNi$_2$ (starting alloy) MgH$_2$+Mg$_2$NiH$_4$+YH$_2$ (hydride) | 200 | 30 | |

Since the reaction of hydrogen with magnesium is a hetero-phase transformation which includes several stages, *viz.* transport of $H_2$ molecule to the surface of magnesium, chemisorption of $H_2$ molecule to H atoms, migration of the H atoms from the surface into the bulk, hydrogen diffusion in the solid and formation and growth of the hydride phase [57], a number of factors may limit the hydrogenation kinetics. These factors include (i) formation of surface oxide which inhibits hydrogen penetration into the material [45], (ii) slow dissociation of hydrogen molecules on the Mg surface [59], (iii) low rate of movement of the MgH$_2$ / Mg interface [60], and (iv) slow diffusion of hydrogen through magnesium hydride [60,61]. Most of these obstacles can be overcome by nanostructuring Mg by several methods including mechanical alloying (MA) and reactive ball milling (RBM) in the presence of catalytic additives.

Apart from the practical necessity to improve the kinetics of hydrogenation and dehydrogenation in Mg-based hydrogen storage materials, poor stability of hydrogen sorption performance during cyclic dehydrogenation / re-hydrogenation at high temperatures is another important problem for establishing applications. Though literature data concerning cycle stability of Mg-based hydrides are quite contradictory (most probably due to different experimental conditions applied by different authors), it has been shown that the observed degradation effects during cycling are related to (i) partially irreversible loss of hydrogen storage capacity [62] and (ii) deterioration of hydrogen absorption and desorption kinetics, particularly at lower temperatures [63]. The origin of the first effect is related to a passivation of $Mg(H_2)$ surface due to a chemical reaction with gas impurities in $H_2$ [64,65]; while the second effect, taking place during the operation in pure $H_2$ at T≥350 °C[I] and mostly pronounced for nanostructured Mg-based composites, is associated with re-crystallisation of $Mg(H_2)$ particles which result in the decrease of specific surface area and longer H diffusion pathways [66,67].

Analysing the data presented above, it can be concluded that the most significant problem in the development of efficient magnesium based H storage materials is in improvement of the hydrogenation / dehydrogenation kinetics.

The second problem is to reduce the high operational temperature required for dehydrogenation of Mg-based H storage materials, due to the thermodynamic limitations. The solution of this problem is much more difficult than for the hydrides of intermetallic compounds of transition metals, which allow a significant variation of their thermodynamic properties by variation of the component composition of starting alloy. It requires either radical revision of the pathways of hydrogenation / dehydrogenation reaction, or creation of thermodynamically metastable materials, by application of special physical methods.

## Historical Overview

Magnesium hydride ($MgH_2$) was first synthesised by pyrolysis of ethyl magnesium halides in 1912 [68]. Further study of the synthesis of $MgH_2$ using the pyrolysis of di-alkyls of Mg was undertaken by Wiberg and Bauer [69] in 1950. Two solvent-based synthesis methods were demonstrated to produce magnesium hydride, the pyrolysis of magnesium dialkylene and hydrogenation of magnesium dialkylene with diborane. At a similar time, the hydrides of beryllium, lithium and magnesium were prepared using lithium aluminium hydride in ethyl ether or diethyl ether solution [70], with a yielded purity of about 75 % with respect to magnesium hydride production. Subsequently, synthesis methods involving the thermal decomposition of magnesium diethyl at 200 °C in high vacuum [71], the hydrogenolysis of Grignard reagents ($2RMgX + 2H_2 \rightarrow 2RH + MgX_2 + MgH_2$) [72], catalytic magnesium hydrogenation with anthracene in THF via anthracene-magnesium as an intermediate [73] and the reaction of phenylsilane and dibutylmagnesium [74] were employed.

The first direct hydrogenation of magnesium was performed in 1951 by heating elemental magnesium in an atmosphere of hydrogen gas [25]. At a temperature of 570 °C and a hydrogen pressure of 200 bar, the reaction yielded 60 % $MgH_2$ - provided a catalyst ($MgI_2$) was used. Later, Ellinger et al [24] measured the desorption pressures for $MgH_2$ for a range of temperatures and showed that the catalyst was only required because the desorption pressure at 570 °C was higher than the applied pressure of 200 bar. They also determined reaction rates at 68 bar $H_2$ for various temperatures and calculated the heat of formation and activation energy. Stampfer et al [75] synthesised magnesium hydride and magnesium deuteride at 300 bar and 500 °C in order to measure desorption isotherms and determine the enthalpy and entropy.

Despite the success of direct hydrogenation and the elimination of the problem of residual solvent material, reaction times were typically slow and quite low yields were observed [75]. An initial problem with magnesium is its strong affinity to oxygen, resulting in a formation of a thin oxide layer on the surface of the Mg particles which limits the hydrogen diffusion [76]. Similarly, a small presence of water leads to the surface formation of $Mg(OH)_2$. Annealing the material can create cracks in the oxide layer (above 400 °C) [77] as well as decompose $Mg(OH)_2$ [45]. Alternatively, ball-milling can break the oxide layer and provide fresh metal surfaces [44]. Regardless of the initial material, once the magnesium starts to hydride, the layer of $MgH_2$ impedes the diffusion of hydrogen to the remaining unreacted metal [21].

---

[I] As pronounced sintering / re-crystallisation of a metal takes place at the temperature above 2/3 of its melting point (in K), for Mg having the melting point at 923 K = 650 °C, its calculated sintering temperature is 2/3 * 923 = 615 K (342 °C).

Reactive ball-milling or mechanical milling is a more recent technique for the synthesis of metal hydrides. In 1961, Dymova et al. [44] applied ball milling of Mg under hydrogen pressure (200 bar) at $T = 350$–400 °C. The yield of $MgH_2$ after 5–6 hours long ball milling was about 75%. Addition of small amounts of catalysts ($I_2$, $CCl_4$, $Mg_2Cu$; 0.5–3 wt%) allowed the milling process to reach a yield of $MgH_2$ above 97% at the same conditions. Chen and Williams [78] ball-milled magnesium powder under 3.4 bar of hydrogen for 23.5 h. They reported full absorption at room temperature, but subsequent studies found the need for additives or catalysts such as transition metals, including Ni [79], V, Zr [45] or carbon [66], or allyl-iodide or multi-ring aromatic compounds of transition metals [44,73,80] to significantly soften conditions of Mg hydrogenation from the gas phase and to increase the yield of $MgH_2$.

Zhao et al [81] proposed a combination of ball-milling and direct hydrogenation termed the second hydrogenation method. In this technique, magnesium powder was hydrogenated for 6 h at 350 °C under 60 bar of hydrogen. This was then milled under argon for 9 h before returning to the reactor for further hydrogenation under pressure and temperature, with best results at 45 bars and 380 °C for 6 h.

Other techniques for the synthesis of magnesium hydride including hydriding chemical vapour deposition (HCVD) [82], laser ablation [83] and hydrogen plasma metal reaction (HPMR) [84] can produce high purity nanostructured magnesium hydride, but these techniques are difficult to scale up to large production quantities.

The first detailed study of the phase equilibria during interaction of Mg metal with gaseous hydrogen and deuterium (PCT diagrams were built in the temperature range $T=314$–576 °C, at gaseous pressures of up to 250 bar) was performed by Stampfer et al. in 1960 [75]. This study showed that Mg–H system has a number of distinct features, including its very small incline in the two-phase region (Figure 4).

Early studies of hydrogenation – dehydrogenation kinetics in $H_2$ – Mg system without catalysts were reported by Ellinger et al. [24], Kennelley et al. [86], Stander [87] and Vigeholm et al. [88]. Hydrogenation / dehydrogenation kinetics of Mg-rich alloys was also studied by Douglass [89], Mintz et al. [53,90], Karty et al. [91], Luz et al. [61] and other authors. It was found that at $T$~250 °C the formation of $MgH_2$ virtually stops when reaching reacted fraction about 75% [87] due to very slow H diffusion through growing $MgH_2$ layer which was found to be the rate-limiting step of the reaction [61,87]. However, at the higher temperatures (~400 °C) the hydrogenation becomes reasonably fast achieving almost complete transformation of Mg to $MgH_2$ in 2.5 hours for the first hydrogenation (Mg powder, ≤75μ in the particle size) and in less than one hour for the subsequent hydrogenations [88].

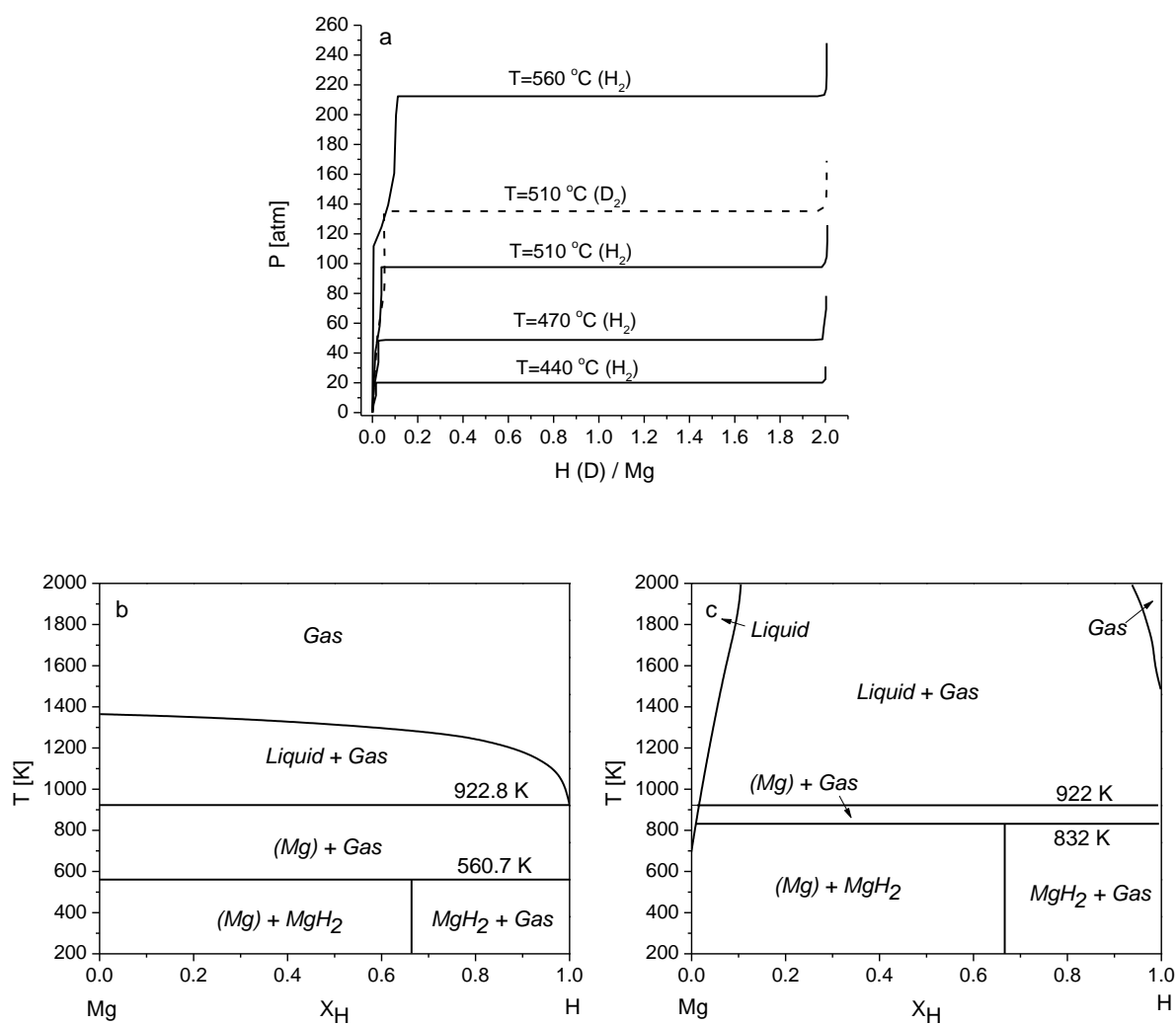

*Figure 4. Phase equilibria in H – Mg system: Top (a) – pressure – composition isotherms [75]; Bottom: phase diagrams at 1 bar (b) and 250 bar (c) [85].*

Since the 1960s, many researchers have attempted to improve the hydrogen sorption properties of magnesium, by alloying with other metals. Mikheeva *et al* have shown the possibility of gas phase hydrogenation for magnesium at room temperature and atmospheric pressure, using Ce–Mg and Ce–Mg–Al alloys [92,93]. Later, the works by Reilly and Wiswall demonstrated an improvement in hydrogenation kinetics of magnesium by hydride-forming intermetallics ($Mg_2Cu$ and $Mg_2Ni$) [30,31] and have stimulated research in this field.

An important milestone achieved at the end of the 1970s – beginning of 1980s was a discovery of catalytic effect of easily-hydrogenated additives which significantly accelerate hydrogenation of Mg. The additives can be introduced either by alloying with Mg (rare-earth and transition metals), or by mixing of Mg powder with hydride-forming intermetallic alloy (e.g. $LaNi_5$) followed by compacting of the mixture [54,57].

An enormous variety of chemical and physical preparation techniques (chemical reaction of a precursor [94], thermolysis [95], gas phase growth [96], reactive mechanical milling [97], etc.) has been undertaken to synthesize a wide arrange of $Mg/MgH_2$ micro and nanostructured morphologies to overcome the main drawbacks related to the thermodynamic and kinetic limitations which prevent its use as a solid hydrogen storage material.

Since end 1980s – 1990s, significant progress was achieved in the improvement of hydrogenation / dehydrogenation kinetics of nanostructured Mg and Mg-based composites employing various catalytic additives [45,98,99]. Nanostructuring, particularly via mechanical alloying and reactive ball milling, is still of a great interest and will be considered in more detail in the following sections of this review.

During recent years, much effort has been focused at reducing the thermodynamic stability of $MgH_2$. One strategy was in reducing the high absolute value of hydrogenation enthalpy by alloying Mg with other elements to form less stable hydrides. Among the different investigated additives (Co, Fe, Mn) the archetypal one is Ni [100] that alloys with Mg to form $Mg_2Ni$ with an absorption enthalpy of 60 $kJ/molH_2$ to become $Mg_2NiH_4$ i.e. thus lowering the value exhibited by $MgH_2$. A similar pathway is achieved by adding a reactive element (Si, Ge) to $MgH_2$ to form a stable Mg-based compound and, therefore to reduce the enthalpy of the hydrogenation reaction. As an example, when adding silicon [101] the reaction enthalpy decreases by more than 30 $kJ/molH_2$, and this effect can become even more pronounced when using other hydrides i.e. $LiBH_4$ because of the formation of reactive hydride composites and changing the hydrogen release pathway [102]. However, such a scenario faces drawbacks related to the reduction of hydrogen capacity as well as slowing down the hydrogenation-dehydrogenation kinetics mainly due to the segregation processes. Currently, the efforts to solve the kinetic problems show that the most promising results are achieved by utilising nanostructuring and by catalyst additions.

More recently, reduction of the particle size to the nanoscale has been considered an alternative approach to modify the enthalpy due to the non-negligible contribution of the surface free energy caused by the high surface area to volume ratios. Theoretical calculations [103,104] predict a drastic reduction of enthalpy for particle sizes smaller than 2 nm and, subsequently, a wide variety of preparation techniques have been employed to achieve this size reduction. These are classified by two approaches, a top-down (divide the matter to the nanoscale) and bottom-top (assemble the atoms to form clusters or nanolayers). Whereas the first approach is mainly based on mechanical milling [105], the second one includes the methods of gas-phase growth [106], scaffold infiltration [106], colloidal methods [108] and intercalation into 2D-materials [109].

Even though a broad range of techniques was applied, giving a variety of studied nanomaterials, the studies failed to show a clear correlation between the particle size and the formation/decomposition enthalpy of $MgH_2$. Nevertheless when in a range below 10 nm, Mg nanoparticles show lowering of the decomposition temperature of magnesium dihydride [21]. Discrepancies between the results of theoretical calculations and experimental data are usually attributed to the complexity of the small nanoparticles where several parameters such as surface state of the studied material [110] strongly affect thermodynamic properties. Furthermore, enthalpy-entropy correlation affects $\Delta H$ value, similar to the situation in other metal-hydrogen systems [111] and in various studies related to the effect of nanosizing [112]. Thus, because of the mentioned correlation, as the temperature of decomposition is derived as $T=\Delta H/\Delta S$, there are challenges in reaching the goal of a decrease in desorption temperature. Efforts to decouple enthalpy and entropy are required in order for nanosizing to become a viable mechanism to reduce the reactions temperature.

Concerning the kinetic constraints, the formation of $MgH_2$ involves several reactions ($H_2$ physisorption, $H_2$ dissociation, chemisorption, H-diffusion) and a successive chain of kinetic barriers has to be overcome. As a consequence, the H-absorption and desorption rates are too low for most applications even at moderate temperatures i.e. at 250-300 °C [45]. To resolve the problem, a huge effort has been invested to uncover the mechanism and to determine activation energies that control the H-absorption/desorption process for $MgH_2$. Different theoretical approaches have been utilized. Although the obtained results are not completely conclusive, due to a huge diversity in experimental conditions and different types of studied materials and their morphologies, the rate limiting steps are generally considered to be nucleation and growth processes (NG) controlled by H-diffusion in the $MgH_2$ phase for absorption and interface reaction for desorption. Activation energies of 170±20 and 220±20 have been respectively reported [58,113] for these two processes. These energies decreased following reduction of the grain size, nanostructuring and removal of oxides/hydroxides from the surface of $MgH_2$ by mechanical milling [45]. Further enhancement was achieved using nanocomposites as in the $MgH_2$-$TiH_2$ system. The latter composite exhibits outstanding kinetic properties (H-absorption is around 20 times faster than for a pure nanometric $MgH_2$ powder) by combining effect of grain growth inhibition and H-gateway mechanism provided by $TiH_2$ [114] revealing the possibilities to explore the acceleration of the kinetic process.

Kinetic improvements can also be achieved by adding catalysts without affecting the thermodynamics. Pd, as an excellent catalyst to dissociate the molecular hydrogen improved the hydrogenation by the spillover effect [115] but finding less expensive but still efficient compounds as catalysts remains a challenge. Transition metals have been habitually considered as the archetypal catalyst elements and, in particular, several elements (Ti, V, Mn, Fe and Ni) have been shown to reduce the activation energy of hydrogen absorption/desorption [116] (Figure 5). A similar diminution of the activation energy of hydrogen desorption was obtained using transition metal oxides [117], metal halides [118], etc. However the mechanism of the catalyst's influence could be very different, since they may involve the formation of intermediate phases (that act as an H-pathways) [119], the refinement of the nanostructure and prevention of agglomeration [120]. Therefore, identifying novel catalyst compositions and processes of their use is crucial to increase the hydrogenation/desorption rates. During the recent years, nanosizing of magnesium has also promoted the overall kinetic improvements of processes of hydrogen exchange by enhancing individual steps such as H-diffusion and also $H_2$-dissociation by the increased effect of the surfaces via reactive role of steps, corner and edge atoms as compared to the bulk [121] and, also, because of achieving a more accurate control of the catalysis such as exploring chemistry of core-shell nanostructures [122]. Thus, importantly, issues related to the reactivity of the surfaces need to be solved.

The cycling behaviour of magnesium hydride is also very important to allow its efficient performance on a reversible hydrogen storage system. It is commonly accepted that cycling negatively affects the H-kinetics due to the degradation of the nanostructures and catalysts redistribution [123]. Hydride confinement [124] and prevention of recrystallization [67,125], are promising pathways that are currently being explored.

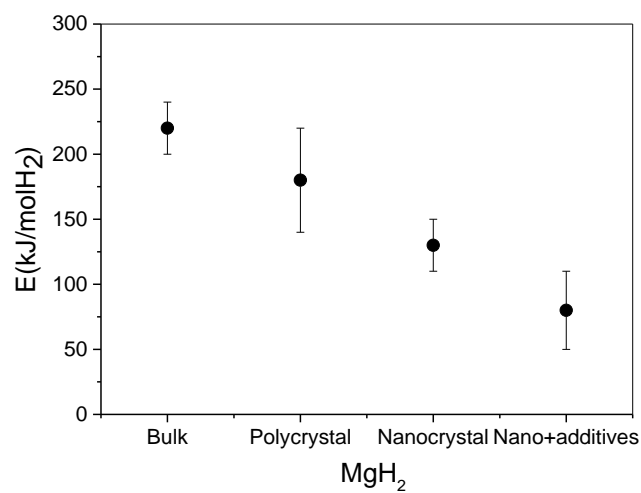

*Figure 5. Influence of different micro-nanostructures and additives on H-desorption activation energy of $MgH_2$ [45,116,117].*

The permanence of $MgH_2$ as an interesting hydrogen related materials is not just related to H-storage properties but also with the changes in electronic state via a metal-isolator transition which takes place during the hydrogenation of magnesium films [126,127]. This opens the possibilities for use of thin films of Mg-rich hydrides as optical/electrical sensors. Moreover, control over the morphology can offer novel ways to destabilize the hydrides. As an example, strain effects able to reduce the stability of the hydride [128] could be tuned by the control over the interface between the different layers in magnesium films. Other applications of magnesium hydride include use of $MgH_2$ in thermal storage [129] or as anodes in lithium ion batteries [130].

# NANOSTRUCTURED MAGNESIUM HYDRIDE

A distinctive feature of nanoobjects such as nanoparticles, nanowires and thin films is the high ratio $A/V$ between the total interface area and volume. Surfaces contribute to $A$ as special solid-vapour interfaces. This scenario can change the thermodynamics of hydride formation if the specific interface free energy per unit area $\gamma$ of the metallic state differs from that of the hydride. For a nanoobject surrounded by different interfaces identified by the index $i$, the free energy change for the Reaction (1) compared to bulk Mg is given by [121,131,132]:

$$\delta\Delta G^0 = (\delta\Delta G^0)^{int} + (\delta\Delta G^0)^{el} = \frac{\bar{V}_{Mg}}{V}\left(\sum_{int} A_{MgH_2|i} \cdot \gamma_{MgH_2|i} - \sum_{int} A_{Mg|i} \cdot \gamma_{Mg|i}\right) - 2B\bar{V}_H\varepsilon. \quad (7)$$

In the first term $(\delta\Delta G^0)^{int}$ that describes the effect of interface area, $\bar{V}_{Mg}$ is the molar volume of Mg and the sums extend over all interfaces. $A_{MgH_2|i}$ and $\gamma_{MgH_2|i}$ denote the area and specific free energy of the $i$-th interface in the hydride phase, respectively (with corresponding notation for the metal).

The second term $(\delta\Delta G^0)^{el}$ describes the effect of the elastic strain $\varepsilon$ (it is negative during compression). $B$ is the bulk modulus of Mg and $\bar{V}_H$ the partial molar volume of hydrogen in $MgH_2$.

$(\delta\Delta G^0)^{int}$ takes on a very simple form whenever, upon hydride formation, the total interface area $A$ does not vary and the change of the specific free energy is the same for all interfaces, i.e. $\gamma_{MgH_2|i} - \gamma_{Mg|i} \equiv \Delta\gamma \;\forall\; i$. In this case, one has: $(\delta\Delta G^0)^{int} = \bar{V}_{Mg}\Delta\gamma\, A/V = \bar{V}_{Mg}(\Delta h - T\Delta s)\, A/V$, where the separation into enthalpic $\Delta h$ and entropic $\Delta s$ contributions has been made explicit. Moreover, $(\delta\Delta G^0)^{int}$ vanishes at the so-called compensation temperature $T_{comp} = \Delta h/\Delta s$, where the equilibrium pressure for hydride formation/decomposition in the nano-object equals the bulk value. Experimental studies on $Mg/MgH_2$ thin films of varying thickness sandwiched between $TiH_2$ layers gave $\Delta\gamma = 0.33$ J/m² at 393 K [131]. This result was in fair agreement with the outcome of Density Functional Theory (DFT) calculations for selected $Mg|TiH_2$ and $MgH_2|TiH_2$ interfaces, which yielded $\Delta h$ between 0.58 and 0.69 J/m² [133]. Notably, there is a real missing knowledge about the interface entropy term $\Delta s$, which has neither been experimentally measured nor determined by model calculations.

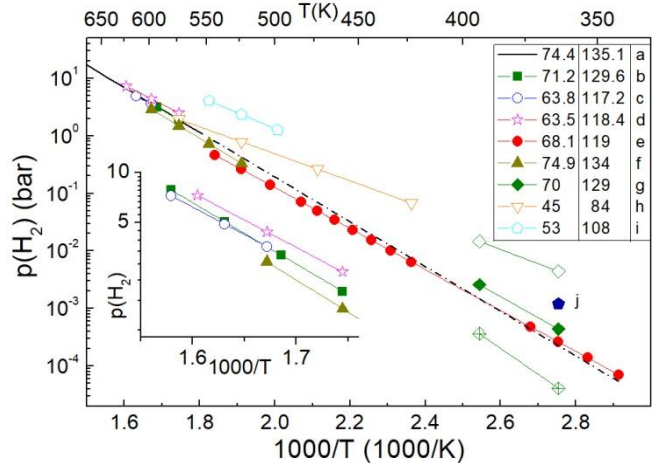

*Figure 6. Compilation of van 't Hoff plots calculated from $\Delta H^0$ and $\Delta S^0$ data for Mg-based nanomaterials confronted to bulk Mg (curve **a** from [75]). The black dash-dotted line is the low temperature extrapolation of bulk Mg data. The data in the legend denote the corresponding absolute values of $\Delta H^0$ (left, in kJ/mol $H_2$) and $\Delta S^0$ (right, in J/K mol $H_2$). The number of symbols represent how many points were actually measured and the temperatures of the measurements. Empty symbols denote absorption pressures $p_{abs}$, filled symbols equilibrium pressures $p_{eq}$. **b**: 2-7 nm Mg nanocrystallites in LiCl matrix [136]. **c**: < 3 nm Mg NPs in carbon scaffold [134]. **d**: 15 nm Mg NPs by electroless reduction [135]. **e**: $MgH_2$-$TiH_2$ composite NPs, 10-20 nm in diameter (6-30 at.% Ti) [125]. **f**: $MgH_2$-$TiH_2$ ball-milled nanocomposite (30 at.% Ti) [114]. **g**: Mg/Ti/Pd nanodots on silica, diameter 60 nm: here, $p_{abs}$ and $p_{des}$ are also plotted separately using empty and crossed symbols, respectively, to highlight the strong pressure hysteresis; the reported enthalpy-entropy values were calculated from $p_{eq}$ data [137]. **h**: Mg-Ti-H NPs, 12 nm in diameter (30 at.% Ti) [140]. **i**: Magnesium-naphtalocyanine nanocomposite with Mg NPs of about 4 nm supported on TTBNc [138]. **j**: ultra-thin (2 nm) Mg film sandwiched between $TiH_2$ layers [131]. The inset represents a zoomed view of the high-temperature region using the same symbols and units as for the main plot.*

Relatively few studies of the equilibrium thermodynamics have been published [131,134-138]. A collection of van 't Hoff plots taken from the literature is presented in Figure 6 and the obtained enthalpy ($\Delta H^0$) and entropy ($\Delta S^0$) data are reported in the legend. Empty symbols indicate absorption pressures $p_{abs}$, whereas filled symbols denote equilibrium pressures $p_{eq} = (p_{abs} \cdot p_{des})^{1/2}$ where $p_{des}$ is the desorption pressure. It is clear that $p_{eq}$, which is the real indicator of thermodynamic stability, exhibits only small changes compared to bulk Mg. The enthalpies $\Delta H^0$ obtained from the fit of $p_{eq}$ data are also quite close to the bulk value $\approx 74$ kJ/mol $H_2$. This scenario is consistent with the predictions that can be cast using Equation (7) and with theoretical calculations [103,139] showing that enthalpy variations are very small in the nanoparticles (NPs) size range explored by the experiments, and that particles of less than 1.3 nm would be needed to have considerable shifts in the equilibrium temperature. On the other hand, the fit of $p_{abs}$ data suggest significantly reduced enthalpies and entropies (see legend of Figure 6). This could be an artefact caused by hysteresis, as low-temperature experiments on encapsulated Mg nanodots (green diamonds in Figure 6) demonstrated very clearly that only by measuring both absorption and desorption branches it is possible to determine the thermodynamic parameters correctly [137]. Other possible sources of errors are the limited temperature range spanned by many investigations, the slow kinetics, particularly at lower temperatures, and the fact that, at temperatures around 600 K, coarsening and sintering lead to microstructural instability.

A few new studies on the preparation of small particles have appeared in recent years. Particles of a size of about 4 nm have been produced in magnesium-naphtalocyanine composites [141]. Very small particles, a series from 3.0 nm down to 1.3 nm in size, were made by Zlotea et al., using solution impregnation of a carbon support [142]. The particles of only 1.3 nm in the carbon support started desorbing hydrogen at 50-75 °C. Of fundamental interest is also the synthesis of magnesium hydride clusters by an organometallic approach using ligands. The largest clusters of this type reported until today contain 12 Mg atoms and are synthesized using phosphor and nitrogen containing ligands [143]. They are also very interesting systems for fundamental studies on cluster size, because the experimental cluster size is small enough to allow DFT calculations and therefore is suitable for a direct experiment-theory comparison. At the same time, all these sophisticated methods to prepare small and stable clusters and particles inevitably imply low interface-energy systems, hence one would not expect a large impact on the equilibrium conditions, while at the same time they are very susceptible to oxidation.

Other studies looking at ways to destabilize $MgH_2$ include investigations of the less thermodynamically stable phase γ-$MgH_2$. Interestingly Shen et al. [144] produced $MgH_2$ particles (60-100 nm) with 5 nm Ni as catalyst by electrochemical reduction of $Mg(BH_4)_2$ using Ni as a sacrificial electrode, which contained almost 30 % γ-$MgH_2$. The enthalpy determined seemed very low (57.5 ± 5.3 kJ/mol), but a concomitant decrease in entropy was reported causing no large shift in equilibrium. Very fast kinetics were observed, possibly related to the presence of the additional γ-$MgH_2$ phase. However, this metastable phase was converted into α-$MgH_2$ within 5 cycles.

Another strategy that has been proposed to destabilize $MgH_2$ is through elastic constraints, i.e. by exploiting the second term $(\delta \Delta G^0)^{el}$ in Equation (7) [145,146]. However, this has not yet been successful, even though simple calculations suggest that a significant hydride destabilization should occur, for instance, in a Mg nanoparticle of 30 nm diameter surrounded by a MgO shell [146]. The main reason behind the failure lies in the strong plastic deformation that develops in constrained systems due to the volume expansion upon hydrogen absorption. Plastic deformation strongly increases the pressure hysteresis and may even lower the desorption plateau pressure below the bulk Mg value [137]. Zhang et al [147] took a different approach by forming a native MgO shell around $Mg_2NiH_4$ hydride nanoparticles encapsulated on the surface of graphene sheets. The transformation to the metallic state shrinks the inner volume while the MgO shell remains intact, also serving as a natural microencapsulation method that prevents coarsening and sintering of the nanoparticles. To summarize, currently the experimentally observed $p_{eq}$ values in nanostructured Mg phases are very close to those of macrocrystalline Mg. This is also due to the experimentally observed enthalpy-entropy correlation, which could be better understood by measuring $p_{eq}$ down to sufficiently low temperatures where $\Delta h$ overwhelms $-T\Delta s$ (<275 °C)  Furthermore a systematic study on a series of well-defined feature sizes would contribute greatly to a more thorough understanding of the effect of particle size on thermodynamic parameters.

However, most studies nowadays focus on another important beneficial effect of nanostructuring namely enhancing the kinetics. Clearly, nanosizing increases the rates of hydrogen desorption and absorption, whether this is related to enhanced surface limited reactions or decreased diffusion distances [148]. Only very few fundamental studies exists in which kinetics is studied systematically as a function of particle size with all other factors equal, one of the few is the study by Yuen et al [149], who investigated carbon-supported magnesium particles of 6 to 20 nm. However, many fundamental questions remain, such as what is the rate limiting step during hydrogen absorption and desorption.

The overall diffusion coefficient obtained for hydride formation in a thin Mg film is very low, of the order of $1\cdot10^{-20}$ m$^2$ s$^{-1}$ at $T \sim 300$ °C [150-152]. More recently, Uchida et al. determined in more detail the diffusion coefficients of hydrogen in thin films. They found that the diffusion of hydrogen as a solute in the metallic Mg phase is relatively fast, i.e. $D_H^{Mg} = 7\cdot10^{-11}$ m$^2$ s$^{-1}$ [153]. However, at higher hydrogen concentrations when the hydride MgH$_2$ phase forms, the diffusivity decreases to $D_H^{tot} = 10^{-18}$ m$^2$ s$^{-1}$. This value is about two orders of magnitude larger than in bulk MgH$_2$, probably due to the contribution from a fast diffusion along the grain boundaries.

The diffusivity drop in macrocrystalline MgH$_2$ leads to the so called "blocking layer effect" where, upon hydrogenation of Mg, first a surface layer of MgH$_2$ is formed, which acts as a diffusion barrier for hydrogenation of inner parts of the material. In a beautiful TEM study, Nogita et al [154] visualized the desorption of hydrogen from macrocrystalline MgH$_2$ and identified the mechanism to be related to the growth of different pre-existing cores/nuclei of Mg in the MgH$_2$, which had not been fully converted upon hydrogenation. On the other hand for nanostructured MgH$_2$ (samples thinned to a few tens of nanometres), they found no blocking layer effect, instead hydrogenation occurred from an outside layer moving to the inside of the material. Indeed, in general, one might expect nanostructures to be too small to support a blocking MgH$_2$ layer. However, many open questions remain, such as why in nanostructures the desorption of hydrogen is generally so much slower than the absorption process at a given temperature [155].

Most recent studies have focused on the practical challenge of maximizing the hydrogen (and hence magnesium) content in the system, by minimizing the weight of the scaffold and additives [156]. An alternative to ball milling, for instance high-pressure torsion, can be used to produce almost pure MgH$_2$ systems, leading to stable materials lacking long-range crystallinity by inducing strain and grain boundaries in the material as well as well-dispersed additives. For instance, Akiba et al [157] used a high-pressure torsion method, proposed 10 years ago, to make ultra-fine grained structures (mostly amorphous, with many grain boundaries and lattice defects), but with V, Ni, Sn as additives. Enhanced hydrogenation kinetics were found, as with ball milling, but interestingly the resistance against deactivation in air was also improved, removing the need to store under protective atmosphere. In an alternative approach, filing was used to make very small chips, which did deactivate upon air exposure, but could very easily be regenerated. There is often an important role for bcc-structured Mg, which is slightly unstable with respect to the hcp structure. It is known that the bcc phase can be stabilized by adding transitions metals such as Ti, or Nb and/or by having an interface with these particles in nanostructures. The bcc structure generally has much higher hydrogen diffusion coefficients than the hcp phase, as confirmed by recent DFT calculations [158]. As argued by several authors, high-pressure treatment to create nanocrystalline materials is more suitable for fundamental studies than ball-milled materials, as the technique avoids contamination from milling balls and vial.

Some authors explored solution phase reduction to obtain freestanding nanostructures. For example, Sun et al. [141] obtained Mg nanofibres of 400-4000 nm length and 40 nm thickness by reduction of dibutylmagnesium with Ca. Unfortunately, this also led to the formation of a ternary phase, Ca$_{19}$Mg$_8$H$_{54}$, which released hydrogen only at much higher temperatures. Extending a more traditional approach, Huen et al [124] tried to achieve high loadings in carbon aerogels by multiple impregnation with dibutylmagnesium. However, due to the large volume of this precursor only up to 17-20% of the pore volume could be filled. Furthermore, it became very challenging to completely convert the dibutylmagnesium, leading to butane being released together with the hydrogen. Capacity loss upon cycling was observed due to reaction of Mg with oxygen still contained in the system, which can however be prevented by using a purer carbon scaffold, or by treating it to remove the oxygen before loading it with magnesium [159].

One of the most successful strategies followed by several groups, is using very thin sheets of a "support" material (wrapping rather than supporting), such as graphene or graphene oxide [109,160,161], in combination with a small amount of catalyst for efficient hydrogen dissociation and association, such Ni or Pd. In this case the structure of the carbon support, which also determines the efficiency of the interaction with the co-catalyst (mostly either Ni or Pd), is very important [162]. Nanocomposites were made at once with wet-chemical and reduction techniques, or alternatively first the co-catalyst, for instance Ni, was deposited on the graphene, after which a composite with Mg was formed by ball milling [160]. Particle sizes varying from 3.0 to 6.0 nm were reported based on TEM analysis, but at the same time sharp reflections observed in the X-ray diffraction patterns evidenced the heterogeneity of these samples and the fact that larger crystallites were also present.

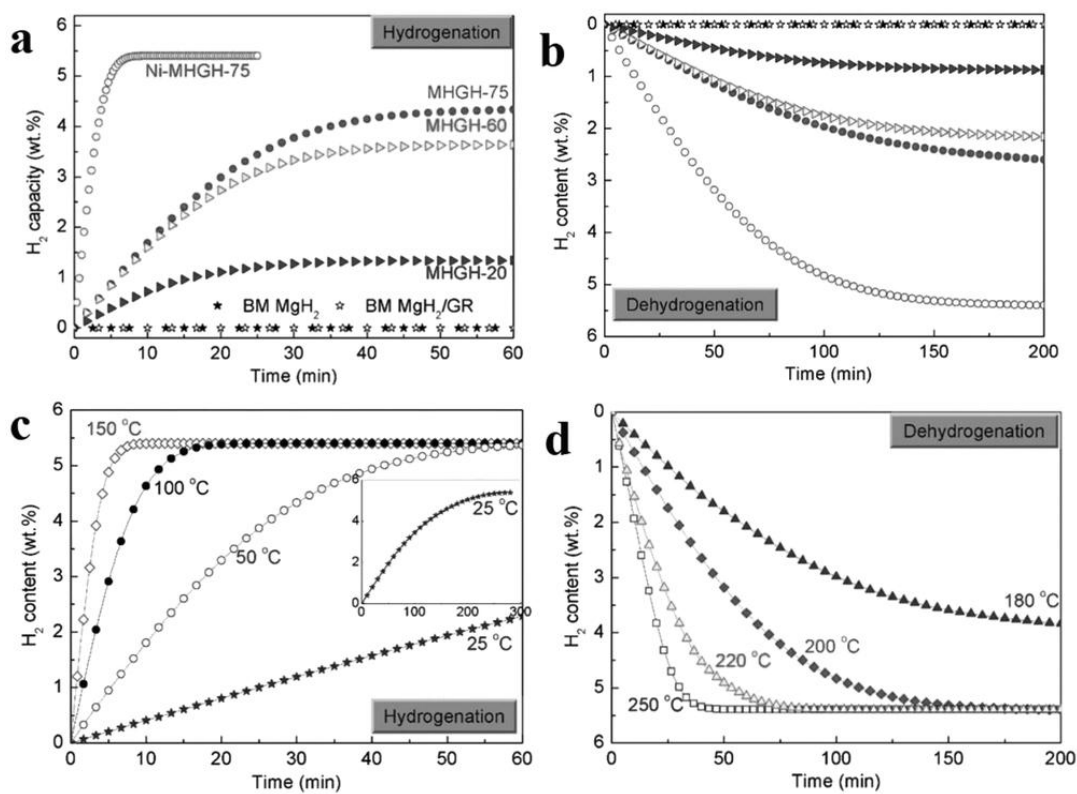

*Figure 7. a) Hydrogenation and b) dehydrogenation of (Ni)-MgH$_2$-graphene nanocomposites at 200 $^{o}C$, including ball-milled MgH$_2$ (BM MgH$_2$) and ball-milled MgH$_2$/GR composite (BM MgH$_2$/GR) for comparison. c) Hydrogenation and d) dehydrogenation of Ni-conMHGH-75 at various temperatures. Hydrogenation was measured under 30 atm hydrogen pressure and dehydrogenation under 0.01 atm.[163] 75wt% 5-6 nm (from TEM) MgH$_2$ on graphene. Mind that even at room temperature there is appreciable hydrogen absorption. The capacity retention was over 98.4% after 30 full cycles.*

In general, these nanocomposites showed favourable hydrogen storage characteristics. Due to the low amount of added graphene (oxide) (typically a few wt%), close to 6 wt% reversible hydrogen storage capacity was achieved. Kinetics were very fast; with absorption generally occurring within minutes at 100 $^{o}C$, and desorption within minutes at 250 $^{o}C$. Figure 7 gives an example of the performance of these materials [163]. Additionally, the formation of nanocomposites with graphene or graphene oxide was reported to "protect" against oxidation. Without significant capacity loss, tens or hundreds of cycles were achieved [109,163], with it being postulated that the graphene (oxide) layer protected the active material by being permeable for hydrogen, but much less so for oxygen and nitrogen. Xia et al. [161] added also LiBH$_4$, which led to a higher capacity (8.9 wt%), but also to the need for much higher desorption temperatures (e.g. 350 $^o$C instead of 250 $^o$C). An open question is whether the use of complicated synthesis techniques and expensive materials like graphene can ever allow large scale practical application, but as a technique to optimize kinetics and capacity this approach seems one of the most successful next to the techniques based on ball milling.

# NANOSTRUCTURED Mg-BASED HYDROGEN STORAGE MATERIALS PREPARED BY MECHANICAL ALLOYING AND REACTIVE BALL MILLING

The low melting temperature and high vapour pressure of magnesium have restricted applications of the conventional methods of sintering or melting to the preparation of magnesium-based composites for hydrogen storage. Mechanical alloying remains the main method of producing magnesium-based hydrogen storage materials after two decades of its intensive application. The obtained materials can be obtained in nanocrystalline or in amorphous state. This is an efficient way to add other elements to magnesium, improving the hydrogenation and dehydrogenation kinetics [45,98,99,164-166].

Mechano-chemical methods including mechanical alloying (MA), mechanical grinding (MG) and reactive ball milling (RBM) consist of mechanical treatment of metal powders in various types of mills which are routinely used both in laboratories and in large-scale synthesis. This treatment results mainly in plastic deformation of the material when applying conditions that yield metastable materials. During MA, MG and RBM, a combination of a repeated cold welding and fracturing of the particles will define the ultimate structure of the powder. These methods can be applied to the synthesis of the metastable phases: amorphous phases, supersaturated solid solutions, non-stoichiometric intermetallic compounds, quasi-crystals, composites with different microstructure and composition including those with non-interacting components of the binary or even more complex metal system. These phases often exhibit unusual physico-chemical properties and show enhanced reactivity.

The facilities for mechano-chemical treatment of the materials (ball mills) may be classified as follows [167]:

- *Low-energy (tumbling) mills* contain cylindrically shaped shells, which rotate around a horizontal axis. Loads of balls or rods are charged into the mill to act as milling tools. The powder particles of the milled materials meet the abrasive and / or impacting forces which reduce the particle size and enhance the solid-state reaction between the elemental powders. The tumbling mills are simple in design and operation and are easy to upscale. However, this type of low-energy mill requires an increased milling time to complete the mechano-chemical process.
- *High-energy ball mills*:
  - *Attritors (attrition ball mills)*. The material in this kind of mills is loaded into a stationary container / vial and comminuted by free moving balls, which are set in motion by a rotating stirrer / impeller. Some attrition mills can operate at a rotation speed up to 2000 rpm. The volume of vials used in attrition ball mills can be up to several litres.
  - *Shaker / vibration type mills*. In shaker mills, a vial oscillates along several axes at high frequency, causing intensive vibration agitation the charge and the balls in three mutually perpendicular directions at a frequency from 180 to 1200–1800 rpm.
  - *Planetary mills*. The planetary ball mills are the most frequently used laboratory mills and are used for MM, MA and RBM. In this type of mill, the milling tools have a considerably high energy, and the effective centrifugal acceleration reaches up to 200–600 m s$^{-2}$ [98,167]. Centrifugal forces caused by the rotation of the supporting disc and autonomous turning of the vial act on the milling charge (balls and powder). A variety of vials with different capacities (12–500 ml) and balls of different diameters (5–40 mm) made of different materials, are available for the commercial planetary ball mills characterised by the disc rotation speed up to 600–1100 rpm.
  - *The uni-ball mill*. This is a special type of ball mill where the ball movements can be confined to the vertical plane by the cell walls and controlled by an external magnetic field generated by adjustable permanent magnet placed close to the vial. Changing the magnet's position affects the mode of the movement of the ferromagnetic balls from shearing (low-energy) to impact (high-energy) mode.

The use of several alternative modifications of the mechano-chemical treatment results in variation of the structure and morphology of the final product as related to a variety of parameters, such as the nature of milling machine, materials of balls and vial, the ball to powder weight ratio, the milling atmosphere, the milling duration, temperature, and so on [99,166,167].

Even without additives, mechano-chemical treatment of Mg-based composites essentially changes the morphology, structure and hydrogen absorption / desorption characteristics of the material. It was shown that ball milling of MgH$_2$ [168] or Mg in H$_2$ [169,170] at room temperature yields a mixture of the usual tetragonal α-MgH$_2$ (called in [168,169] as β-MgH$_2$) and orthorhombic high pressure modification of γ-MgH$_2$ which is also obtained under GPa-level hydrostatic pressures (see the data presented in the review later). The presence of γ-MgH$_2$ destabilises the MgH$_2$ phase reducing H$_2$ desorption temperature and improves kinetics of H desorption.

A variety of additives has been used during the mechano-chemical treatment (mostly, RBM) to improve H sorption / desorption kinetics: 3d-transition metals and alloys [45,63,171-177, etc.], oxides [117,172,178-180, etc.], halides [181-183], binary and complex hydrides [114,184-188], carbon materials [66,67,162,189-192, etc.], and other compounds. A systematic experimental survey of the influence of various additives on hydrogen desorption performance of Mg-based nanocomposites prepared by high-energy RBM under H$_2$ pressure (HRBM) was published in [193]. More data on the influence of catalytic additives on hydrogenation / dehydrogenation performance of Mg-based nanocomposites will be presented in the chapter "Catalysis for de/hydriding of Mg based alloys".

Preparation and characterisation of Mg-based hydrogen storage composites by MA, MG and RBM have been a subject of numerous studies which, until now, represent a major part of the publications on the topic of this review. About ~2/3 of these works were published recently, between 2010 and 2018.

The following parameters were varied during the course of these studies:

- Type of the main starting component in the charge: individual Mg [44,45,98,99,175,194-197], Mg alloys [175,198-203], or MgH$_2$ [63,184,188,203-206];
- Catalytic additives (see present review);
- Milling medium: inert gas (mostly, argon [45,98,99,187,194,206], less frequently N$_2$ [164,207,208] or He [199]), hydrogen [44,175,185,197], or organic liquid [164,209]. Sometimes the milling was carried out in vacuum [210] or even in air without any protective atmosphere [211,212].
- Type of milling machine: mostly planetary [34,98,175,197] and shaker [45,195,196] mills, uni-ball mill operating in various modes [199,213], attritor [195], low-energy ball mill (rotating autoclave) [44].
- Other process parameters: rotating speed / vibration frequency, ball-to-powder ratio, pressure of reactive medium (H$_2$), etc.
- Pre- or post-processing of the charge including; rapid solidification [214], heat treatment in inert or hydrogen atmosphere [34,194,205,215], pre-milling of the components at different conditions [172,196,216, etc.].

Depending on the combination of the parameters listed above, various hydrogen storage composites with improved performance have been synthesized. As a rule, the improvements relate to hydrogenation / dehydrogenation kinetics, though sometimes minor altering of thermodynamic characteristics (destabilisation of Mg–H bonding or changing reaction pathway) has been achieved [217-221].

The best kinetic improvements were observed for the materials prepared by HRBM in planetary mills, as well as for the composites prepared by RBM in inert atmosphere when MgH$_2$ was taken as a starting material. It was noted [222] that the use of MgH$_2$ instead of Mg improves the kinetics of Mg interaction with the additives during the ball milling process.

Examples of the best composites exhibiting high hydrogen sorption / desorption rates are $MgH_2+NbF_5$ and $MgH_2 + Nb_2O_5$ ball milled in inert or hydrogen atmosphere [204,205], $MgH_2 + TiH_2$ (or Mg+Ti) and Mg + bcc-V alloy ball milled in $H_2$ [63,67,175].

Selected results on hydrogenation kinetics of Mg during HRBM with and without catalytic additives are presented in Figure 8 and Table 3. As it can be seen, the main factors accelerating hydrogenation of Mg are (i) introduction of the catalyst, (ii) increase of the rotation speed and ball-to-powder weight ratio (BPR), (iii) increase of $H_2$ pressure.

Catalytic additives exert the strongest influence on the improvement of hydrogenation kinetics of Mg during HRBM (compare curves 1a and 1 b, 5a and 5b), as well as on the kinetics of dehydrogenation and re-hydrogenation (see the chapter "Catalysis for de/hydriding of Mg based alloys" for the details). The second strongest factor seems to be milling energy supplied by the balls to the charge. This energy depends on the vial geometry and increases with the increase of rotation speed and BPR (compare curves 4a, 4b and 4d), as well as milling time. Quantification of this factor for the process of milling graphite in a planetary mill was presented in work [224] and references therein. Influence of $H_2$ pressure on the hydrogenation kinetics of Mg during HRBM is the weakest as can be seen from comparison of curves 3 and 6, as well as 4b and 4c in Figure 8.

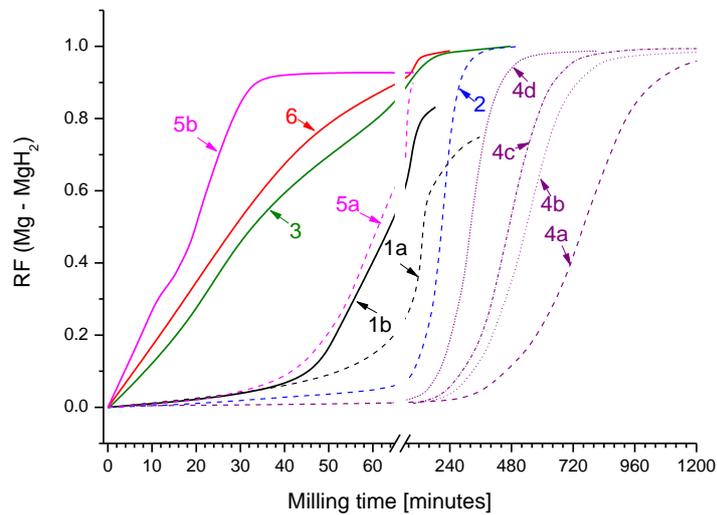

*Figure 8. Hydrogenation of Mg during its ball milling in $H_2$*

*Table 3. Process parameters of hydrogenation of Mg during its ball milling in $H_2$ (Figure 8)*

| Curve # | $H_2$ pressure [bar] | Type of ball mill | Milling parameters | Catalyst | Ref |
|---|---|---|---|---|---|
| 1a | 200 | Low-energy (rotating autoclave) | 150 rpm, BPR~20:1 | None | [44] |
| 1b | | | Heating to 350–400°C | $I_2$ (0.7%) | |
| 2 | 30 | Planetary | 500 rpm, BPR=80:1 | None | [170] |
| 3 | 300 | Planetary | 400 rpm, BPR=50:1 | $TiH_2$ (10 mol%) | [185] |
| 4a | 5 | Planetary | 300 rpm, BPR=60:1 | 5.5–6 wt%Zn, 0.4–0.5 wt%Zr (commercial Mg alloy) | [223] |
| 4b | 5 | | 400 rpm, BPR=60:1 | | |
| 4c | 10 | | 400 rpm, BPR=60:1 | | |
| 4d | 5 | | 400 rpm, BPR=120:1 | | |
| 5a | 80 | Planetary | 400-800 rpm, BPR=60:1 | None | [114] |
| 5b | | | | Ti (30 mol%) | |
| 6 | 20 | Planetary | 500 rpm, BPR=40:1 | Ti (25 mol%) | [67] |

More details about interaction of Mg with $H_2$ under HRBM conditions will be given in the next chapter

# MECHANOCHEMISTRY OF MAGNESIUM UNDER HYDROGEN GAS

Hydrogen sorption kinetics in bulk coarse-grain Mg are extremely slow. This results from kinetic limitations on both dissociation of hydrogen molecules at the Mg surface and diffusion of hydrogen atoms in bulk Mg [14,16, 45,225]. Surface limitations can be overcome by decoration of the metal surface with catalysts. The use of $Nb_2O_5$ as additive is a paradigmatic example of this approach [16,226,227]. Bulk limitations are attributed to the formation of a blocking hydride shell over the Mg metal. Indeed, the diffusion coefficient of hydrogen in magnesium hydride is very low: $10^{-18}$ m$^2$/s at 300 °C [150,152]. Thus, full hydrogenation of 90 μm in diameter of Mg particles requires about 1 day at 400 °C and $P(H_2) = 2$ MPa [60]. Slow bulk kinetics can be faced by Mg nanostructuring. The idea is to shorten the bulk diffusion lengths (i.e. the grain size) and to take advantage of the fast hydrogen diffusivity along Mg grain boundaries [228-230]. To achieve Mg nanostructuring at a large scale, mechanical milling is a very attractive method. However, milling of Mg metal under inert gas (usually argon) is problematic. Ductility as well as significant atomic mobility of magnesium lead, respectively, to powder agglomeration induced by cold welding and limited grain size reduction (~ 40 nm) resulting from enhanced recovery rate [231]. Therefore, to get loose powders of low crystallinity (~ 10 nm), ball milling of brittle magnesium hydride is preferred to that of the pure metal [168,232].

As a step forward, mechanochemistry under hydrogen gas can be used as a one-pot synthesis method to get surface-catalyzed Mg in nanocrystalline form [233]. To this aim, Mg powder is milled under hydrogen atmosphere with possible addition of one or several catalytic species (metal oxides, carbon, early or last transition metals TM…) [114,229,235,236]. As indicated in the previous section, this technique is also denoted as reactive ball milling under hydrogen atmosphere (HRBM). The proof-of-concept was given by Chen and Williams [78] motivating further studies on $MgH_2$ synthesis [169,233,237]. In the pioneering studies, formation of $MgH_2$ was generally not completed as a result of insufficient hydrogen supply in the vials. In 2007, Doppiu et al. were able to completely transform Mg into $MgH_2$ in less than 10 hours of milling using a large milling vial, V = 0.22 l, operated at high pressure, $P(H_2) \leq 9$ MPa [234]. Moreover, the vial was equipped with pressure and temperature sensors and a telemetric system for data acquisition. By a proper calibration of this device, which is commercialized by Evico Magnetics (Germany), *in-situ* hydrogenation kinetics can be analysed [97]. This provides sound information on the mechanism and hydrogenation kinetics of $MgH_2$ formation as well as on the influence of milling additives.

Figure 9a displays the sorption curve of Mg under hydrogen gas ($P(H_2) = 8$ MPa). Milling was performed in an Evico Magnetics vial using a Fritsch P4 planetary mill [114]. Disk and vial rotation speeds were 400 and -800 rpm, respectively with a ball to powder mass ratio fixed to 60. Hydrogen uptake follows a sigmoidal shape, which is characteristic of Avrami's nucleation and growth equations [238], with a complete formation of $MgH_2$ in 2 hours of milling time $t_m$. The obtained product consists of two phases: the rutile-type α-$MgH_2$ phase (S.G.: $P4_2/mnm$, 76 wt%) and the metastable high-pressure γ-$MgH_2$ phase (S.G. : $Pbcn$, 24 wt%). Both phases are nanocrystalline with grain size of ~ 6 nm.

The same synthetic procedure was conducted under deuterium gas [239]. The sorption curve follows a similar trend (Figure 9a) with slower kinetics. This can be better visualized by plotting the time-evolution of the *in-situ* absorption rate, showing a maximum at $t_m = 150$ min for $D_2$ absorption as compared to $t_m = 60$ min for $H_2$ one (Figure 9b). Such a high difference evidences that $MgH_2$ formation by mechanochemistry under hydrogen gas is controlled by the diffusion of hydrogen atoms through the freshly formed magnesium hydride over-layer [239]. According to the classical rate theory, deuterium diffusion is expected to be slower than hydrogen one due to its higher atomic mass, $D_H/D_D = \sqrt{m_D/m_H} = \sqrt{2}$. Indeed, it has been observed that absorption kinetics in Mg deuteride are slower than in Mg hydride [240].

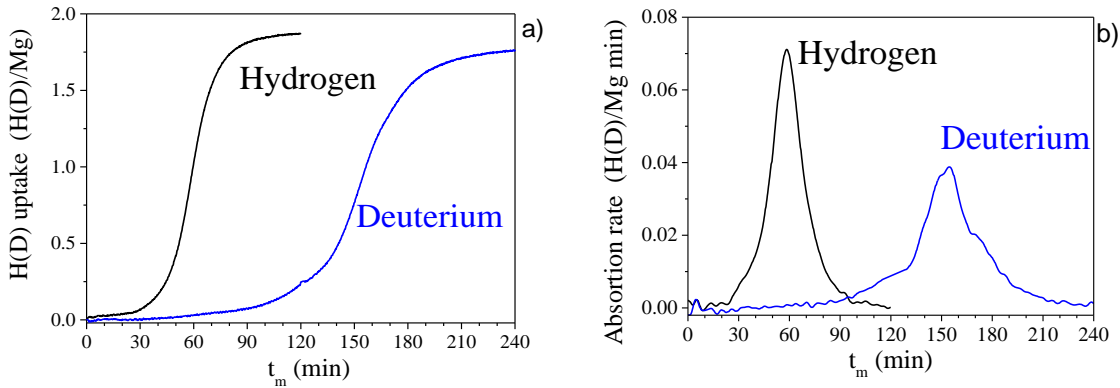

*Figure 9. Mechanochemistry of Mg powder under hydrogen and deuterium gas. a) In-situ hydrogen uptake curves as a function of milling time $t_m$, b) In-situ absorption rate (derivative curves of Figure 9a)*

$MgH_2$ formation kinetics during mechanochemical synthesis can be accelerated by addition of TMs (Figure 10). By using last transition metal (LTM) additives such as Fe, Co and Ni, the time needed to achieve the maximum absorption rate decreases by a factor of two, from $t_m = 60$ down to $t_m = 30$ min (Figure 10a) [97]. It is claimed that decoration of the Mg surface with LTM particles facilitates hydrogen dissociation and that LTM/Mg interfaces may act as active nucleation sites for the hydride formation. It should be noted that, by prolonged milling, LTMs react with $MgH_2$ through a solid-solid reaction leading to the formation of ternary (e.g. $Mg_2NiH_4$) and quaternary (e.g. $Mg_2(FeH_6)_{0.5}(CoH_5)_{0.5}$) Mg-based complex hydrides [97,241-243].

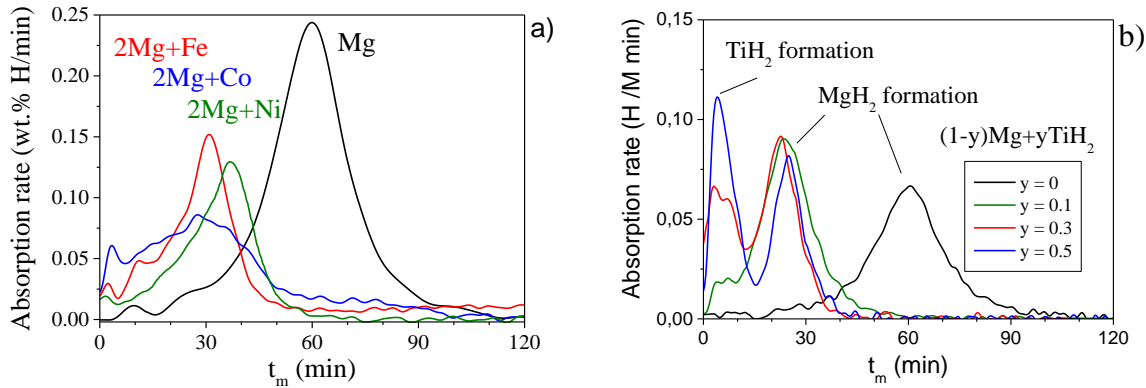

*Figure 10. Mechanochemistry of Mg under hydrogen gas using several transition metals TM as additives. a) In-situ absorption rate with LTM = Fe, Co and Ni (atomic ratio Mg/LTM = 2), b) In-situ absorption rate with ETM = Ti for different titanium contents y.*

Early transition metals (ETMs) can be also used as milling additives with the peculiarity to form stable hydride phases at the starting of the milling process. Reaction kinetics for ETM = Ti with $10 \leq$ Ti (at.%) $\leq 30$ at.% are displayed in Figure 10b [114]. Titanium hydride is formed at short milling time ($t_m \leq 15$ min) and it helps to accelerate $MgH_2$ formation rate by a factor of three, from $t_m = 60$ min to $t_m = 23$ min. This effect is mainly attributed to the abrasive properties of $TiH_2$ that during milling scrapes off the surfaces of pristine Mg and freshly formed $MgH_2$ [239]. In addition, it has been suggested that $TiH_2$ hydride phase may also act as a gateway for hydrogen chemisorption and diffusion toward the Mg phase [114]. In contrast to Mg-LTMs systems, prolonged milling of Mg-Ti under hydrogen gas results in the formation of composite materials, and not ternary hydrides, consisting of $MgH_2$ (mixture of $\alpha$ and $\gamma$ polymorphs) and $TiH_2$ phases. Those are nanostructured materials with grain size around 5 nm for $MgH_2$ and 10 nm for $TiH_2$ [244]. These materials are characterized by fast hydrogenation kinetics, even for hydrogen absorption at room temperature, with synergetic kinetic effects between both phases. They are of high interest both for hydrogen storage at moderate temperatures [114,245] and as negative electrodes of Li-ion batteries [246-249].

# CATALYSIS FOR DE/HYDRIDING OF Mg BASED ALLOYS

The dehydrogenation of Mg based hydrides requires rather high temperature, even above 300 °C, which must be overcome for practical applications. To enhance the hydrogen absorption/desorption rate and reduce the reaction temperature of Mg based hydrogen storage alloys, different improving strategies, such as alloying, nanosizing, nanoconfinement, catalyzing and compositing, have been applied to tune the dehydriding/hydriding thermodynamics and kinetics [14,18,21,156,251-253]. It should, however, be noted that the beneficial effect of tuning a certain characteristic of the hydrogen storage material can be accompanied by a deterioration in its other service characteristics. Thus, tailoring is usually aimed at improving one specific property without destroying other features relevant for hydrogen storage.

A large variety of different types of metals and compounds, including carbon materials, metals and intermetallics, transition-metal compounds (oxides, halides, hydrides, carbides, nitrides, and fluorides) have been added as catalytic additives by different material preparation processes. Some of them exhibit excellent catalytic activity on the de-/hydrogenation behaviour of $MgH_2$, leading to faster hydrogen sorption rate and lower reaction temperature. In general, the catalytic effect is determined by several key factors: (1) The type of additives with specific catalytic mechanism; (2) size and distribution of catalysts, which is related with the preparation process; and (3) the structural stability of catalyst in the de/hydrogenation cycles.

In this part of the review, the earlier performed catalysis work on the de/hydrogenation reactions of Mg-based alloys has been summarized with the emphasis on the effect of different type of catalysts and their catalyzing mechanism. The preparation processes to obtain catalyzed Mg-based composites are also discussed in view point of the catalyzing effect and the convenience for large scale fabrication.

## Non-metal additives

Carbon-based materials, including graphite (G), carbon nanofibers (CNFs), carbon nanotubes (CNTs), and graphene (GN), are the most effective non-metal additives to show prominent catalytic effect on the de-/hydrogenation of $Mg/MgH_2$ system. Imamura *et al.* first demonstrated the greatly enhanced hydrogen storage properties of Mg/G composites [254], which were prepared by mechanical milling of Mg and G with different organic additives such as benzene, cyclohexane or tetrahydrofuran. They suggested that mechanical milling in liquid organic additives resulted in highly dispersed cleaved lamellae of graphite and the generation of large amounts of dangling carbon bonds in graphite, which could act as active sites for hydrogen absorption. The hydrogen uptake of the Mg/G nanocomposites is in the form of C–H bonds and hydrides in the graphite and magnesium, respectively. The hydrogen absorption amount by the graphite was estimated to be ~1.4 wt% of $H_2$ per gram of carbon. Wu *et al.* found various carbon additives exhibited advantage over the non-carbon additives [255], such as BN nanotubes or asbestos, in improving the hydrogen storage capacity and kinetics of Mg. The hydrogen storage capacities of all mechanically-milled Mg/C composites at 300 °C exceeded 6.2 wt% within 10 min, about 1.5 wt% higher than that of pure $MgH_2$ at the identical operation conditions. The remarkable improvement in the hydrogen capacity and absorption/desorption kinetics of Mg is attributed to the incorporation of carbon that increases the area of phase boundaries and hydrogen diffusion driving force. The milling results in many carbon fragments with $sp$ and $sp^2$ hybridizations and unhybridized electrons delocalized, which may have strong interaction with hydrogen molecules. Physically adsorbed hydrogen molecules can be concentrated around the carbon fragments and act as hydrogen source for the further chemical dissociation on Mg. In addition, the specific nanostructure of single-walled carbon nanotubes (SWNTs) may facilitate the diffusion of hydrogen into Mg grains, and thus exhibited the most prominent "catalytic" effect over other carbon materials such as graphite, activated carbon, carbon black and fullerene. The onset dehydrogenation temperature of $MgH_2$ with the addition of SWNTs could be reduced by 60 °C in comparison with non-carbon additives.

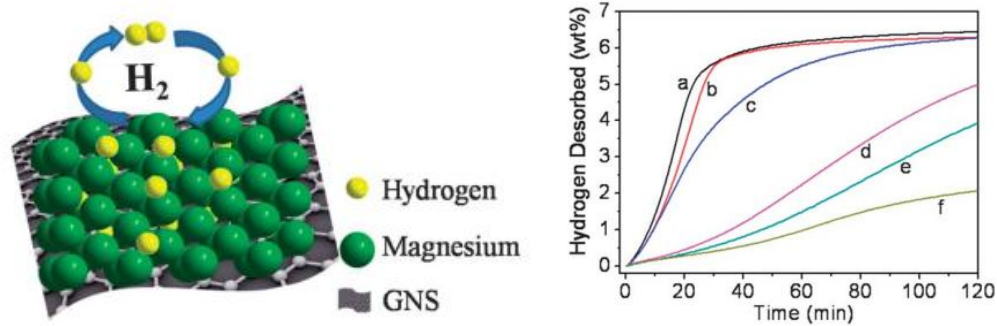

*Figure 11. Schematic of hydriding/dehydriding reaction in MgH₂–GNS composite, and hydrogen desorption curves of the sample at 300 ℃: (a) MgH₂–5GNS-20 h, (b) MgH₂–5GNS-15 h, (c) MgH₂–5GNS-10 h, (d) MgH₂–5GNS-5 h, (e) MgH₂–5GNS-1 h, and (f) MgH₂–20 h [256].*

Recently, a two-dimensional carbon material, graphene, has also been found to significantly improve the hydrogenation/dehydrogenation reaction of MgH₂. Y.J. Wang et al. synthesized highly crumpled graphene nanosheets (GNS) with a BET surface area as high as 1159 m² g⁻¹ by a thermal exfoliation method [256], and further composited graphene nanosheets with Mg by ball milling. As schematically shown in Figure 11, the smaller GNS dispersed in an irregular and disordered manner in the composite after milling, providing more edge sites and hydrogen diffusion channels, preventing the nano-grain (5–10 nm) of MgH₂ sintering and agglomeration, thus leading to enhanced hydrogen storage properties. The 20 h-milled composite MgH₂–5%GNS can absorb 6.3 wt% H within 40 min at 200 °C and 6.6 wt% H within 1 min at 300 °C, even at 150 °C, it can also absorb 6.0 wt% H within 180 min. It was also shown in Figure 11 that 6.1 wt% H at 300 °C within 40 min could be released from MgH₂–GNS composite. The mechanism investigation found that the GNS served as both dispersion matrix and catalyst for hydrogen diffusion.

In addition, graphene nanosheets are a good support for nanoscale catalysts and Mg nanoparticles. Yu *et al*. reported a bottom-up self-assembly of MgH₂ from the organic solution of dibutylmagnesium [163], obtaining a large number of monodispersed MgH₂ nanoparticles (~5 nm) distributed uniformly on the graphene nanosheets (Figure 12a). Moreover, the loading percentage of MgH₂ nanoparticles on graphene could be increased up to 7.5 wt%, and the maximum hydrogen capacity of Mg/GNS system is 5.7 wt%, both of which are much higher than the nanoconfinement method for the preparation of Mg nanoparticles [257]. By further incorporation of Ni catalyst into the Mg/GNS composite, the Ni-catalyzed MgH₂/GNS system exhibited superior hydrogen storage properties and cycling performances. A complete hydrogenation could be achieved within 60 min at 50 °C, and 2.3 wt% hydrogen uptake even at ambient temperature within 60 min. Moreover, the hydrogenation capacity at room temperature reached up to ~5.1 wt% within 300 min. In the dehydrogenation, the Ni-catalyzed MgH₂/GNS system can completely desorb 5.4 wt% H within 30 min (Figure 12b). It was also shown that the apparent activation energy ($E_a$) for hydrogenation and dehydrogenation, which is based on the isothermal kinetic curves at different temperature and Arrhenius equation, was calculated to be 22.7 kJ mol⁻¹ and 64.7 kJ mol⁻¹, respectively, which are drastically lower than the corresponding values (99.0 kJ mol⁻¹ for hydrogenation and 158.5 kJ mol⁻¹ for dehydrogenation) for the bulk Mg. More importantly, there is almost no capacity retention for the Ni-catalyzed MgH₂/GNS system after 100 hydrogenation/dehydrogenation cycles (Figure 12c), with no loss in kinetic performance. This is attributed to the nature of graphene that could act not only as a structural support for loading MgH₂ nanoparticles, but also as a space barrier to prevent the sintering and growth of MgH₂ nanoparticles during hydrogenation/dehydrogenation cycles.

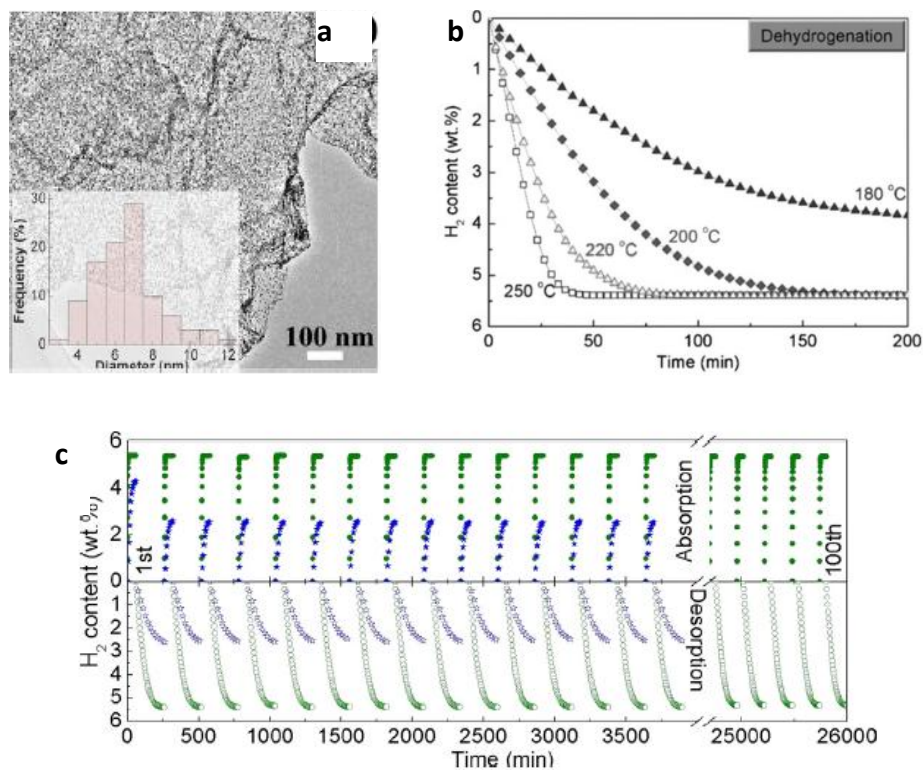

*Figure 12. TEM image, dehydrogenation kinetic curves, and reversible H₂ absorption (under 3 MPa H₂) and desorption (under 0.001 MPa H₂) of Ni catalyzed 75 wt% MgH₂ [163].*

The introduction of carbon additives was shown to result in the improvement of hydrogen sorption / desorption performance of Mg-based nanocomposites including a significant increase of their stability during H absorption-desorption cycling at high temperatures. This effect was associated with distribution of carbon in between nanoparticles of Mg(H₂) during RBM preventing their coalescence and surface oxidation [258,259], although a study of the kinetics of MgH₂ with C₆₀ buckyballs as an additive showed little or no improvement [260].

Recently, Lototskyy et al [66] suggested that the effect of $sp^2$-hybridized carbon additives to Mg is related to the formation of graphene layers during their RBM in H₂ (HRBM) which then encapsulated the MgH₂ nanoparticles and prevented the grain growth on cycling. This results in an increase of absorption–desorption cycle stability and in a decrease in the MgH₂ crystallite size in the re-hydrogenated Mg–C materials as compared to Mg alone. Recent experimental studies of composite materials containing MgH₂ with graphene / graphene derivatives and exhibiting improved and stable dehydrogenation kinetics [224,261,262] confirmed the correctness of this hypothesis.

In [67] it has been shown that introduction of 5 wt% of graphite into the MgH₂ − TiH₂ composite system prepared by HRBM results in outstanding improvement of the hydrogen storage performance when hydrogen absorption and desorption characteristics remained stable through the 100 hydrogen absorption / desorption cycles and were related to an effect of the added graphite. A TEM study showed that carbon is uniformly distributed between the MgH₂ grains covering segregated TiH₂, preventing the grain growth and thus keeping unchanged the reversible storage capacity and the rates of hydrogen charge and discharge (Figure 13).

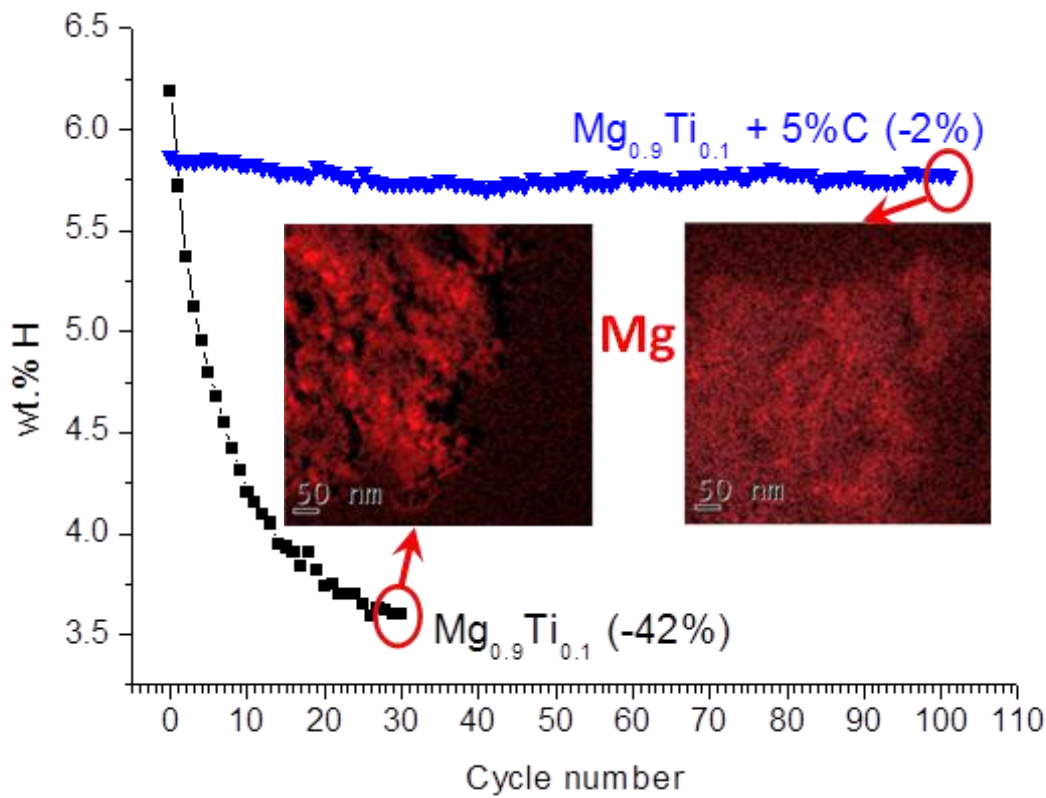

*Figure 13. Reversible hydrogen storage capacity of HRBM $MgH_2$–$TiH_2$ at T=350 °C. The values in brackets specify the capacity losses throughout the cycling. The insets show elemental maps of Mg in the cycled composites clearly indicating its grain refinement in the graphite-modified material [67].*

It has to be noted that kinetic improvements in the formation / decomposition of $MgH_2$ in the carbon-containing materials are more pronounced when minor amounts ($\leq$5-10 wt%) of the carbon additives were introduced together with catalytic additives of transition metals or oxides [66,67,263-267]. This synergetic effect was explained by the facilitation of hydrogen dissociation / recombination on the surface of the catalyst while the carbon species played the role of an efficient mediator of the H atoms between the catalyst and $Mg(H_2)$ [66,265]. In addition, carbon may also inhibit oxidation of $Mg(H_2)$ during cyclic H absorption / desorption thus preventing the deterioration of the reversible hydrogen storage capacity of the material [180].

## The effect of microstructure on the hydrogenation properties of Mg-carbonaceous additives composites

Initial reports on the super-high gravimetric hydrogen storage capacity of the single wall carbon nanotubes (CNTs) created a wave of enthusiasm and initiated a number of studies on the subject. However, it soon became clear that early estimates of hydrogen storage capacity of CNTs were highly exaggerated [268]. It was shown that various structural modifications of carbon store hydrogen primarily by physisorption, and that the maximum hydrogen storage capacity scales with the specific surface area of the material, similarly to the other nanoporous materials storing hydrogen at low temperatures [268]. Yet it is now universally accepted that the single- and multiwall CNTs, as well as other carbonaceous nanomaterials are very efficient catalytic agents enhancing the hydrogenation kinetics of Mg and its hydride-forming alloys. In spite of a high number of works already published on this subject and continuing flow of new publications, presenting a self-consistent picture of the effect of carbonaceous nanomaterials on hydrogenation of Mg-based alloys is hardly possible, mainly because of the differences in the processing methods employed by different groups and resulting differences in the microstructures of the Mg-carbonaceous material composites. In this respect, the recent work of Ruse et al. [269] should be particularly mentioned. They performed a systematic study of the effect of various carbon allotropes on the hydrogenation behaviour of Mg-carbonaceous material composites. It was found that hydrogen absorption and desorption are both accelerated in presence of carbon allotropes, with the allotropes of lower dimensionality exhibiting stronger catalytic effect (i.e. one-dimensional CNTs exhibited the strongest catalytic effect, followed by the 2-D graphene nano-platelets and activated carbon) [269]). Moreover, the catalytic effect increased with decreasing density of defects in carbon allotropes, as determined by the ratio of intensities of the disorder-related D-band and G-band in Raman spectra [269]. These results of Ruse et al. [269] are in a good agreement with the earlier findings of Skripnyuk et al. [270] that severe plastic deformation and concomitant increase of defect density in the multiwall CNTs (MWCNTs) cancels their beneficial effect on the increase of the equilibrium hydrogen plateau pressure observed in the as-processed Mg - 2 wt% MWCNTs composite [270].

The mechanisms of the catalytic effect of carbon allotropes on hydrogen interaction with Mg remain poorly understood. One of the possible mechanisms is the spillover effect reducing the role of carbon allotropes to that of efficient transport paths for dissociated hydrogen atoms [162,269]. However, the role of most anisotropic carbon allotropes in modifying the microstructure of the two-phase Mg-MgH$_2$ composite material during hydrogen absorption/desorption remains largely ignored. Surprisingly, there are very few metallographic studies of the morphology of a two-phase Mg- MgH$_2$ mixture, in spite of the fact that it plays a crucial role in kinetics of Mg interaction with hydrogen [271]. For example, high nucleation rate of MgH$_2$ during hydrogenation leads to accelerated formation of the continuous hydride shell on the surface of Mg particles, inhibiting further hydrogenation due to the sluggish diffusion of hydrogen through MgH$_2$. The well-known beneficial effect of nanostructuring on hydrogenation kinetics of Mg and its alloy can be interpreted in terms of transition from core-shell metal-hydride morphology in coarse grained powders to the Janus-type morphology in nanopowders.

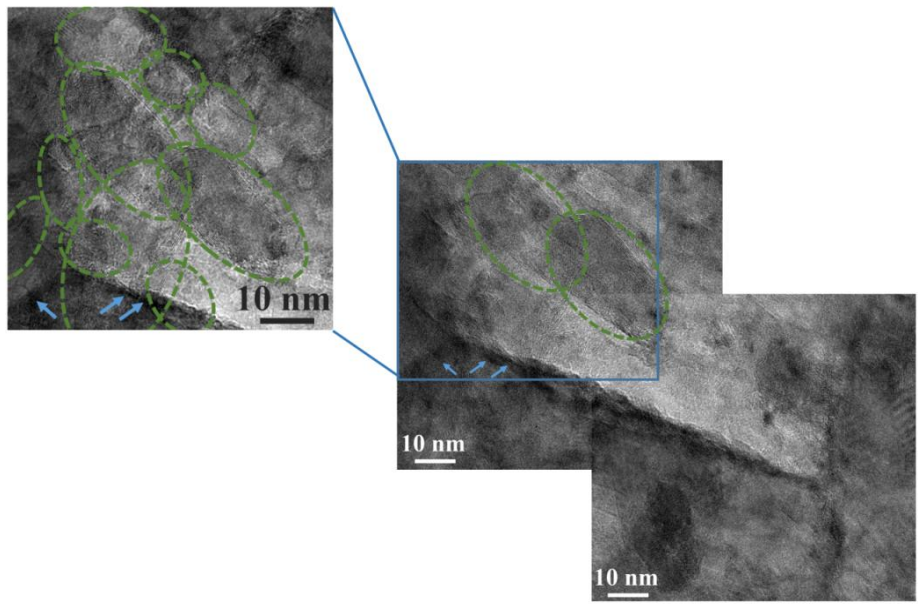

*Figure 14. Scanning Transmission Electron Microscopy (STEM) High Angle Annular Dark Field (HAADF) micrographs of MWCNTs segments/carbon nanoparticles (marked by the circles) located in close proximity to each other and forming a "chain" within Mg grains. Arrows point on grain boundary in Mg, indicating that carbon nanoparticles are located inside the Mg grains rather than along the grain boundaries. The sample was prepared by co-milling of Mg powder with 2 wt% MWCNTs in the Pulverisette - 7 planetary micro mill in hexane for 4 h at 800 rpm using the stainless steel balls of 10 mm in diameter. BTP ratio was 20:1 [273].*

Recent works [272, 273] have uncovered the role played by MWCNTs and the products formed after prolonged ball milling in modifying the morphology of two-phase Mg-MgH$_2$ mixture. It was found that high-energy co-milling of Mg powder with 2 wt% of MWCNTs results in partial destruction of MWCNTs which transform into anisotropic chains of carbon nanoparticles (see Figure 14). These nanoparticles, and their interface with Mg matrix serve as preferential nucleation sites of hydride phase during hydrogenation of the composite. As a result, the two-phase Mg-MgH$_2$ mixture exhibits highly anisotropic morphology, with the bicontinuous intertwined networks of the metallic and hydride phases (see Figure 15). This morphology results in high thermal conductivity of the partially hydrogenated samples [273], and ensures continuous supply of hydrogen to the metal-hydride interface due to the fast diffusion of former through the continuous metallic phase. The porous pellets with such bi-continuous microstructure exhibited thermal conductivity approx. 50% higher than those made from pure Mg powder and were hydrogenated to achieve a comparable fraction of their maximum hydrogen storage capacity. These examples demonstrate that optimising the microstructure of the two-phase Mg- MgH$_2$ mixture with the aid of anisotropic additives (such as MWCNTs [272,273] and nanoplatelets [224]) represent a promising path to improving hydrogenation properties of Mg and its alloys.

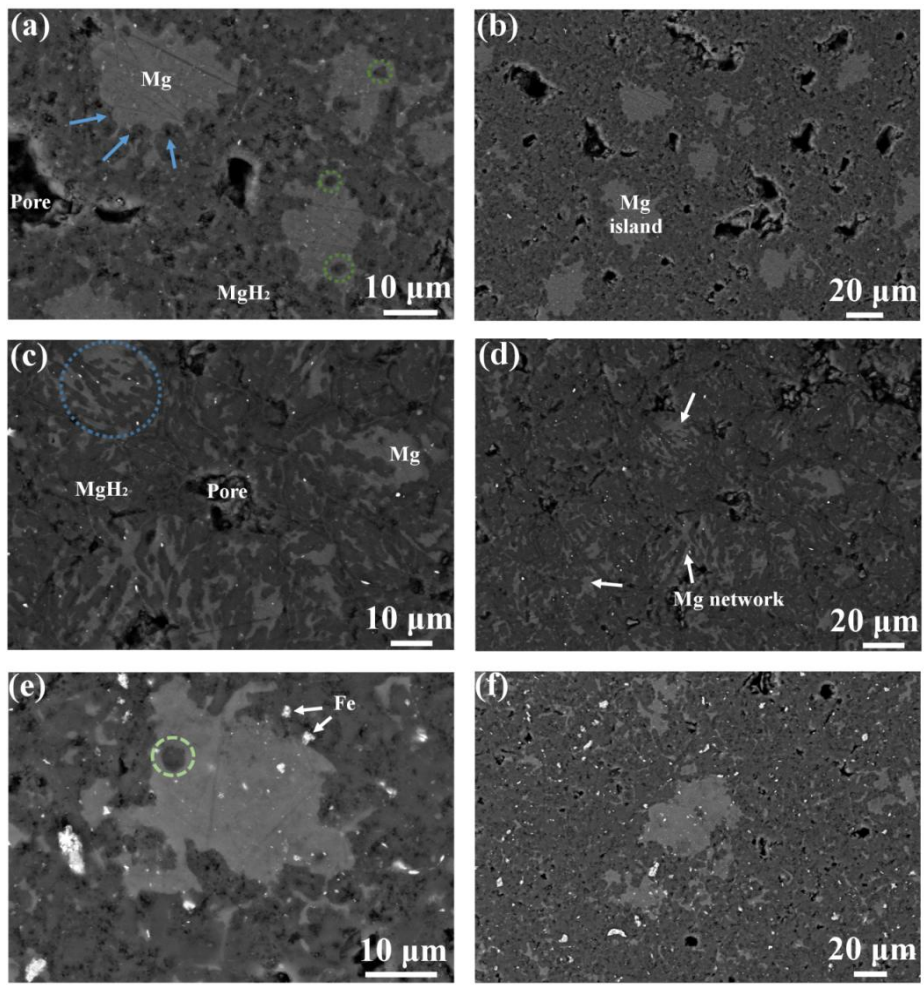

*Figure 15. Backscattered electrons (BSE) scanning electron microscopy micrographs of the pellets hydrogenated to 80-90% of maximum theoretical hydrogen storage capacity (a,b- Mg pellet; c,d- Mg-2wt% MWCNTs, and e,f- Mg-2wt% Fe). The view plane is perpendicular to the compression axis. (a)-Circles mark individual unimpinged isotropic MgH$_2$ nuclei. Arrows point on impinged MgH$_2$ nuclei forming wavy Mg/MgH$_2$ interface. (b) Lower magnification micrograph showing the isolated pockets of unreacted Mg surrounded by the MgH$_2$ phase. (c) Elongated anisotropic MgH$_2$ nuclei (marked by the circle), (d) Micrograph showing the developed Mg network along the sample. (e) Symmetrical MgH$_2$ nucleus formed next to the Fe particle. (f) Increased number of hydride nucleation sites results in smaller size of metallic Mg islands in comparison with the reference Mg pellet [273].*

## Transition metals and their compounds

Among various additives, the transition metals and their compounds showed superior catalytic performances on $MgH_2$ [16]. For example, Liang et al. first reported different catalytic effects of 3d-TMs (TM=transitional metal Ti, V, Mn, Fe, Ni, etc.) on the reaction kinetics of Mg-H system [116]. The ball-milled Mg-Ti composite exhibited the most rapid hydrogen absorption rate, followed by the Mg-V, Mg-Fe, Mg-Mn and Mg-Ni composites, while the most hydrogen desorption rate was attributed to the $MgH_2$-V composite, followed by the $MgH_2$-Ti, $MgH_2$-Fe, $MgH_2$-Ni and $MgH_2$-Mn composites. By reactive milling under $H_2$ atmosphere, the Mg-M (M=Co, Ni and Fe) systems showed better hydrogen storage properties due to the dual effects of catalyzing and particle refining.

The transitional metal oxides are also extensively studied as effective catalysts for hydrogen storage in $MgH_2$. Bormann et al. prepared the $MgH_2/Me_xO_y$ nanocomposites ($M_xO_y=Sc_2O_3$, $TiO_2$, $V_2O_5$, $Cr_2O_3$, $Mn_2O_3$, $Fe_3O_4$, CuO, $Al_2O_3$ and $SiO_2$) powders via ball milling, and found that the catalytic effect of $TiO_2$, $V_2O_5$, $Cr_2O_3$, $Mn_2O_3$, $Fe_3O_4$, and CuO on the hydrogenation of Mg are similar [274]. The $Fe_3O_4$ showed the best effect in the dehydrogenation reaction, which was followed by $V_2O_5$, $Mn_2O_3$, $Cr_2O_3$ and $TiO_2$. Especially, Barkhordarian et al. reported the superior catalytic effect of $Nb_2O_5$ and the fast hydrogen sorption kinetics of 0.2 mol.% $Nb_2O_5$-doped Mg with nanocrystalline structure [226]. The absorption of 7 wt% H at 300 °C was reached within 60 s and the desorption was completed within 130 s. The hydrogen absorption at 250 °C was almost as fast as at 300°C, while the desorption required only 10 min. The authors also investigated the effect of $Nb_2O_5$ concentration (0.05, 0.1, 0.2, 0.5, and 1 mol.% $Nb_2O_5$) on the kinetics of the magnesium [117], and found that fastest kinetics were obtained using 0.5 mol.% $Nb_2O_5$. At 250 °C, more than 6 wt% hydrogen were absorbed in 60 s and desorbed again in 500 s. The apparent activation energy for dehydrogenation varied with the $Nb_2O_5$ concentration and reached the minimum value of 61 kJ $mol^{-1}$ $H_2$ at 1 mol.% $Nb_2O_5$. The $Nb_2O_5$ catalyst may act also as nucleation sites for $H_2$ on Mg, since it was also demonstrated that the unusual catalytic effectiveness of $Nb_2O_5$ for the recombination of hydrogen molecule on the Mg surface [275]. To further reveal the catalytic mechanism of $Nb_2O_5$, Jensen et al. performed in situ real time synchrotron radiation X-ray diffraction experiments under dynamic hydrogenation and dehydrogenation reactions of $MgH_2$ ball milled with 8 mol.% $Nb_2O_5$ [276], and the results indicated that the $Nb_2O_5$ reacted with Mg forming ternary solid solution $Mg_xNb_{1-x}O$ with a composition in the range 0.2~0.6 in the heating, which facilitated the surface reaction and chemisorption or the recombination of hydrogen molecule.

In addition to oxides, the transition metal-hydrides are also proven to be effective catalysts. Pelletier et al. performed the time-resolved X-ray scattering measurement on the hydrogen desorption in the milled $MgH_2$-Nb system [277], and revealed an intermediate niobium hydride phase with an approximate composition of $NbH_{0.6}$ and a small crystallite size of ~8-10 nm formed during dehydrogenation (Figure 16). The intensity of metastable niobium hydride increased simultaneously with the intensity of metallic magnesium and only after the magnesium was fully dehydrided. This short-lived metastable niobium-hydride phase may be the real catalytic species, and this is direct evidence of the hydrogenation mechanism in such composite metal hydrides. The authors proposed a model in which the niobium nanoparticles act as gateway for hydrogen flowing out of the magnesium reservoir.

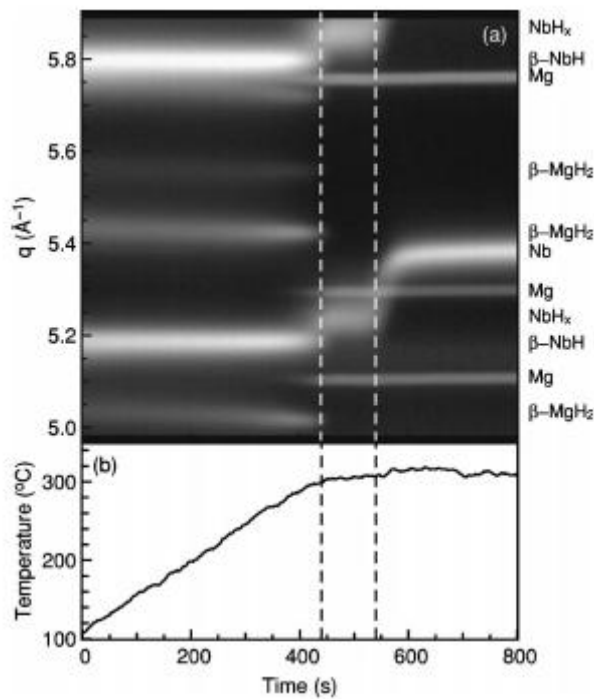

*Figure 16. Time-resolved X-ray scattering data for MgH$_2$-Nb heated to 310 °C. (a) Gray-scale contour plot of the X-ray scattering where intensity increases with lighter tones, (b) temperature profile [277].*

Obviously, these catalysts could effectively overcome the reaction kinetic barriers related to different steps of hydrogenation and dehydrogenation reactions, including the hydrogen physisorption, hydrogen chemisorption, hydrogen diffusion, or the recombination of hydrogen molecules, hydride formation and decomposition. The extremely slow hydrogen diffusion rate in MgH$_2$ at ambient temperature [153] will dramatically slow down the hydrogenation rate upon the formation of magnesium hydride layer around the Mg particles. Therefore, the hydrogen diffusion in the MgH$_2$ is the rate-limiting step for both the hydrogenation and dehydrogenation reactions. Especially, the reaction rate for the dehydrogenation of MgH$_2$ is generally slower than that for the hydrogenation of Mg, and the apparent activation energy for the dehydrogenation is estimated to be ~160 kJ mol$^{-1}$, much higher than ~120 kJ mol$^{-1}$ for the hydrogenation.

As a generally adopted catalyst doping process, the simple ball-milling of the additives with Mg brings beneficial structural modifications, including the particle and grain refinement of the Mg and additives, the breaking of any surface oxide layer (MgO), the introduction of a large number of structural defects on the surface of Mg, and the homogeneous distribution of active sites for the hydrogenation reaction. The reduction of particle/grain size in MgH$_2$ leads to a pronounced kinetic enhancement due to the increased surface area of the interfaces and shortened diffusion paths, while the surface structural modifications caused by milling are necessary for the gas-solid reaction between Mg and hydrogen gas because of increased rates of the processes at the surface. Even though tremendous efforts have been devoted to improving the hydrogen storage properties of Mg by ball milling with a large variety of additives, however, it is still difficult to obtain a very homogeneous distribution and to obtain greatly refined catalysts, even when using very long milling times. Unfortunately, this may cause severe contamination and negatively affect hydrogen storage performance. One solution to this problem is to use a liquid additive which disperses through the magnesium hydride more readily. Alsabawi et al. [278] used a titanium-based organic liquid and found it to be just as effective as Nb$_2$O$_5$. Some other methods have been developed for the preparation of nanocomposites of hydrides and rare-earth metal or transition metal catalysts, including the chemical solution [183,279], the *in situ* decomposition of Mg-based multi-component alloys [280,281], and the crystallization of amorphous Mg-based alloys [114,282,284]. Excellent hydrogen storage properties especially hydrogen sorption kinetics have been achieved.

One problem to overcome is that the passivated MgO layer covering magnesium metal is considered almost impermeable to hydrogen and decreases the kinetics significantly. Recently, it was found that the transition metal catalysts could greatly enhance hydrogen absorption/desorption of $MgH_2$. By the reaction of Mg powder in THF solution with $TMCl_x$ (TM: Ti, Nb, V, Co, Mo, or Ni), micro-sized Mg particles were coated by different transition metals, forming a continuous TM shell with a thickness of less than 10 nm [183]. All composites released hydrogen at a low temperature of 225 °C. It was noted that the nano-coating of a TM around the micro-sized $MgH_2$ particles is much more effective than milling Mg with the corresponding TMs. The catalytic effect on the dehydrogenation is in the sequence Mg–Ti, Mg–Nb, Mg–Ni, Mg–V, Mg–Co and Mg–Mo. This may be due to the decrease in electro-negativity (c) from Ti to Mo. The nano-composite of Mg with a nano-coating of Ti-based catalysts can release 5 wt% $H_2$ within 15 min at 250 °C. However, Ni is a special case with a high catalytic effect in spite of the electro-negativity. It was supposed that the formation of the $Mg_2Ni$ compound may play an important role in enhancing the hydrogen de/hydrogenation of the Mg–Ni system. It was also found that the larger the formation enthalpy, the worse the dehydrogenation kinetics.

In addition, Cui *et al.* also coated different Ti-based nanoscale catalysts [279], including Ti, $TiH_2$, $TiCl_3$ and $TiO_2$, on the surface of ball-milled Mg powders. Firstly, the mixed powders of Mg and $MgH_2$ with a weight ratio of 9:1 were pre-milled on a vibratory milling apparatus for 6 h, and then the pre-milled powder were reacted with $TiCl_3$ in THF solution under electromagnetic stirring for 5 h. This resulted in dehydrogenation properties much better than those of the conventionally ball-milled sample. Hydrogen release started at about 175 °C and reached 5 wt% $H_2$ within 15 min at 250 °C, with the dehydrogenation $E_a$ reduced to 30.8 kJ mol$^{-1}$. It was suggested that the multiple valence Ti sites facilitated electron transfer among them, and thus acted as the intermediate for electron transfer between $Mg^{2+}$ and $H^-$. Figure 17 illustrates the catalytic mechanism of multi-valence Ti-compounds on the dehydrogenation of $MgH_2$. Because the electronegativity of Ti (1.54) is between Mg (1.31) and H (2.2), it facilitates the weakening the Mg–H bond. Also, a splitting of the 3d state of Ti ions can induce the Ti ions to gain electrons (e$^-$) easier than Mg ions and lose e$^-$ easier than H$^-$ ions. As shown in Figure 17, there exist a large number of interfaces among $MgH_2$, high valence and low valence Ti compounds in the hydrogenated composite. The electron transfer between $Mg^{2+}$ and $H^-$ is proposed in the following steps: (1) H$^-$ at the interface of $MgH_2$/Ti-compounds donates e$^-$ to high valence Ti ($Ti^{3+/4+}$) which transforms into low valence Ti ($Ti^{2+}$) simultaneously; (2) with the breakage of the weakened Mg–H bond, the dissociative H is produced and dehydrogenation reaction occurs; (3) H atoms are recombined into $H_2$; and (4) Mg nucleates and grows coupled with $H_2$ recombination. Thus, the dehydrogenation of $MgH_2$ is promoted due to the lowered barrier in the above inferred catalytic process.

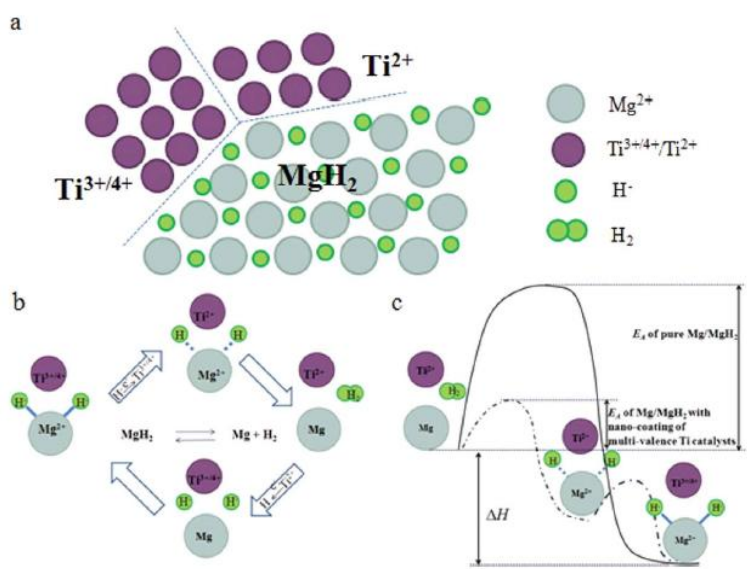

*Figure 17. Schematic for the catalytic mechanism of muti-valence Ti doped $MgH_2$ [279]*

To obtain a stable and pronounced catalytic effect in long-term de-/hydrogenation cycling, it is very important to maintain the nanocrystalline structure of $MgH_2$ composite containing homodispersed nanosized catalysts, namely keeping the particle/grain of Mg-based nanocomposite from growing in repeated dehydrogenation/hydrogenation cycles. Cuevas *et al.* reported that the $0.7MgH_2–0.3TiH_2$ composite, which was obtained by reactive ball milling of elemental powder under 8 MPa hydrogen pressure [114], exhibited outstanding kinetic properties and cycling stability. At 300 °C, hydrogen uptake took place in less than 100 s, which is 20 times faster than for a pure nanosized $MgH_2$ powder by milling. Further, it was demonstrated that the $TiH_2$ phase inhibited the grain coarsening of Mg, which allows extended nucleation of the $MgH_2$ phase in Mg nanoparticles prior to the formation of a continuous and blocking $MgH_2$ hydride layer. The hydrogenation process of $0.7MgH_2–0.3TiH_2$ composite also follows a gateway mechanism for hydrogen transfer from the gas phase to Mg.

It is thus suggested that *in situ* formed catalyst would show higher catalytic activity and superior stability than those of the externally added catalyst due to the better homogeneity and finer particle size. Gross *et al.* observed that the hydrogen absorption/desorption of $La_2Mg_{17}$ was far more rapid than Mg due to the catalytic effect of *in situ* formed $LaH_{3−x}$ phase upon hydrogenation [285]. It has been widely accepted that 3d transition metal Ni could significantly lower the dissociation barrier. For that, Ouyang *et al.* synthesized the $MgH_2$-based composites containing $CeH_{2.73}$ and Ni catalysts [286], which were *in situ* formed from a $Mg_3Ce$ structured $Mg_{80}Ce_{18}Ni_2$ alloy in the hydrogenation process. This nanocomposite exhibited excellent hydrogen absorption/desorption performance and cycling stability, absorbing hydrogen at room temperature and desorbing hydrogen at 232 °C with a high capacity of ∼4 wt% and fast kinetics. The apparent activation energy ($E_a$) is only 63 kJ mol$^{−1}$ H$_2$, which is far below that of milled $MgH_2$ (∼158 kJ mol$^{−1}$ H$_2$) or that of $Mg_3Ce$ alloy (∼104 kJ mol$^{−1}$ H$_2$). The existence of Ni is a key role for refining the microstructure of $CeH_{2.73}$-$MgH_2$ composites, the Ni not only demonstrated good catalytic effect on the hydrogen desorption of $MgH_2$ but also promoted the transformation of $CeH_{2.73}$ to $CeH_2$, increasing the practical hydrogen storage capacity. Transmission electron microscopy analysis revealed the combinational catalytic effect of *in situ* formed extremely fine $CeH_2/CeH_{2.73}$ and Ni to Mg/$MgH_2$. As shown in Figure 18, the grain size of $CeH_{2.73}$ and $MgH_2$ phases is much smaller in the hydrogenated $Mg_{80}Ce_{18}Ni_2$ alloy compared to that for the hydrogenated $Mg_3Ce$ alloy. The very fine $CeH_{2.73}$ and Ni nanophases could effectively inhibit the growth of Mg. In the hydrogenation process, the $Mg_{80}Ce_{18}Ni_2$ alloy initially transformed into Mg-$CeH_{2.73}$-Ni composite, the Mg then reacted with hydrogen to form $MgH_2$, and the $MgH_2$-$CeH_{2.73}$-Ni nanocomposite was finally obtained. The great improvement in the hydrogenation kinetics for the $MgH_2$-$CeH_{2.73}$-Ni nanocomposite is due to the presence of Ni nanoparticles, high-density interfaces between $CeH_{2.73}$ and $MgH_2$, and plenty of grain boundaries in nanocrystalline $MgH_2$. A large amount of interfaces and boundaries act as hydrogen diffusion channels and nucleation sites of hydrides. In addition, the *in situ* formed Ni and $CeH_{2.73}$/$CeH_2$ nanophases act as dual catalysts for the hydriding/dehydriding reactions of Mg. It is noted that pure Ni instead of $Mg_2Ni$ existed in the $MgH_2$-$CeH_{2.73}$-Ni composites, which is totally different from the findings of other *in situ* hydrogenated Mg-RE-Ni alloys. In the dehydrogenation process, it is assumed that Mg nuclei preferentially nucleate along the surface of $CeH_{2.73}/CeH_2$ and Ni phase at the starting transition stage of $MgH_2$ to Mg. More importantly, this nanocomposite structure can effectively suppress Mg/$MgH_2$ grain growth and enable the material to maintain its high performance for more than 500 hydrogenation / dehydrogenation cycles. As shown in Figure 18, the capacity retention exceeds 80% after 500 cycles, and the main reason for the capacity loss is by slight oxidation. In summary, the enhanced hydrogen storage kinetics and stability of $MgH_2$ is attributed to the synergetic effect of *in situ* formed $CeH_{2.73}$ and Ni.

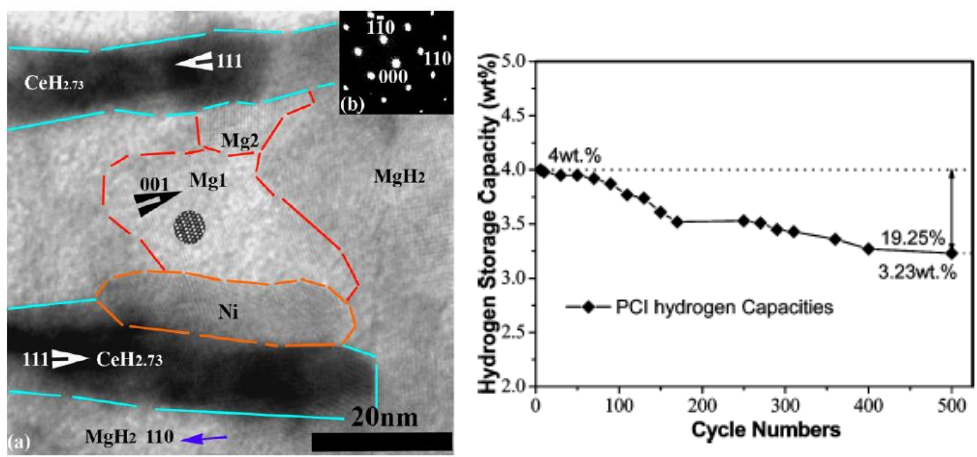

*Figure 18. Left: TEM images of the microstructure of the partially dehydrogenated MgH₂-CeH₂.₇₃-Ni nanocomposites demonstrate the catalyst effect of CeH₂.₇₃ and Ni on MgH₂ dehydrogenation process. (a) Bright field image and (b) selected area diffraction patterns of MgH₂ (zone axis [011̄]). Right: Evolution of the maximum hydrogen sorption capacities versus cycle times of MgH₂-CeH₂.₇₃-Ni composite. [286]*

To achieve the synergetic role of different catalysts, Lin et al developed a simple method to induce a novel symbiotic $CeH_{2.73}/CeO_2$ catalyst into Mg-based hydrides, which was prepared via controllable hydrogenation and oxidation treatments of as-spun $Mg_{80}Ce_{10}Ni_{10}$ amorphous ribbons [287]. Namely, crushed $Mg_{80}Ce_{10}Ni_{10}$ amorphous powder was first activated under 10 MPa hydrogen pressure at 300 °C for 3 h and then treated by 15 cycles of dehydrogenation/hydrogenation at 300 °C. Then, the sample was controllably oxidized under specific condition. The obtained sample has a nanocomposite microstructure of $MgH_2$–$Mg_2NiH_4$–$CeH_{2.73}/CeO_2$ and displays remarkable reduction in the dehydrogenation temperature. Maximum hydrogen desorption temperature reduction of $MgH_2$ could be reduced down to ~210 °C for the composite with the molar ratio of $CeH_{2.73}$ to $CeO_2$ being 1:1. The dynamic boundary evolution during hydrogen desorption was observed in the symbiotic $CeH_{2.73}/CeO_2$ at atomic resolution using *in situ* high-resolution transmission electron microscope, as shown in Figure 19. Combining the *ab-initio* calculation, which shows significant reduction in the formation energy of hydrogen vacancy at the $CeH_{2.73}/CeO_2$ interface boundary in comparison to those for the bulk $MgH_2$ and $CeH_{2.73}$, it was proposed that the outstanding catalytic activity can be attributed to the spontaneous hydrogen release effect at the $CeH_{2.73}/CeO_2$ interface as efficient "hydrogen pump".

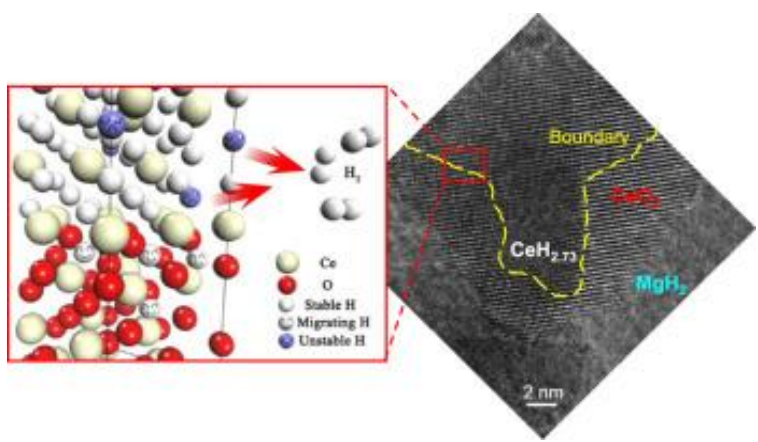

*Figure 19. Schematic illustration and TEM image showing the enhanced hydrogen release at the interface of $CeH_{2.73}/CeO_2$ [287].*

## Metal halides and fluorides

The catalytic role of the anion in metal additives on the hydrogenation of Mg was also investigated. Suda *et al.* first claimed that the fluorination treatment in an aqueous solution containing F⁻ could provide a highly reactive and protective fluorinated surface for hydrogen uptake on different alloys [288,289]. Therefore, as a large variety of additives containing both transition metal and fluorine anion, transition metal fluorides such as $TiF_3$, $TiF_4$, $NbF_5$, $ZrF_4$ etc. have been widely used to designed catalysed Mg-based hydrogen storage system. Xie et al. milled $MgH_2$ nanoparticles with 5 wt% $TiF_3$ under hydrogen atmosphere, and obtained the composite of $MgH_2$, $TiH_2$, and $MgF_2$ [290]. The resultant composite desorbed 4.5 wt% H within 6 min at 300 °C and absorbed 4.2 wt% H in 1 min at room temperature. The authors suggested that decrease of particle size was beneficial for enhancing absorption capacity at low temperature, but has no effect on desorption. Moreover, the hydrogen storage properties of $MgH_2$-$TiF_3$ composite are superior over the $MgH_2$-$TiH_2$ composite. Here, the $TiF_3$ helps to dissociate the hydrogen molecule at low temperature as the Ti species is introduced into magnesium. Wang et al. investigated the influence of $NbF_5$ as an additive on the H-sorption kinetics of $MgH_2$ [291], and found that fast kinetics were obtained in the 5h-milled $MgH_2$ + 2 mol.% $NbF_5$ composite. At 300 °C, the $MgH_2$ + 2 mol.% $NbF_5$ could absorb 5 wt% hydrogen in 12 s and 6 wt% in 60 min, and desorb 4.4 wt% in 10 min and 5 wt% in 60 min. The structural analysis indicated that $MgH_2$ reacted with $NbF_5$ to form $MgF_2$ and $Nb^{x+}$–containing compounds. It was suggested that in situ formed $MgF_2$ has a highly reactive and protective effect for hydrogen uptake (later confirmed by the observation of improvement of hydrogen desorption kinetics in ball milled $MgH_2$–$MgF_2$ composites [292]), which may further combine with the catalytic function of Nb species to produce a synergetic effect. Alternatively, $F^-$ anion may also directly participate in the generation of the catalytically active species.

Another origin of the improvement of H sorption performance of $MgH_2$ by fluorine substitution can be in the formation of $Mg(H_xF_{1-x})_2$ solid solutions, isostructural to rutile-type $MgH_2$. These solid solutions have been shown to absorb hydrogen with a practical hydrogen capacity of 4.6 wt% H for x = 0.85 and 5.5 wt% H when x = 0.9 [293,294]. Additionally, with the formation of $Mg(H_xF_{1-x})_2$ solid solutions, stabilization is found to occur with respect to $MgH_2$. The thermodynamics for $Mg(H_{0.85}F_{0.15})_2$ were determined, with an enthalpy of decomposition of 73.6 ± 0.2 kJ mol$^{-1}$ $H_2$ and entropy of 131.2 ± 0.2 J K$^{-1}$ mol$^{-1}$ $H_2$ [294]. These values are decreased in comparison with $MgH_2$ from 74.06 kJ mol$^{-1}$ $H_2$ and 133.4 J K$^{-1}$ mol$^{-1}$ $H_2$ [136].

The effects of typical titanium compounds ($TiF_3$, $TiCl_3$, $TiO_2$, $TiN$ and $TiH_2$) on hydrogen sorption kinetics of $MgH_2$ were also compared by Wang et al. [182]. Among them, adding $TiF_3$ resulted in the most pronounced improvement of both hydrogen absorption and desorption kinetics. At 150 °C, 3.8 wt% H can be absorbed within 30 s for the milled $MgH_2$ + 4 mol.% $TiF_3$ composite, leading to the absorption rate at least 10 times higher than others. Further, the favourable kinetic performances persisted well in the absorption/desorption cycles, and the cyclic capacity loss is <10% after 15 cycles. Comparative studies indicate that the $TiH_2$ and $MgF_2$ phases in situ introduced by $TiF_3$ were not responsible for the superior catalytic activity. For other titanium compounds, the catalytic activity follows the order of $TiO_2$ > $TiN$ > $TiH_2$, which may be understandable from the electron transfer associated with multi-valence of Ti cation.

Further comparison between $TiF_3$ and $TiCl_3$ additives indicated that $TiF_3$ showed superior catalytic effect over $TiCl_3$ in improving the hydrogen sorption kinetics of $MgH_2$ (Figure 20), this result suggested the specific catalytic role of F anion [295]. It was shown that both titanium halide and fluoride reacted with $MgH_2$ in a similar way during milling or hydrogen cycling processes, forming $TiH_2$ and $MgCl_2$ or $MgF_2$. It was thus assumed that the $F^-$ anion may result in catalytically active species, while the chlorine anion may not. This assumption was confirmed by X-ray photoelectron spectroscopy studies. It was revealed in Figure 21 that the incorporated fluorine (F) showed significantly different chemical bonding state from its analogue chlorine (Cl). The asymmetry of F 1s spectra and the sputtering-induced peak shift suggested that a new and localized Ti–F–Mg bonding was formed in the $TiF_3$-doped $MgH_2$. In contrast, the stable binding state of Cl was assigned to the $MgCl_2$ for the $TiCl_3$-doped $MgH_2$. Therefore, the generation of active F-containing species well explains the advantage of $TiF_3$ over $TiCl_3$ in improving the hydrogen sorption kinetics of $MgH_2$. The functionality of F anion in tuning the activity of compound catalyst was also found in other hydrogen storage materials [296].

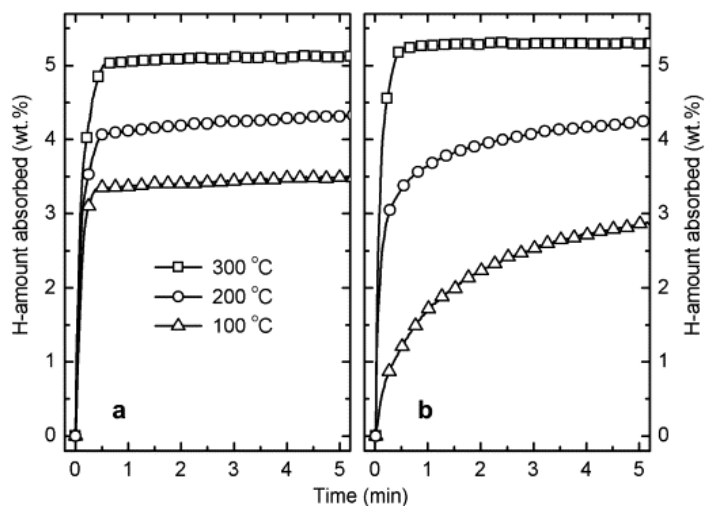

*Figure 20. Hydrogen absorption kinetics of (a) $MgH_2 + 4\,mol.\%\,TiF_3$; (b) $MgH_2 + 4\,mol.\%\,TiCl_3$. [295].*

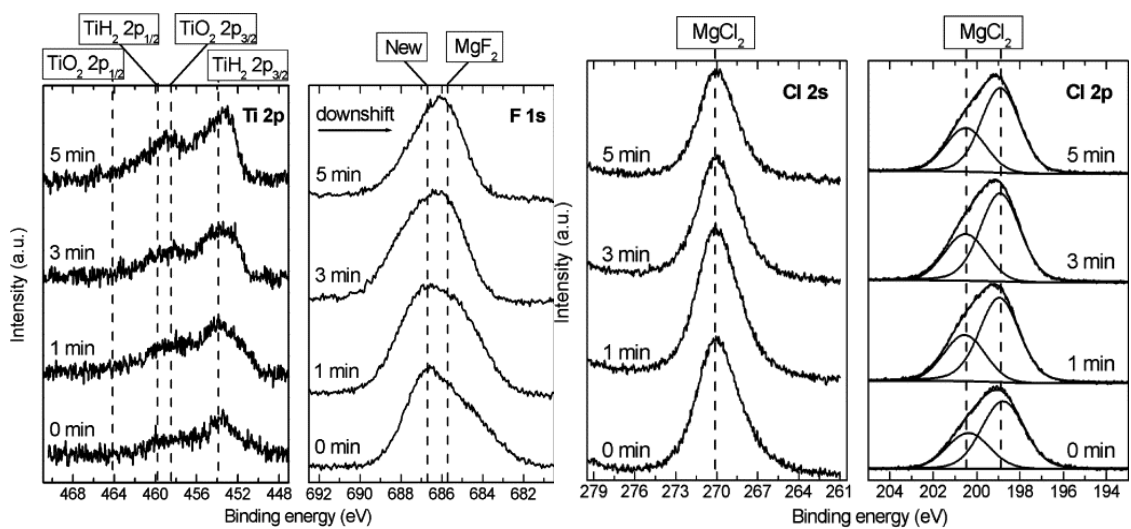

*Figure 21. Left: Evolution of Ti 2p and F 1s photoelectron lines for the dehydrogenated $MgH_2 + 4\,mol.\%\,TiF_3$ sample as a function of sputtering time; Right: Evolution of Cl 2s and Cl 2p photoelectron lines for the dehydrogenated $MgH_2 + 4\,mol.\%$ $TiCl_3$ sample as a function of sputtering time. [295]*

Recent experimental and computational studies of the reactivity of bulk and ball-milled $TiCl_3$ and $TiF_3$ with hydrogen gas [297] showed that these compounds themselves cannot accelerate H exchange reactions (by promoting $H_2$ dissociation) due to the endergonic thermodynamic driving forces. However, these additives can potentially promote $H_2$ dissociation at interfaces where structural and compositional varieties or chemical transformations are expected that probably take place in $Mg/MgH_2$ doped by Ti halides and fluorides.

## Summary

The search for and identification of effective catalysts is a subject of great importance for the development of Mg-based hydrogen storage system. Numerous efforts in the catalysts optimization have brought remarkable improvement on hydrogen absorption/desorption kinetics of $MgH_2$. The multi-valence transition metal Ti, Nb and their oxides, hydrides and halides have been proven to show superior catalytic effect over other additives. The combination of different types of catalysts may improve the overall kinetics [16], and especially combined with the light carbon materials, could offer multiple advantages with regard to kinetics, capacity as well as cyclability owing to their roles of supporting and confining catalysts. Regarding the catalyst doping, the incorporation of catalytic species by *in situ* generation chemical reaction are strongly recommended. Despite the great achievements by catalysing, the thermodynamic properties of $MgH_2$ do not change with additives.

This explains that the dehydrogenation kinetics of catalysed Mg-based materials is always inferior to the hydrogenation kinetics. Thus, the combination of thermodynamic tuning strategies, such as nanoconfinement and destabilization by reactive compositing, and catalysing should be given key consideration in the future study of Mg-based hydrogen storage materials.

## ROLE OF DRIVING FORCE ON KINETICS OF HYDROGEN SORPTION REACTIONS

The reaction rate in $H_2$ absorption-desorption experiments is often described on the basis of the Johnson-Mehl-Avrami-Kolmogorov (JMAK) equation, which has been developed for phase transformations based on nucleation and growth mechanism. Several papers showing a JMAK kinetic analysis are available in the literature, providing a wide range of values for the activation energy for hydrogen sorption reactions. Nevertheless, in order to determine the effect of additives or particle size reduction on kinetic parameters for hydrogen absorption and desorption reactions in $MgH_2$, the thermodynamics of phase transformations has to be considered in detail.

Nucleation and growth model for phase transformations include both thermodynamic and kinetic parameters. In particular, for the determination of the homogeneous nucleation barrier ($\Delta G^*$) in the classical nucleation theory [298] (ClNT), interfacial energy and driving force are necessary. In case of heterogeneous nucleation, the reduction of $\Delta G^*$ is defined on the basis of catalytic effects of nucleants, which act on the interfacial energy, promoting nucleation. In the frame of ClNT, the nucleation frequency can be obtained from $\Delta G^*$ via a diffusion coefficient. On the other hand, the diffusion coefficient, which is also related to the growth rate, is composed of a kinetic contribution (i.e. mobility) and a thermodynamic factor, related to the first derivative of the chemical potential with respect to composition. So, it is clear that a careful kinetic analysis of hydrogen sorption reactions requires a deep knowledge of corresponding thermodynamics.

According to the van 't Hoff equation, the relationship between the equilibrium temperature ($T_{eq}$) and pressure ($P_{eq}$) is given by $T_{eq}=\Delta H/(RlnP_{eq}+\Delta S)$, where $\Delta H$ and $\Delta S$ are, respectively, the enthalpy and entropy of the hydrogen sorption reactions and $R$ is the gas constant. From the pressure dependence of the Gibbs free energy, the driving force for the phase transformation at temperature $T$ and pressure $P$ can be obtained from $\Delta G= \Delta H - T \Delta S + RTln(P/P_0)$, where $P_0$ is a reference pressure, often taken as 1 bar.

The driving forces for hydrogen absorption and desorption in Mg, according to the $Mg + H_2 \leftrightarrow MgH_2$ reaction, can be calculated from available thermodynamic databases [299]. Considering [21] $\Delta H = -74.5$ kJ/mol$_{H2}$ and $\Delta S = -135$ J/mol$_{H2}$/K, lines connecting constant values of free energy difference for hydrogen sorption reactions have been calculated as a function of pressure and temperature and the results are shown in Figure 22. Negative values of driving force correspond to a spontaneous hydrogenation reaction of magnesium to form magnesium hydride. A more accurate description of driving forces for hydrogen sorption in Mg can be developed by the CALPHAD method [85], where the temperature dependence of thermodynamic properties of solid phases is taken into account.

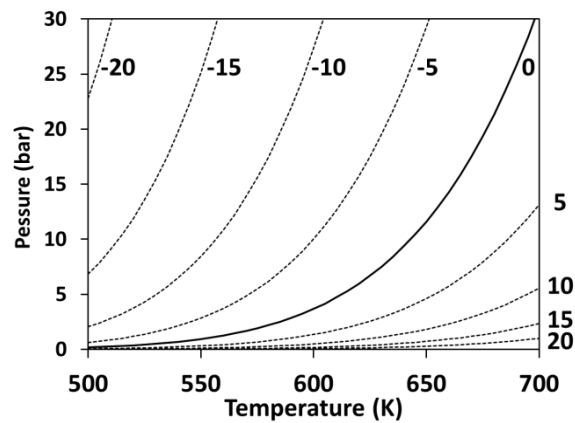

*Figure 22. Calculated driving forces for hydrogen absorption/desorption in Mg as a function of temperature and pressure. Lines connect constant values, as indicated in kJ mol⁻¹$_{H2}$. Thick continuous line corresponds to equilibrium conditions.*

Experiments aimed at determining the kinetic parameters for hydrogen sorption reactions are often performed following the phase transformation as a function of time at different temperatures. This approach requires a constant driving force for nucleation, which implies a change of $H_2$ backpressure at the various temperatures. From Figure 22, it turns out that, in order to maintain a constant driving force for hydrogenation of magnesium equal to –5 kJ $mol_{H_2}^{-1}$ in a temperature range of 20 K (i.e. from 600 K to 620 K), a corresponding change in pressure of about 5 bar is necessary. On the basis of a careful analysis of possible rate determining steps of the hydrogen sorption mechanism, a relationship between the temperature of experiments and backpressure of hydrogen has been obtained [300], suggesting to maintain, during experiments, a constant value proportional to $T[1-(P_{eq}/P)^{1/2}]$ and $T[1-(P/P_{eq})^{1/2}]$ for absorption and desorption experiments, respectively. This approach has been applied to determine kinetic parameters for hydrogen sorption reactions in Mg nanoparticles [301].

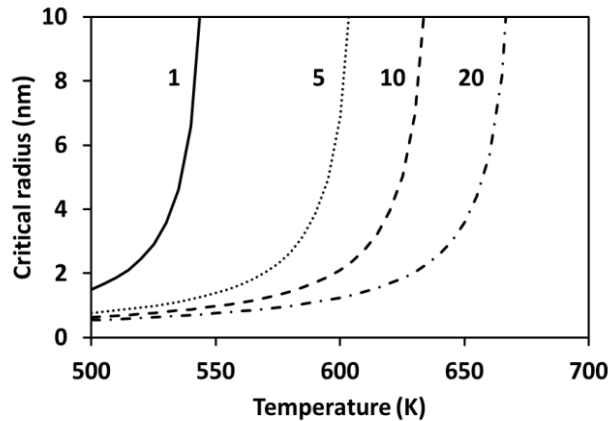

*Figure 23. Critical radius for various values of P. Continuous line: 1 bar; dotted line: 5 bar; dashed line: 10 bar; dot-dashed line: 20 bar.*

Applying the ClNT to hydrogen sorption reactions [153], the critical radius for nucleation ($r_c$) can be estimated according to $r_c=-2\gamma/\Delta G_v$, where $\gamma$ is the interfacial energy between the metal and hydride phases and $\Delta G_v$ is the driving force per unit volume. Because of the pressure and temperature dependence of the driving force, the value of $r_c$ can be estimated for the Mg/MgH$_2$ system as a function of $T$ for different values of $P$. $\Delta G_v$ has been obtained from $\Delta G * V_m$, where $V_m$ is the molar volume of the nucleating phase, which has been considered as the average between $V_{Mg}$ = 1.38×10$^{-5}$m$^3$·mol$^{-1}$ and $V_{MgH2}$= 1.81 ×10$^{-5}$m$^3$·mol$^{-1}$ [300]. An experimental or calculated value of $\gamma$ for the Mg/MgH$_2$ interface is not available, but it can be estimated considering the difference between the free energy of Mg/TiH$_2$ and the MgH$_2$/TiH$_2$ interfaces. So, it has been taken equal to 0.33 J·m$^{-2}$, as obtained experimentally from thin film experiments [131]. As an example, the values of $r_c$ obtained as a function of temperature for various values of $P$ are reported in Figure 23. It turns out that, as expected, an increase in $P$ and a reduction in $T$ decreases the value of the critical radius, which is in the range of 1-10 nm. Of course, constant values of driving force in kinetics experiments will maintain a constant value for the critical radius, allowing kinetic information from isothermal experiments to be obtained.

As a conclusion, from simple thermodynamic arguments, it turns out that the backpressure of hydrogen is a key parameter for kinetic experiments, because it drives the selection of temperatures for isothermal measurements. If experiments are performed at various temperatures maintaining a constant pressure, the driving force is changed and, as a consequence, the $r_c$ value turns out to be different. In case of the presence of nanoparticles or stress, interface and/or strain contributions to the free energy should be also considered, as described previously in the section concerning nanostructured magnesium hydride.

# Mg-H SYSTEM AT HIGH PRESSURES

Magnesium dihydride powder supplied by Sigma–Aldrich consisting of 92 wt% $\alpha$-MgH$_2$, 7 wt% Mg(OH)$_2$ and 1 wt% Mg metal according to X-Ray diffraction, was used to investigate the Mg-H reaction at high H$_2$ gas pressures.

In order to determine the equilibrium pressure of the $\alpha \leftrightarrow \gamma$ transformation in the range of a smaller baric hysteresis, the high-pressure experiments were carried out at temperatures as high as 700 °C (973 K).

Each powder sample was enclosed in a Cu container with an inner diameter of 5 mm, a height of 4 mm and the walls 0.5 mm thick. The container was filled with 90 mg of compacted MgH$_2$ and tightly plugged with a copper lid in an Ar glove box. The container was then placed into a lens-type high pressure chamber [302]; compressed to a pre-selected pressure up to 6 GPa, and heated to 973 K using a graphite heater electrically insulated from the copper container by a layer of mica. The pressure was measured with an accuracy of ±0.2 GPa, the temperature ±50 K. After exposure to these conditions for 1 hour the sample was quickly cooled (quenched) to room temperature and restored to ambient pressure. The sample was removed from the container under an Ar atmosphere, crushed, placed on an X-ray sample holder, hermetically sealed with a Mylar film on a grease ring to further protect it from contact with air and examined by X-ray diffraction at room temperature with a Siemens D500 powder diffractometer using Cu K$\alpha$ radiation. Along with MgH$_2$, all quenched samples contained about 25 wt% MgO. No signs of any reaction between the MgH$_2$ and copper container were observed at 973 K.

Results of the X-ray investigation of the quenched MgH$_2$ samples are presented in Figure 24. There were two series of experiments.

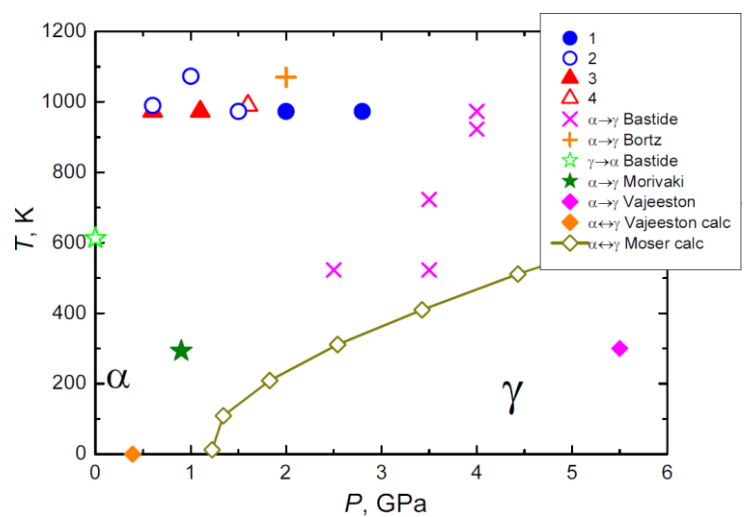

Figure 24. Phase diagram of MgH$_2$ near the $\alpha \leftrightarrow \gamma$ equilibrium line. 1 – $\alpha$ transformed to $\gamma$; 2 – $\alpha$ did not transform to $\gamma$; 3 – $\gamma$ transformed to $\alpha$; 4 – $\gamma$ did not transform to $\alpha$. The other symbols show literature data (Bastide [26], Bortz [303], Morivaki [29], Vajeeston [28], Moser [304]).

To determine the starting pressure of the $\alpha \rightarrow \gamma$ transition, the initial $\alpha$-MgH$_2$ was compressed. As shown by the filled circles in Figure 24, the formation of $\gamma$-MgH$_2$ started at P = 2 GPa. This gives the upper limit $P_{eq}$ < 2 GPa for the pressure of the $\alpha \leftrightarrow \gamma$ equilibrium at 973 K.

The quenched MgH$_2$ samples produced at 2.0 and 2.8 GPa (solid circles in Figure 24) were composed of a mixture of the $\alpha$ and $\gamma$ phases and contained, respectively, 25 and 40 wt. % $\gamma$- MgH$_2$, the rest being $\alpha$-MgH$_2$ and MgO. These samples were used in the second series of experiments aimed at determining the starting pressure of the reverse $\gamma \rightarrow \alpha$ transition. As the experiments showed, the relative content of the $\alpha$ and $\gamma$ phases did not change in the sample exposed to 1.6 GPa (open triangle in Figure 24) whereas the $\gamma$ phase disappeared in the samples exposed to $P \leq 1.1$ GPa (solid triangles in Figure 24). This sets the lower limit $P_{eq}$ > 1.1 GPa.

Estimating from the obtained measurements, we arrive at $P_{eq} = 1.5\pm0.5$ GPa at $T = 973$ K. The literature data [303] are consistent with this estimate.

Raman scattering study of the $\gamma$-MgH$_2$ to HP1-MgH$_2$ and reverse phase transformations at room temperature in diamond anvil cell were performed, both without pressure transmitting medium and in helium. The pressures of the phase transformations were found to be 8.7(10) GPa and 7.4(3) GPa respectively with no pressure transmitting medium, and 10.3(3) GPa and 8.0(10) GPa respectively in helium. No traces of pyrite-type MgH$_2$ were found in both cases, in contrast to the data of Vajeeston et al [28]. Compression of the HP1 phase to higher pressures in helium revealed transformation to the HP2 phase at 17.2(3) GPa and a reverse phase transformation at 10.0(10) GPa, in accordance with the data of Moriwaki et al. [29].

# MODELLING THE MgH$_2$ HYDRIDE WITHIN HIGH PRESSURE MODEL

## Methodology

DFT calculations [305,306] were carried out using the Vienna ab initio simulation package (VASP) [307,308]. The exchange and correlation (XC) functional was considered within the generalized gradient approximation (GGA) in the frame of Perdew-Burke-Ernzerhof (GGA-PBE) [309,310]. An energy cut-off of 800 eV was used for the projector augmented-plane wave basis set (PAW) [311], and a dense grid of k-points in the irreducible wedge of the Brillouin zone (k-points spacing less than $2\pi\cdot0.05$ Å$^{-1}$) was used with the sampling generated by the Monkhorst-Pack procedure [312]. Both the internal atomic coordinates and the lattice parameters were fully relaxed so that the convergence of Hellmann-Feynman forces was better than $10^{-6}$ eV/Å in order to eliminate any residual strain. Electron Localization Function (ELF) was plotted using VESTA software [313].

The zero-point energy (ZPE) and finite-temperature properties, arising from vibrational displacements around equilibrium positions were determined in the frame of the theory of lattice dynamics, with phonon calculations using the frozen phonon (supercell) method [314]. The vibrational modes of the different compounds were computed using the phonopy code [315,316] coupled with VASP calculations by using the quasi-harmonic approximation (QHA), i.e. repeating the harmonic calculation at several volumes $V$ to get the minimum of $F(V,T)$, the heat capacity at constant pressure, thermal expansion and bulk modulus.

Using the same methodology as in previous work on Ni-H [317], the thermodynamic high pressure model of Lu [318] was applied. The temperature $T$ and pressure $P$ dependencies of the molar volume of solid phases are described on the basis of an empirical relationship between molar volume $V_m$ and bulk modulus $B = -V\left(\frac{\partial P}{\partial V}\right)_{T,c}$:

$$V_m(T,P) = x + y \ln\left(\frac{B}{P_{ref}}\right) \qquad (8)$$

with $x$ and $y$ functions of temperature, characteristic of the considered material. Knowing values of $V_m$ and $B$ at $P_{ref}=10^5$ Pa, the previous equation of state could be written in the following way:

$$V_m(T,P) = V_0 + c(T) \ln\left(\frac{\kappa(T,P)}{\kappa_0}\right) \qquad (9)$$

where we defined 4 values such as $c(T) = -y(T)$, the compressibility $\kappa(T) = 1/B$, $\kappa_0 = \kappa(T,P_0)$ and $V_0 = V_m(T,P_0)$. These parameters are fitted using the data obtained by the QHA phonon calculations ($V_m(T)$ at $P = 0$, $V_m(P)$ at $T = 0$, $B(T)$ at $T = 0$ and thermal expansion at $P = 0$). After integration, the Gibbs energy derived from this equation of state (Equation (9)) can be written as:

$$^{comp}\Delta G = \int_P^{P_0} V_m\mathrm{d}P = \frac{c(T)}{\kappa_0}\left[\exp\left(\frac{V_m(T,P) - V_0}{c(T)}\right) - 1\right] \qquad (10)$$

The equilibrium is calculated by a Gibbs energy minimization for each pressure and temperature using the Thermo-Calc software [319].

## Crystal and electronic structures

We defined α, β and γ polymorphic phases of magnesium dihydride as α−TiO₂, β−FeS₂ and γ−PbO₂, respectively, as described in Table 4. The δ and ε phases reported in other works have not been calculated [320,321]. The rutile α−TiO₂ is known to be stable under ambient condition. The β−FeS₂ form (sometimes called β−PdF₂) is derived from the fluorite CaF₂-type by a displacement of the hydrogen atoms from the tetrahedral sites of a *fcc* host structure to triangular interstitial positions along the [111] direction. Indeed, H is located in the triangular position in all polymorphic phases of MgH₂, with a small deviation from the centre in γ−PbO₂.

It is possible to describe the structures as a network of Mg-centred H-octahedra. Depending on the polymorphic form, the arrangement and density of the corner-linked octahedra is slightly different as shown in Figure 25.

*Table 4. Crystallographic description and calculated heat of formation of the 3 polymorphic forms of MgH₂.*

| Prototype | Space group | Pearson symbol | $\Delta H_{for}$ (kJ/mol-fu) | $\Delta ZPE$ | $\Delta H_{for}^{corrected}$ (kJ/mol-fu) |
|-----------|-------------|----------------|------------------------------|--------------|------------------------------------------|
| TiO₂ | *P4₂/mnm* (136) | *tP6* | -52.1 | 9.8 | -42.3 |
| PbO₂ | *Pbcn* (60) | *oP12* | -52.0 | 9.8 | -42.1 |
| FeS₂ | *Pa*-3 (205) | *cP12* | -43.1 | 9.4 | -33.7 |

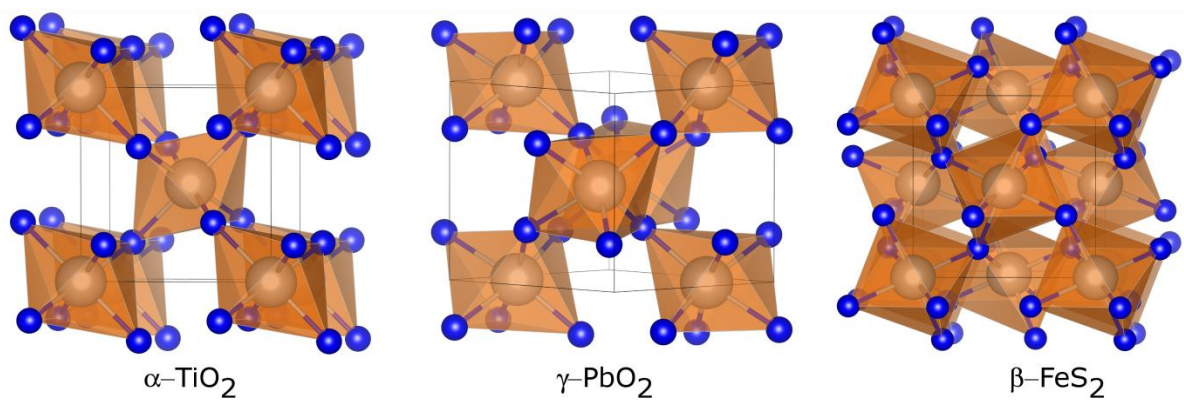

α−TiO₂          γ−PbO₂          β−FeS₂

*Figure 25. Crystal structure of the 3 polymorphic forms of MgH₂: α−TiO₂, γ−PbO₂ and β−FeS₂. Mg and H atoms are in grey and blue respectively.*

The electronic structure of MgH₂ has been published in length [243,320,322], with a calculated band gap underestimated with classical GGA around 3.7 and 4.2 eV for the α and γ−forms. Only calculations using PAW in GW approximation (Green's Function) are able to predict indirect band gap values of 5.5 and 5.2 eV respectively [322], in agreement with experiments [323]. The β−form presents a smaller gap between 2.7 and 3.9 eV depending on the choice of the XC functional.

The GGA Bader calculations give almost equivalent charge transfer of 0.8 e- to the hydrogen atom, and, as discussed in the review [21], the bonding character is not fully of ionic type. A representation of the ELF surface is given in Figure 26, where the electrons localized around H form a high density area inside a regular triangle for α−TiO₂. The case of γ−PbO₂ is different since H is not exactly in the plane formed by the 3 Mg atoms. Because of the remarkable compactness of the *fcc* host structure in (111) plane, the electron localized areas are denser for the β−FeS₂.

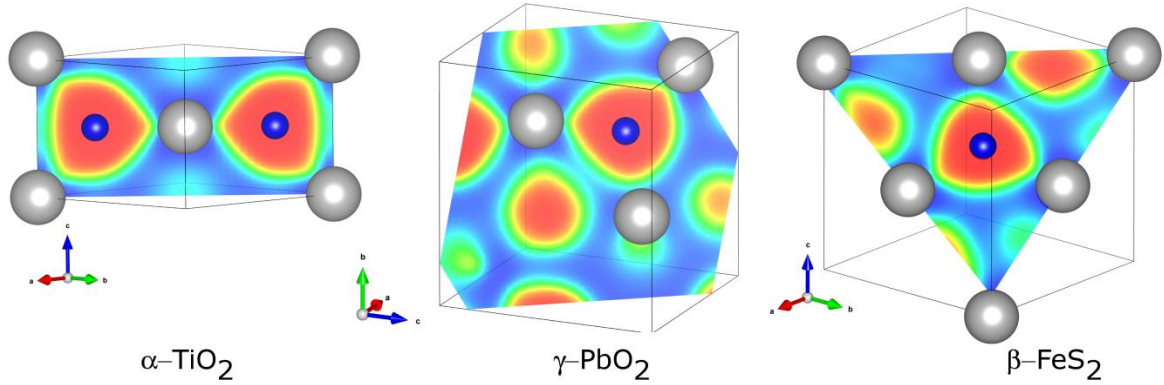

*Figure 26. Electron Localization Function (ELF) of MgH$_2$ in α–TiO$_2$, γ–PbO$_2$ and β–FeS$_2$ forms in the plane containing the triangular H interstitial site. Warm colours indicate a localized valence electron region (high probability), whereas the cold colours shows electron-gas like region (low probability).*

## High pressure modelling

The enthalpies of formation of α–TiO$_2$ and γ–PbO$_2$-types are very similar (less than 0.1 kJ/mol difference). This relates to the close nature of the crystallographic structures and the same octahedral coordination for the metal atom. The calculated cell volumes are also very close as indicated in Figure 27. Experimentally, a density increase of 1.6% is observed at the α→γ transition. The calculations predict a slightly larger increase of 2.7 %.

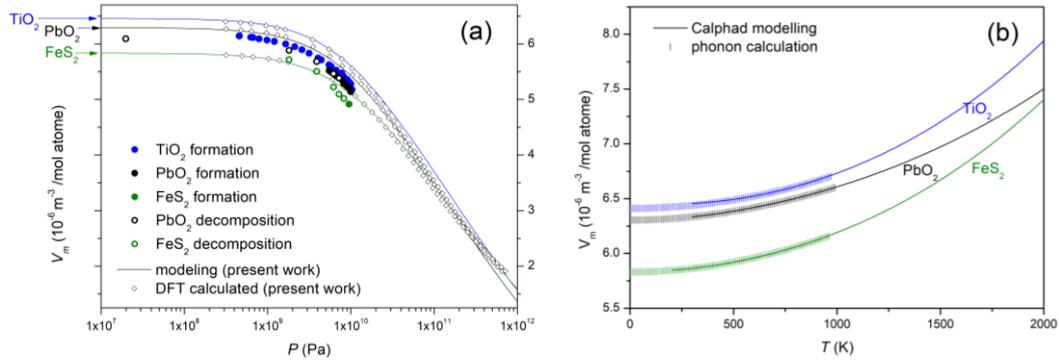

*Figure 27. Cell volume as a function of pressure (a) and temperature (b).*

The difference between enthalpies of formation of PbO$_2$ and FeS$_2$ types is clearly more important, reaching 8.4 kJ/mol according to the DFT calculations corrected with the ZPE contribution. This relates to the very big differences in crystal structures with the latter structure derived from the CaF$_2$ type. The calculations as already reported in Ref. [324] actually predict FeS$_2$ to be more stable than CaF$_2$ by 13.2 kJ/mol, without the ZPE corrections: the ideal CaF$_2$ structure is showing large imaginary frequencies in phonon bands. This shows an energetic and mechanical stabilization of the structure by the displacement of hydrogen atoms from the ideal tetrahedral to triangular positions. The increase of the Mg atom density in the *fcc* structure compared to TiO$_2$ and PbO$_2$ is associated with the decrease of the molar volume as shown in Figure 27. This increase is calculated to be 4.9% at 650 K according to our calculations, in agreement with the calculations of Cui et al. [325], see Table 5.

*The results described above refer to ambient conditions. Figure 27 shows the calculated volume as a function of pressure in excellent agreement with experimental data. A good agreement is obviously observed, too, for the bulk modulus as shown in Table 6. The pressure tends to stabilize the phases with lower cell volumes.*

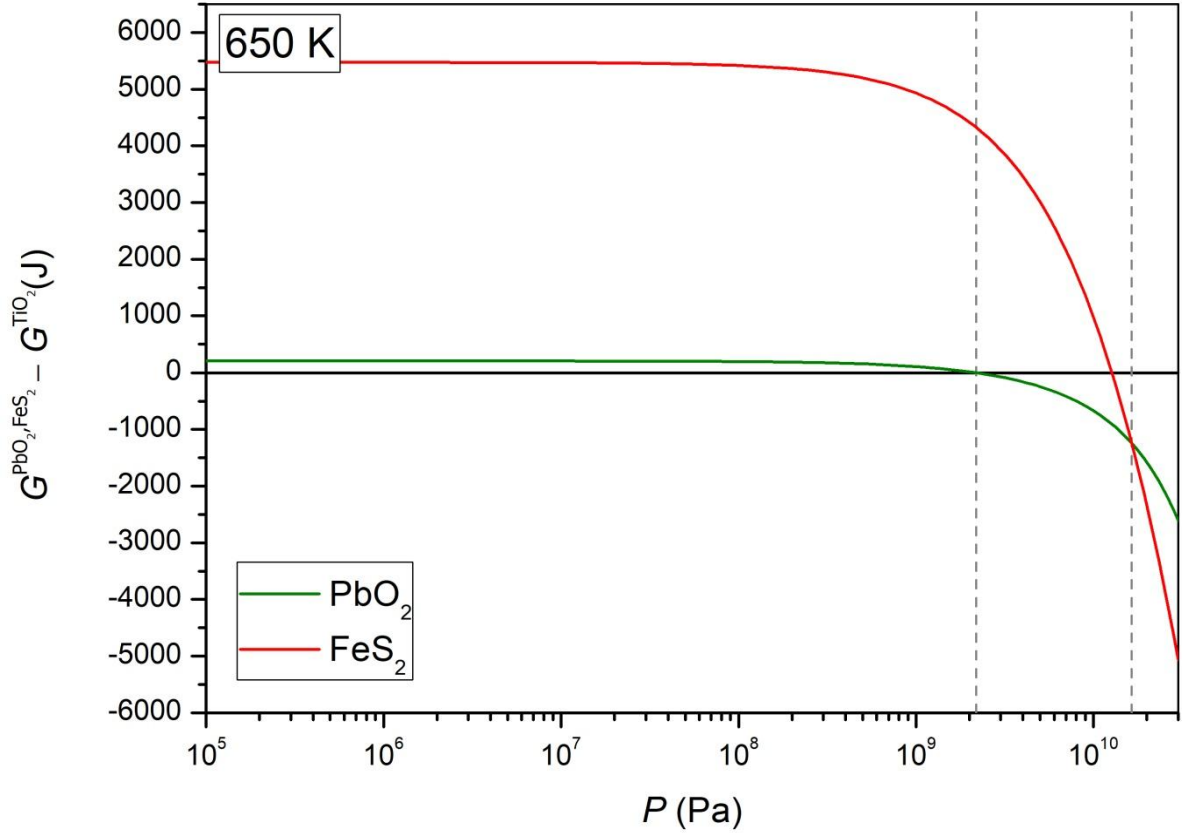

Figure 28 plots the Gibbs energies as a function of pressure and shows two phase transitions at 650 K.

At room temperature, the transitions $TiO_2/PbO_2$ and $PbO_2/FeS_2$ are calculated at pressures of $1.0 \times 10^9$ Pa and $1.5 \times 10^{10}$ Pa. These values, reported in Table 5, are in agreement with the data calculated by Cui et al. [325] and Vajeeston et al. [28].

Table 5. Calculated phase transition pressure at ambient temperature.

| Phases transition | P (Pa) present work | Other calculations |
|---|---|---|
| $TiO_2$ / $PbO_2$ | $2.1\ 10^9$ | $0.39\ 10^9$ [320] |
| | | $1.2\ 10^9$ [325] |
| | | $6.1\ 10^9$ [304] |
| | | $2.4\ 10^9$ [321] |
| $PbO_2$ / $FeS_2$ | $1.6\ 10^{10}$ | $3.9\ 10^9$ [320] |
| | | $9.7\ 10^9$ [325] |
| | | $7.1\ 10^9$ [304] |

Table 6. Bulk modulus at room temperature and pressure calculated from our model and measured experimentally.

| Prototype | B (Pa) present work | B (Pa) experimental | Ref |
|---|---|---|---|
| $TiO_2$ | $4.3 \cdot 10^{10}$ | $4.3 \pm 0.2 \cdot 10^{10}$ | [325] |
| | | $4.9 \cdot 10^{10} (MgD_2)$ | [326] |
| $PbO_2$ | $4.3 \cdot 10^{10}$ | $4.4 \pm 0.2 \cdot 10^{10}$ | [325] |
| $FeS_2$ | $4.7 \cdot 10^{10}$ | $4.7 \pm 0.4 \cdot 10^{10}$ | [325] |

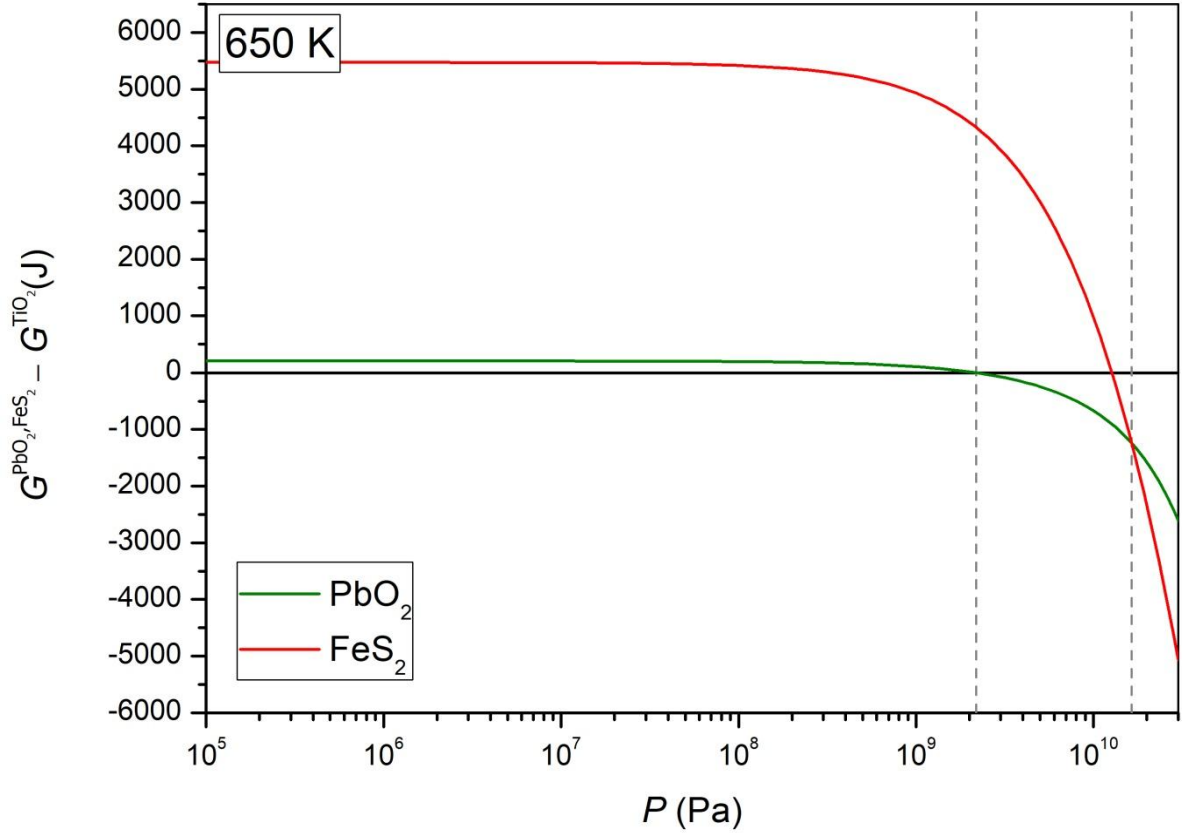

*Figure 28. Gibbs energy as a function of pressure at 650 K. The phase transitions are shown with vertical dotted lines.*

Finally, the complete *P-T* phase diagram could be obtained from our approach. It is presented in Figure 29 in comparison with experimental data and calculations by Moser et al. [304] and AlMatrouk et al. [321] who employed a similar approach and the quasi-harmonic approximations. While a good agreement between Moser and our model is observed for the $PbO_2$-$FeS_2$ transition, the results differ considerably for the $TiO_2$-$PbO_2$ transition at high temperatures. This latter transition is however well described by AlMatrouk et al. who did not predict a transition to the $FeS_2$ type because they took into account the δ and ε phases, which we did not. Figure 29 also compares the calculations with the available experimental data at room temperature (shown as bars) and at high temperature including the experimental data from the present work (see previous Section). The presence of $PbO_2$ at 450 K and $4.3 \times 10^9$ Pa [327] is in agreement with the three models. However, our calculations are in a closer agreement with the experimental value $P \approx 1.5$ GPa for the $TiO_2$/$PbO_2$ equilibrium at $T = 973$ K (open green right-triangle in Figure 29).

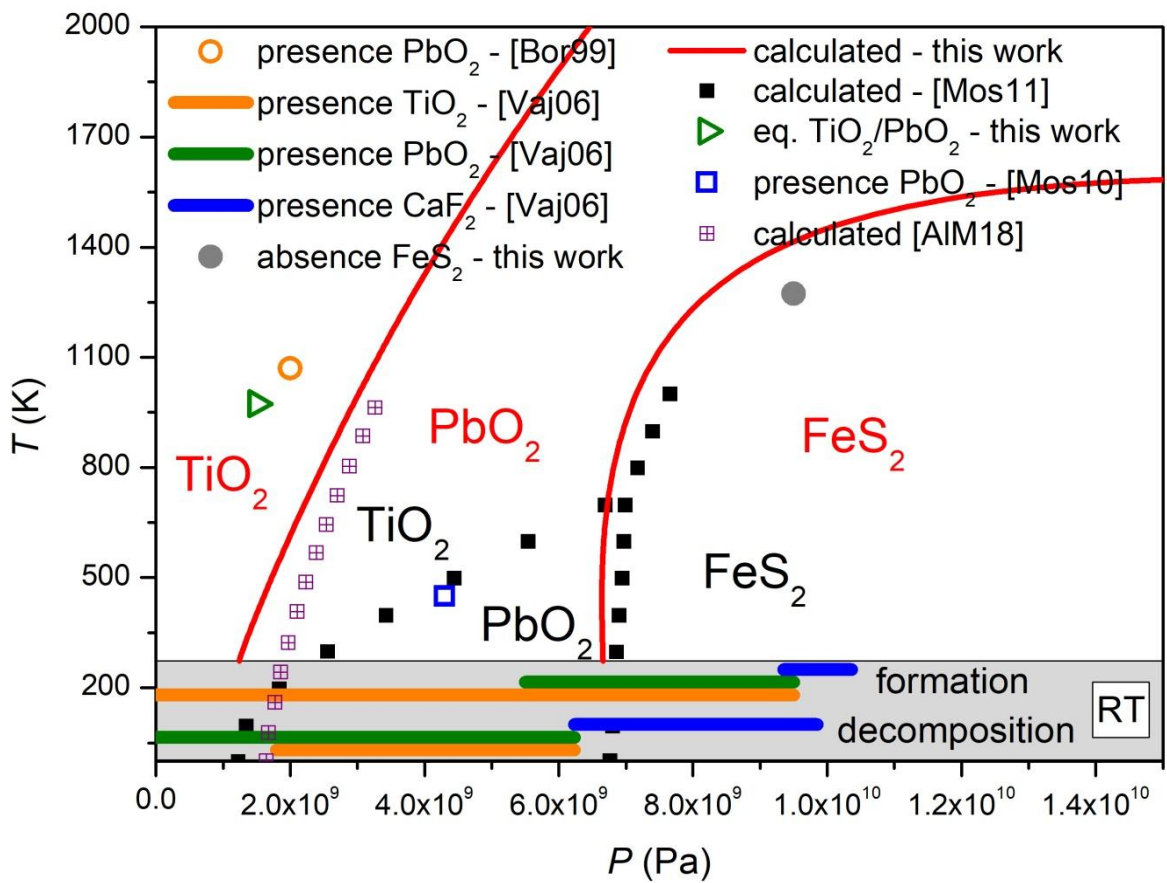

*Figure 29. P-T phase diagram of MgH₂ according to model (continuous lines), compared with literature data (Bor99 [303], Vaj06 [28], Mos10 [327], Mos11 [304], AIM18 [321]).*

$FeS_2$ has not been evidenced in our high pressure/high temperature experiments although our calculation predicts it to be stable (however, close to the limit of the studied pressure interval). Work is still in progress to identify the reasons for this discrepancy. In particular, anharmonic contributions could be responsible for a slight modification of the phase diagram.

## MIXED TRANSITION-METAL COMPLEX Mg-BASED HYDRIDES

In Mg-based transition-metal (TM) complex hydrides, the TM is covalently bonded to hydride ligands, $H^-$, forming a complex anion $[TMH_x]^{\delta-}$ that obeys the 18-electron rule [328-330]. $Mg^{2+}$ stabilizes the anion by electron transfer. $Mg_2NiH_4$, $Mg_2CoH_5$ and $Mg_2FeH_6$ have been investigated extensively. One question was whether different TM complex anions can coexist in the same crystal structure?

Samples of mixed TM complex Mg-based hydrides have been synthesized from the elemental powder mixtures by HRBM at $P(H_2) = 50$ bar using a Fritsch P6 planetary ball mill using the Evico Magnetics high-pressure vial [234]. Neutron diffraction has been used for structural characterization, and a number of samples was then prepared in a deuterium atmosphere.

A mixture with nominal composition $Mg_2Fe_{0.5}Co_{0.5}$ has been prepared by reactive ball milling at 50 bar $D_2$ atmosphere [242]. During milling the temperature steeply increased from room temperature to about 320 K, and consequently the pressure increased at the very beginning, but then after some time, decreased because of hydrogen absorption. The formation of the hydride proceeds in two steps, with the highest hydrogen absorption rate for the first step. It has been proposed that $MgH_2$ is an intermediate reaction product associated with the first reaction step, followed by the slower reaction with the formation of the complex hydride [241]. However, in [242] it is also suggested that a direct formation of the complex hydride from the elements is possible via H-absorption at the Mg/Fe-Co interfaces.

From powder neutron and X-ray diffraction studies it was found that $Mg_2(FeH_6)_{0.5}(CoH_5)_{0.5}$ takes the $K_2PtCl_6$-type structure (space group *Fm-3m*) with $a = 6.426$ Å, and is consequently similar to $Mg_2FeH_6$ with $[FeH_6]^{4-}$ anions exhibiting octahedral geometry and the high-temperature modification of $Mg_2CoH_5$ with disordered distribution of H atoms Co atoms in a square-pyramidal $[CoH_5]^{4-}$ arrangement [242]. This results in coexistence of $[FeH_6]^{4-}$ and $[CoH_5]^{4-}$ ions in the formed complex hydride (Figure 30). The presence of both types of ions in $Mg_2(FeH_6)_{0.5}(CoH_5)_{0.5}$ has been also confirmed by IR spectroscopy [331]. Synchrotron powder X-ray diffraction (SR-PXD) showed that hydrogen is desorbed in one step at temperatures between 500 and 600 K with Mg and a FeCo solid solution as desorption products. The thermal stability is similar for $Mg_2FeH_6$, $Mg_2CoH_5$ and the mixed complex hydride. DFT calculations and inelastic neutron scattering have been performed in order to better understand the properties of $Mg_2(FeH_6)_{0.5}(CoH_5)_{0.5}$ [332,333].

Reactive milling of Mg, Co and Ni gives the $Mg_2Ni_{0.5}Co_{0.5}H_{4.4}$ hydride [243]. This compound takes a tetragonal *P4/nmm* structure that is isostructural with $Mg_2CoH_5$ at room temperature.

Recently obtained data on the formation and hydrogen desorption properties of $Mg_2Fe_xCo_{1-x}H_y$ shows that the cubic $K_2PtCl_6$ structure-type is formed for samples with increased amount of Fe, but for smaller x the data indicates a tetragonal distortion. Recent results show that $Mg_2(FeH_6)_{0.5}(CoH_5)_{0.5}$ is partly reversible at 30 bar $H_2$ [Barale J, Deledda S, Dematteis EM, Sørby MH, Baricco M, Hauback BC, to be submitted].

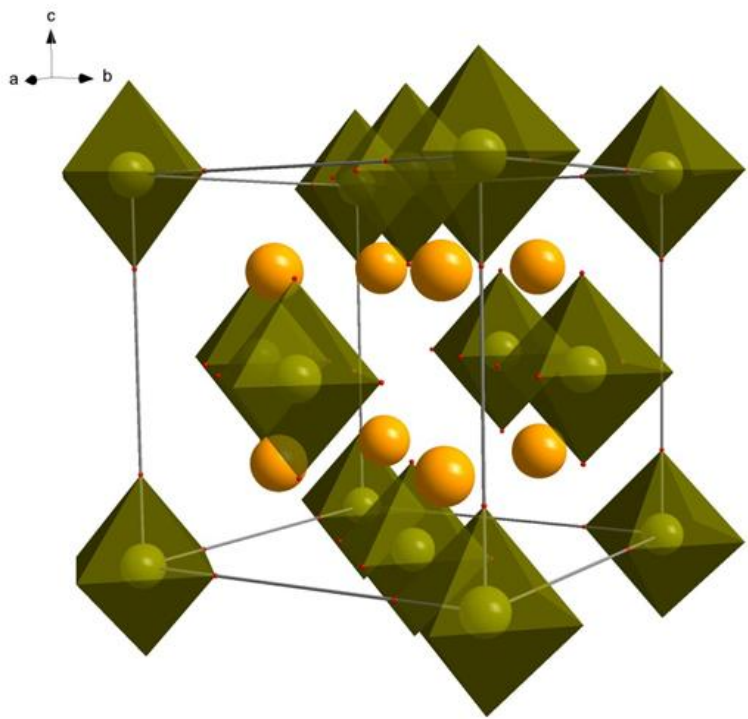

*Figure 30. The crystal structure of $Mg_2(FeH_6)_{0.5}(CoH_5)_{0.5}$ determined from Rietveld refinements of powder neutron and X-ray diffraction data. Mg yellow, Fe/Co green and H/D atoms - red dots in corners of the octahedra (D used for neutron diffraction).*

# NON DIRECT THERMAL DESORPTION METHODS

In most research work, to overcome the thermodynamic and kinetics barriers, thermal energy is used to drive the absorption/desorption hydrogenation processes in magnesium. The use of non-direct thermal energy sources is starting to be considered to open new pathways to destabilize magnesium hydride. Albeit a systematic study of all possible non-direct thermal heating methods has not been performed yet, the following non-direct thermal energy sources have been attempted:

    (i)        electrochemical sources,

    (ii)      electromagnetic (microwave and light driven) and ultrasonic radiation and

    (iii)     mechanical bias.

Electrochemical charging/discharging of hydrogen in metals is the most common alternative to the direct gas-solid hydrogenation driven by heat. This method is widely used in Ni-metal hydrides batteries [334] and its utilization in magnesium and alkaline metals.

Using electromagnetic waves to replace the heat source has been performed by using different ranges in the spectrum i.e. microwave, visible and ultraviolet radiation. Concerning microwave radiation, some works [335,336] show the effectiveness of this radiation in decomposition on different types of hydrides such as metal and complex hydrides (LiH, NaH, LiBH$_4$) using microwave irradiation at 2.45 GHz and 500 W for 30-60min. In the particular case of MgH$_2$, only a small amount of hydrogen (<1%) is released [335] because of the non-metallic character of the hydride as well as low microwave penetration depth compared to metallic hydrides such as TiH$_2$, the latter exhibits rapid heating (T $\sim 600$ K) due to the conductive losses. Improvements have been obtained by the inclusion of MgH$_2$ particles into metallic supports such as Ni acting as heating media [336].

A similar approach has been recently proposed in [337] but using visible radiation. In this case, the plasmonic effect is used to promote light to heat conversion allowing the decomposition of different hydrides with visible light. To this aim, Au-nanoparticles were added to MgH$_2$ nanoparticles and exposed under illumination at the Au-resonance frequency. Desorption and absorption process are enhanced due to the local increase of temperature ($\sim 100$ K) by plasmonic effect and partial decomposition of MgH$_2$ is observed. Further improvements are needed to extend this effect for larger particles.

Investigations of the effect of UV-radiation on MgH$_2$ were performed by [338,339]. Unlike lower energy radiation, no light to thermal conversion is observed. Those works show the formation of MgH$_2$ by applying UV-light on matrix-isolated Mg atoms under H$_2$ atmosphere at low T (< 30K). Decomposition of magnesium hydride was also investigated using UV-light (0.3 W/cm$^2$) onto MgH$_2$ powder at RT detecting a very small amount of hydrogen released (<0.1 %) because the effect is essentially circumscribed to the surface (creation of excited sites at the surface which energy transfer). A more extensive investigation is needed to overcome those problems.

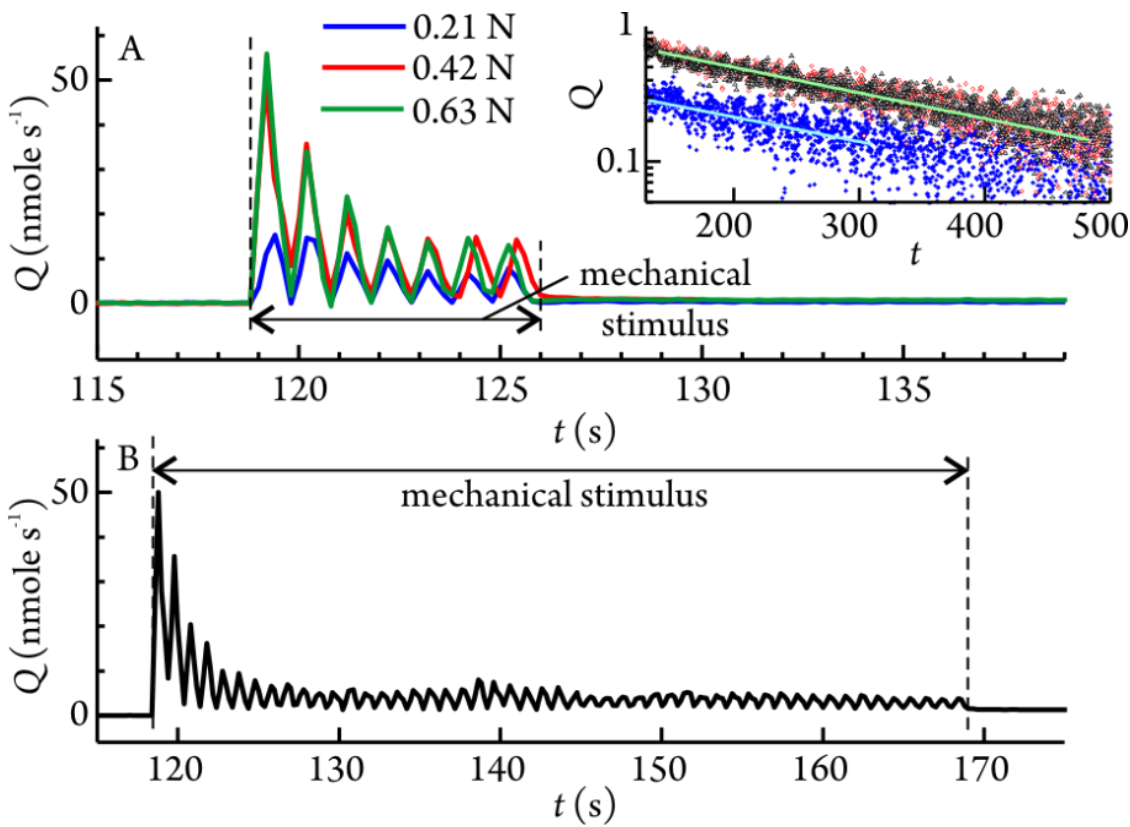

*Figure 31. MSGE from MgH₂. (A) Emission rate (nanomol H₂/s.) under different loads (0.21 N, 0.42 N, and 0.63 N). The inset shows slow desorption decay after the end of deformation. Lines are linear fit of the experimental data. (B) Evolution of the emission rate (H₂ eq.) for 50 sliding cycles. Normal load 0.42 N [341].*

Finally, mechanical bias is a technique that has been habitually applied to drive different reactions [340] i.e. tribochemistry. In particular, it was recently reported that $H_2$ desorption was associated with tribochemical decomposition under mechanical micro-deformation of $MgH_2$ [341]. The phenomenon was characterized in situ and in real time during deformation of $MgH_2$ on the micrometric scale using a novel technique of Mechanical Stimulated Gas Emission (MSGE) spectrometry [341]. Results shows that $MgH_2$ decomposition occurs by an instantaneous (t ~ s) non-thermal H-release at RT during mechanical treatment (Figure 31A and B). This process must provide a huge increase in the driving force to explain the release of hydrogen at such low temperatures, resembling thermal decomposition mediated by catalyst. Although some explanations have been offered to describe this non-thermal decomposition (triboelectric effects, hydrogen directly released in the molecular state), a full concept of the tribodesorption mechanism is challenging and is still under development.

In summary, a wide variety of non-direct thermal desorption methods are starting to be investigated. All of them provide different formation/decomposition paths than usually tested by thermal methods and are worthy of investigation in the quest to overcome the well-known drawbacks for the use of Mg in hydrogen storage applications.

## CYCLING Mg AT ELEVATED TEMPERATURES

Although the behaviour of Mg under elevated temperatures is well documented, the true nature of the dominant elementary processes occurring during cycling are not fully understood [342-344]. It is possible that the relatively low melting point of Mg, 650 $^{o}$C, is an important contributing factor to the complex behaviour of Mg as sintering between individual particles has been observed to occur at elevated temperatures [345,346]. Past investigations have noted the occurrence of sintering as a barrier to its utilisation as a thermal storage medium as it leads to the formation of an expanding structure that may place additional stresses on the walls of reactors and impede the material kinetics [347,348], e.g., in Figure 32 a Mg structure formed during cycling through sintering has continued to grow out through 0.5 mm diameter holes in the Cu gaskets and expand up the 1/4" gas transfer connection. Alternatively, it was speculated that a porous structure of this kind could potentially deliver a more stable thermal storage medium with enhanced kinetic and thermal properties. Consequently steps were taken to 1) optimise an activation process through which the highly porous structure could be reliably fabricated, 2) test the structure's resilience to various cycling conditions and 3) characterise the structure's thermal performance.

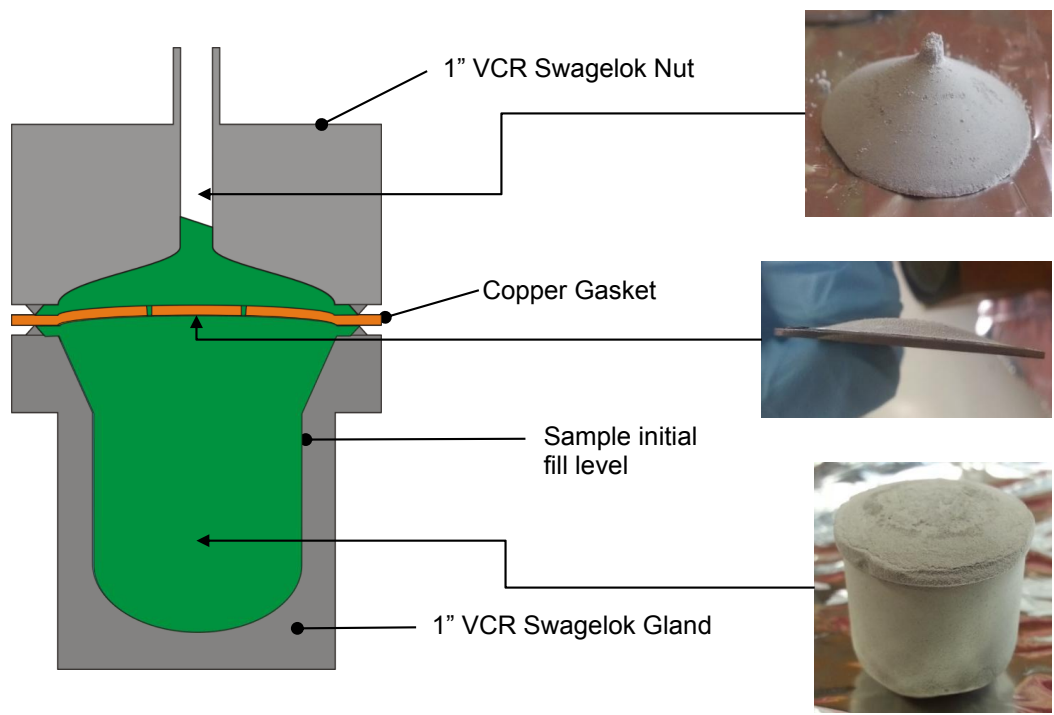

*Figure 32. Following 150 rapid cycles at 400 °C the Mg powder has sintered into porous structure growing through 0.5 mm diameter holes in a Cu gasket and expanding up the ¼" gas transfer tube.*

## Optimisation of the activation process

Initial fabrication of a continuous porous structure was achieved by 280 successive 30 min hydrogenations and dehydrogenations of a 0.2 g sample of ball milled $MgH_2$ powder. The result of this rapid cycling process is shown in Figure 33 (a), where it can be seen that significant sintering has occurred and there is no loose powder remaining. Micrographs of a cross-section taken from the cylindrical sample reveal that there appears to be a high degree of porosity resulting in a continuous sponge like structure, Figure 33 (b).  Following a process of optimisation, it was found that, at a temperature of 400 °C and starting pressures of 40 bar and 0.1 bar, the number of successive cycles required to form the porous structure could be reduced to ca. 50.

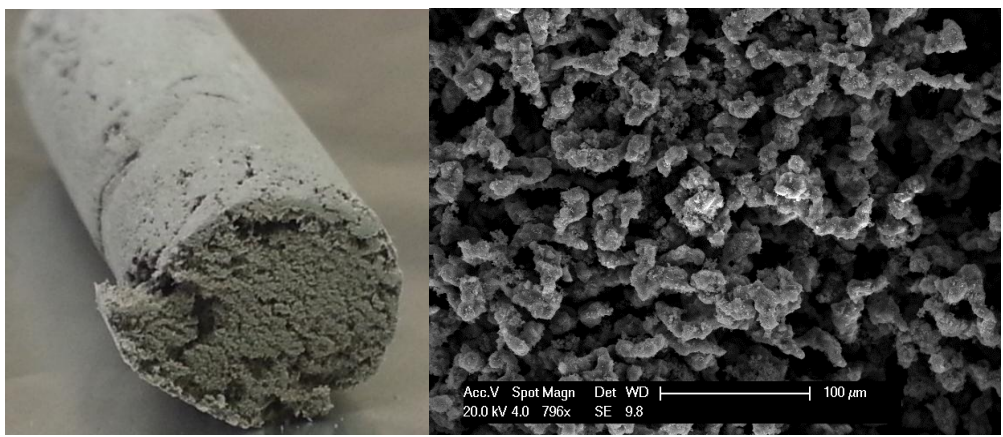

*Figure 33. a) Mg Powder activated and then fabricated into the continuous porous structure. b) A micrograph of a cross-section of the cylindrical shaped sample shows that the porous structure at the surface continues through the sintered sample.*

## Testing of the Storage Materials Resilience – Exposure to Air

In order to investigate the effects of exposure to air, a sample of the fabricated porous structure was passivated in air for several weeks. Following re-activation at 400 °C and 40 bar $H_2$, the maximum capacity of the sample after three hours of cycling was found to be 7.4 wt% $H_2$, which is comparable to pre-passivation capacity of 7.5 wt% H.

## Characterisation of the Structures Thermal Performance

The thermal conductivity under ambient conditions in air was measured for both ball milled powder and the fabricated porous structure as 0.1 and 0.6 $Wm^{-1}K^{-1}$ respectively. This improvement in thermal conductivity is due to the presence of the continuous structure formed by the sintering process, which significantly reduces the number of boundaries that heat encounters whilst passing through the material. The lower value of thermal conductivity obtained for the ball milled powder is due to the fact that it comprises numerous individual particles which leads to a higher degree of contact resistance and a lower rate of heat transfer.

In conclusion, a method by which a continuous porous storage medium can be fabricated from inexpensive Mg powder without additional costly processing has been discovered and optimised. Through experimentation the continuous porous structure has been demonstrated to have a degree of handleability (can be exposed to air and the capacity does not degrade when stored in the metallic state for prolonged periods at elevated temperatures), and increased thermal conductivity (increased from ca. 0.1 to ca. 0.6 W $m^{-1}K^{-1}$). In the context of large scale thermal energy storage the improvements in handle-ability and thermal performance in comparison to other MH powders are valuable characteristics.

# MAGNESIUM COMPOUNDS FOR THERMAL ENERGY STORAGE

Magnesium based hydrides have been under consideration as high temperature metal hydrides for energy storage applications since at least 1967 when Brookhaven National Laboratory suggested the use of Mg, Mg-5%Ni, or $Mg_2Ni$ in automotive applications [349]. The heat required to release hydrogen from these compounds is sequestered from the exhaust gas of the vehicle. Since this time, the use of these materials has been proposed for a number of applications including hydrogen storage, waste heat recovery and thermal batteries for concentrating solar thermal power (CSP) stations [20,349-352]. For application as thermal battery for CSP applications, the metal hydride must have a thermal efficiency greater than that of the molten salts currently employed in CSP plants. Currently molten salts are limited to a maximum operating temperature of 565 °C, at which point they decompose [350]. In addition, this technology uses sensible heat storage, which is 153 kJ $kg^{-1}$ $K^{-1}$ [353]. Current targets for implementing metal hydrides as thermal batteries would require these thermo-chemical storage materials to operate at temperatures of > 600 °C.

In the late 80's, Bogdanović et al. began investigating $MgH_2$ and Ni-doped Mg to determine their application as heat energy storage systems [352,354]. In 1995 a process steam generator was subsequently built containing 14.5 kg of $MgH_2$ + 1 - 2 wt% Ni and cycled 1000 times [345]. Since then other Mg-based hydrides have been investigated as heat storage materials including $Mg_2NiH_4$, $Mg_2CoH_5$ and $Mg_2FeH_6$ of which have been shown to cycle in excess of 1000 times between a temperature range of 250 and 350 °C for $Mg_2NiH_4$ and up to 550 °C for the Fe and Co analogues [352,355,356].

Recently, the $Na_2Mg_2NiH_6$ hydride related to $Mg_2NiH_4$ has been studied as a thermal energy storage material with 10 cycles being conducted between 315 and 395 °C without any loss in capacity [357]. This material decomposes in two steps (Reactions (11),(12)) with only the first step being of use due to the evaporation of Na if NaH were to be decomposed. As a result, the practical reaction enthalpy $\Delta H_{des}$ of this material is 83 kJ/mol. $H_2$, which is higher as compared to $Mg_2NiH_4$ (64 kJ/mol. $H_2$) and $Mg_2FeH_6$ (77 kJ/mol. $H_2$) [31,358,328]. The main drawback towards technological application of $Na_2Mg_2NiH_6$ (and $Mg_2NiH_4$) is the cost of the Ni constituting component and having to inhibit the decomposition of NaH by strict control of temperature and pressure during absorption and desorption of hydrogen [357]. This material has a theoretical operating temperature range of 318 - 568 °C (1 - 150 bar $H_2$) with a thermal storage capacity of 1042 kJ/kg.

$$Na_2Mg_2NiH_6 \longrightarrow Mg_2NiH_x + 2NaH + (2 - x)H_2 \ (x < 0.3, > 2.15 \ wt\% \ H_2) \qquad (11)$$

$$2NaH \longrightarrow 2Na + H_2 \quad (1.27 \ wt\% \ H_2) \qquad (12)$$

Besides the aforementioned materials, there is also a number of other Mg-based complex transition metal hydrides containing $[NiH_4]^{4-}$, $[CoH_5]^{4-}$ and $[FeH_6]^{4-}$ anions including $LaMg_2NiH_7$, $Na_2Mg_2FeH_8$ and $Ca_4Mg_4Co_3H_{19}$ [328]. To date, the decomposition temperature of most of these compounds has been determined although only a few have had their thermodynamic properties measured. These include $LaMg_2NiH_7$ ($\Delta H_{des}$ = 94 kJ/mol $H_2$), $YbMgNiH_4$ ($\Delta H_{des}$ = 111 kJ/mol $H_2$), $CaMgNiH_4$ ($\Delta H_{des}$ = 129 kJ/mol $H_2$), $Na_2Mg_2FeH_8$ ($\Delta H_{des}$ = 95 kJ/mol $H_2$), $Ca_4Mg_4Fe_3H_{22}$ ($\Delta H_{des}$ = 122 kJ/mol $H_2$) and $Ca_4Mg_4Fe_3H_{22}$ ($\Delta H_{des}$ = 137 kJ/mol $H_2$) [328]. There have also been a variety of theoretical studies carried out on this class of complex hydrides to determine their thermodynamic properties [359-361]. The large $\Delta H_{des}$ of these compounds would potentially allow them to be ideal thermal energy storage materials with operating temperatures well above the 565 °C maximum of molten salts currently employed in CSP plants [350]. Important requirements for metal hydrides in CSP plants are low costs of the materials, long-term cycle stability, a high reaction enthalpy and good heat conductivity for solar heat in- and output. Many of the thermodynamic properties of these compounds are still yet to be determined although cost of raw materials and the potential for multiple decomposition steps may hinder their technological application. A cheaper magnesium-based complex hydride is $Mg(BH_4)_2$, that releases hydrogen near 300 °C [362-365] However, re-hydrogenation requires extreme pressure [366,367] or much lower temperatures so that it does not fully decompose to $MgB_2$ [368,369]. As such, the re-hydrogenation pressure should be decreased and kinetics of $Mg(BH_4)_2$ must be improved if it is to be considered for technological applications.

Despite sintering at above 450 °C [62], $MgH_2$ ($\Delta H_{des}$ = −74 kJ/mol) [136] has continued to be of interest as a thermal energy storage material with a host of theoretical modelling studies being compiled towards the optimisation of containment vessels and conditions [370]. As discussed earlier in the paper, sintering results in the grain and particle growth, decreasing surface area and restricting the pathway for hydrogen to unreacted Mg. It is possible to reduce the effect of sintering by including a small quantity (a few wt%) of an impurity phase, such as Ni, $Nb_2O_5$ or $TiB_2$ [354,371], which act to separate Mg grains from one another and restrict sintering. $MgH_2$ mixed with exfoliated natural graphite (ENG) has been trialled as a thermal energy storage (TES) material in two laboratory prototype TES systems towards integration into CSP systems [129,372]. Super-critical $H_2O$ was employed as the heat transfer fluid which would bring heat to the system from the solar concentrators and also remove the heat from the system at low periods of solar irradiation in order to produce electricity [350]. A volumetric hydrogen source was used to store the hydrogen during the day cycle, whereas optimally and if cost effective, a low temperature hydride such as $NaAlH_4$ would be used on a larger scale due to the greater volumetric hydrogen storage density of the metal hydride compared to high pressure gas storage [350,373-376]. These prototypes have shown the feasibility of using hydrogen in a large scale CSP plant although thermal management was shown to be difficult to manage on a small scale system (40 g of material) primarily due to radiative heat loss [129,372].

In contrast to $MgH_2$ the ternary hydride $Mg_2FeH_6$ shows no sintering process also at temperatures up to 550 °C. One reason for this unusual behaviour might be a complete phase separation of Mg- and Fe-metal after the decomposition (13).

$$Mg_2FeH_6 \rightleftharpoons 2Mg + Fe + 3H_2 \quad (5.5 \ wt\% \ H_2) \qquad (13)$$

Both metals are completely immiscible and produce no alloys at these temperatures. During the hydrogen uptake and release a complete reconstruction of the complex structure is carried out which prevents particle growth. This results in superior stable cycling properties also after 1000 cycles of re- and dehydrogenation.

The macroscopic description of the formation of this ternary hydride was discovered with an electron microscopy study [377]. $MgH_2$ is formed initially with enhanced kinetics, because of the presence of Fe. This is followed by nucleation of $Mg_2FeH_6$ between $MgH_2$ and Fe, which grows with columnar morphology. As the $Mg_2FeH_6$ columns grew, both the capping Fe-particles and the $MgH_2$ are consumed. Even though this describes the production of $Mg_2FeH_6$ on a macroscopic scale, the processes on an atomic scale producing an octahedral $[FeH_6]^{4-}$ complex anion are still unclear.

Recently, the use of $Mg_2FeH_6$ for heat storage at temperatures around 500 °C was demonstrated [355,356]. Heat transfer for the heat storage (dehydrogenation) and heat release process (hydrogenation) was done with molten salts. Thermal oils cannot be used at these high temperatures because of decomposition processes. The heat storage unit was performed using a tube bundle reactor with 13 tubes for the heat storage material. Molten salt flows around the tubes and supplies heat during the decomposition (heat storage) or takes up heat during the hydrogenation (heat release). The heat storage material was easily prepared by mixing Mg- and Fe-metal powder in a 2:1 stoichiometric ratio. Practically, the material can absorb up to 5 wt% of hydrogen with an equilibrium pressure of 30 - 77 bar in a temperature range of 450 - 510 °C. An overall amount of 5 kg of $Mg_2FeH_6$ was used as the heat storage material. The storable heat is 2.7 kWh if a gravimetric hydrogen storage density of 5 wt% $H_2$ can be reached. In the original experiments only 1.5 to 1.6 kWh could be stored, because of heat losses and the non-optimal flow of the heat transfer fluid. The overall performance of the system should be optimized in a forthcoming project.

$NaMgH_3$ is another promising Mg-based material for high temperature thermal energy storage applications due to its $\Delta H_{des}$ of 87 kJ/mol $H_2$ [378]. Its thermodynamic properties allow for an impressive theoretical operating temperature range of 382 - 683 °C (1 - 150 bar $H_2$), with a theoretical thermal storage capacity of 1721 kJ/kg. The decomposition of this material occurs over 2 steps (Reactions (14),(15)) and as such the first step is only useful due to the evaporation of Na if the second step were to occur [379].

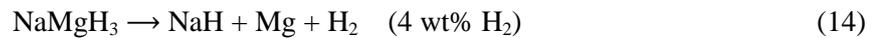
$$NaMgH_3 \longrightarrow NaH + Mg + H_2 \quad (4 \text{ wt\% } H_2) \quad\quad\quad (14)$$

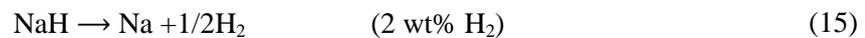
$$NaH \longrightarrow Na + 1/2 H_2 \quad\quad (2 \text{ wt\% } H_2) \quad\quad\quad (15)$$

Recently, partial substitution of fluorine for hydrogen has been investigated to stabilise metal hydride compounds [373,380-381]. This is in contrast to using $F^-$ based compounds as an additive to destabilise metal hydride compounds [381-382]. Due to the comparable ionic size of the hydride and fluoride ions and the structural similarity of their compounds new metal hydrides based on known fluoride structures and hydride – fluoride solid solution systems have been explored [379,380,383,384]. The thermodynamics and cycling capabilities of $NaMgH_2F$ have shown a $\Delta H_{des} =$ 97 kJ/mol $H_2$, a significant increase when compared to pure $NaMgH_3$ (87 kJ/mol $H_2$) providing a theoretical thermal storage capacity of 1416 kJ/kg [373]. The decomposition pathway of this material is interesting, as below 478 °C a two-step pathway is observed, whereas a one-step pathway occurs above this temperature. Over 10 hydrogenation cycles of this material at 500 °C (1 bar desorption, 45 bar absorption), there was a marked decrease in hydrogen capacity from 2.6 wt% $H_2$ to ~0.8 wt% $H_2$ [376]. This is due to the formation of some Na and Mg rather than $NaMgF_3$ as would be expected.

The concept of producing hydride-fluoride solid solutions has also been extended to $MgH_2$ with the formation of $Mg(H_{x-1}F_x)_2$ (x = 0.95, 0.85, 0.7, 0.5) compounds [294]. Ball milling and annealing produces single phase solid-solutions and are shown to be thermodynamically stabilised by increasing F content, a result that is in accord with the analogous Na-H-F system [379]. In contrast, previous studies where $MgF_2$ has been ball-milled and used as an additive have shown a destabilisation and increased kinetics [385]. The thermodynamics of the $Mg(H_{0.85}F_{0.15})_2$ system was determined by PCI measurements to be 73.6 ± 0.2 kJ/mol $H_2$ and an entropy of 131.2 ± 0.2 J/K.mol $H_2$ [294]. In comparison with $MgH_2$, these values are decreased from 74.06 kJ/mol $H_2$ (enthalpy) and an entropy of 133.4 J/K.mol $H_2$ [136]. The decrease in entropy is the key factor in the increased stability of the system. Cycling of this system has shown that this material can operate at ~80 °C higher than bulk $MgH_2$.

## Other Mg-based systems

Porous Mg scaffolds have been synthesised to simultaneously act both as a confining framework and a reactive destabilizing agent for infiltrated metal hydrides [386]. The scaffolds were synthesised by sintering a pellet of $NaMgH_3$ under dynamic vacuum. The pores were created by the removal of $H_2$ and Na vapour from the body of the pellet during sintering. The majority of the pores are in the range of macropores (> 50 nm) with only a small number of mesopores (2 – 50 nm) (Figure 34). $LiBH_4$ was melt-infiltrated into the scaffold allowing for the formation of a small quantity of $MgH_2$. Temperature Programmed Desorption (TPD) experiments, showed a $H_2$ desorption onset temperature ($T_{des}$) at 100°C, which is 250 °C lower than bulk $LiBH_4$ and 330 °C lower than the bulk $2LiBH_4/MgH_2$ composite. LiH that was formed during the decomposition of the $LiBH_4$ was fully decomposed at 550 °C. These novel Mg scaffolds formed from sintered $NaMgH_3$ are a promising reactive containment vessel for metal hydrides for stationary or mobile applications.

The potential in finding new multicomponent Mg containing intermetallic hydrides is far from being exhausted. One example is thermodynamically stable ternary $LaMgPdH_5$ hydride formed at high, 66.7 at % content of La+Mg [387] with high H content of 1.67 at.H/Me. Absorption-desorption isotherms, measured at 400 – 600 °C, show two distinct single plateau regions. Estimated enthalpy of hydride formation are around −130 kJ mol $H_2^{-1}$, significantly smaller than the corresponding value of about −208 kJ mol $H_2^{-1}$ for $LaH_2$ formation. Vacuum Thermal Desorption Spectroscopy (TDS) of $LaMgPdH_5$ shows presence of two major H release events with the peaks at 233 and at 510 °C (see Figure 35), both located at significantly lower temperatures than the desorption peaks of $MgH_2$ (300 °C) and $LaH_2$ (800 °C) thus clearly showing a destabilization effect of Pd.

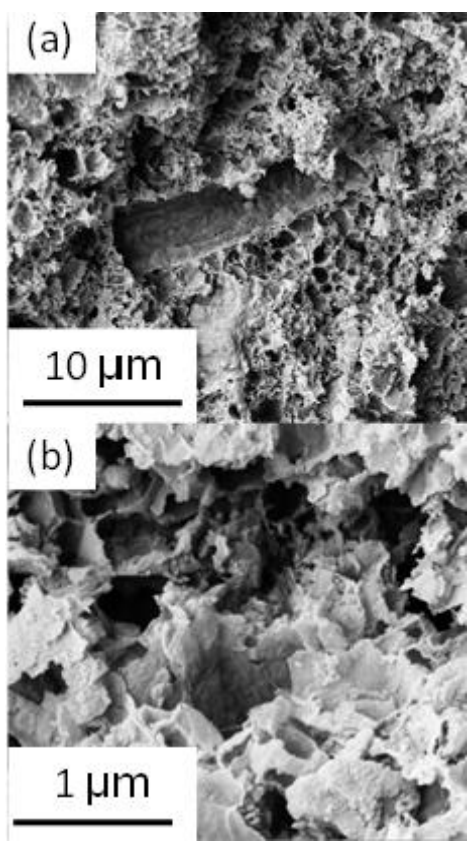

*Figure 34. SEM micrographs of the as-prepared porous Mg scaffold with two different magnifications (a) low and (b) high magnification.*

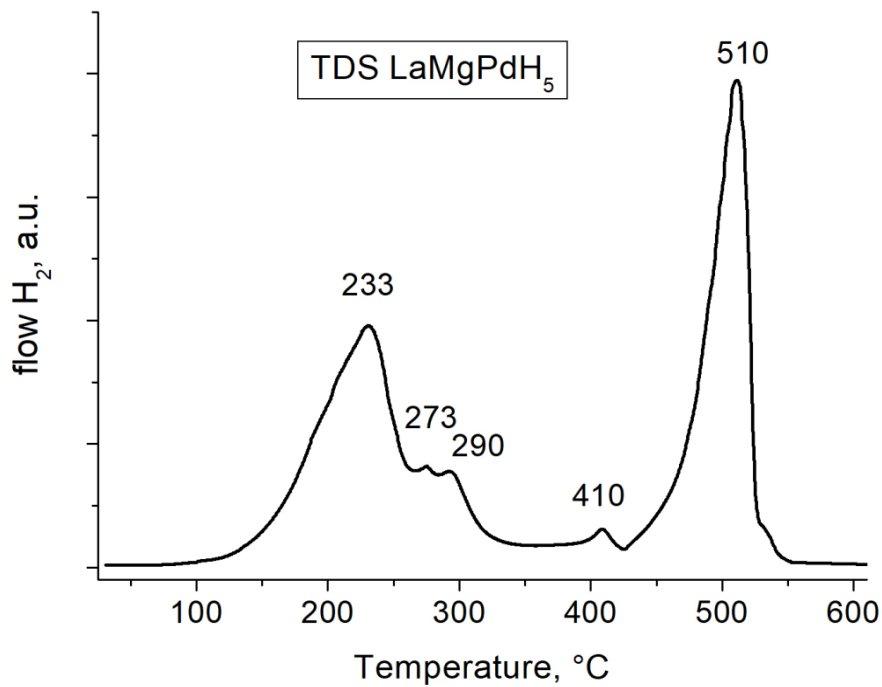

*Figure 35. Vacuum Thermal Desorption Spectra of hydrogen from LaMgPdH₅ measured at a heating rate of 2 °C/min*

## SYSTEM DEVELOPMENT

Magnesium based materials, as has been mentioned earlier, present an ideal combination of high capacity, straightforward reaction pathways and fast kinetics, provided the synthesis and the heat transfer necessary for their use are adequately carried out. It is this combination, as well as the long time over which the materials have been known and studied, that makes them excellent candidates for use in hydrogen or, in general, energy storage systems.

In the past decades, several systems using Mg or its alloys/compounds have been developed. Importantly, they were technically very successful. The fact that they were not widely accepted for usage in their chosen fields was due more to economic factors than any deficiencies that could be said to be attached to the technology of these systems.

As early as in the nineteen seventies, Mg was tried out as one of the hydrogen storage materials in a binary system to supply hydrogen to a vehicle [388]. In this case, a room temperature hydride was used to supply hydrogen to an internal combustion engine. However, the exhaust gases were not hot enough to achieve desorption in the magnesium hydride at lower power settings. This changed for the full-power operation, where the magnesium hydride provided the hydrogen to operate the vehicle. This concept was shown to work in practice; however, the lowering of oil prices during the late seventies and early eighties seemingly led the company in question to abandon development of hydrogen-powered vehicles as a viable alternative to petrol- and diesel-powered ones. Only years later, with the resurgence of fuel cells, was hydrogen as a power source for vehicles considered in earnest again.

Despite intensive activity in the development of metal hydride hydrogen storage systems, most of them use "low-temperature" $AB_5$ and $AB_2$ intermetallic hydrides [389] with much lower storage capacities than $MgH_2$. Only a few publications considered development of H storage systems where "high-temperature" Mg-based H storage materials were used. The reason for that is the high operating temperatures ($\geq$300 °C) which, together with high amount of heat required for $H_2$ desorption from $MgH_2$, cause a high energy consumption for the operation of Mg-based hydrogen storage units. In addition, the high operating temperatures pose certain limitations on the material of containment (usually, only steels can withstand so high temperatures) that, in turn, results in the decrease of gravimetric hydrogen storage density at the system level [390], as well as increased cost.

Nevertheless, magnesium hydride was the candidate of choice for a number of other applications, mostly related to "medium"-temperature (300–450 °C) heat management including thermal energy storage (see previous chapter). A solar power station with thermochemical Mg / MgH$_2$ energy store was developed by Groll et al in 1994 [391]. In 1995, Bogdanović et al. designed and built a process steam generator based on MgH$_2$ [345]. Both units used the Mg / MgH$_2$ H storage material: Ni-doped Mg hydrogenated in THF in the presence of MgCl$_2$ as a co-catalyst; the material's H storage capacity reached up to 7 wt% [80].

The schematic layout of the MgH$_2$ tank for the steam generator reported in [345] is presented in Figure 36 (left). The unit (19.4 dm$^3$ in the volume, empty weight 26 kg, MH load 14.5 kg, ~1 kg H / 10 kWh hydrogen / heat storage capacity) had the weight hydrogen storage efficiency (the ratio of weight H storage capacity of the material to the one for the system as a whole) of ~35%, which is lower than the typical values for H storage units using heavier "low-temperature" MH (50–60% [392]). The operation cycle of the unit coupled with a "low-temperature" MH container included H$_2$ charge and discharge at $P$=20–40 bar and $T$=350–450 °C during ~2 hours. The unit successfully operated for 1.5 years, and then it was opened for inspection, see Figure 36 (right). As it can be seen, during operation, the H storage material had sintered forming a ductile, porous, chalk-like mass. Although it was reported that the sintering had no influence on the H storage capacity of MH bed and its reaction kinetics, the practical usage of Mg-based H storage materials can nevertheless be challenging, as it requires careful temperature control to avoid overheating. On the other hand, as it was noted in [345], the sintering improved heat transfer performance of the MH bed.

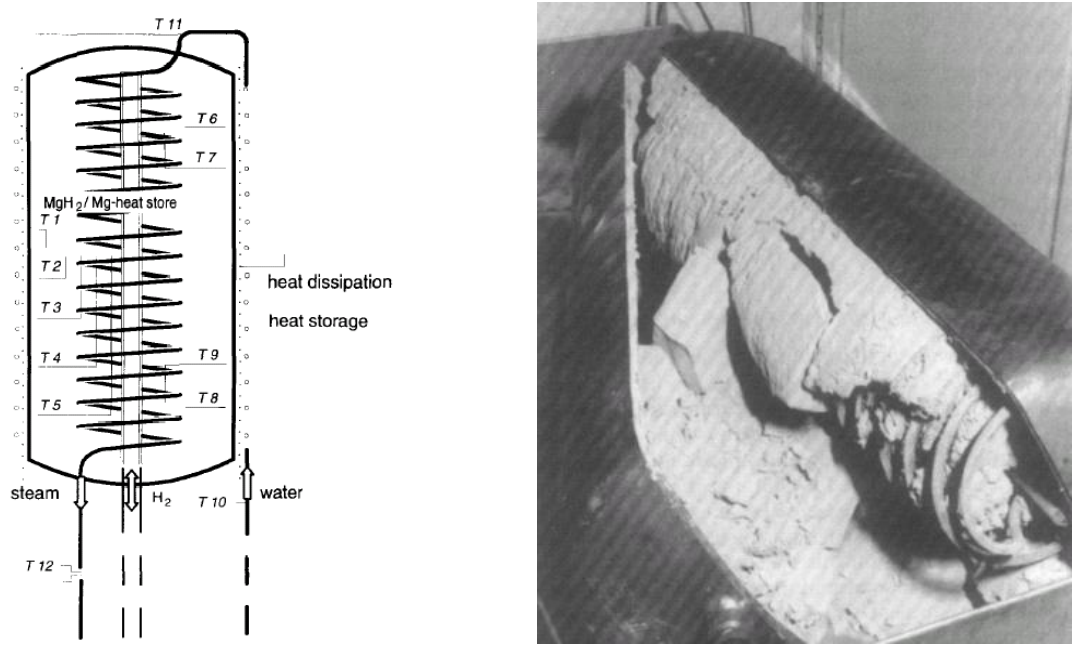

Figure 36. Left: layout of MgH$_2$ tank for a process steam generator. Right: The tank opened after 1.5 years of operation [345].

Considering the examples presented above, we can conclude that magnesium hydride and its related compounds seem to be promising for heat storage and heat management applications. Recently, the authors (ML, VAY) presented a concept of combined cooling, heating and power system utilising solar power and based on reversible solid oxide fuel cell (R-SOFC) and metal hydrides [393]. The work considered utilisation of waste heat losses (T~600 °C) released during the operation of R-SOFC using three types of metal hydride based system components: (i) MH hydrogen and heat storage system (MHHS) on the basis of magnesium hydride; (ii) MH hydrogen compressor (MHHC) on the basis of AB$_5$-type hydrogen storage intermetallic compound; and (iii) compressor-driven MH heat pump (MHHP) on the basis of multi-component $AB_2$-type hydrogen storage alloy. The MH-assisted production of the useful heating and cooling leads to an improvement of 36% in round-trip energy efficiency as compared to that of a stand-alone R-SOFC.

The quest for a magnesium hydride-based system to be used in vehicular (and other) applications is, however, still ongoing. One creative approach in this area is presented in [348,394,395], with a tank based on magnesium hydride operating together with a starter tank based on room temperature hydrides. Later, a similar system was proposed where a lithium amide and magnesium hydride were used in the main storage system instead of $MgH_2$ alone. In the latter case, the realized double-hydride concept allows the operation of a high-temperature hydride starting from room temperature [396,397].

Large scale hydrogen storage systems based on the use of magnesium hydride have been developed by the company McPhy and have a potential to be commercially deployed as offering a mature operation [398]. In these systems, large diameter tank tubes are filled with the H storage material assembled as a stack of a mechanically stable (on cycling) pellets used instead of the powder as in most of the metal hydride tanks. The scientific foundation for this approach was laid at CNRS [399-403].

The MH material used in the McPhy's hydrogen storage tanks (storage capacity 8 kg $H_2$ in the standard configuration [398]) is a composite containing $MgH_2$ powder ball-milled with 4 at% of Ti–V–Cr alloy and further compacted with ~5 wt% of expanded natural graphite (ENG) at a pressure ~ 200 MPa [399]. Apart from an increase in the filling density of the MH material in the containment resulting in an increase of hydrogen storage efficiency and improvement of effective thermal conductivity of the MH bed, the $MgH_2$–ENG compacts are not flammable when exposed to open air (see Figure 37) thus providing exceptional safety during assembling the MH tanks and eliminating a possibility of the accidents (rupture of the containment) during their operation.

A complementary feature of the McPhy's hydrogen storage tanks based on $MgH_2$ is in the storage of the heat of hydrogenation / dehydrogenation reaction by using a phase-change material (PCM) based on Mg–Zn eutectic alloy as a heat transfer medium [403]. The use of latent heat of solidification of the PCM allows mitigation of the problem of too high energy consumption for $H_2$ release from $MgH_2$ (more than 30% of the lower heating value of hydrogen).

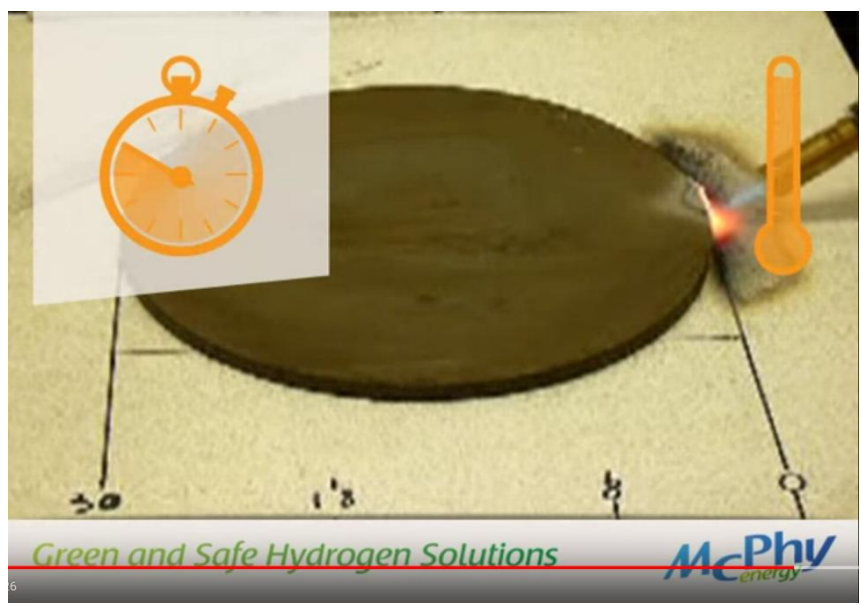

*Figure 37. Fire testing of McPhy's $MgH_2$ composite. No explosion or fire propagation observed during one minute long test. The video is available at https://www.youtube.com/watch?v=RohMt2-UKQI*

In conclusion, the large variety of Mg-based materials that have been explored as thermal energy storage materials indicate that sintering of the Mg is unavoidable at high temperatures. The use of additives including $TiB_2$, which act as particle reducing agents, are effective but do not eradicate the problem [371], although this problem has not appeared to have inhibited $Mg_2FeH_6$, which has been shown to operate effectively at ~550 °C for at least 1000 cycles [352]. This generally limits Mg-based compounds to operate at a maximum operating temperature of ~450 °C in order to prolong the life time of these compounds. In terms of industrial application as thermal energy storage materials, Mg-based compounds are likely to find a market in the medium temperature range (230 – 650 °C) [404]. This would provide application potential for thermally regenerative electrochemical systems, industrial waste heat recovery or preheating applications.

While a great deal of research has been published on Mg-based compounds, many materials have yet to be characterised, such as complexes containing $[NiH_4]^{4-}$, $[CoH_5]^{4-}$ and $[FeH_6]^{4-}$ anions [328,359]. For the few that have been identified by practical and theoretical studies, their cycling and thermodynamic properties are still largely unknown.

The story told by the developments described above shows that, while magnesium-based materials have a long and technically successful history in the field of hydrogen (and energy) storage, it has been also a challenging one, especially when seen from the perspective of a fruitful commercialization into the products required by the consumers. It seems that in the future, these materials will play an increasingly important role, first in applications like heat storage, but hopefully also in transportation.

## SUMMARY, OUTLOOK AND FUTURE PROSPECTS

The present review article summarises the work done in the recent years by the participants of the IEA Task 32 "Hydrogen Based Energy Storage" and their collaborating institutions. 19 research units from Australia, China, Denmark, France, Germany, Japan, Italy, Israel, Netherlands, Norway, Russia, South Africa, Spain and United Kingdom shared their expertise and results in the topic Magnesium hydride based materials for hydrogen based energy storage. Significant if not all the activities are based on collaborations between two and more research groups on a particular topic. This resulted in a high impact of the group on the work in the area of solid state hydrogen storage in total.

The present paper covers a broad variety of topics of fundamental and applied studies of Mg-H based systems, covering fundamental properties of $MgH_2$-based systems, historical overview of the activities, and a review of the ongoing experimental and theoretical studies. The latter include nanostructured $MgH_2$, kinetics, thermodynamics and catalysis of the Mg based hydrides, mechanochemistry, reactive ball milling, metallurgy, composite materials, further to the novel experimental characterisation techniques, novel proccesion techniques to influence and tailor properties of the hydride systems and theoretical and experimental studies of Mg-H system at high pressures. Finally, applied oriented properties, including cycling of hydrogen charge and discharge, storage of thermal energy are considered and reviewed aimed at large scale applications of $MgH_2$.

The research field remains very dynamic with prospects of wide-scale future implementation subject to developments in key areas. These key areas guide current and future research towards solving specific issues.

The main problem of applied oriented work on magnesium hydride remains in addressing three main problems;

(a)     How to decrease thermodynamic stability of $MgH_2$;

(b)     How to achieve fast kinetics of hydrogen absorption and desorption;

(c)     How to extend the cycle life and life time to build and utilise efficient systems for hydrogen and thermal storage.

Various complementary approaches can be used and the proposals and expectations of their use are summarized below.

Fundamenal studies: thermodynamics, kinetics, properties under special conditions (high pressures, when influenced by interfaces):

Future activities may be focused on the determination of the free energy difference $\Delta\gamma$ for the interfaces between Mg-based nanoparticles and porous scaffolds or carbon-based materials, on the modelling of the interface entropy $\Delta s$, and on the experimental / theoretical identification of other phases, beyond $TiH_2$, able to maximize $\Delta\gamma$ and $\Delta h$. In addition, there is a need for synthetic methodologies able to enhance the $A/V$ ratio while keeping a high Mg weight fraction in the material.

Thus, mechano-chemical treatment offers a wide range of different reaction conditions, which facilitate different types of chemical reactions and nanostructuring. The future developments within this field are expected to provide new experimental options for materials treatment. This will likely lead to new materials with a range interesting properties

Mechanical treatment is a generally accepted strategy for nano structuring hydrogen storage materials. This approach is well established, efficient and can be used under a variety of different conditions. Traditionally, solid-solid reactions were conducted under ambient pressure and temperatures, including frictional heating. Varying ball-to-powder ratio, time and intensity of the mechano-chemical treatment have a range of effects, such as reducing particle size, nanostructuring, e.g. grain-boundary formation. Chemical reactions may take place, which can produce new stoichiometric compounds or solid solutions. The latter is a typical reaction among metals, known as alloying. In all these cases mechano-chemical treatment has a significant influence on hydrogen release and uptake properties.

Reactive mechano-chemistry is conducted in a gas atmosphere with elevated pressure or a using a liquid phase. Under such conditions other types of reactions may take place, e.g. solid-gas or solid-liquid reactions. Hydrogen gas is an efficient one-pot synthesis method to obtain catalyzed magnesium hydride in nanostructured state. The screening of catalytic additives for improving sorption properties of $MgH_2$ remains of vivid interest for future research. All sort of additives can be attempted by this technique. Moreover, this method can be extended to the synthesis of Mg-containing hydride materials such as novel complex hydrides (e.g. $Li_3MgN_2H_x$ [250]) and reactive hydride composites.

Catalytic additives can affect the hydrogenation properties of Mg and Mg-based alloys indirectly, by changing the locations of hydride nucleation, and the anisotropy of hydride growth. Thus, understanding the microstructure evolution of the two-phase $Mg-MgH_2$ alloy in the course of hydrogen absorption/desorption is a key to improving the hydrogenation kinetics and thermal conductivity of the alloy. Improving the thermodynamic properties of $MgH_2$ turned out to be much more difficult task than improving the hydrogenation kinetics of Mg. Most attempts to increase the plateau pressure of the $Mg - MgH_2$ equilibrium by nanostructuring and strain engineering did not bear practical fruit. In this respect the most promising approach seems to be hydride destabilization by formation of intermetallic phases and solid solutions [219]. Since most of the binary and even ternary Mg-based alloys have already been tried, the key here is in multicomponent alloying of Mg, in the hope to cause a synergetic effect on parameters of two-phase metal-hydride equilibrium. In this respect, employing the CALPHAD and DFT approaches for modelling the effect of alloying on the thermodynamics of the system, and identifying the most promising combinations of alloying additives for further experimental studies is very promising.

While the effect of various additives or catalysts on $MgH_2$ has been profound, dramatically increasing the rates of hydrogen absorption and desorption, far less is known about the mechanisms by which the additives work. As described above, $Nb_2O_5$, one of the best additives, is believed to form a ternary Mg-Nb oxide which facilitates transport of hydrogen. But this mechanism does not work for the equally effective results of Titanium Isopropoxide [278], nor does it explain the kinetic enhancement of MgO additive [120] which is otherwise expected to reduce hydrogen diffusion. A better understanding of the mechanisms by which these kinetic enhancing additives work might guide the development of better materials. It has also been found that some additives are more effective for desorption than absorption and vice versa, suggesting that combinations of additives may be advantageous [16].

So far as high-pressure modifications of magnesium dihydride are concerned, the synthesis of massive single-phase samples of $\gamma$-$MgH_2$ achieved in the present work opens a good perspective of studying its lattice dynamics by inelastic neutron scattering. A few years ago [405], an INS investigation of the low-pressure α-MgH$_2$ phase made it possible for the first time to experimentally construct its density $g(E)$ of phonon states and further use it to accurately calculate the temperature dependence of the heat capacity $C_P$, which nearly coincided with that determined earlier [406] in a wide temperature interval from 300 to 2000 K. A similar INS investigation of γ-MgH$_2$ would give its $g(E)$ and $C_P(T)$. The dependences $C_P(T)$ for the α and γ dihydrides would give temperature dependences of their Gibbs free energies, and the balance between these energies would determine the line of the α↔γ equilibrium in the $T$–$P$ diagram of MgH$_2$. If the work is a success, this will be the first equilibrium line ever determined for two modifications of any substance on the basis of the experimental spectra of phonon density of states.

From the computational point of view, more theoretical developments are still needed, e.g. phonon calculations using softer functionals also including the anharmonic contributions.

Applications:

Cyclic hydriding and dehydriding at varing pressure, temperature and time can expand magnesium to form a porous 3-D structure confined in the space of the vessel. This offers potential to enable good thermal contact between the hydride bed and the vessel walls and any internal structure. Good thermal conductivity paths are crucial to effective thermal management of Mg hydride beds both for hydrogen stores and thermal stores operating at elevated temperatures typically above 300 ˚C, for example energy scavenging devices of low grade heat from large plant. The 3-D porous structures also have the added benefit of safer handle-ability, such as for maintenance or decommissioning.

Mg$_2$FeH$_6$ is an ideal Mg based material for heat storage applications in the 500 – 600 °C range and future work will be done on this material to optimise it for applications in this temperature range. For thermal energy storage materials to operate at T > 600 °C, it is most likely that materials containing Mg will not be used due to the low vaporisation temperature of Mg, and hence loss of Mg at T > 600 °C. If Mg is used at T > 600 °C encapsulation is required to inhibit Mg vaporisation. It is therefore more likely that for applications requiring T > 600 °C Ca based materials will be used as TES materials.

In the near future, it is to be expected that the increase in use of concentrating solar power will bring with it the need for massive heat storage. Since Mg-based materials are singularly well suited to this, they can be expected to be produced in quantity at low cost for this application. Regarding other uses, the acceptability of high operational temperatures will be key in the implementation of these materials. Where such high temperatures are no obstacle (mainly in industrial processes such as the steel, hydrocarbon and chemical industries, but also other large-scale applications), Mg-based materials could be the materials of choice due mainly to their low prices and excellent availability, among the other qualities mentioned above.

Novel characterisation techniques:

A wide range of well-known characterization techniques (X-ray and neutron diffraction, Raman and visible spectroscopy , scanning and transmision microscopy, thermal gravimetry, differential scanning calorimerry, mass spectrometry, Sieverts systems..etc)  are habitually used by the hydrogen community to determine the structural, compositional as well as thermodynamic and kinetic properties of bulk Mg-based compounds and their respective hydrides. In the next future, those techniques will be still used but they will be frequently complemented by those focused on nanoscale characterization i.e. high resolution microscopies and "in situ" operando techniques that are able to provide direct information about the hydrogenation and dehydrogenation process at moderate temperatures and pressures i.e. μ-Raman, environmental TEM and XPS. The increasing spread of those techniques will hopefully provide a closer understanding of magnesium hydride phenomenology and, therefore, an improvement of their hydrogen related properties.

**CONCLUSIONS**

Here we review an exciting research field of magnesium-based materials, including a great variety of alloys, compounds and composites. This is expected to bring inspiration to new important developments and excitements for the years ahead resulting in important applications. This review contains a broad variety of topics of fundamental and applied studies of magnesium based systems and a review of the fronties of both experimental and theoretical research, including nanostructuring, kinetics, thermodynamics and catalysis of magnesium and the hydrides. Metallurgy, composite materials, advanced experimental characterisation techniques, novel procession techniques to influence and tailor properties towards large scale applications of $MgH_2$ are presented. The review also includes a historical overview of early developments within the research field mainly driven by development of solid state hydrogen storage. Currently, magnesium-based materials attract much attention for storage of concentrated solar heat with much higher energy densities than traditional phase change materials. The thermodynamic dynamic properties and high abundancy provide a potential for development of large scale heat storage systems. In future, novel types of magnesium batteries may replace the currently very successful lithium batteries. These Mg-batteries have a potential in reaching much higher energy densities as compared to current standards and also become safer. This clearly highlights the relevance of continuing research and developments within the area of magnesium based materials.

# ACKNOWLEDGEMENTS

V. Yartys acknowledges a support from Research Council of Norway (project 285146 "IEA Task Energy Storage and Conversion Based on Hydrogen"). V.A. Yartys, R.V. Denys and M.V. Lototskyy acknowledge financial support from EU Horizon 2020 / RISE project "Hydrogen fuelled utility vehicles and their support systems utilising metal hydrides – HYDRIDE4MOBILITY" and EU FP7 ERAfrica program, project RE-037 ''Advanced Hydrogen Energy Systems – HENERGY'' (2014–2018) co-funded by the Research Council of Norway / RCN and the Department of Science and Technology / DST of South Africa).

M.V. Lototskyy acknowledges financial support from DST within Hydrogen South Africa / HySA programme (project KP3-S02), as well as from National Research Foundation / NRF of South Africa, incentive funding grant number 109092.

N. Bourgeois, J.-C. Crivello and J.-M. Joubert acknowledge support from the French GDR CNRS n°3584 TherMatHT and the national program investments for the future ANR-11-LABX-022-01. DFT-QHA calculations were performed using HPC resources from GENCI-CINES (Grant 2017-96175).

C.E. Buckley, T.D. Humphries, M.P. Paskevicius and M.V. Sofianos acknowledge financial support from the Australian Research Council for grants LP120101848, LP150100730, DP150101708, LE0989180 and LE0775551.

J.R. Ares acknowledges financial support from MINECO (Nº.MAT2015-65203R)

The work by V.E. Antonov was supported by the program "The Matter under High Pressure" of the Russian Academy of Sciences.

I. Jacob acknowledges support from Israel Science Foundation Grant 745/15.

## List of the Tables

**Table 1.**

Qualitative analysis of main criteria of solid hydrogen storage families according to DOE 2020 targets for on-board applications. (Color code: Red = deficient ; Yellow = Fair ; Green = Good)

**Table 2.**

Sample compositions and hydrogenation conditions for hydrogen absorption curves presented in Figure 3

**Table 3.**

Process parameters of hydrogenation of Mg during its ball milling in $H_2$ (Figure 8)

**Table 4.**

Crystallographic description and calculated heat of formation of the 3 polymorphic forms of $MgH_2$.

**Table 5.**

Calculated phase transition pressure at ambient temperature.

**Table 6.**

Bulk modulus at room temperature and pressure calculated from our model and measured experimentally.

Figure captions in the manuscript

**MAGNESIUM BASED MATERIALS FOR
HYDROGEN BASED ENERGY STORAGE:
PAST, PRESENT AND FUTURE**

by

V.A. Yartys, M.V. Lototskyy, E. Akiba, R. Albert, V.E. Antonov,
J.-R. Ares, M. Baricco, N. Bourgeois, C.E. Buckley, J.M. Bellosta von Colbe, J.-C. Crivello,
F. Cuevas, R.V. Denys, M. Dornheim, M. Felderhoff,
D.M. Grant, B.C. Hauback, T.D. Humphries, I. Jacob, T.R. Jensen,
P.E. de Jongh, J.-M. Joubert, M.A. Kuzovnikov, M. Latroche,
M. Paskevicius, L. Pasquini, L. Popilevsky, V.M. Skripnyuk, E. Rabkin,
M. V. Sofianos, A. Stuart, G. Walker,
Hui Wang, C.J. Webb and Min Zhu

**Figure 1.**

Number of articles published during the last years 2000-2017 having "hydrogen storage" and the "name" of the respective compound in the **title**, **abstract** and **keyword** fields of the publication. Source: Scopus.

**Figure 2.**

Pressure – composition isotherms (H desorption) for systems of $H_2$ gas with: 1 – Mg-based nanocomposite, T=300 °C [41]; 2 – $Mg_2Ni$, T=300 °C [41]; 3 – Mg – Ni – Mm eutectic alloy , T=300 °C [42];  4 – $MmNi_{4.9}Sn_{0.1}$, T=22 °C [41]. Mm is lanthanum rich mischmetal.

**Figure 3.**

Hydrogen absorption by magnesium and magnesium-based alloys / composites. The legend describing compositions and hydrogenation conditions of the samples is presented in Table 2.

**Figure 4.**

Phase equilibria in H – Mg system: Top (a) – pressure – composition isotherms [75]; Bottom: phase diagrams at 1 bar (b) and 250 bar (c) [85].

**Figure 5.**

Influence of different micro-nanostructures and additives on H-desorption activation energy of $MgH_2$ [45,116,117].

**Figure 6.**

Compilation of van 't Hoff plots calculated from $\Delta H^0$ and $\Delta S^0$ data for Mg-based nanomaterials confronted to bulk Mg (curve **a** from [75]). The black dash-dotted line is the low temperature extrapolation of bulk Mg data. The data in the legend denote the corresponding absolute values of $\Delta H^0$ (left, in kJ/mol $H_2$) and $\Delta S^0$ (right, in J/K mol $H_2$). The number of symbols represent how many points were actually measured and the temperatures of the measurements. Empty symbols denote absorption pressures $p_{abs}$, filled symbols equilibrium pressures $p_{eq}$. **b**: 2-7 nm Mg nanocrystallites in LiCl matrix [136]. **c**: < 3 nm Mg NPs in carbon scaffold [134]. **d**: 15 nm Mg NPs by electroless reduction [135]. **e**: MgH$_2$-TiH$_2$ composite NPs, 10-20 nm in diameter (6-30 at.% Ti) [125]. **f**: MgH$_2$-TiH$_2$ ball-milled nanocomposite (30 at.% Ti) [114]. **g**: Mg/Ti/Pd nanodots on silica, diameter 60 nm: here, $p_{abs}$ and $p_{des}$ are also plotted separately using empty and crossed symbols, respectively, to highlight the strong pressure hysteresis; the reported enthalpy-entropy values were calculated from $p_{eq}$ data [137]. **h**: Mg-Ti-H NPs, 12 nm in diameter (30 at.% Ti) [140]. **i**: Magnesium-naphtalocyanine nanocomposite with Mg NPs of about 4 nm supported on TTBNc [138]. **j**: ultra-thin (2 nm) Mg film sandwiched between TiH$_2$ layers [131]. The inset represents a zoomed view of the high-temperature region using the same symbols and units as for the main plot.

## Figure 7.

a) Hydrogenation and b) dehydrogenation of (Ni)-MgH$_2$-graphene nanocomposites at 200 °C, including ball-milled MgH$_2$ (BM MgH$_2$) and ball-milled MgH$_2$ /GR composite (BM MgH$_2$ /GR) for comparison. c) Hydrogenation and d) dehydrogenation of Ni-conMHGH-75 at various temperatures. Hydrogenation was measured under 30 atm hydrogen pressure and dehydrogenation under 0.01 atm.[163] 75wt% 5-6 nm (from TEM) MgH$_2$ on graphene. Mind that even at room temperature there is appreciable hydrogen absorption. The capacity retention was over 98.4% after 30 full cycles.

## Figure 8.

Hydrogenation of Mg during its ball milling in H$_2$.

## Figure 9.

Mechanochemistry of Mg powder under hydrogen and deuterium gas. a) In-situ hydrogen uptake curves as a function of milling time t$_m$, b) In-situ absorption rate (derivative curves of Figure 9a)

## Figure 10.

Mechanochemistry of Mg under hydrogen gas using several transition metals TM as additives. a) In-situ absorption rate with LTM = Fe, Co and Ni (atomic ratio Mg/LTM = 2), b) In-situ absorption rate with ETM = Ti for different titanium contents y.

## Figure 11.

 Schematic of hydriding/dehydriding reaction in MgH$_2$–GNS composite, and hydrogen desorption curves of the sample at 300 °C: (a) MgH$_2$–5GNS-20 h, (b) MgH$_2$–5GNS-15 h, (c) MgH$_2$–5GNS-10 h, (d) MgH$_2$–5GNS-5 h, (e) MgH$_2$–5GNS-1 h, and (f) MgH$_2$–20 h [256].

## Figure 12.

TEM image, dehydrogenation kinetic curves, and reversible H$_2$ absorption (under 3 MPa H$_2$) and desorption (under 0.001 MPa H$_2$) of Ni catalyzed 75 wt% MgH$_2$ [163].

## Figure 13.

Reversible hydrogen storage capacity of HRBM MgH$_2$–TiH$_2$ at T=350 °C. The values in brackets specify the capacity losses throughout the cycling. The insets show elemental maps of Mg in the cycled composites clearly indicating its grain refinement in the graphite-modified material [67].

## Figure 14.

Scanning Transmission Electron Microscopy (STEM) High Angle Annular Dark Field (HAADF) micrographs of MWCNTs segments/carbon nanoparticles (marked by the circles) located in close proximity to each other and forming a "chain" within Mg grains. Arrows point on grain boundary in Mg, indicating that carbon nanoparticles are located inside the Mg grains rather than along the grain boundaries. The sample was prepared by co-milling of Mg powder with 2 wt% MWCNTs in the Pulverisette - 7 planetary micro mill in hexane for 4 h at 800 rpm using the stainless steel balls of 10 mm in diameter. BTP ratio was 20:1 [273].

**Figure 15.**

Backscattered electrons (BSE) scanning electron microscopy micrographs of the pellets hydrogenated to 80-90% of maximum theoretical hydrogen storage capacity (a,b- Mg pellet; c,d- Mg-2wt% MWCNTs, and e,f- Mg-2wt% Fe). The view plane is perpendicular to the compression axis. (a)- Circles mark individual unimpinged isotropic $MgH_2$ nuclei. Arrows point on impinged $MgH_2$ nuclei forming wavy $Mg/MgH_2$ interface. (b) Lower magnification micrograph showing the isolated pockets of unreacted Mg surrounded by the $MgH_2$ phase. (c) Elongated anisotropic $MgH_2$ nuclei (marked by the circle), (d) Micrograph showing the developed Mg network along the sample. (e) Symmetrical $MgH_2$ nucleus formed next to the Fe particle. (f) Increased number of hydride nucleation sites results in smaller size of metallic Mg islands in comparison with the reference Mg pellet [273].

**Figure 16.**

Time-resolved X-ray scattering data for $MgH_2$-Nb heated to 310 °C. (a) Gray-scale contour plot of the X-ray scattering where intensity increases with lighter tones, (b) temperature profile [277].

**Figure 17.**

Schematic for the catalytic mechanism of muti-valence Ti doped $MgH_2$ [279]

**Figure 18.**

Left: TEM images of the microstructure of the partially dehydrogenated $MgH_2$-$CeH_{2.73}$-Ni nanocomposites demonstrate the catalyst effect of $CeH_{2.73}$ and Ni on $MgH_2$ dehydrogenation process. (a) Bright field image and (b) selected area diffraction patterns of $MgH_2$ (zone axis $[01\bar{1}]$). Right: Evolution of the maximum hydrogen sorption capacities versus cycle times of $MgH_2$-$CeH_{2.73}$-Ni composite. [286]

**Figure 19.**

Schematic illustration and TEM image showing the enhanced hydrogen release at the interface of $CeH_{2.73}/CeO_2$ [287].

**Figure 20.**

Hydrogen absorption kinetics of (a) $MgH_2$ + 4 mol.% $TiF_3$; (b) $MgH_2$ + 4 mol.% $TiCl_3$. [295].

**Figure 21.**

Left: Evolution of Ti 2p and F 1s photoelectron lines for the dehydrogenated $MgH_2$ + 4 mol.% $TiF_3$ sample as a function of sputtering time; Right: Evolution of Cl 2s and Cl 2p photoelectron lines for the dehydrogenated $MgH_2$ + 4 mol.% $TiCl_3$ sample as a function of sputtering time. [295]

**Figure 22.**

Calculated driving forces for hydrogen absorption/desorption in Mg as a function of temperature and pressure. Lines connect constant values, as indicated in kJ $mol^{-1}_{H2}$. Thick continuous line corresponds to equilibrium conditions.

**Figure 23.**

Critical radius for various values of P. Continuous line: 1 bar; dotted line: 5 bar; dashed line: 10 bar; dot-dashed line: 20 bar.

**Figure 24.**

Phase diagram of $MgH_2$ near the $\alpha \leftrightarrow \gamma$ equilibrium line. $1 - \alpha$ transformed to $\gamma$; $2 - \alpha$ did not transform to $\gamma$; $3 - \gamma$ transformed to $\alpha$; $4 - \gamma$ did not transform to $\alpha$. The other symbols show literature data (Bastide [26], Bortz [303], Morivaki [29], Vajeeston [28], Moser [304]).

**Figure 25.**

Crystal structure of the 3 polymorphic forms of $MgH_2$: $\alpha - TiO_2$, $\gamma - PbO_2$ and $\beta - FeS_2$. Mg and H atoms are in grey and blue respectively.

**Figure 26.**

Electron Localization Function (ELF) of $MgH_2$ in $\alpha - TiO_2$, $\gamma - PbO_2$ and $\beta - FeS_2$ forms in the plane containing the triangular H interstitial site. Warm colours indicate a localized valence electron region (high probability), whereas the cold colours shows electron-gas like region (low probability).

**Figure 27.**

Cell volume as a function of pressure (a) and temperature (b).

**Figure 28.**

Gibbs energy as a function of pressure at 650 K. The phase transitions are shown with vertical dotted lines.

**Figure 29.**

P-T phase diagram of $MgH_2$ according to model (continuous lines), compared with literature data (Bor99 [303], Vaj06 [28], Mos10 [327], Mos11 [304], AlM18 [321]).

**Figure 30.**

The crystal structure of $Mg_2(FeH_6)_{0.5}(CoH_5)_{0.5}$ determined from Rietveld refinements of powder neutron and X-ray diffraction data. Mg yellow, Fe/Co green and H/D atoms - red dots in corners of the octahedra (D used for neutron diffraction).

**Figure 31.**

MSGE from $MgH_2$. (A) Emission rate (nanomol $H_2$/s.) under different loads (0.21 N, 0.42 N, and 0.63 N). The inset shows slow desorption decay after the end of deformation. Lines are linear fit of the experimental data. (B) Evolution of the emission rate ($H_2$ eq.) for 50 sliding cycles. Normal load 0.42 N [341].

**Figure 32.**

Following 150 rapid cycles at 400 °C the Mg powder has sintered into porous structure growing through 0.5 mm diameter holes in a Cu gasket and expanding up the ¼" gas transfer tube.

**Figure 33.**

a) Mg Powder activated and then fabricated into the continuous porous structure. b) A micrograph of a cross-section of the cylindrical shaped sample shows that the porous structure at the surface continues through the sintered sample.

**Figure 34.**

SEM micrographs of the as-prepared porous Mg scaffold with two different magnifications (a) low and (b) high magnification.

**Figure 35.**

Vacuum Thermal Desorption Spectra of hydrogen from $LaMgPdH_5$ measured at a heating rate of  2 °C/min

**Figure 36.**

Left: layout of $MgH_2$ tank for a process steam generator. Right: The tank opened after 1.5 years of operation [345].

**Figure 37.**

Fire testing of McPhy's $MgH_2$ composite. No explosion or fire propagation observed during one minute long test. The video is available at https://www.youtube.com/watch?v=RohMt2-UKQI

**Table 1.**

Qualitative analysis of main criteria of solid hydrogen storage families according to DOE 2020 targets for on-board applications.
(Color code: Red = deficient ; Yellow = Fair ; Green = Good)

| Compound families | Gravimetric capacity | Volumetric capacity | Minimum and maximum delivery temperature | Absorption / desorption rates | Toxicity , abundancy |
|---|---|---|---|---|---|
| Metallic hydrides ($AB_2$, $AB_5$..) |  |  |  |  |  |
| Magnesium hydride and alloys |  |  |  |  |  |
| Complex hydrides (alanates, borohydrides) |  |  |  |  |  |
| Chemical hydrides (amides, aminoboranes..) |  |  |  |  |  |
| Adsorbent materials (nanocarbon, MOFS) |  |  |  |  |  |

**Table 2.**

Sample compositions and hydrogenation conditions for hydrogen absorption curves presented in Figure 3

| Curve # | Sample composition | | Hydrogenation conditions | | Notes |
|---|---|---|---|---|---|
| | **Components** | **Phases** | **T, °C** | **P, bar** | |
| 1 | Mg | Mg | 410 | 40 | |
| 2 | Mg | Mg | 410 | 40 | $H_2$ with the admixture of $CCl_4$ (~2%) |
| 3 | $Mg_{0.99}In_{0.01}$ | Mg (solid solution) | 270 | 80 | Alloy |
| 4 | Mg 94.11 Zn 4.01 La 1.24 Cd 0.52 Zr 0.12 (wt%) | Mg (solid solution) Traces of intermetallic phases | 340 | 30 | Industrial Mg alloy |
| 5 | Mg 80 $LaNi_5$ 20 (wt%) | $Mg + LaNi_5$ | 345 | 30 | Compacted mixture of powdered Mg and $LaNi_5$ |
| 6 | $REMg_{12}$ (RE = La, Ce) | $REMg_{12}$ (starting alloy and product of vacuum heating at T>450 °C) $MgH_2 + REH_3$ (hydride) $Mg + REH_2$ (dehydrogenated sample) | 325 | 30 | Alloys |
| 7 | $Mg_{75}Y_6Ni_{19}$ | $Mg + Mg_2Ni + YNi_2$ (starting alloy) $MgH_2 + Mg_2NiH_4 + YH_2$ (hydride) | 200 | 30 | |

**Table 3.**

Process parameters of hydrogenation of Mg during its ball milling in $H_2$ (Figure 8)

| Curve # | $H_2$ pressure [bar] | Type of ball mill | Milling parameters | Catalyst | Ref |
|---|---|---|---|---|---|
| 1a | 200 | Low-energy (rotating autoclave) | 150 rpm, BPR~20:1 | None | [44] |
| 1b | | | Heating to 350–400°C | $I_2$ (0.7%) | |
| 2 | 30 | Planetary | 500 rpm, BPR=80:1 | None | [170] |
| 3 | 300 | Planetary | 400 rpm; BPR=50:1 | $TiH_2$ (10 mol%) | [185] |
| 4a | 5 | Planetary | 300 rpm, BPR=60:1 | 5.5–6 wt%Zn, 0.4–0.5 wt%Zr (commercial Mg alloy) | [223] |
| 4b | 5 | | 400 rpm, BPR=60:1 | | |
| 4c | 10 | | 400 rpm, BPR=60:1 | | |
| 4d | 5 | | 400 rpm, BPR=120:1 | | |
| 5a | 80 | Planetary | 400-800 rpm, BPR=60:1 | None | [114] |
| 5b | | | | Ti (30 mol%) | |
| 6 | 20 | Planetary | 500 rpm, BPR=40:1 | Ti (25 mol%) | [67] |

**Table 4.**

Crystallographic description and calculated heat of formation of the 3 polymorphic forms of $MgH_2$.

| Prototype | Space group | Pearson symbol | $\Delta H_{for}$ (kJ/mol-fu) | $\Delta ZPE$ | $\Delta H_{for}^{corrected}$ (kJ/mol-fu) |
|---|---|---|---|---|---|
| $TiO_2$ | $P4_2/mnm$ (136) | $tP6$ | -52.1 | 9.8 | -42.3 |
| $PbO_2$ | $Pbcn$ (60) | $oP12$ | -52.0 | 9.8 | -42.1 |
| $FeS_2$ | $Pa$-3 (205) | $cP12$ | -43.1 | 9.4 | -33.7 |

**Table 5.**

Calculated phase transition pressure at ambient temperature.

| Phases transition | P (Pa) present work | Other calculations |
|---|---|---|
| $TiO_2$ / $PbO_2$ | $2.1 \cdot 10^9$ | $0.39 \cdot 10^9$ [320] |
| | | $1.2 \cdot 10^9$ [325] |
| | | $6.1 \cdot 10^9$ [304] |
| | | $2.4 \cdot 10^9$ [321] |
| $PbO_2$ / $FeS_2$ | $1.6 \cdot 10^{10}$ | $3.9 \cdot 10^9$ [320] |
| | | $9.7 \cdot 10^9$ [325] |
| | | $7.1 \cdot 10^9$ [304] |

**Table 6.**

Bulk modulus at room temperature and pressure calculated from our model and measured experimentally.

| Prototype | B (Pa) present work | B (Pa) experimental | Ref |
|---|---|---|---|
| $TiO_2$ | $4.3 \cdot 10^{10}$ | $4.3 \pm 0.2 \cdot 10^{10}$ | [325] |
| | | $4.9 \cdot 10^{10} (MgD_2)$ | [326] |
| $PbO_2$ | $4.3 \cdot 10^{10}$ | $4.4 \pm 0.2 \cdot 10^{10}$ | [325] |
| $FeS_2$ | $4.7 \cdot 10^{10}$ | $4.7 \pm 0.4 \cdot 10^{10}$ | [325] |

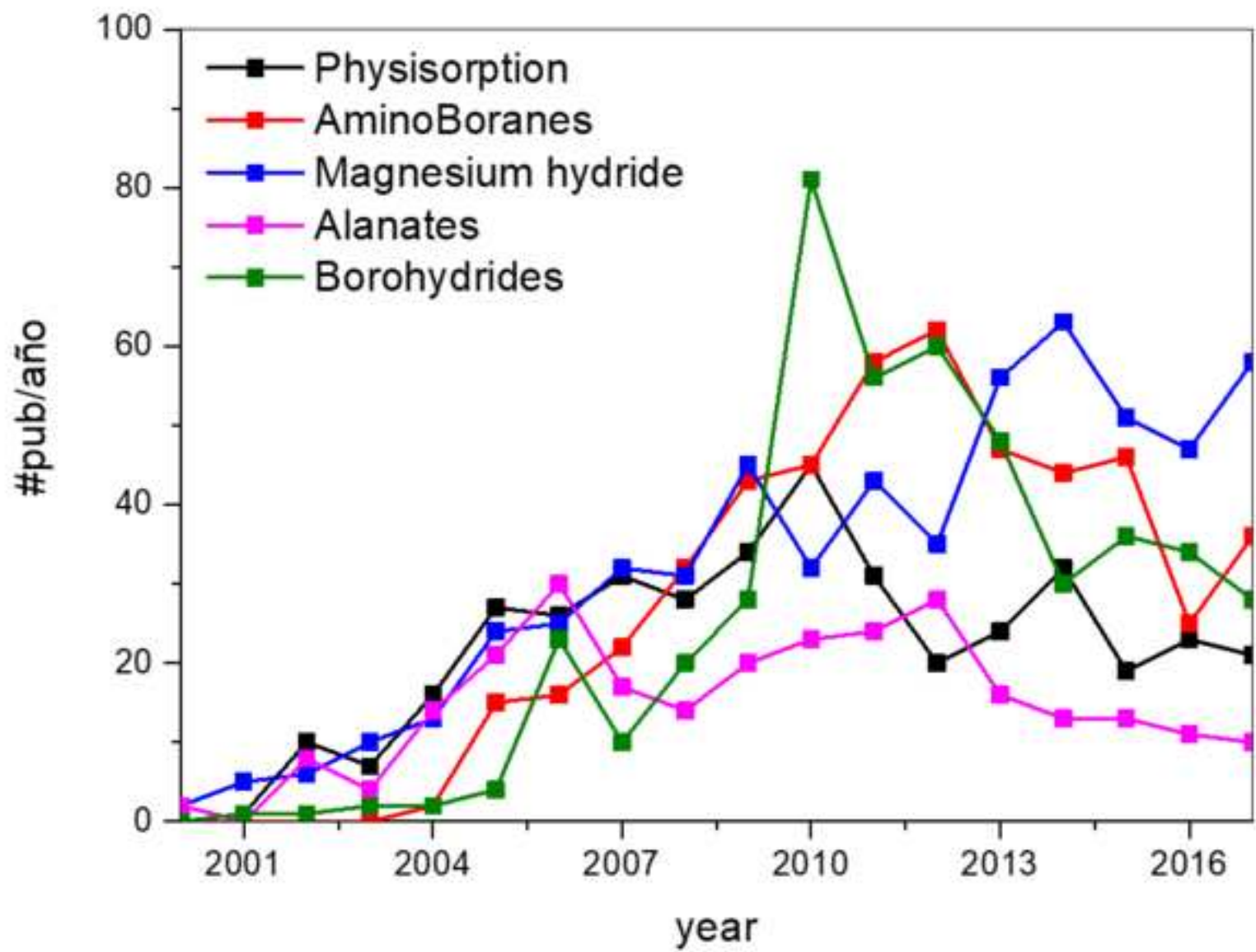

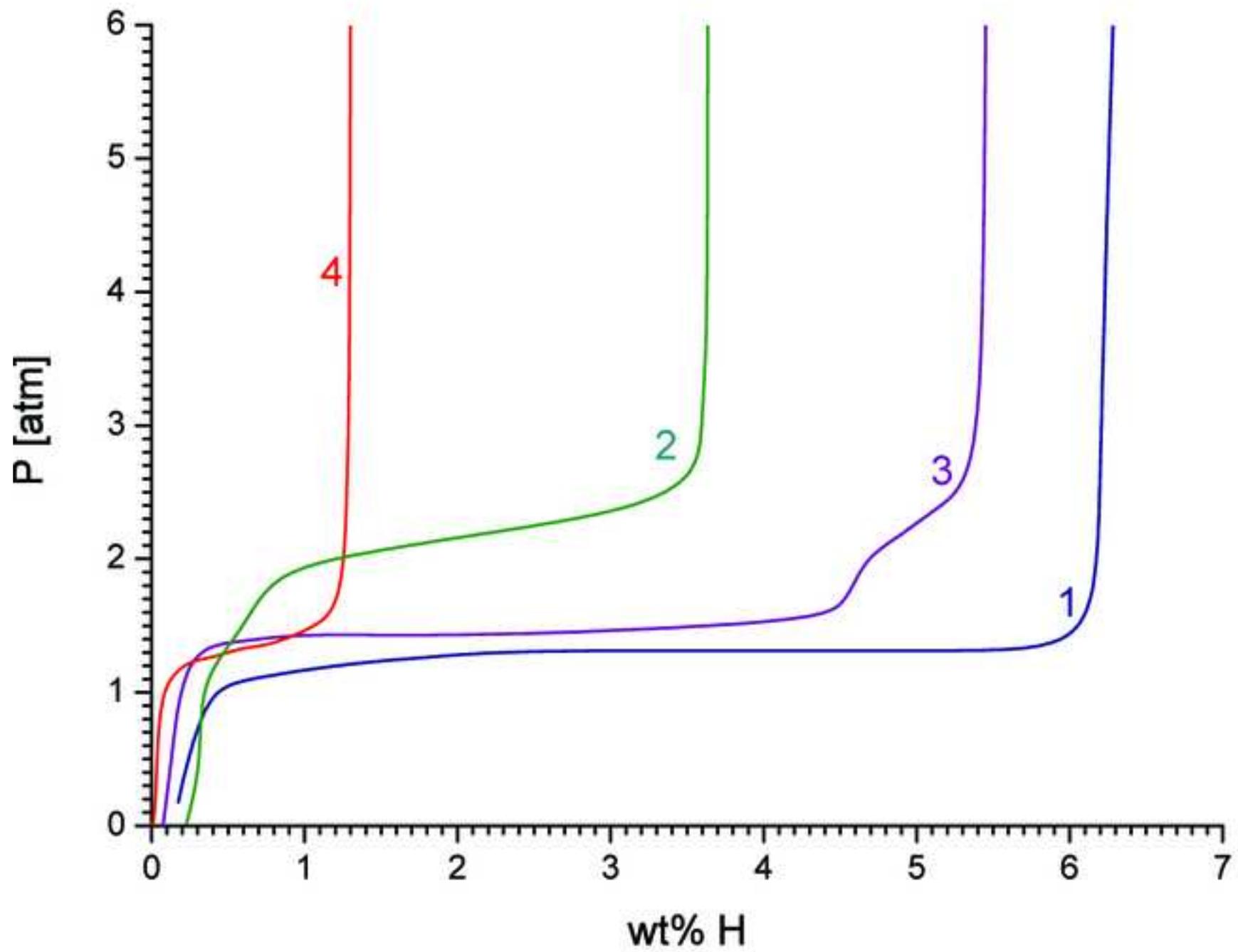

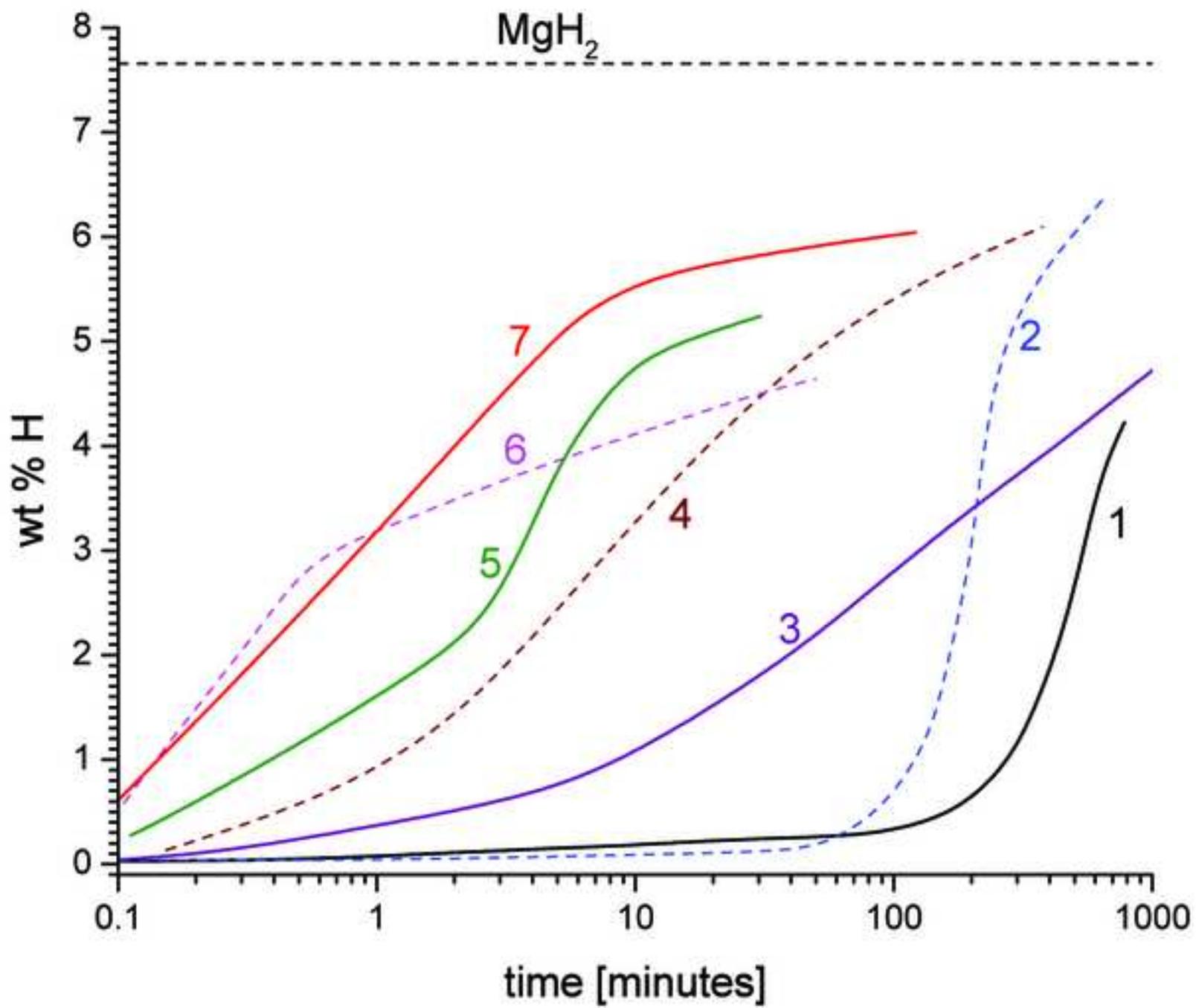

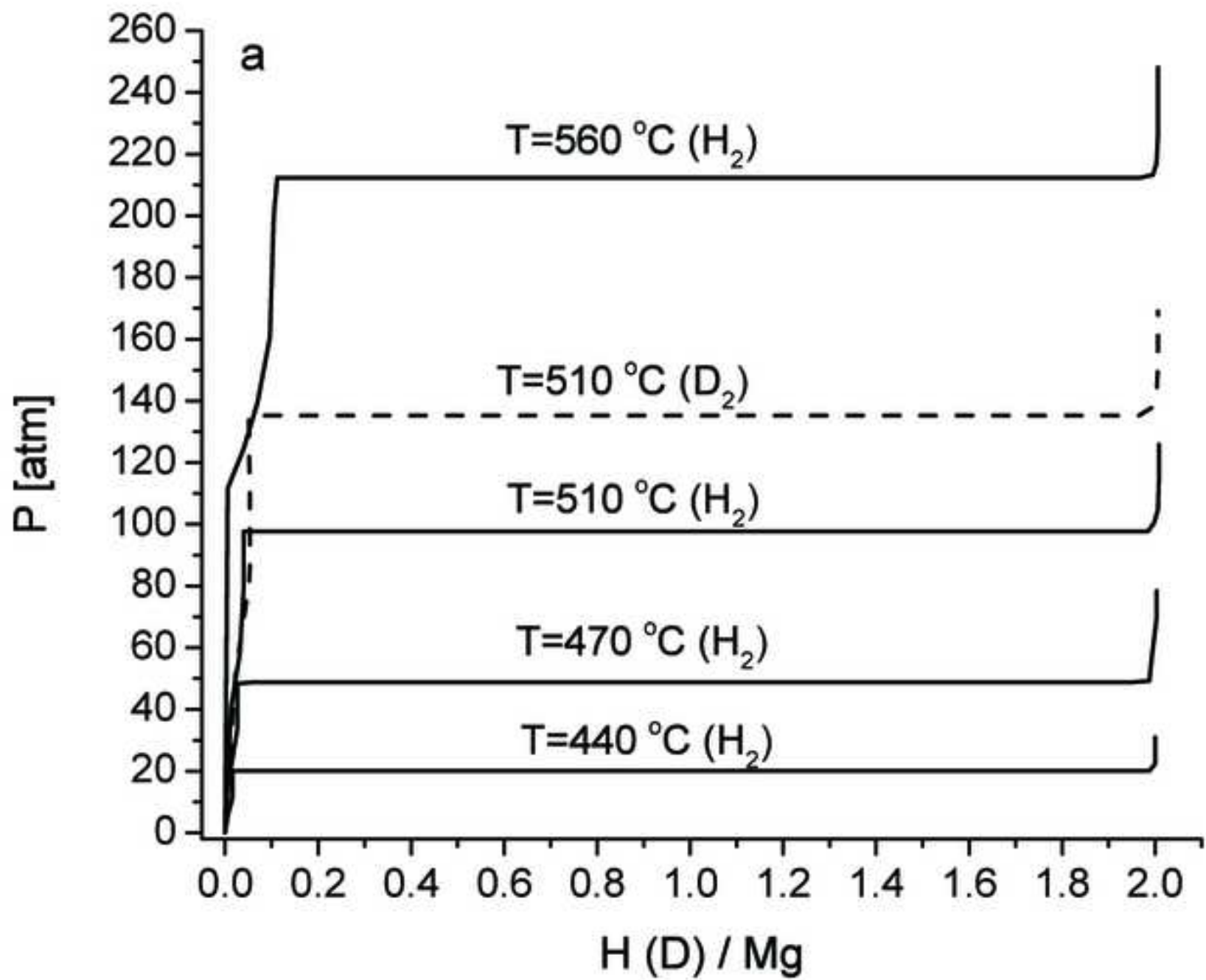

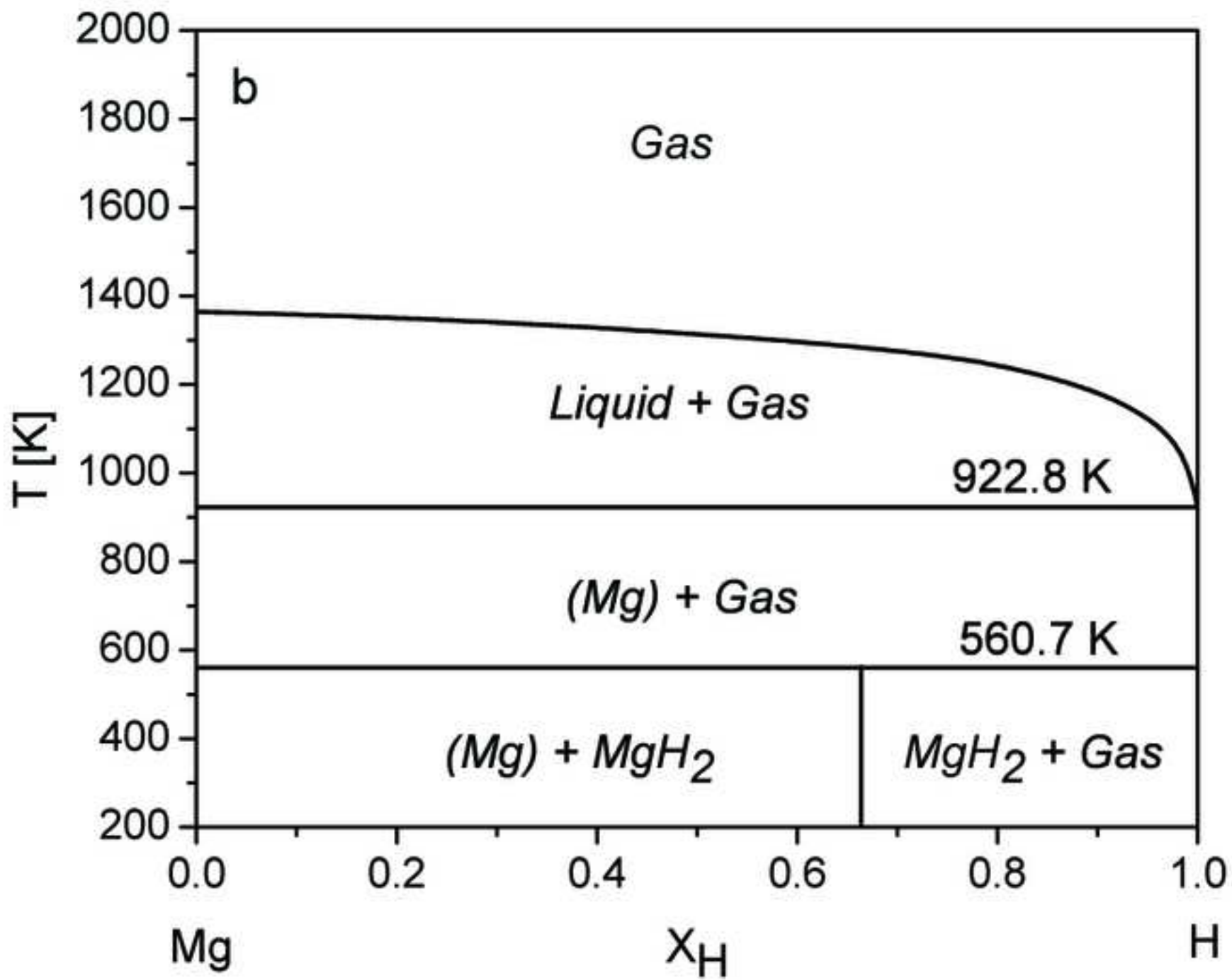

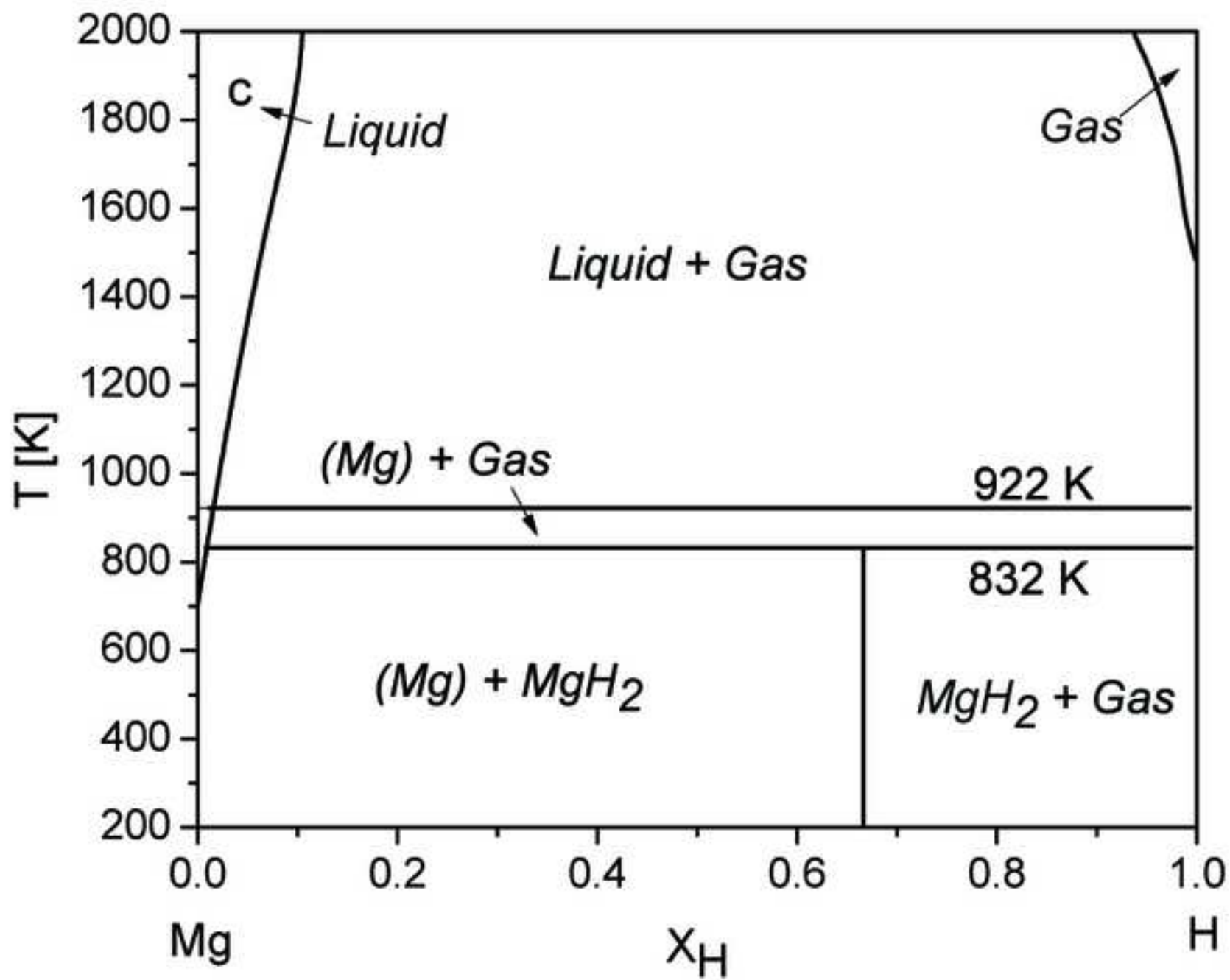

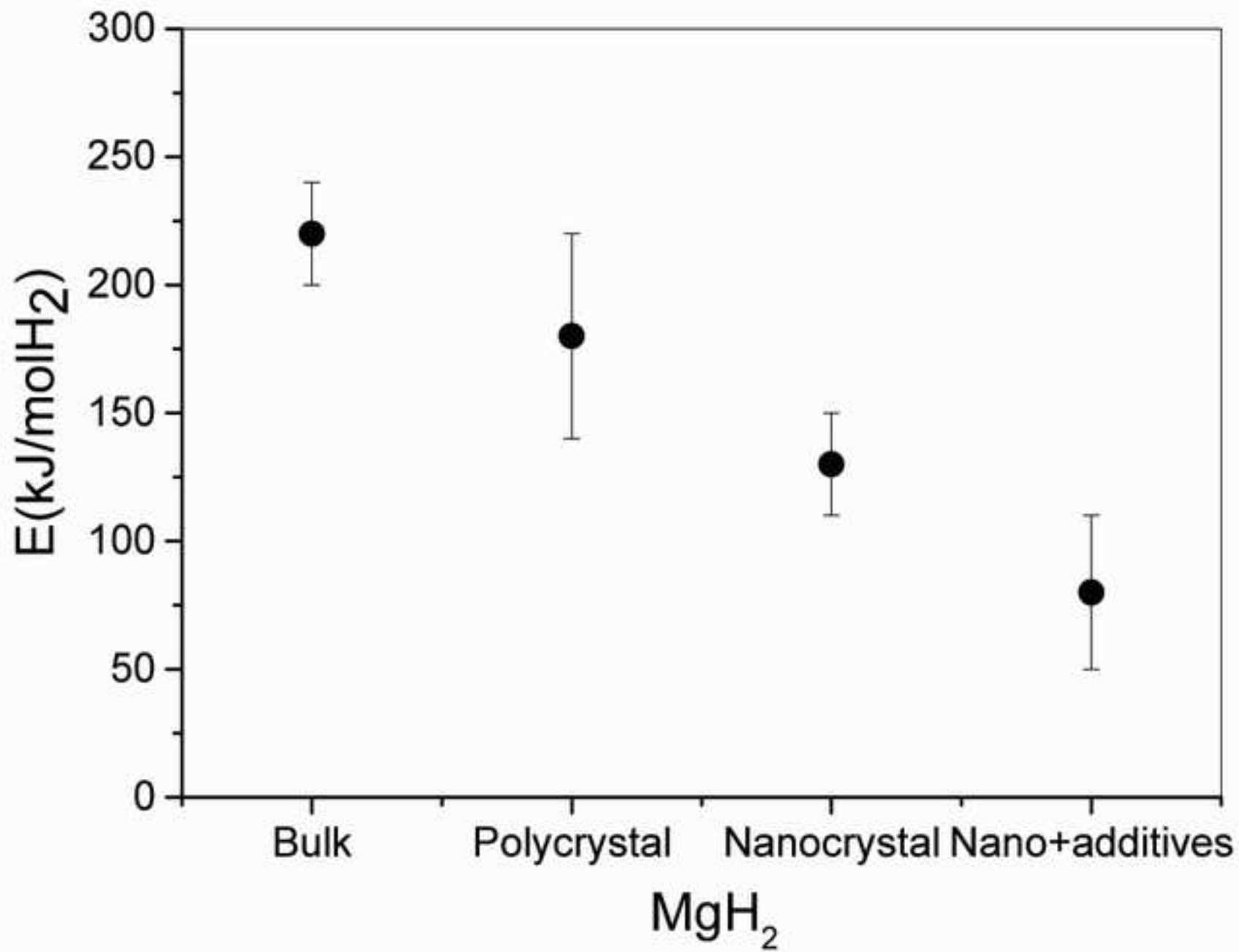

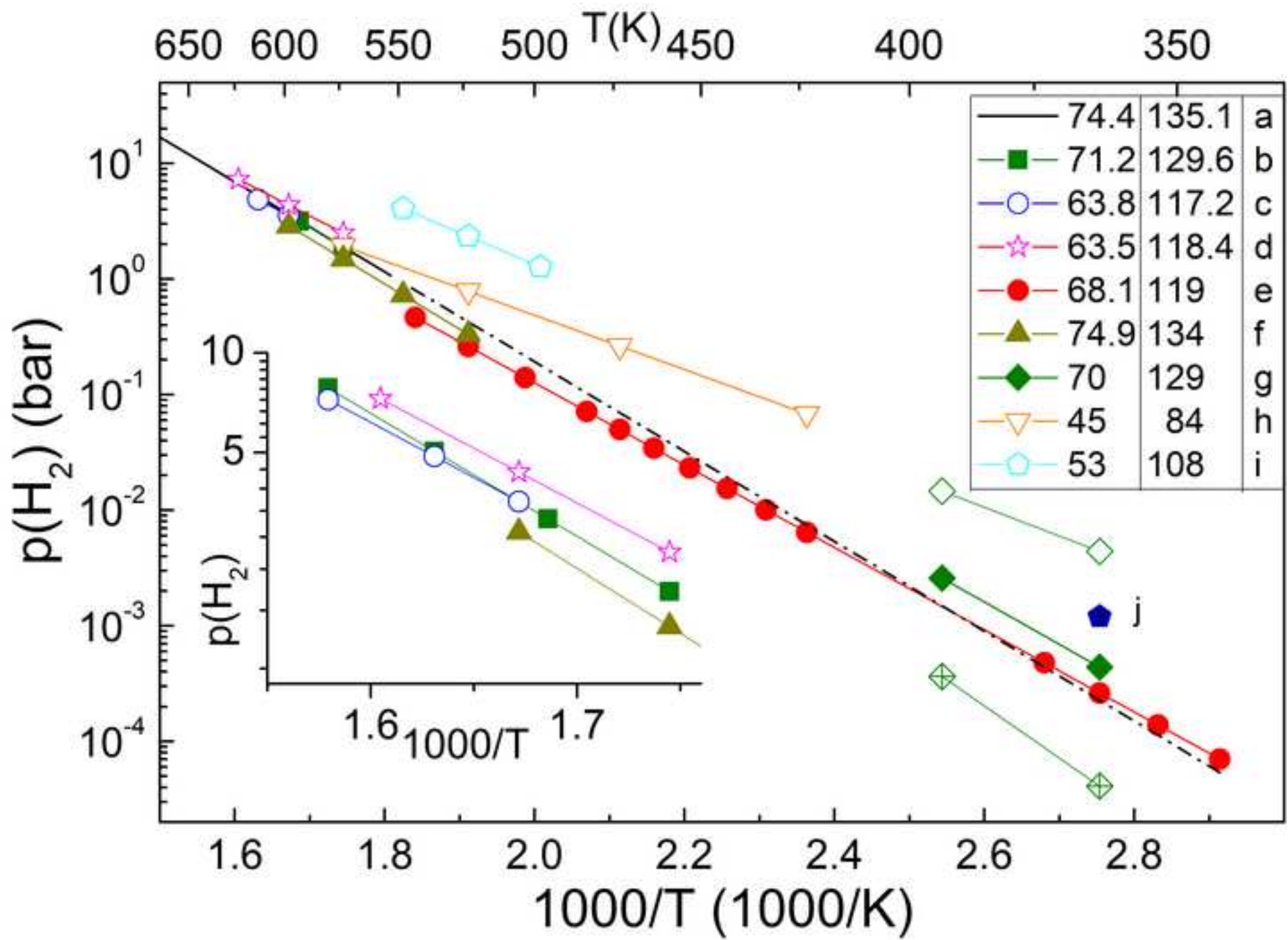

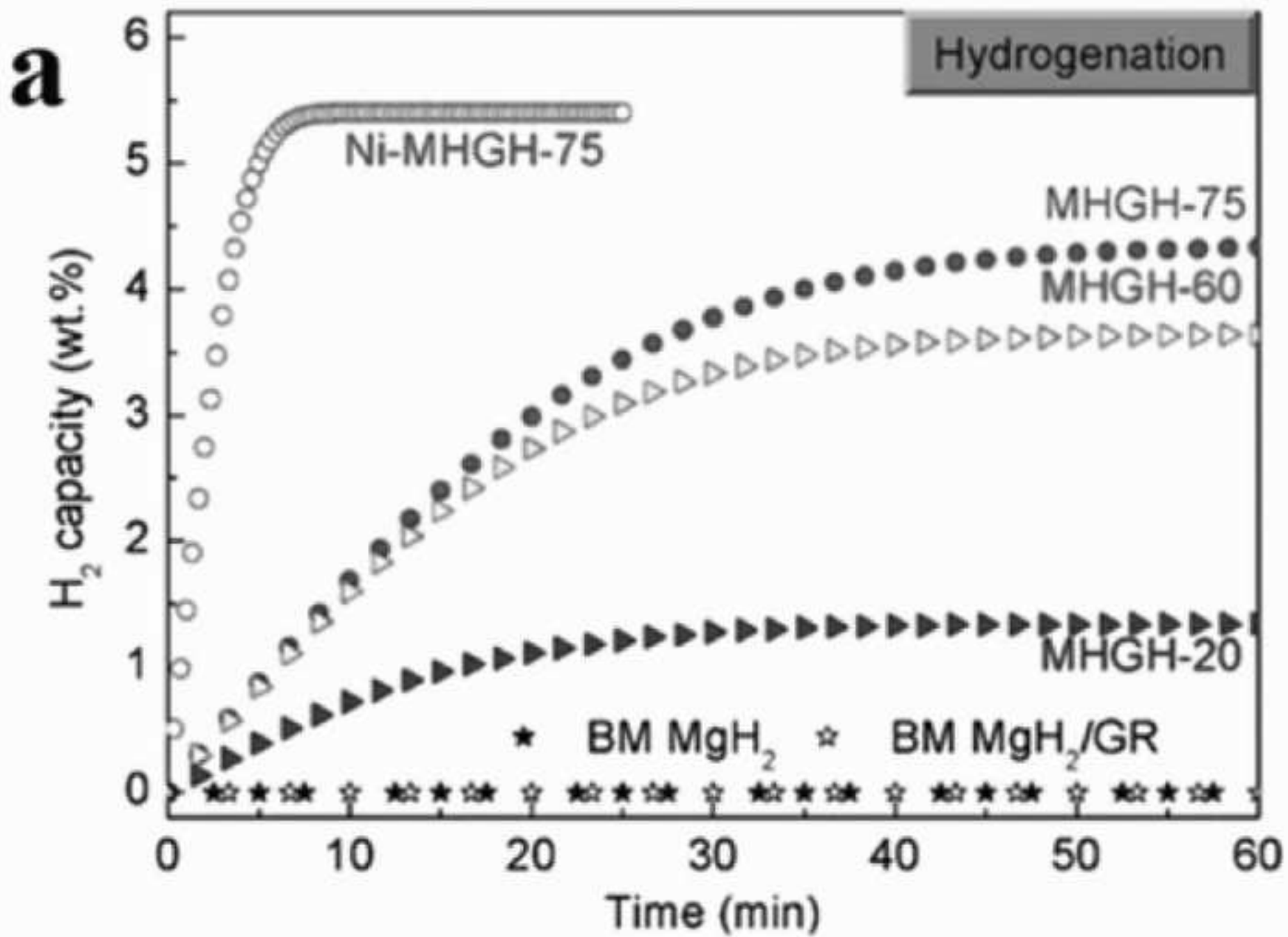

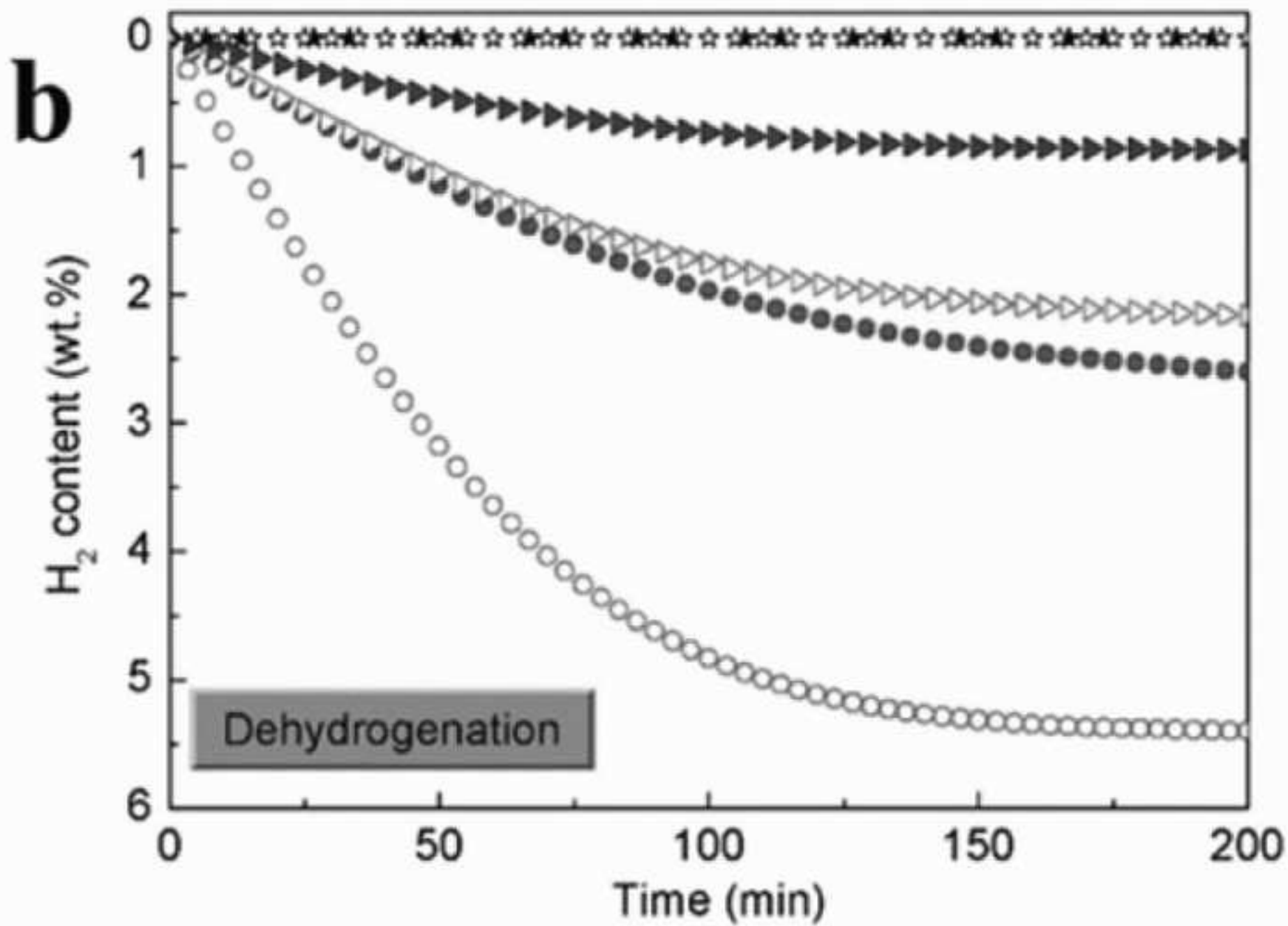

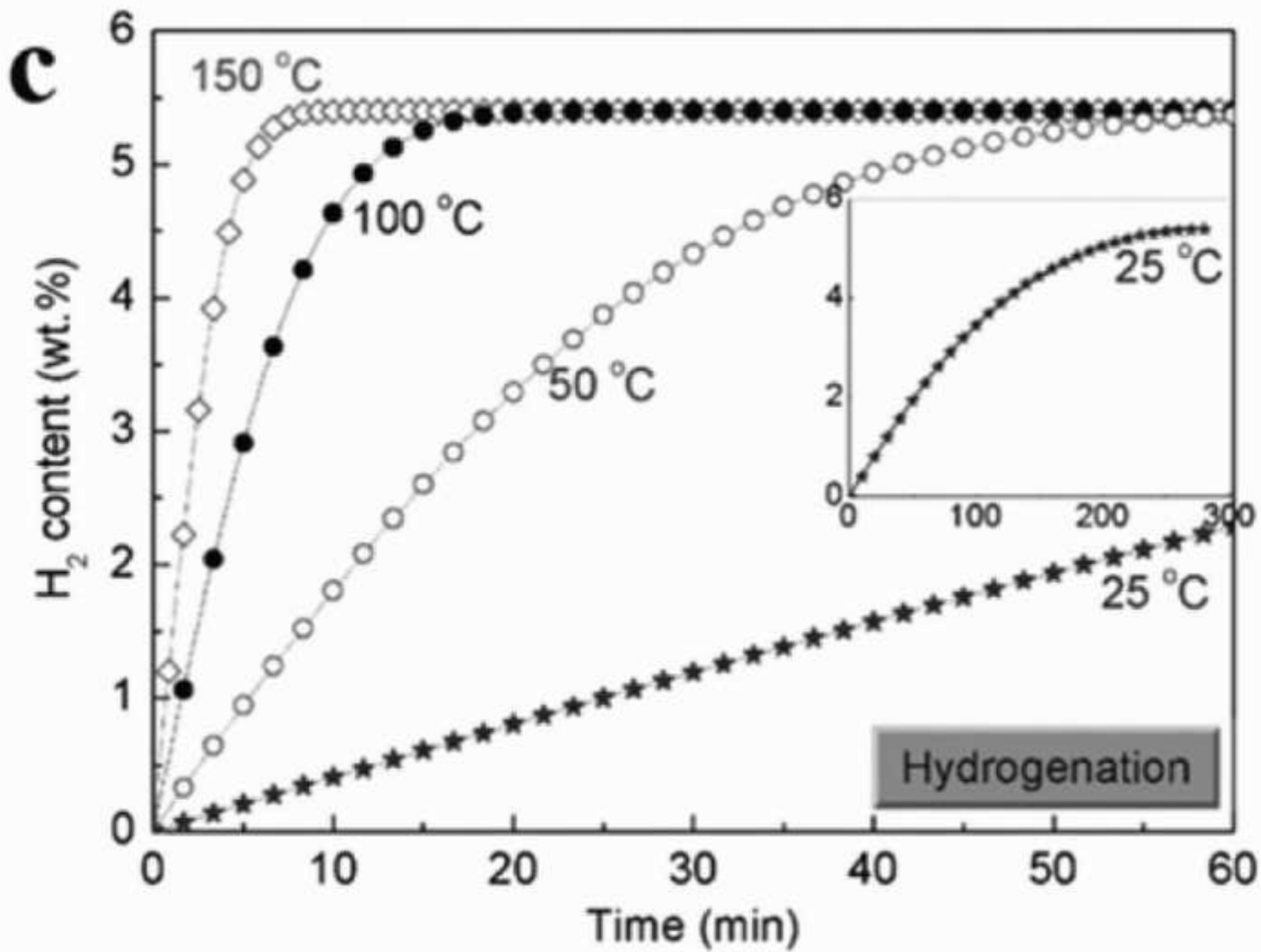

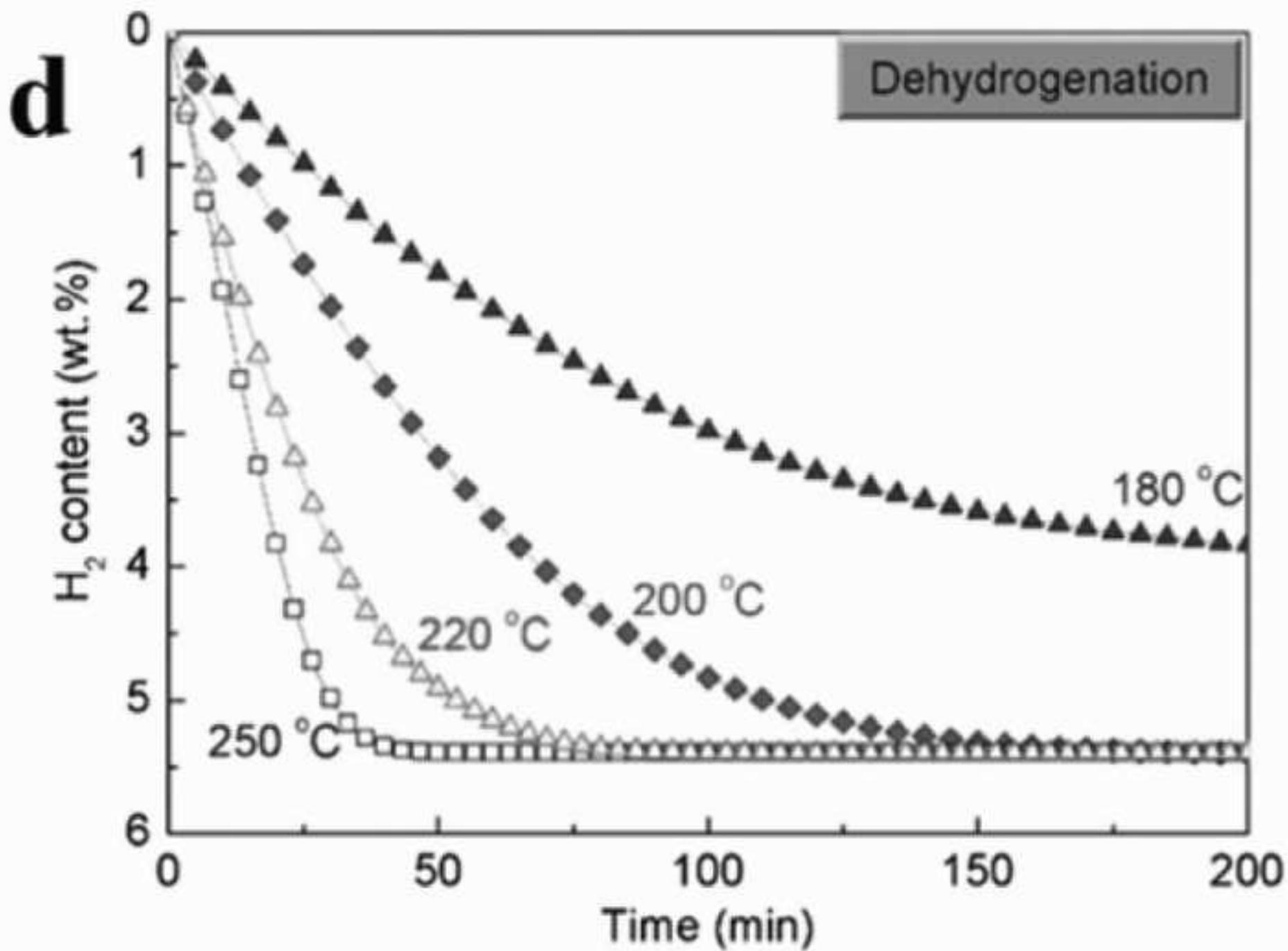

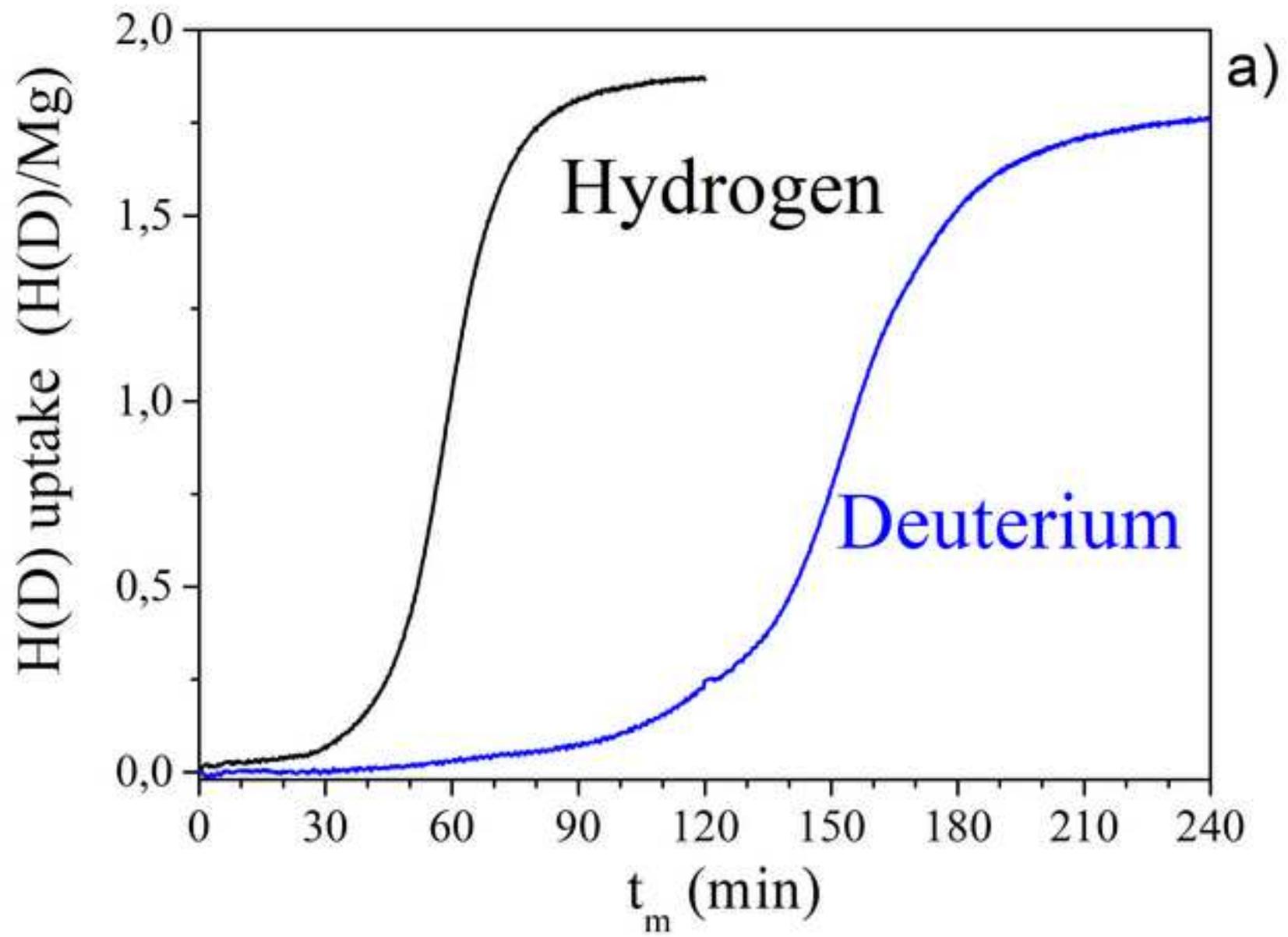

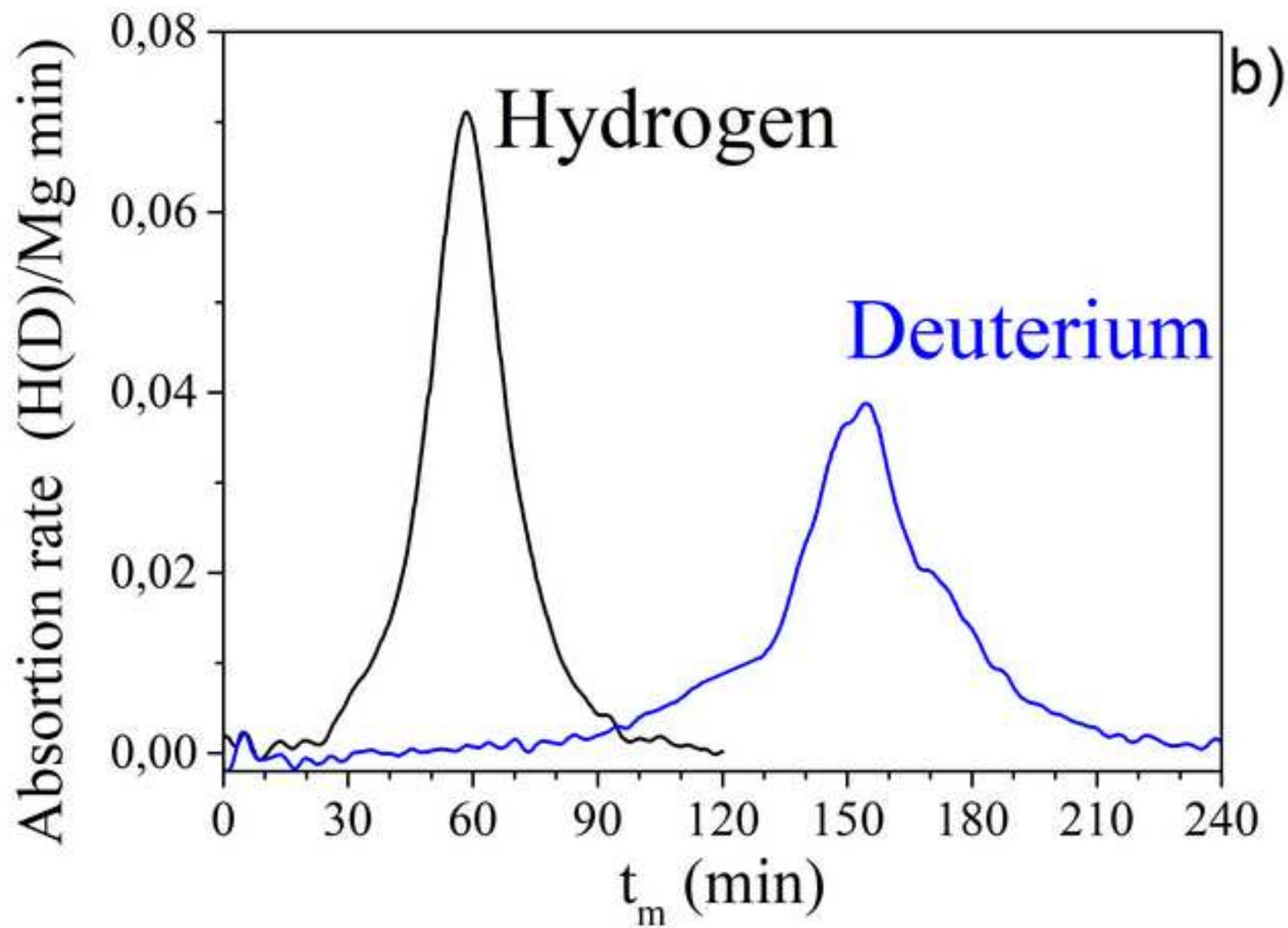

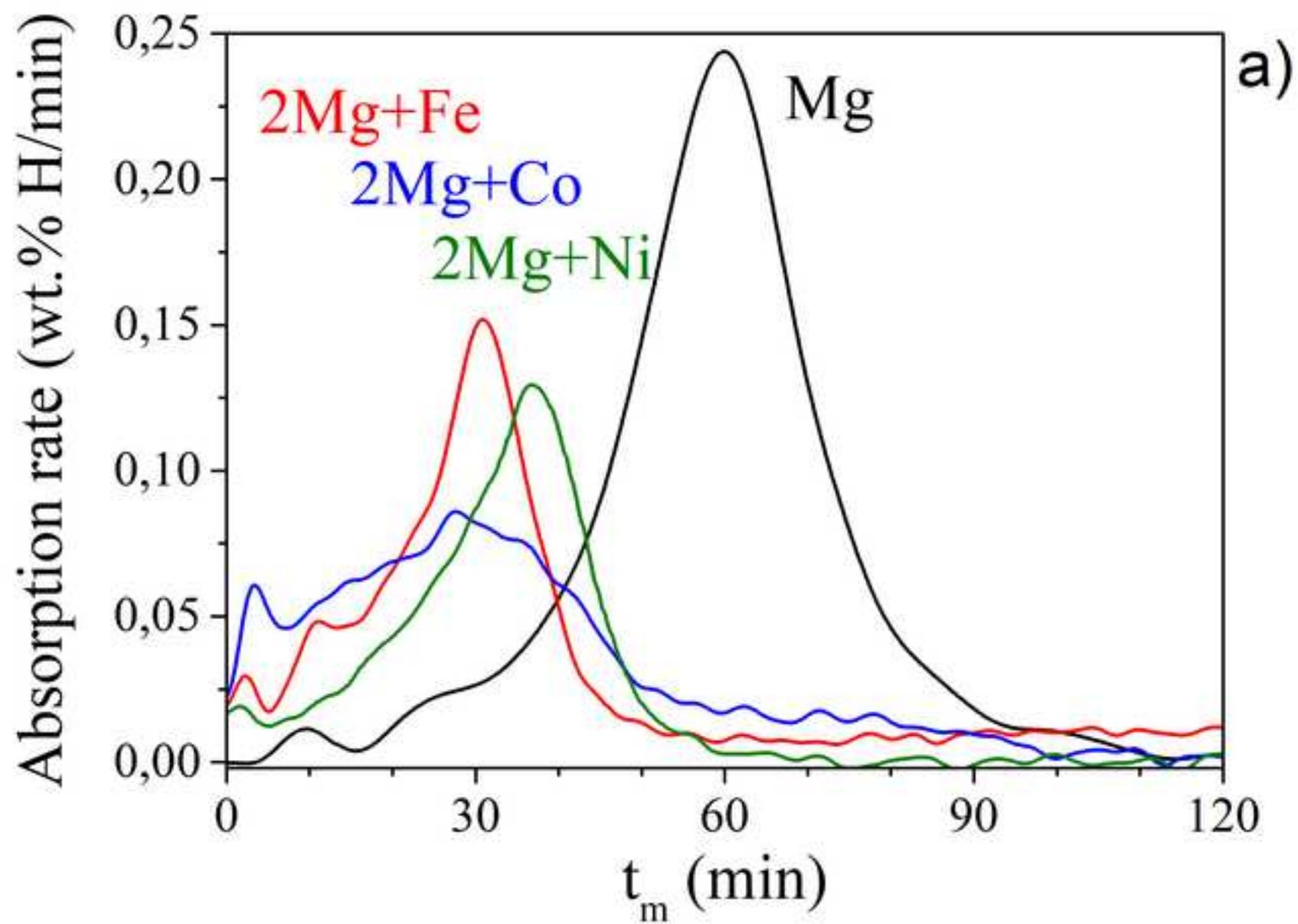

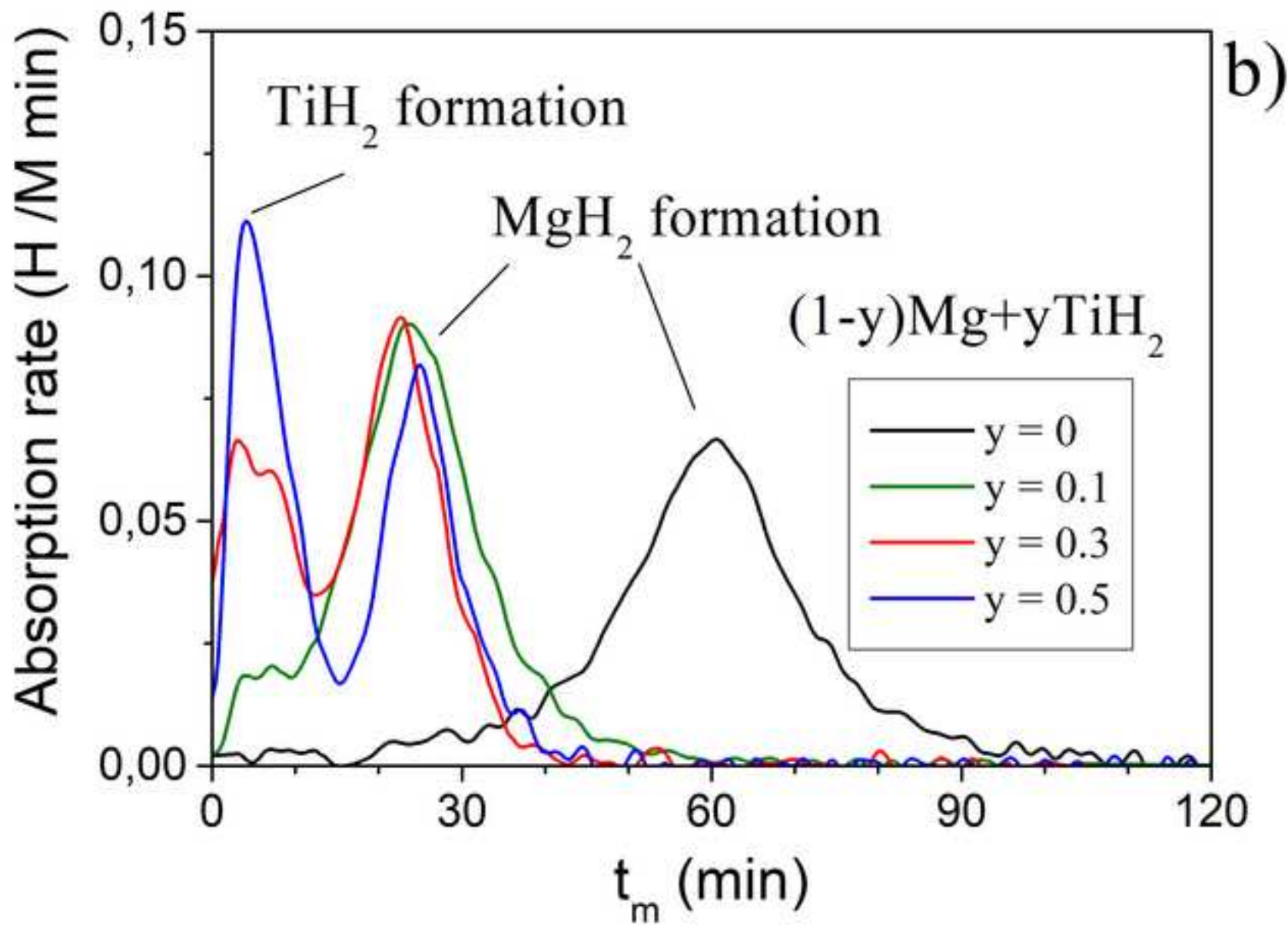

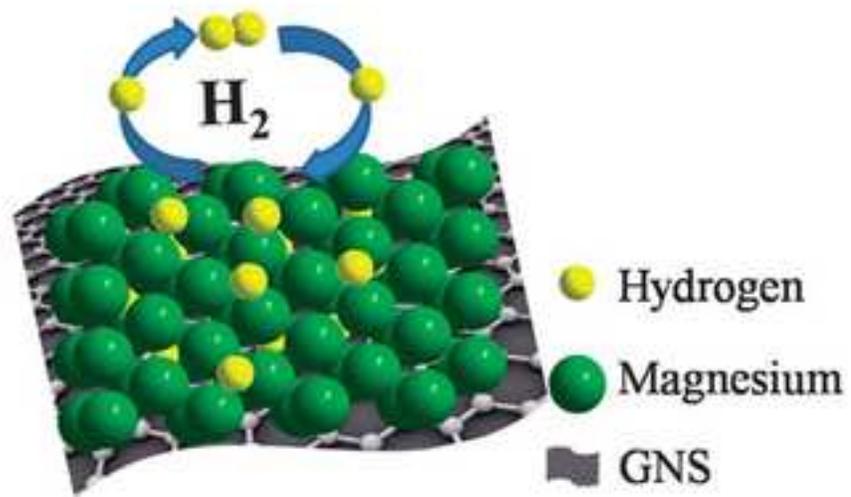

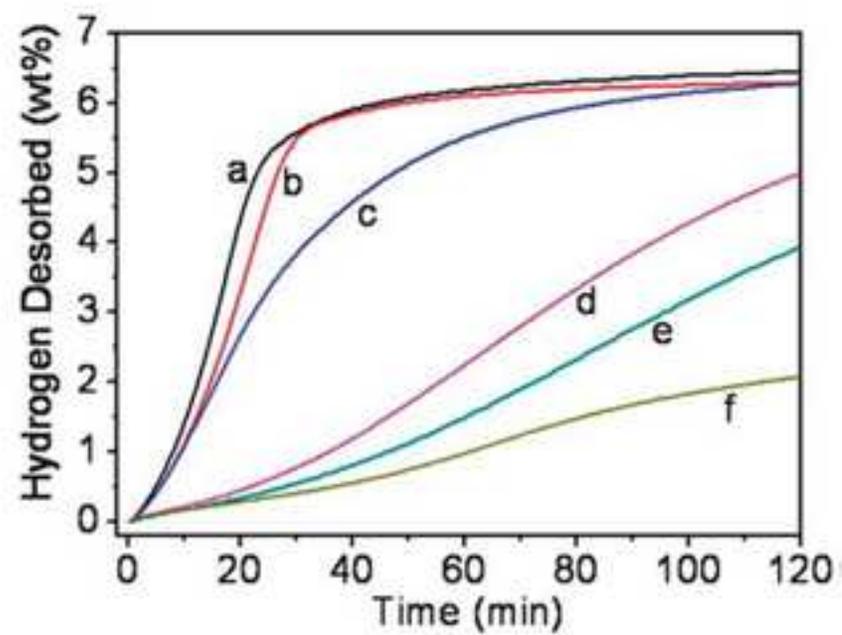

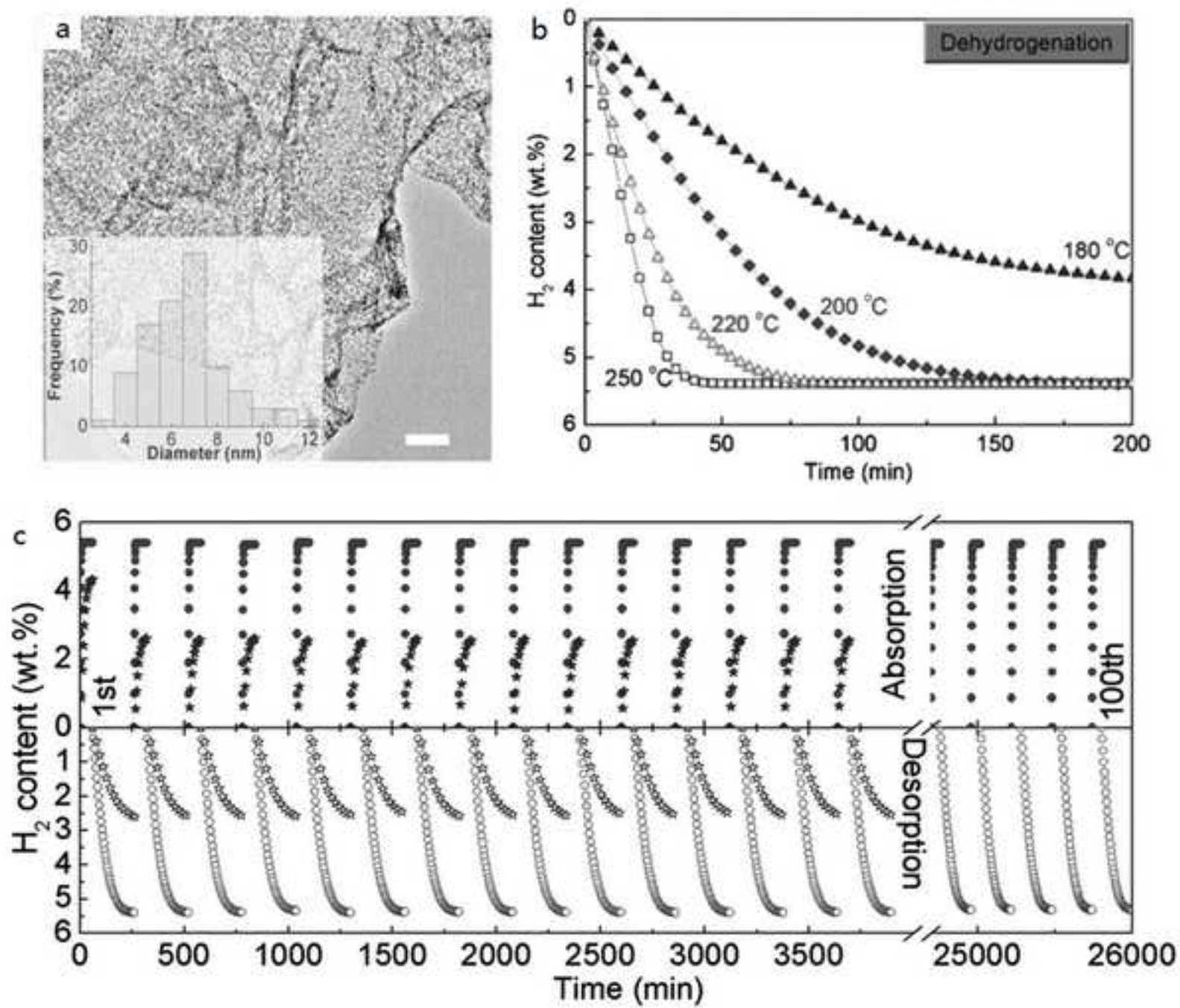

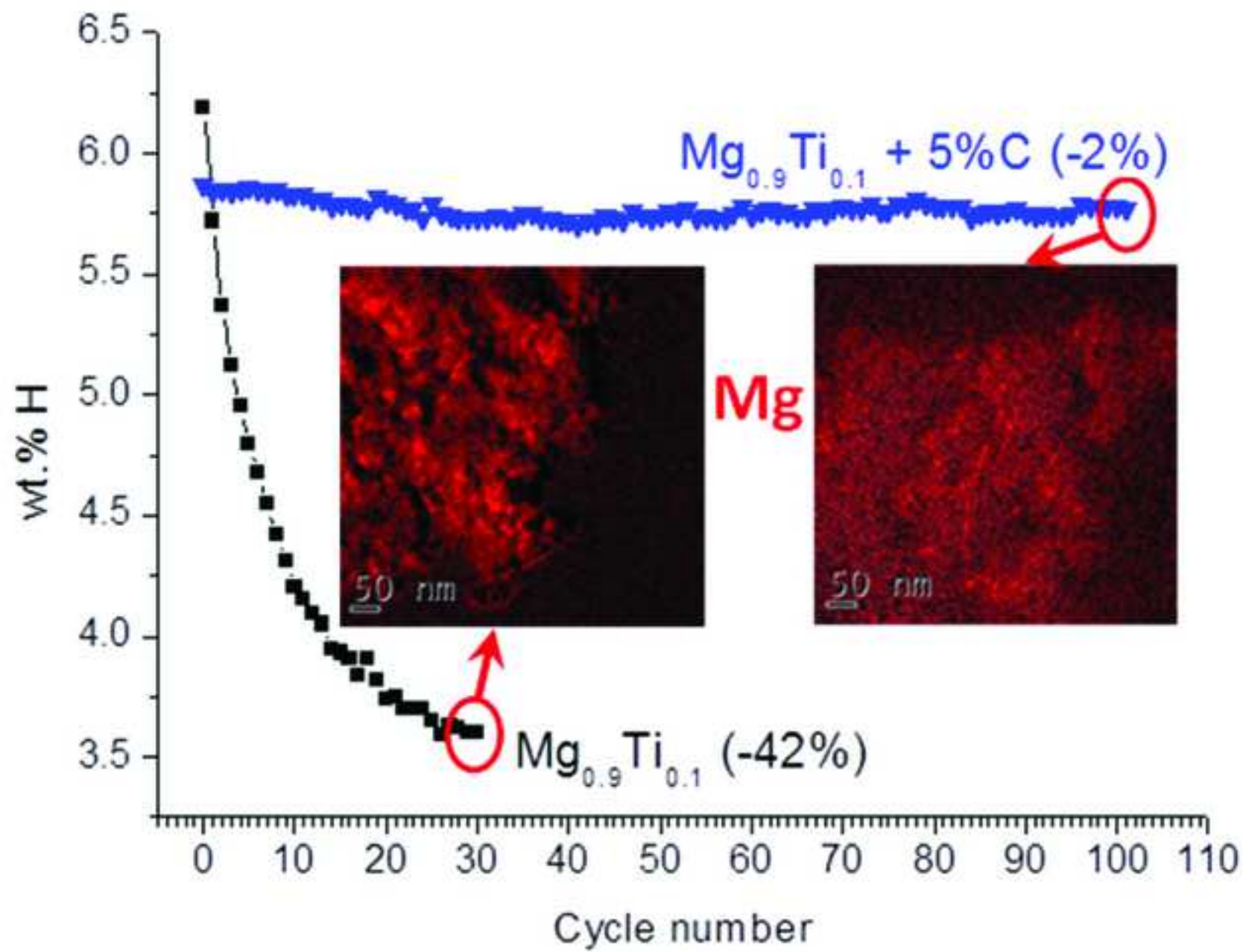

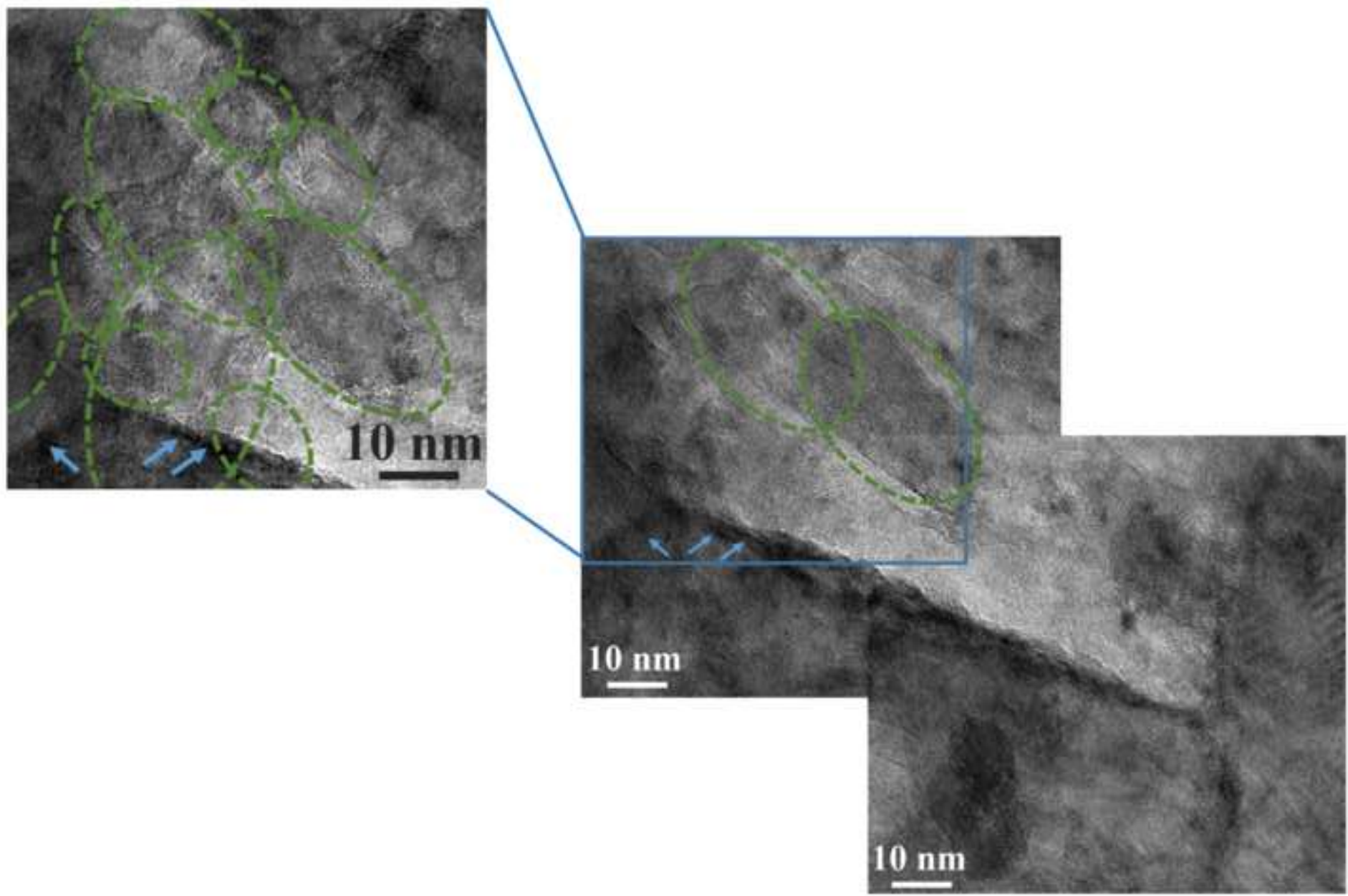

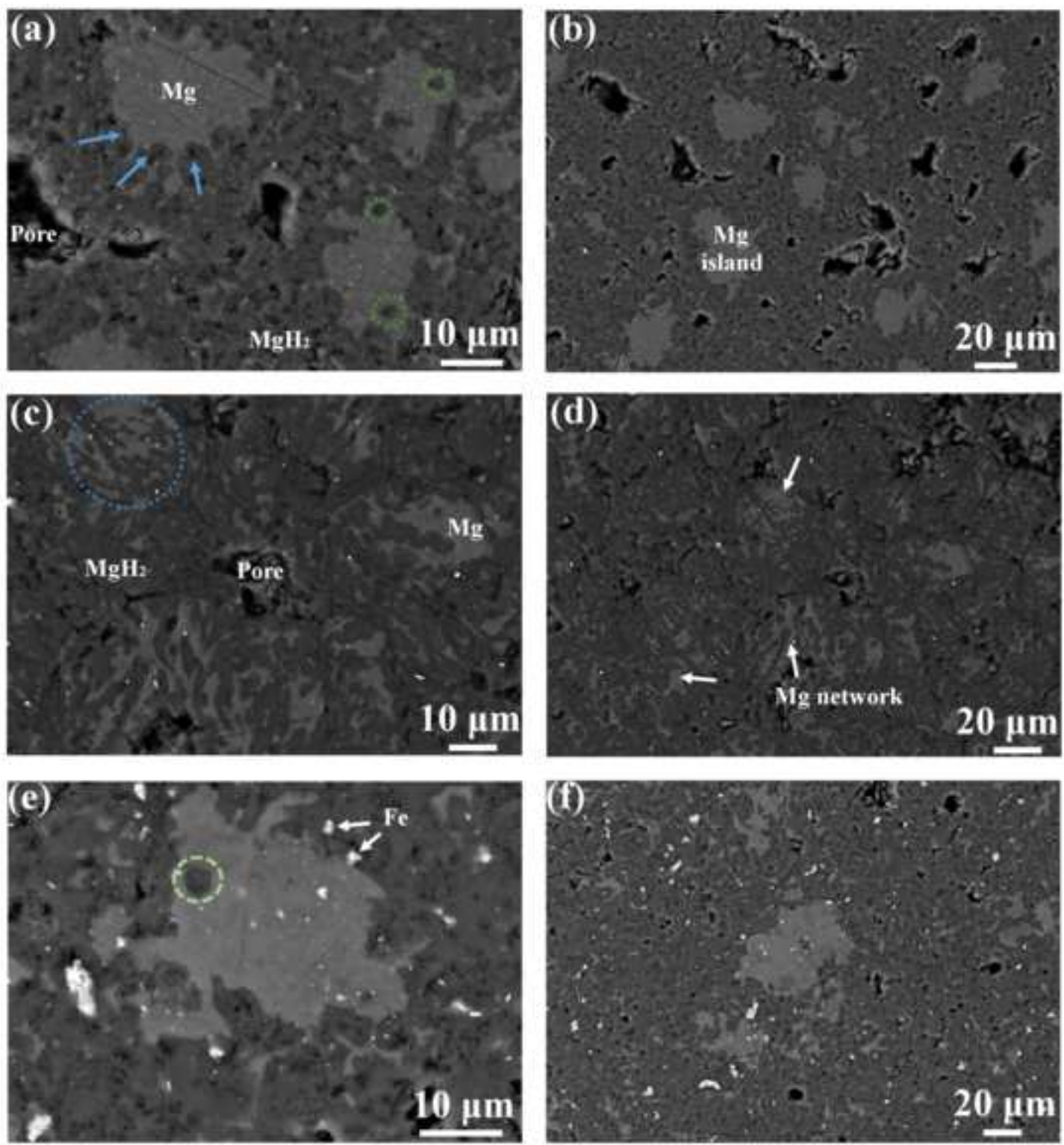

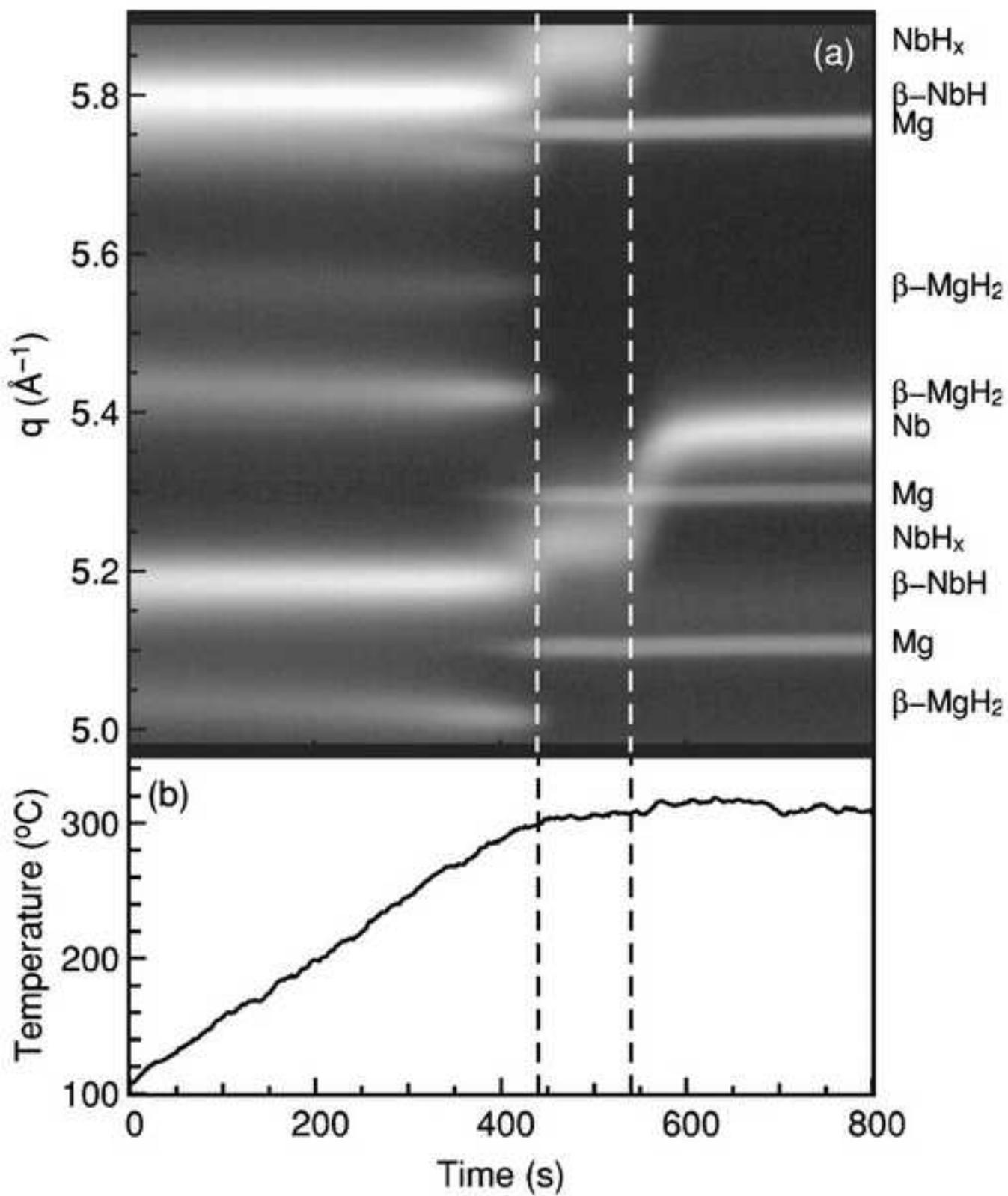

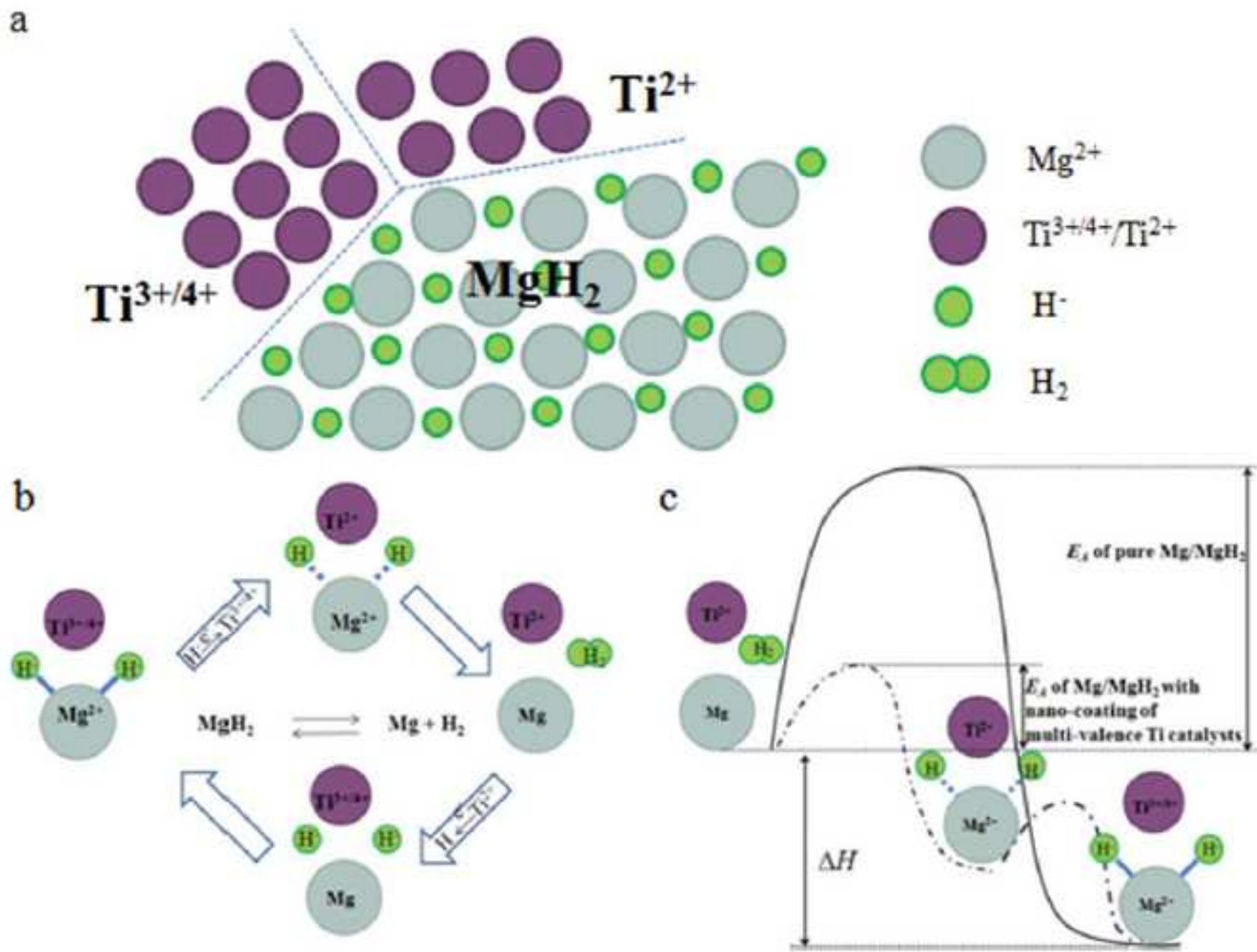

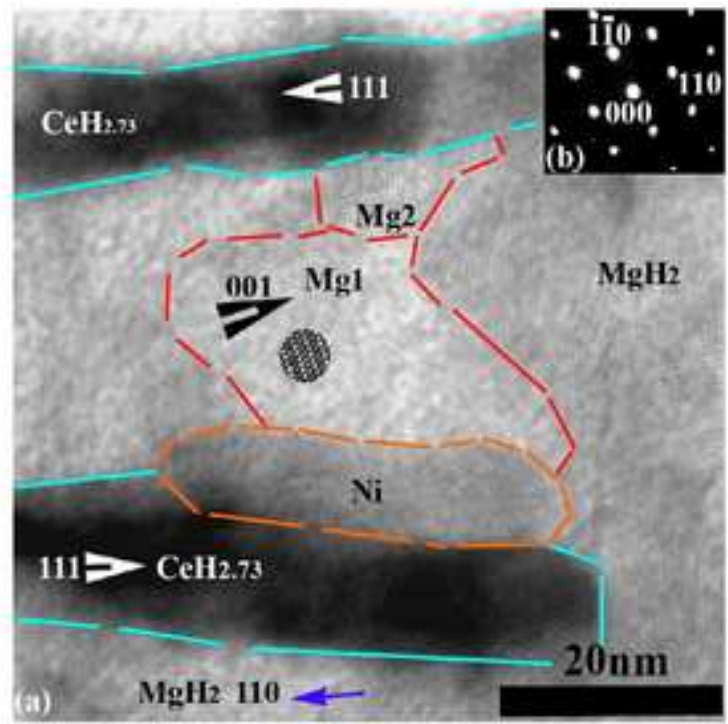

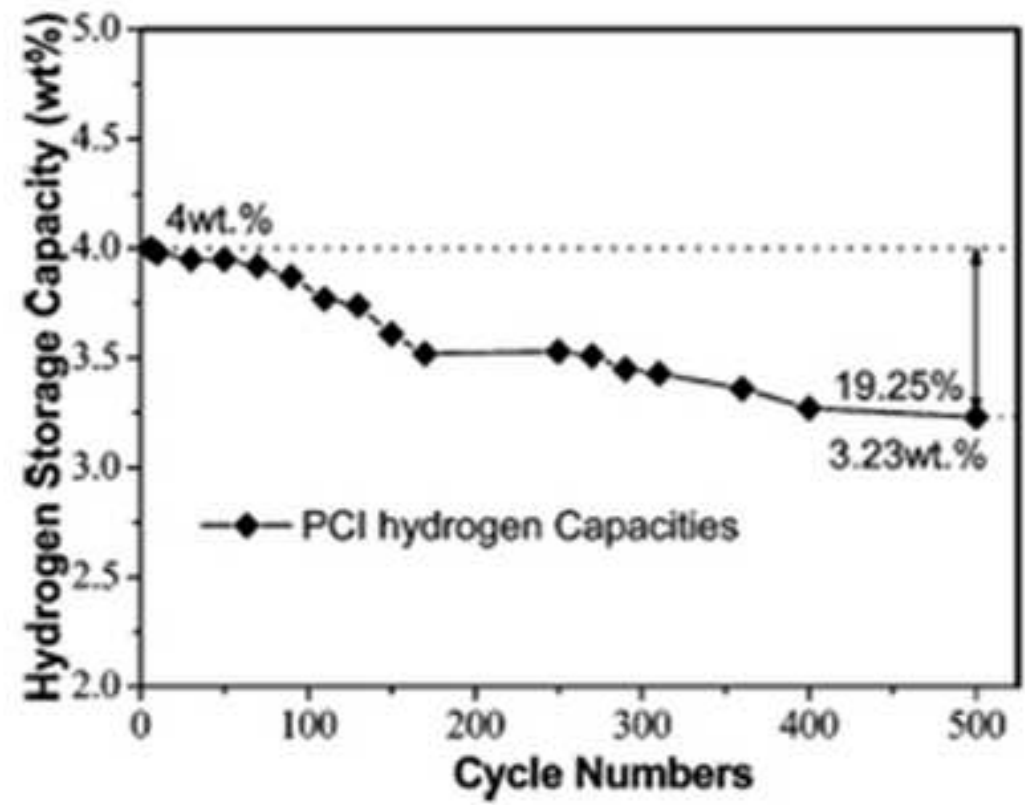

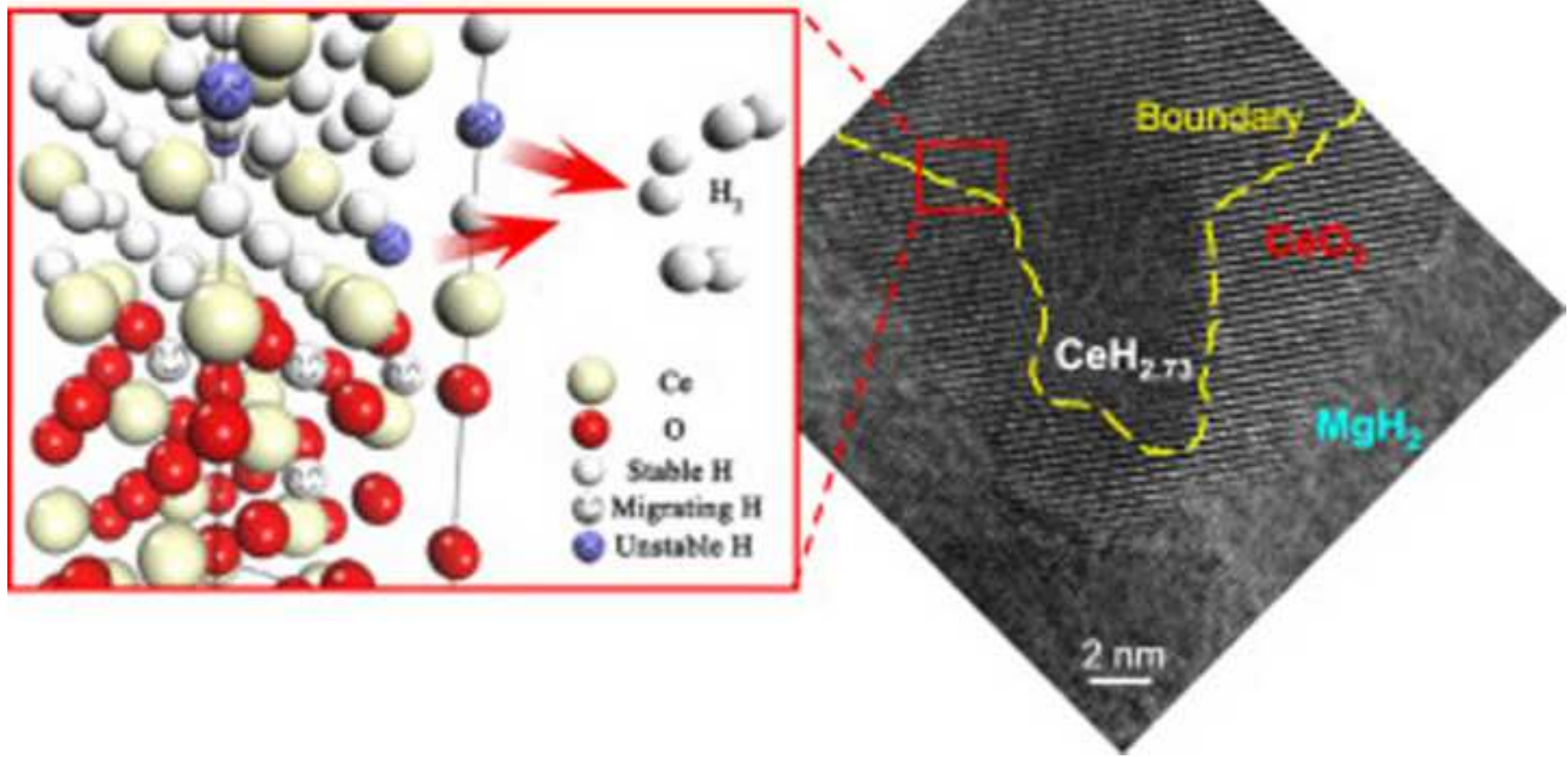

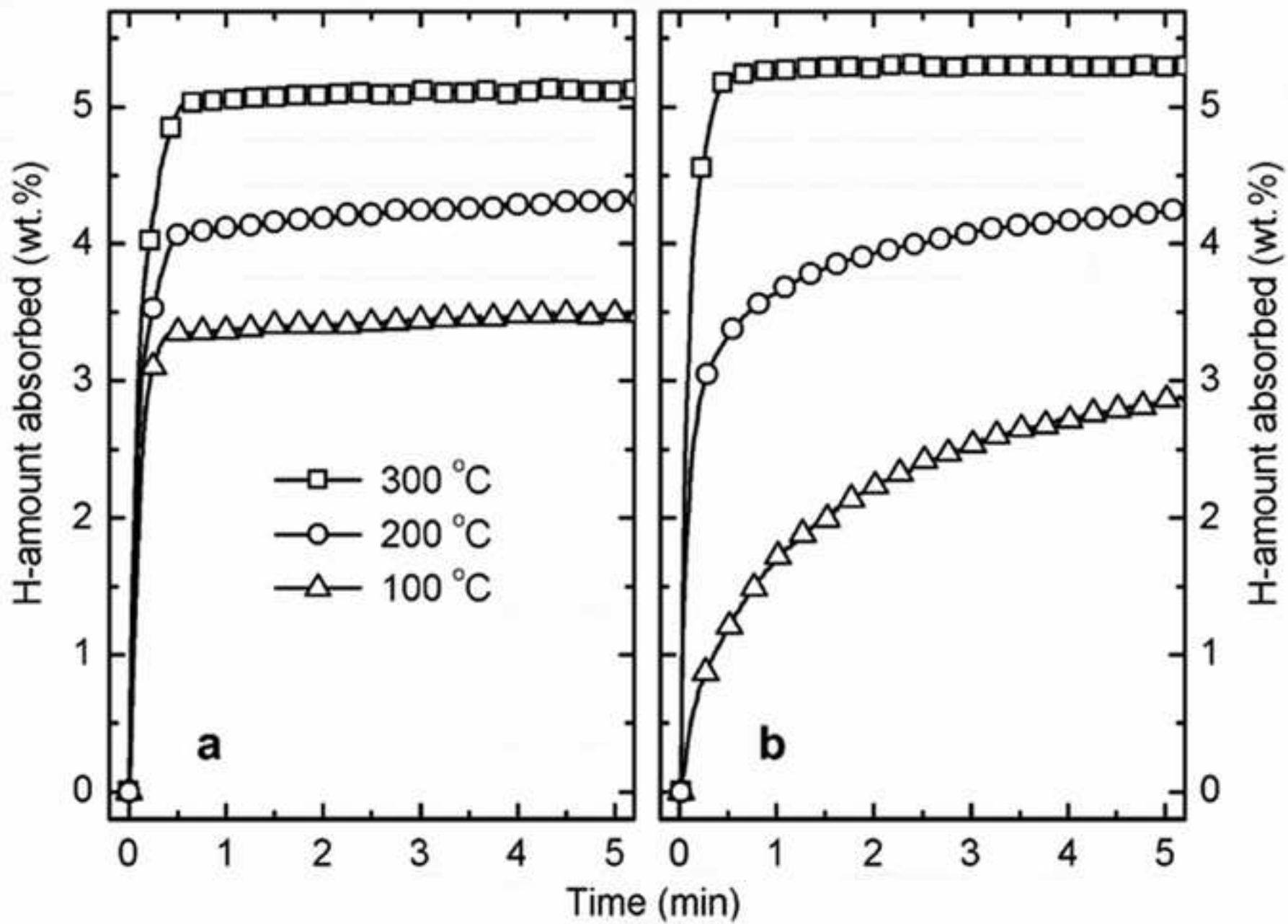

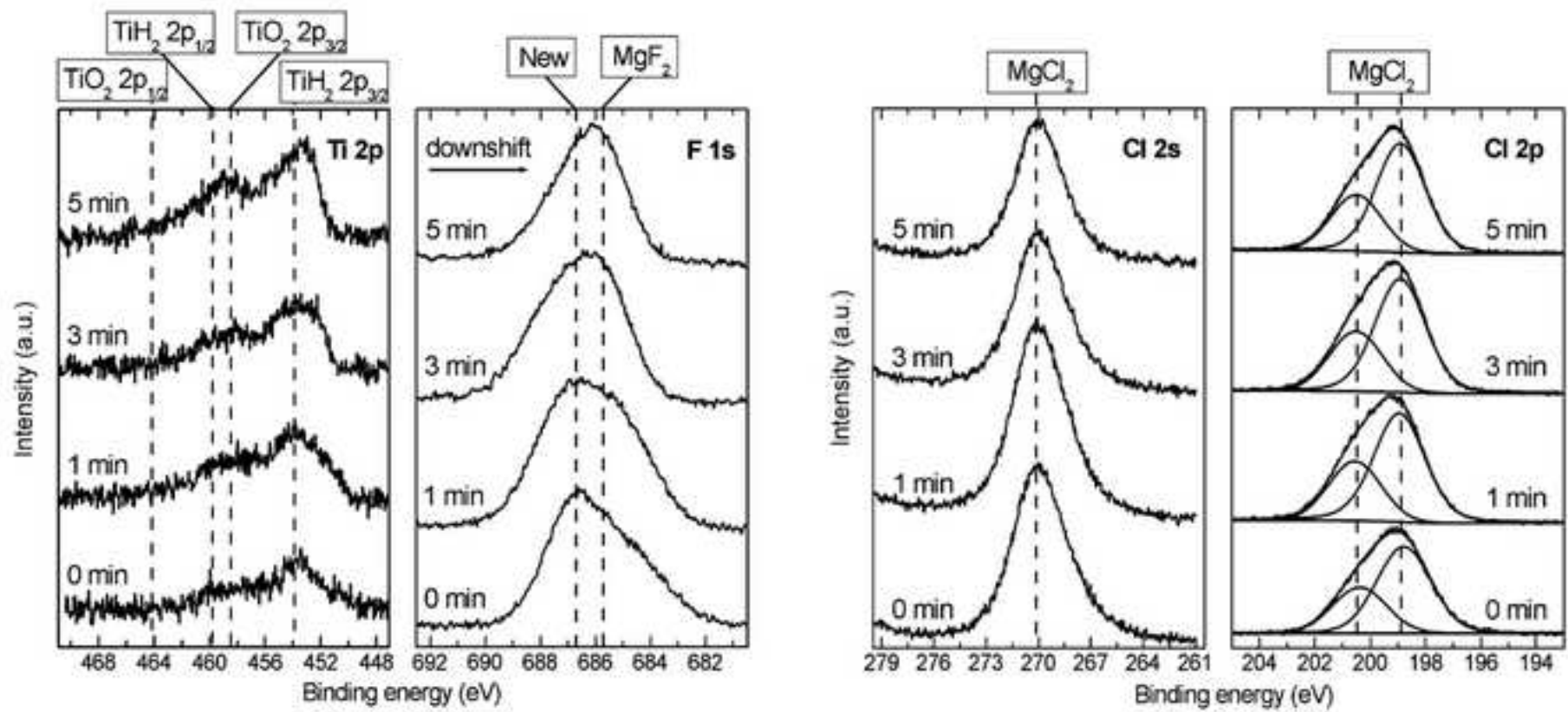

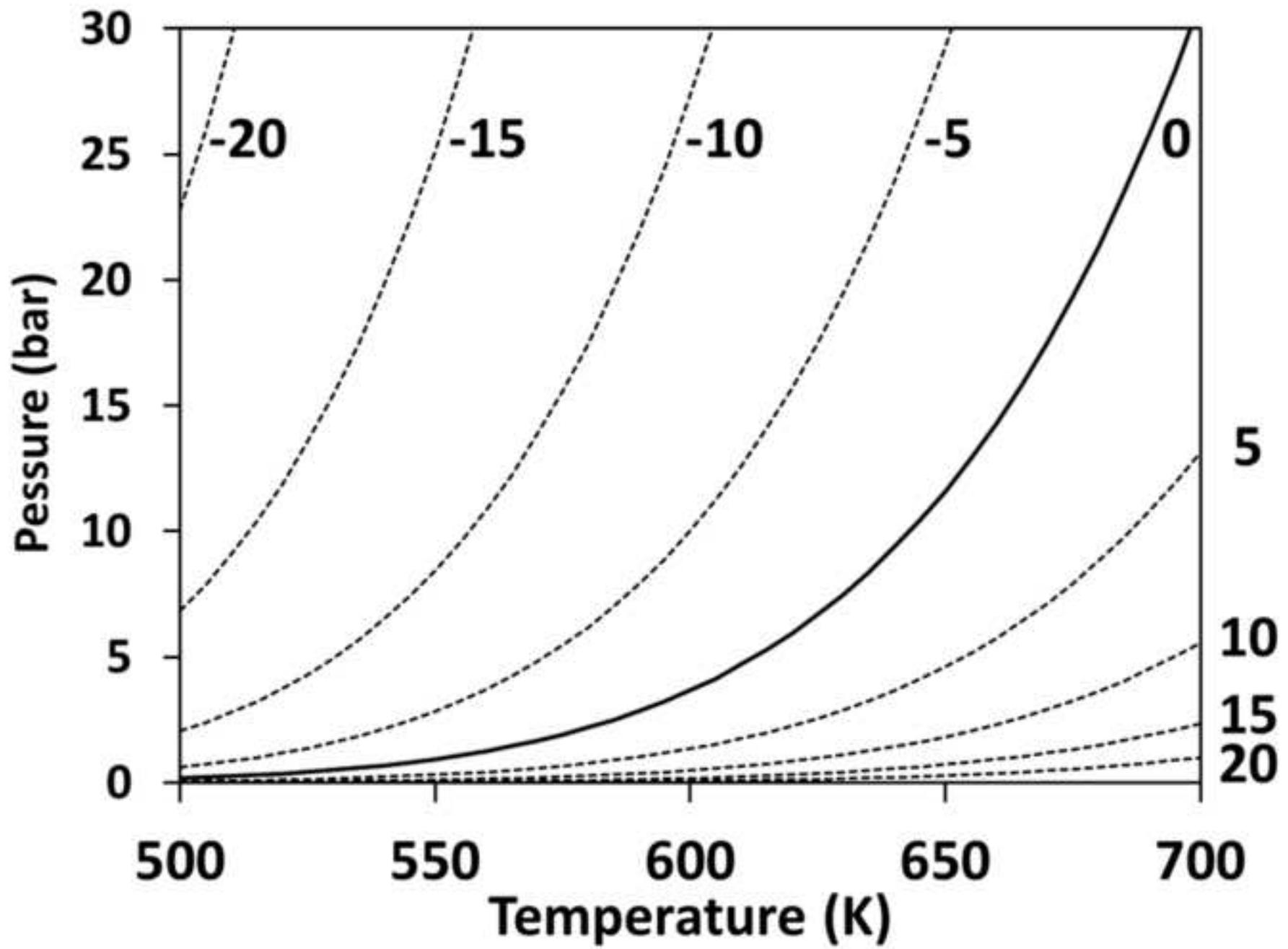

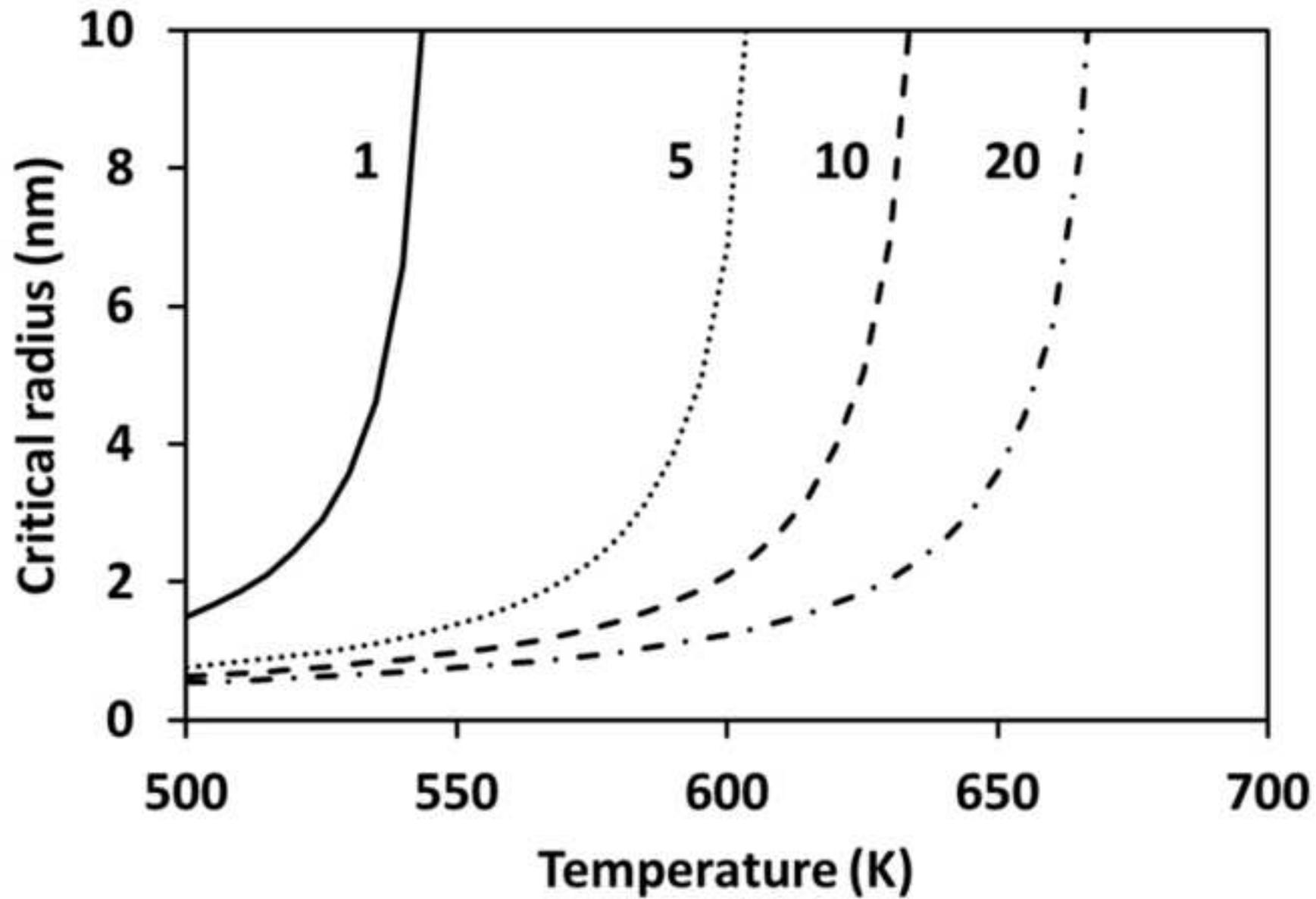

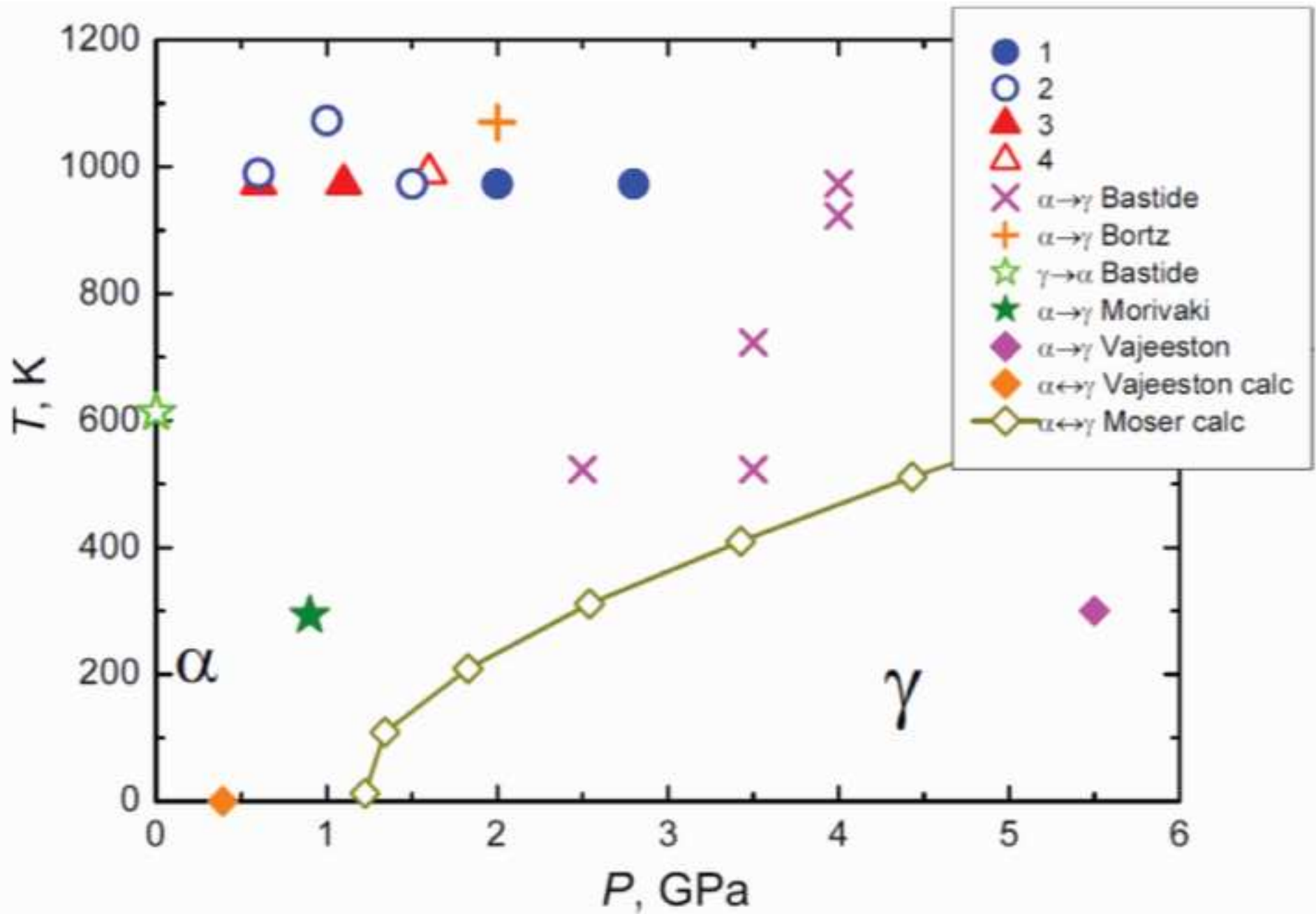

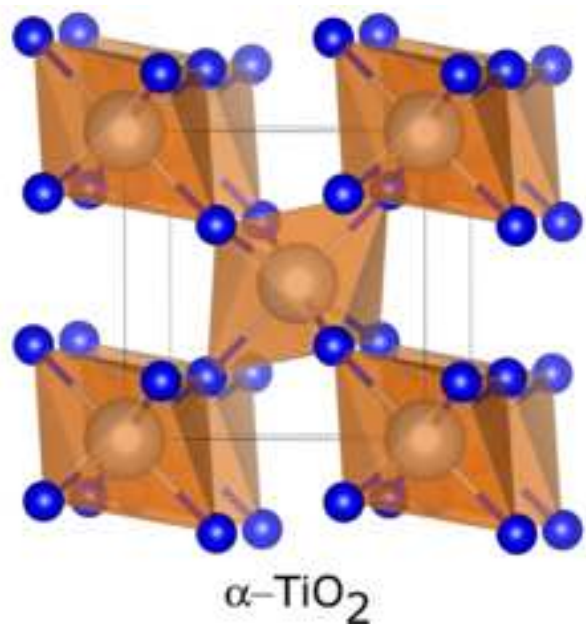
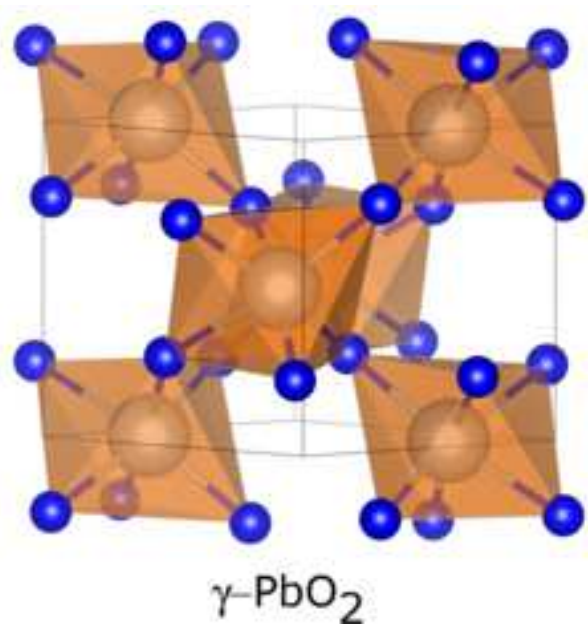
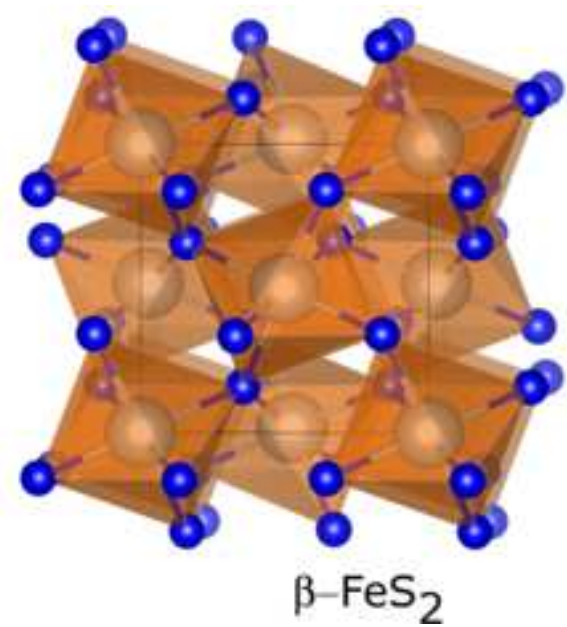

$\alpha-TiO_2$ $\qquad\qquad\qquad$ $\gamma-PbO_2$ $\qquad\qquad\qquad$ $\beta-FeS_2$

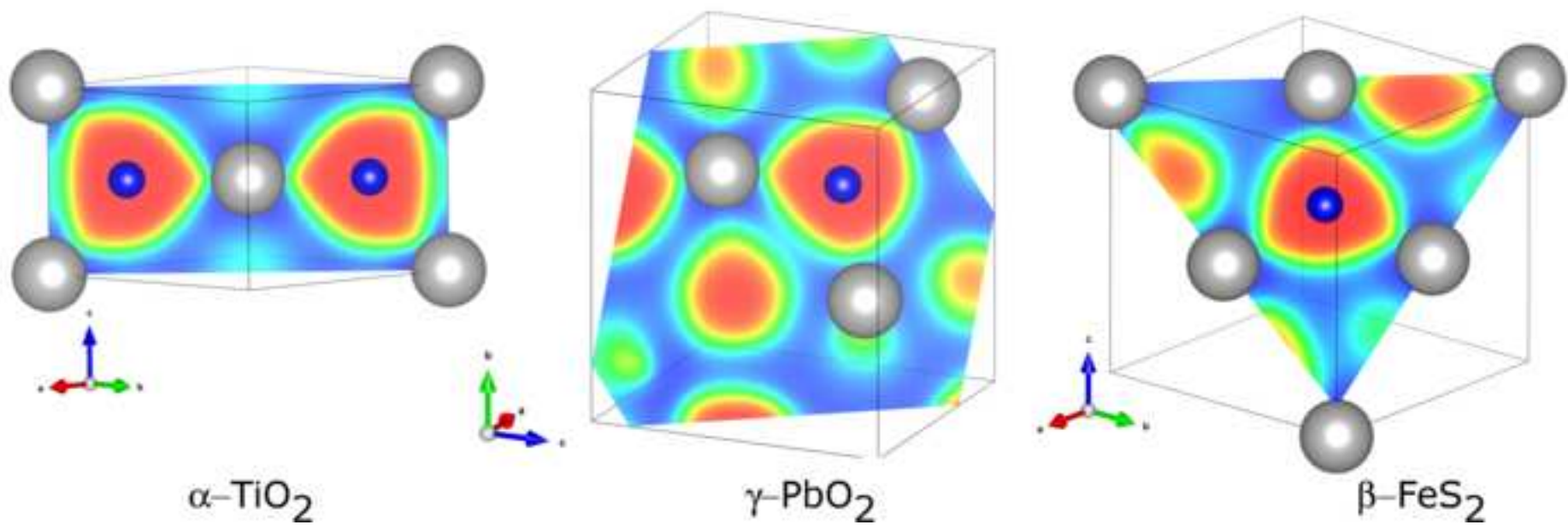

$\alpha$–TiO$_2$ $\qquad$ $\gamma$–PbO$_2$ $\qquad$ $\beta$–FeS$_2$

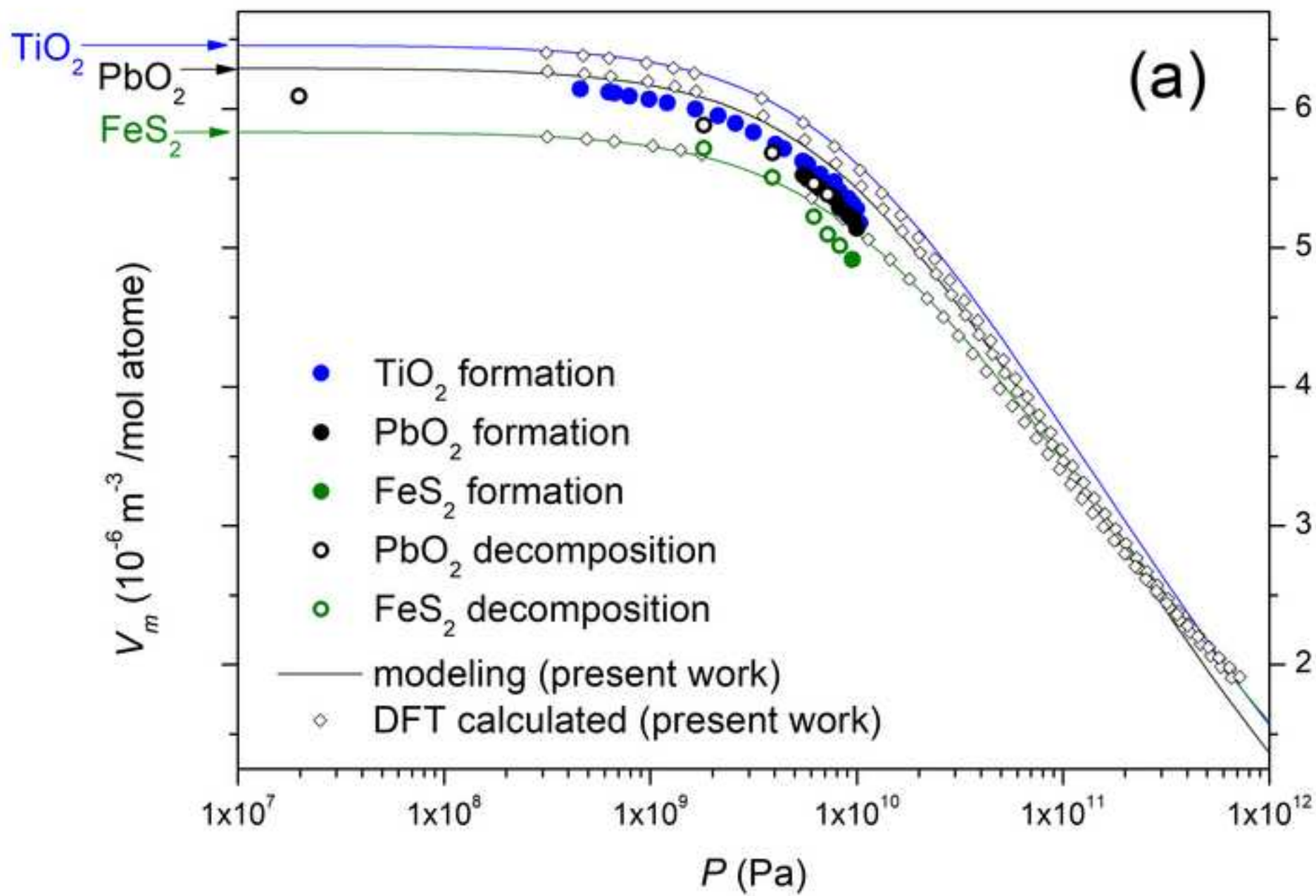

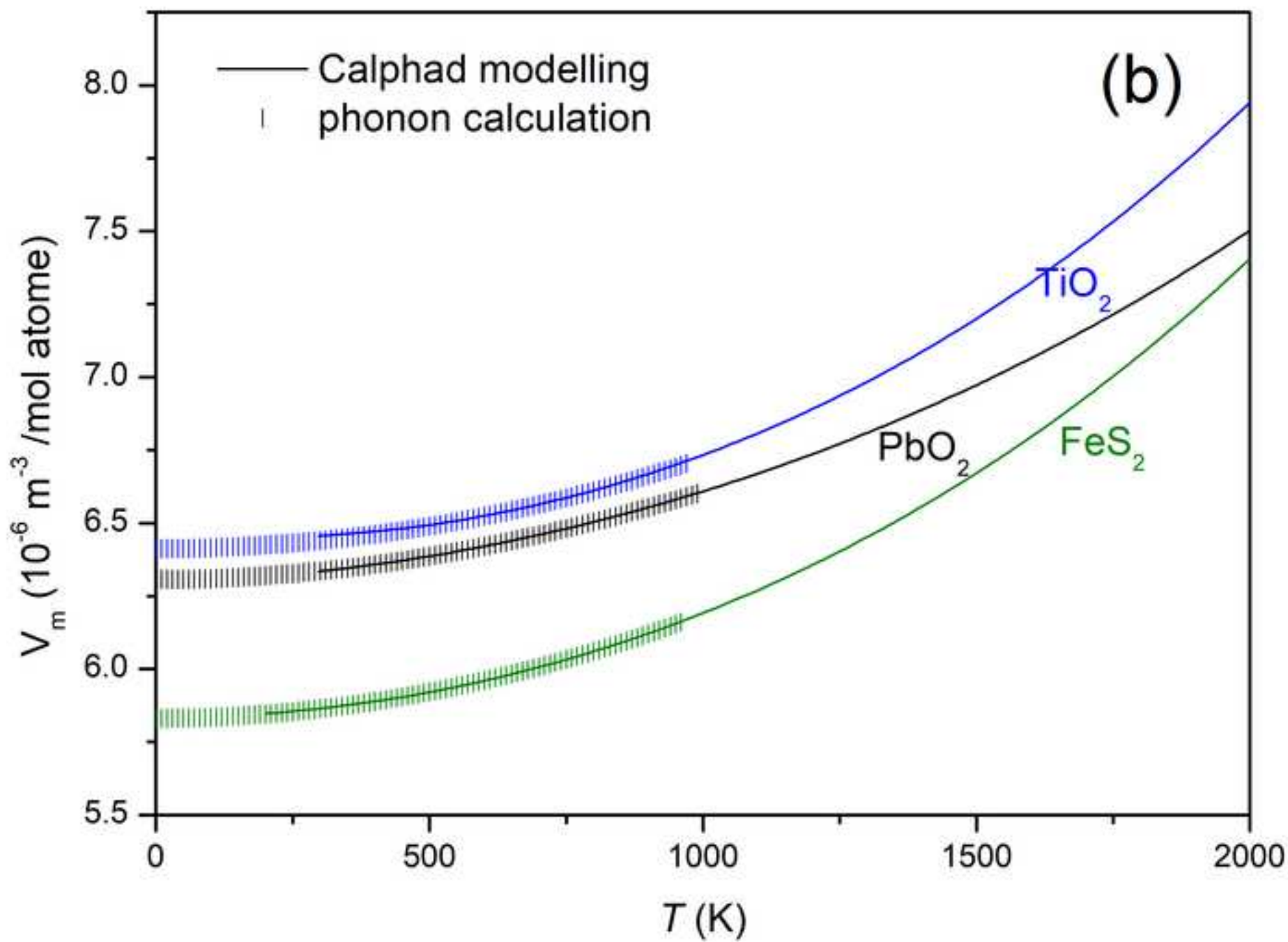

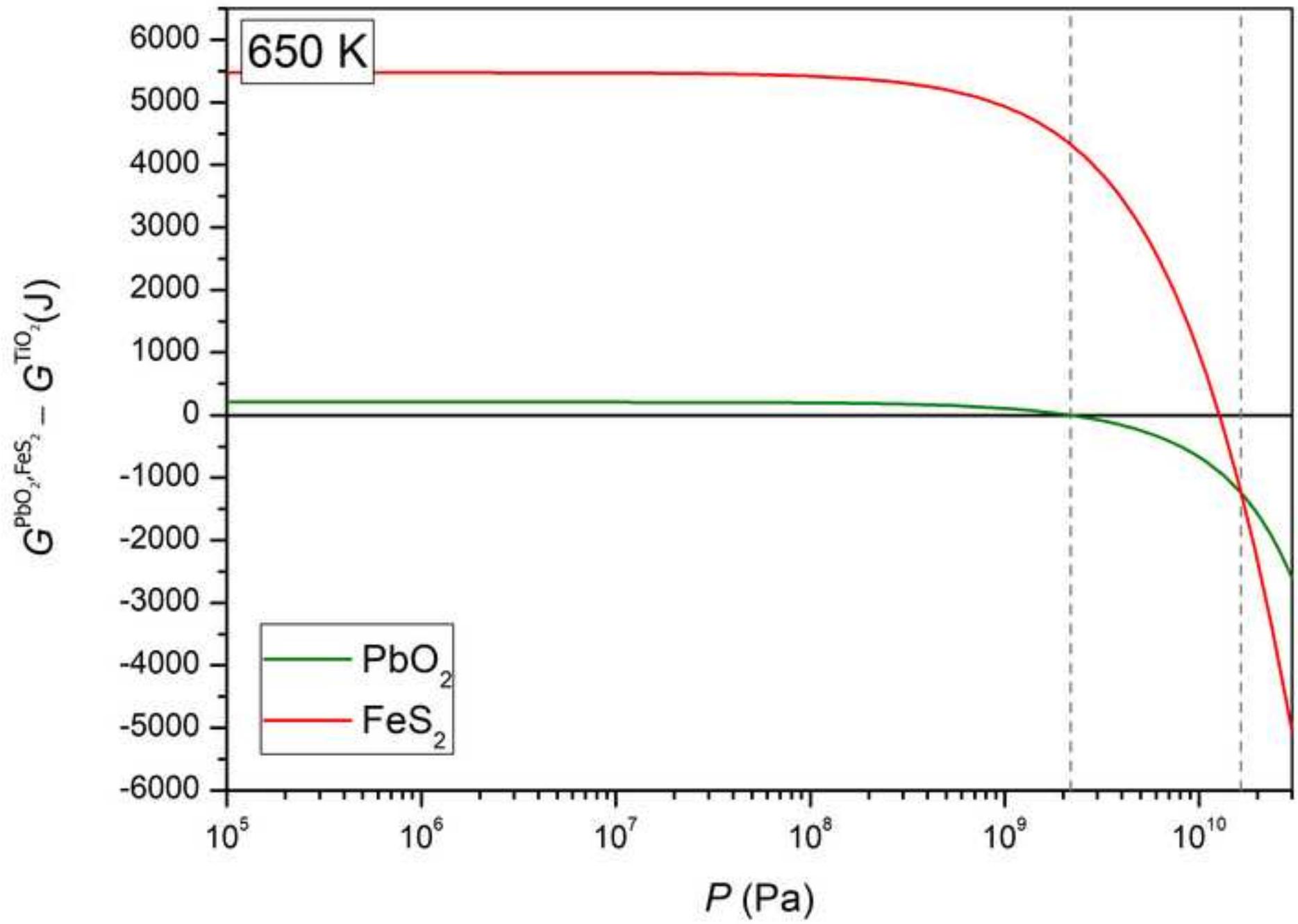

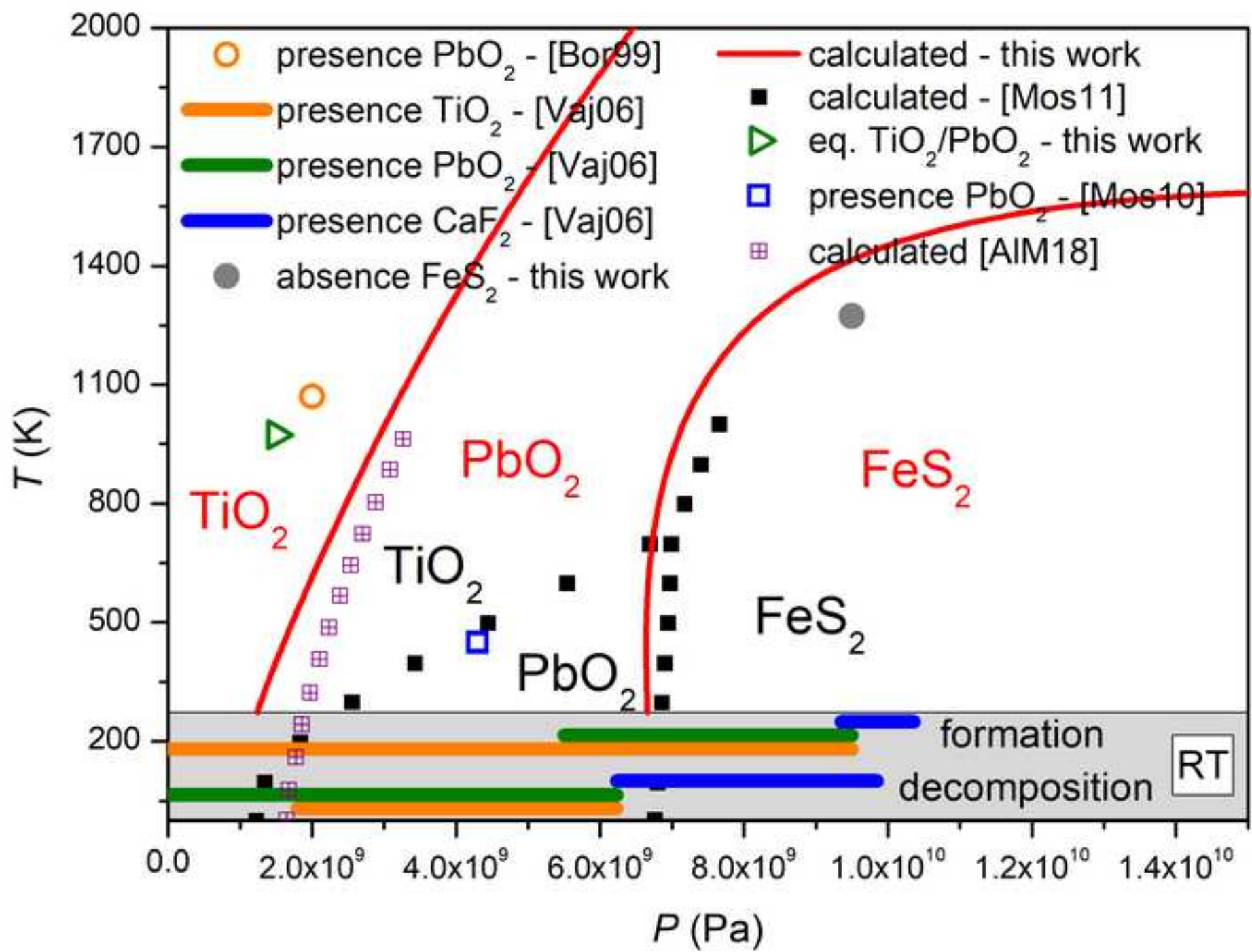

**Figure 30**
Click here to download high resolution image

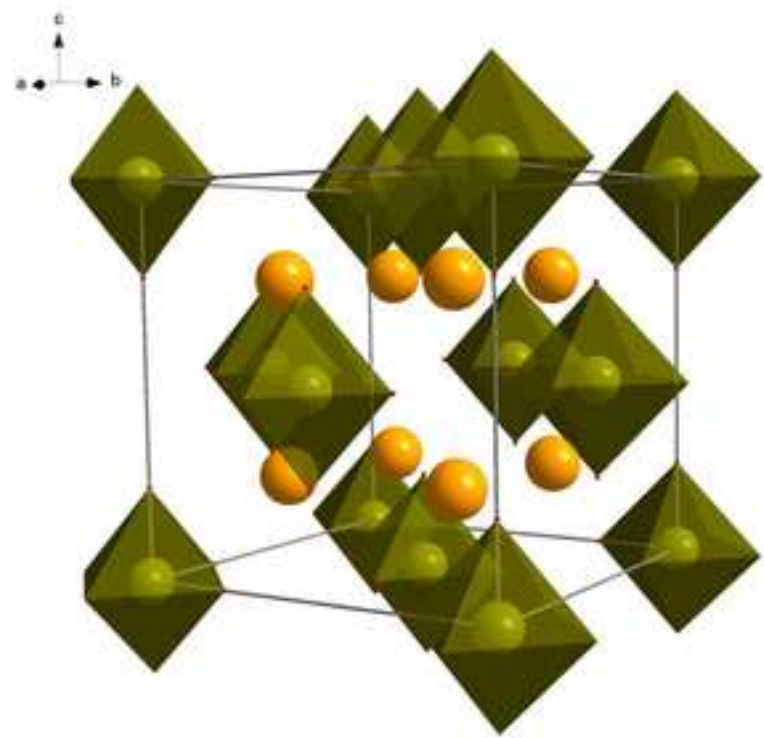

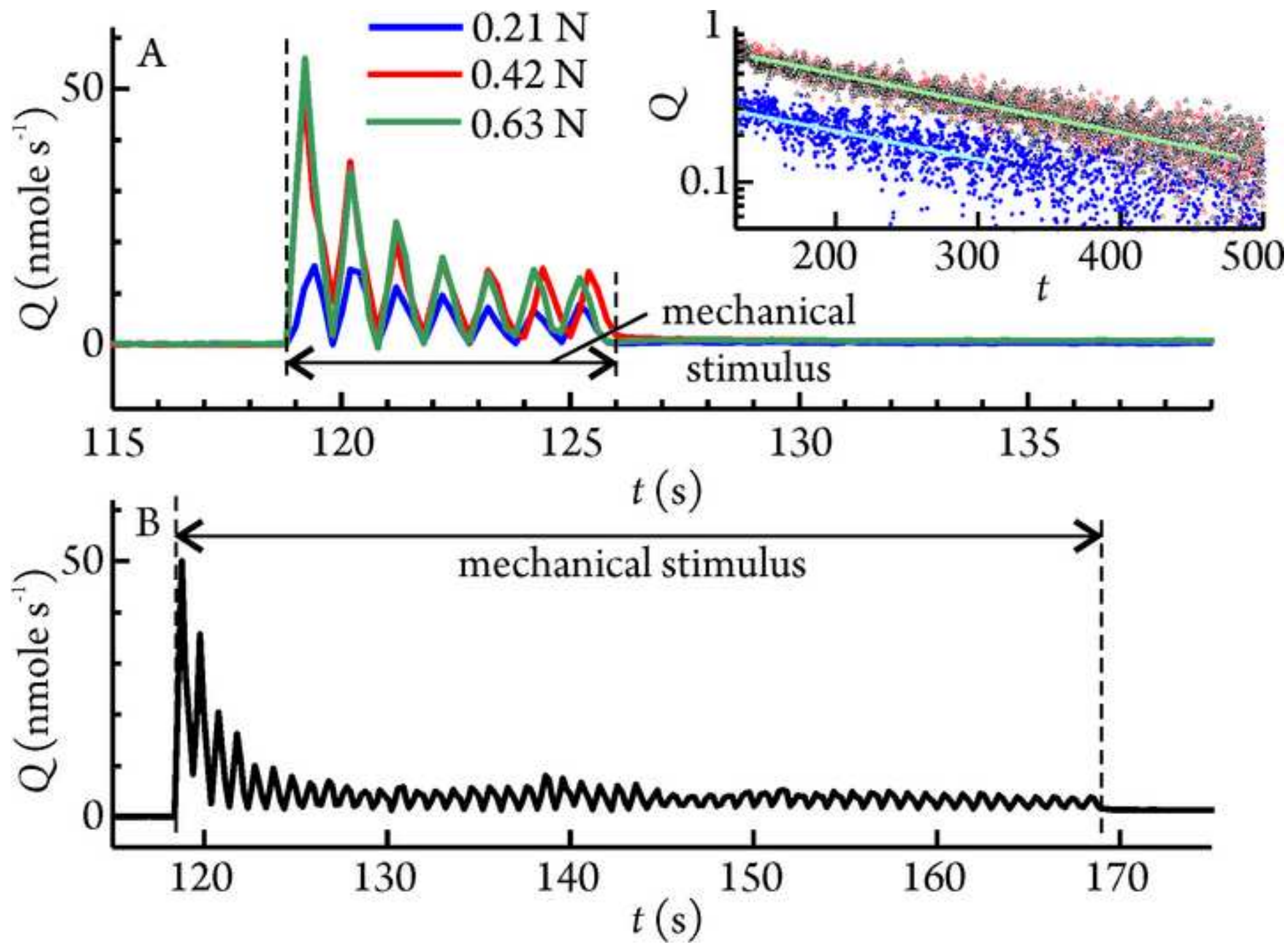

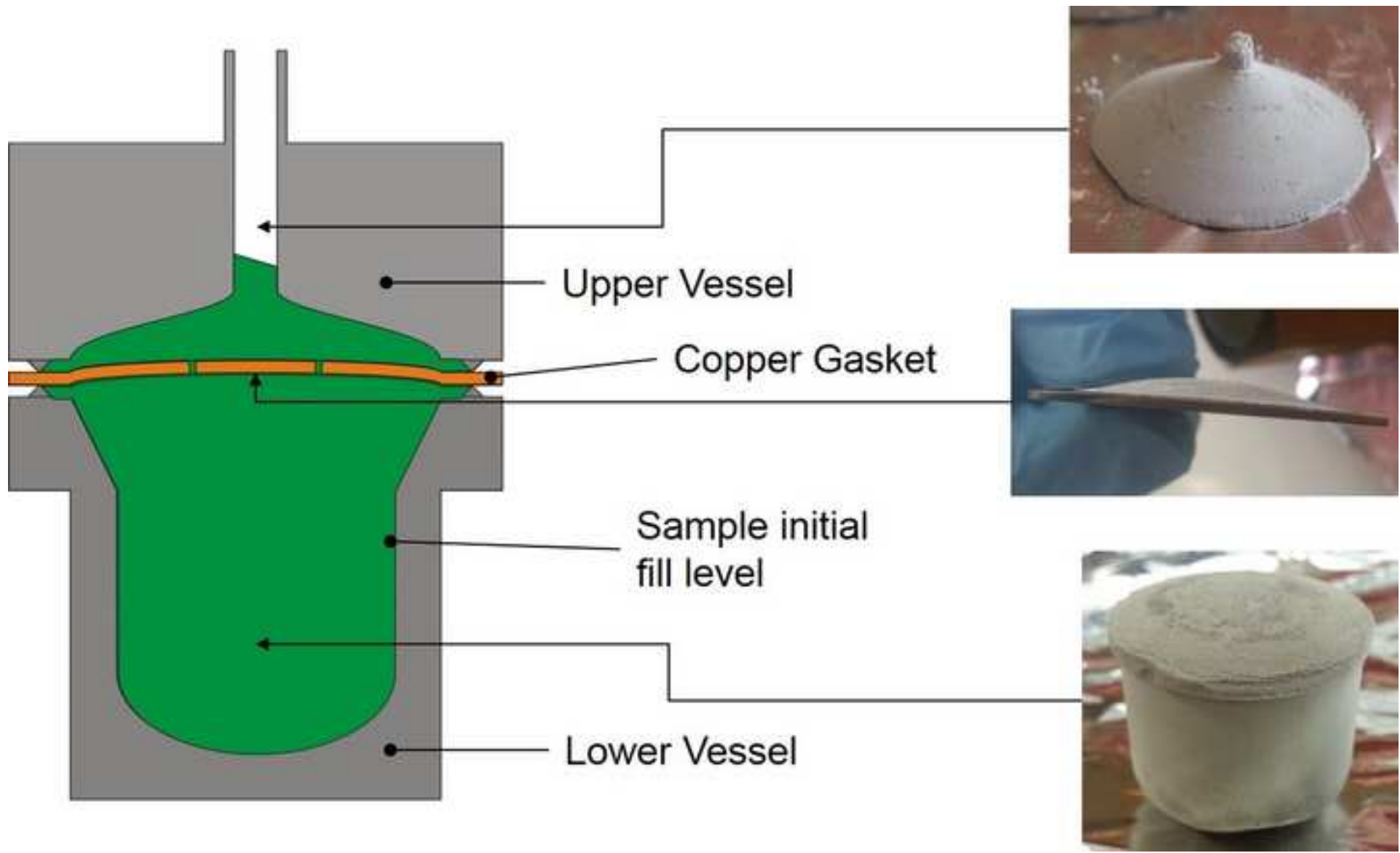

Upper Vessel

Copper Gasket

Sample initial
fill level

Lower Vessel

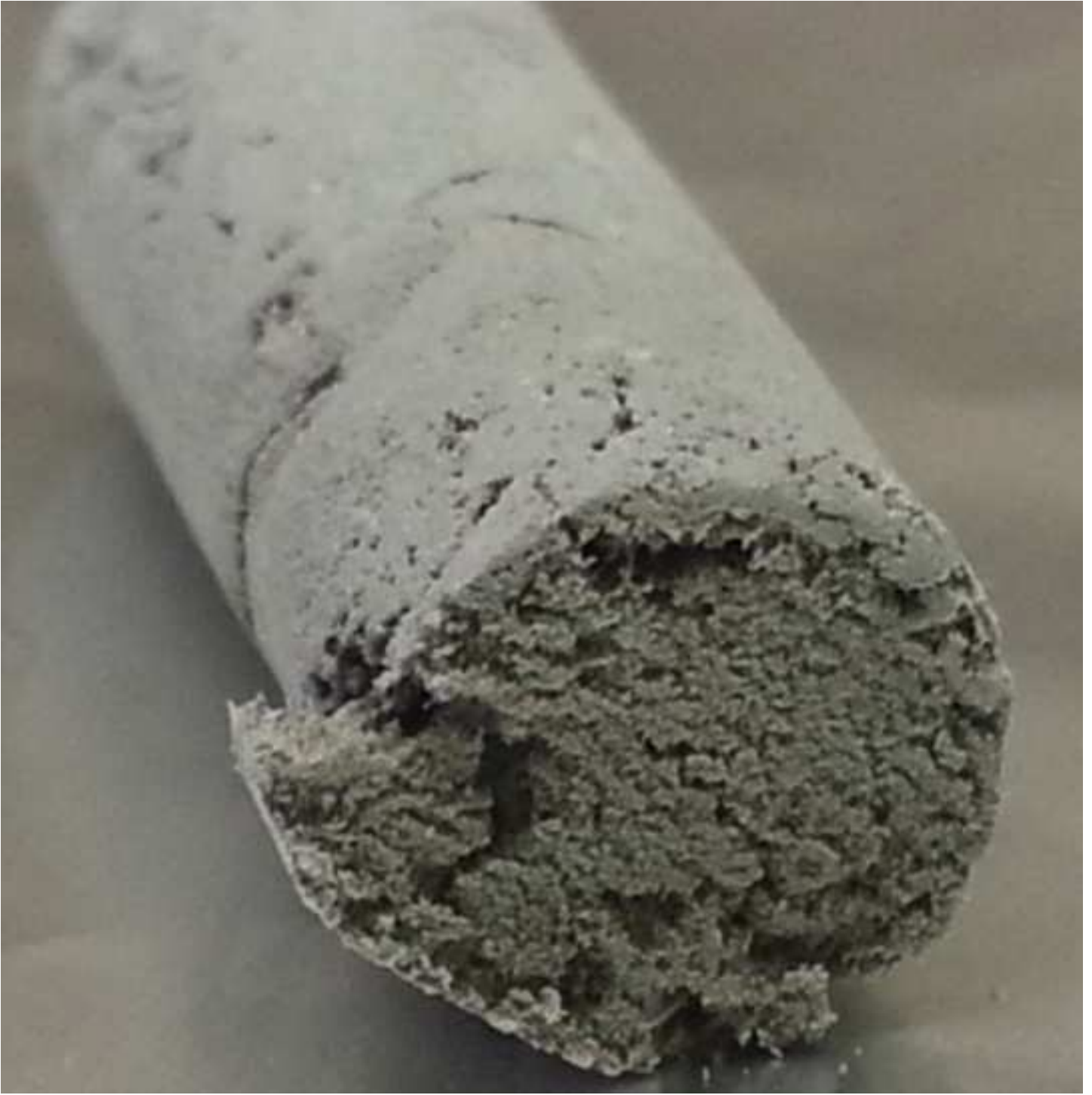

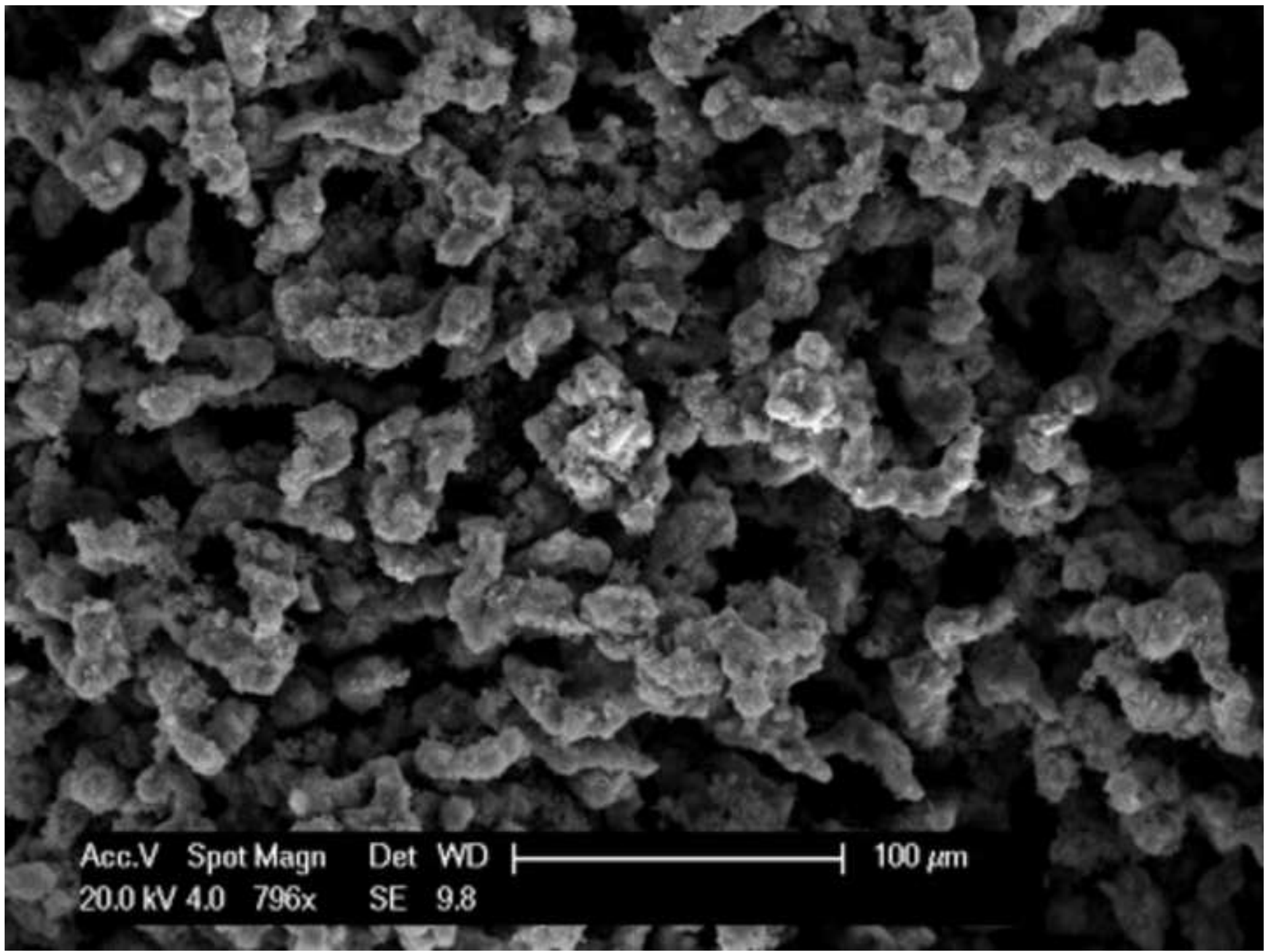

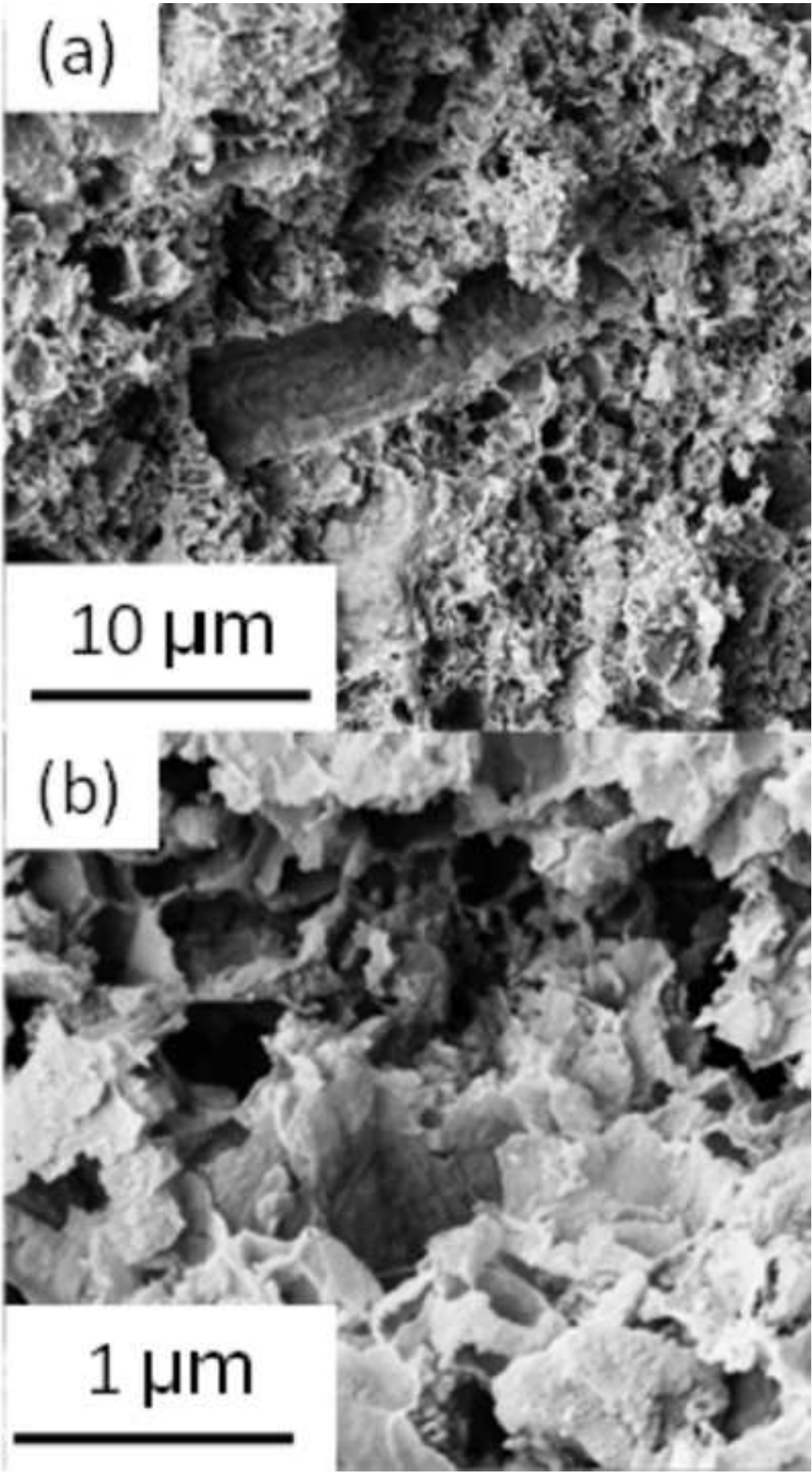

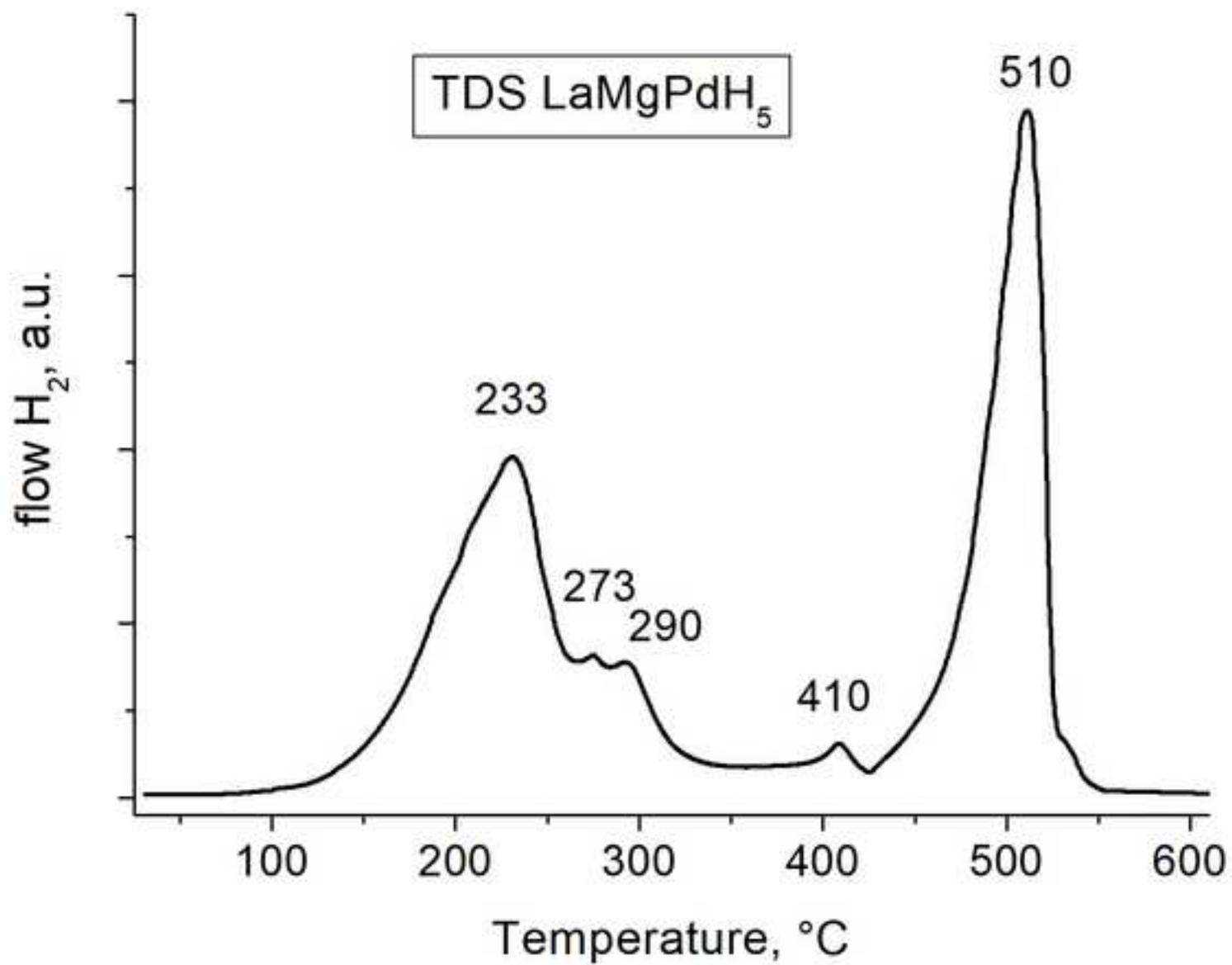

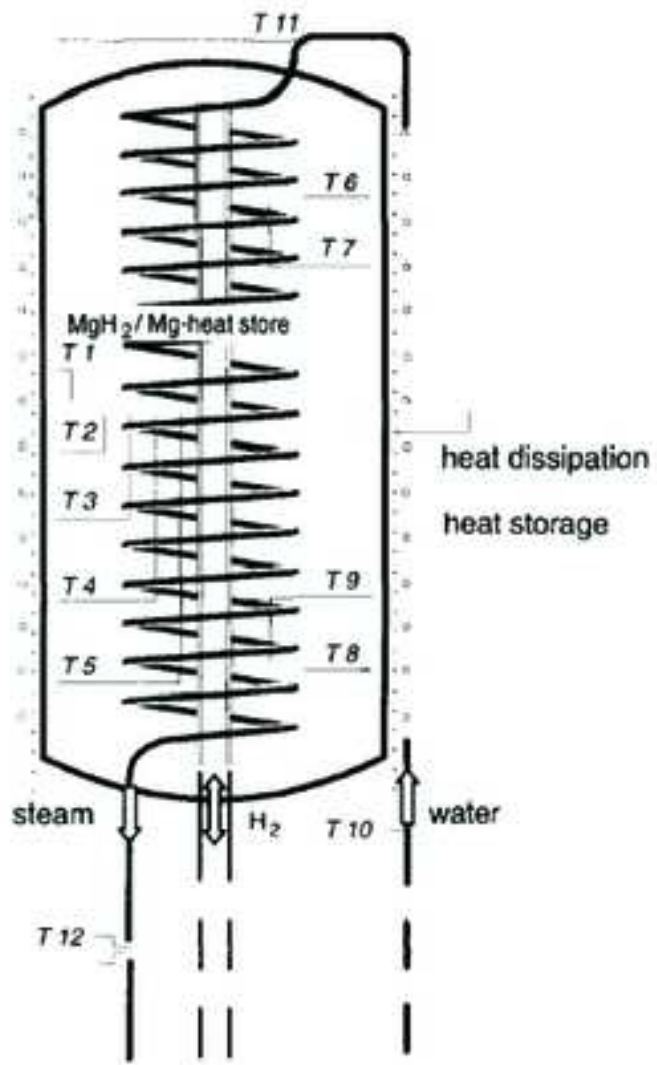

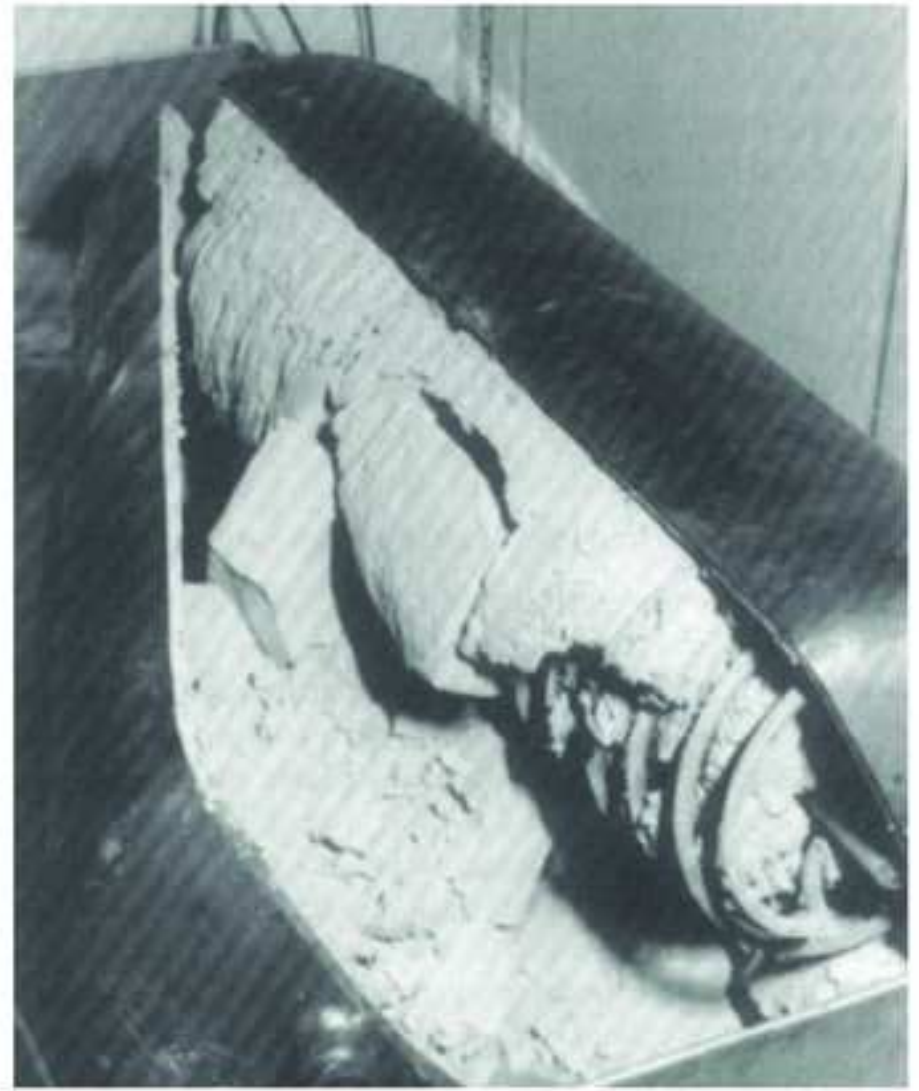

**Figure 37**

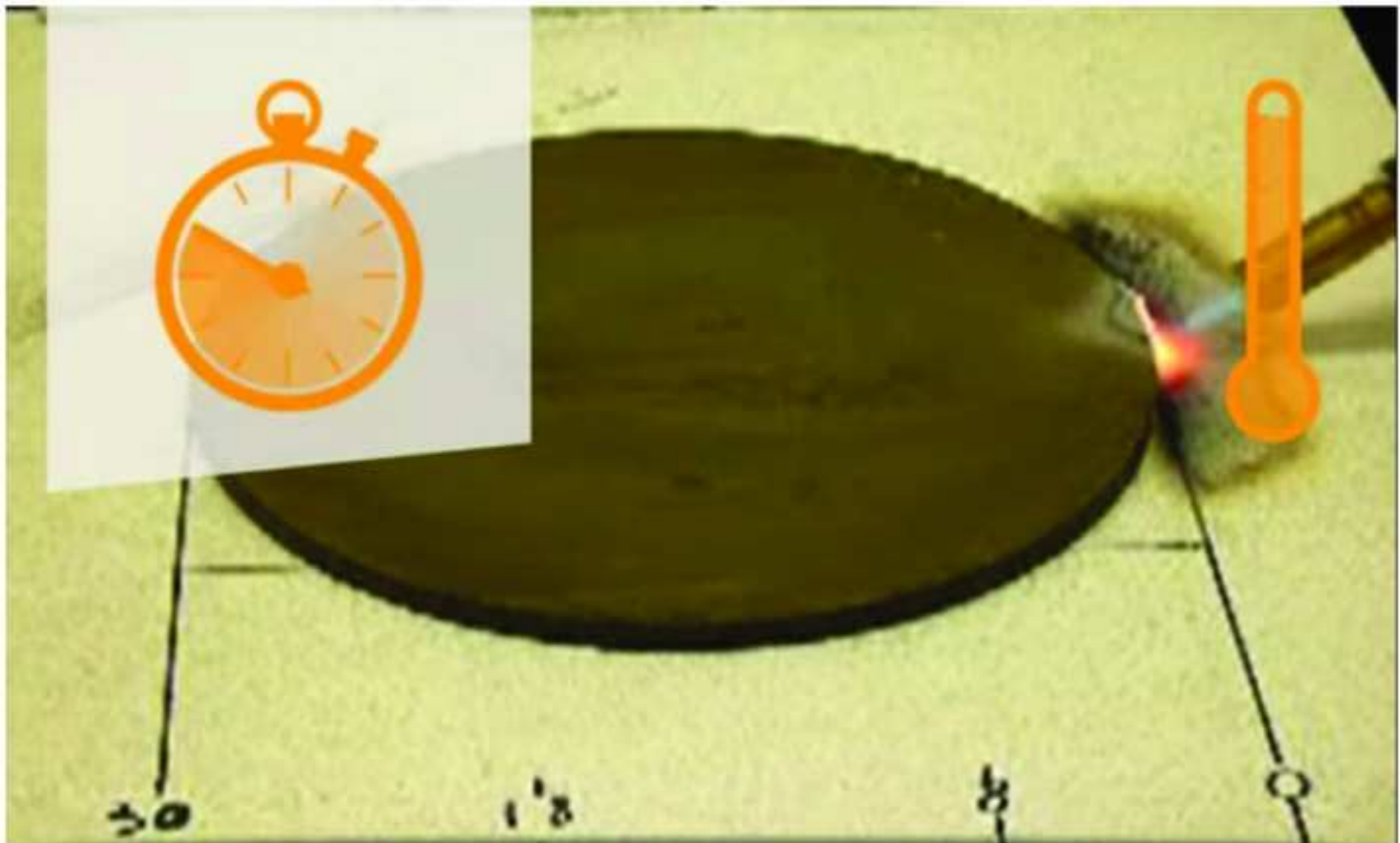